\documentclass[a4paper,11pt,fleqn]{book}

\usepackage{amsmath}
\usepackage{graphicx}
\usepackage[usenames]{color}
\usepackage{hyperref}
\usepackage{amssymb}
\usepackage{latexsym}
\usepackage{dsfont}
\usepackage{picins}
\usepackage{picinpar}

\newcommand{\m}[1]{\mbox{\boldmath{$#1$}}}
\newcommand{\eq}{{(0)}}
\newcommand{\ssssection}[1]{\paragraph{\it $\sim$ #1 $\sim$}$ $\\ }
\newcommand{\re}{\text{Re}}
\newcommand{\im}{\text{Im}}
\newcommand{\sgn}{\text{Sgn}}

\hypersetup{
    colorlinks,%
    citecolor=blue,%
    filecolor=blue,%
    linkcolor=blue,
    urlcolor=blue
}

\usepackage{fancyhdr}
\makeatletter
\def\cleardoublepage{\clearpage\if@twoside \ifodd\c@page\else
	\hbox{}
	\vspace*{\fill}
	\thispagestyle{empty}
	\newpage
	\if@twocolumn\hbox{}\newpage\fi\fi\fi}
\makeatother

\RequirePackage[avantgarde]{quotchap}
\renewcommand\chapterheadstartvskip{\vspace*{-5\baselineskip}}
\RequirePackage[calcwidth]{titlesec}
\titleformat{\section}[hang]{\sffamily\bfseries}
{\Large\thesection}{12pt}{\Large}[{\titlerule[0.pt]}] 
\titlespacing{\section}{0pt}{15pt}{15pt}

\def\resume{\newpage\chapter*{R\'esum\'e}
	\thispagestyle{empty}}
\def\abstract{\newpage\chapter*{Abstract}
	\thispagestyle{empty}}
\def\publications{\newpage\chapter*{Publications \& Orals}
	\thispagestyle{empty}}
\def\dedication[#1]{
  \chapter*{ }
  \thispagestyle{empty}
  \vspace*{\fill}
  \begin{flushright} 
    \textit{\large #1} \\
    \hfill \\  \hfill \\  \hfill\\ 
    \textit{\large To the energy of the future... }\\
    \hfill \\ 
    \begin{minipage}{\linewidth}
      \begin{flushright}
      \includegraphics[scale=3]{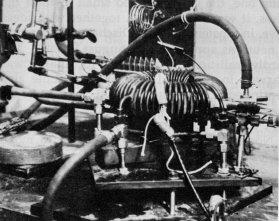}
      \end{flushright}
    \end{minipage}
  \end{flushright}
  \vspace*{\fill}
}
\def\acknowledgements{\chapter*{Acknowledgements}
\addcontentsline{toc}{chapter}{Acknowledgements} 
}

\renewcommand{\maketitle}{
\thispagestyle{empty}
\begin{center}
\begin{minipage}{0.4\linewidth}
\begin{center}
\includegraphics[scale=0.2]{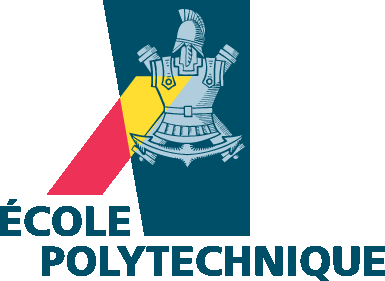}\\
\small \sf
Ecole Doctorale de l'Ecole Polytechnique\\
Route de Saclay\\
F-91128 Palaiseau, France.\\
\end{center}
\end{minipage}
\hfill
\begin{minipage}{0.4\linewidth}
\begin{center}
\includegraphics[scale=0.35]{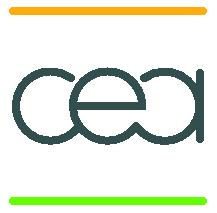}\\
\small \sf
IRFM, Association Euratom-CEA\\
Centre de Cadarache\\
F-13108 Saint Paul-Lez-Durance, France.\\
\end{center}
\end{minipage}\\
\vspace{3.0cm}
{\fontfamily{phv}
\LARGE
Magneto-HydroDynamic Activity and Energetic
Particles \\} 
\vspace{0.2cm}
{\fontfamily{phv}
\LARGE
$\sim $ Application to Beta Alfv\'en Eigenmodes $ \sim$}\\
\vspace{1cm}
{\rmfamily\large  by}\\
\vspace{0.5cm}
{\rmfamily\Large\bfseries Christine Nguyen}\\
\vspace{1.cm}
\vspace{0.5cm}
{\rmfamily \large A dissertation submitted in partial satisfaction\\ of the
  requirements for the degree of}\\
\vspace{1cm}
{\rmfamily \bfseries \large Doctor of Philosophy}\\
{\rmfamily \large in}\\
{\rmfamily \bfseries \large Plasma Physics}\\
{\rmfamily \bfseries \large Ecole Doctorale de l'Ecole Polytechnique}\\
\vspace{1.5cm}
\begin{tabular}{lcl}
  \hline\\ 
  \large
  {\large Dr. Xavier  Garbet}    &\quad&{\large CEA Research Director} -{ PhD advisor}\\
  {\large Dr. Pascale Hennequin} &\quad&{\large CNRS Research Director}\\  
  {\large Dr. Hinrich L\"utjens} &\quad&{\large CNRS Research Scientist}\\  
  {\large Dr. Patrick Mora}      &\quad&{\large CNRS Research Director} -{ Chair of PhD committee}\\  
  {\large Dr. Roland  Sabot}     &\quad&{\large CEA Research Director} -{ CEA advisor}\\
  {\large Dr. Andrei  Smolyakov} &\quad&{\large Professor at the University of Saskatchewan}\\  
  {\large Dr. Laurent Villard}   &\quad&{\large Professor at EPFL} -{ Referee}\\
  {\large Dr. Fulvio  Zonca}     &\quad&{\large Senior Researcher at ENEA} -{ Referee}\\  
\vspace*{-0.2cm}\\
\hline
\end{tabular}\\
\vspace{1.5cm}
{\rmfamily Jan 07 - Dec 09}\\
\end{center}
\cleardoublepage
\pagenumbering{roman}\setcounter{page}{1}
}

\begin{document}

\maketitle
\resume



\vspace*{5cm}
La faisabilit\'e de la fusion magn\'etique est d\'ependante de notre capacit\'e \`a confiner l'\'energie des particules supra-thermiques lib\'er\'ees \'a haute \'energie par les r\'eactions de fusion, dans les meilleures conditions de s\'ecurit\'e et d'efficacit\'e. Dans ce but, il est n\'ecessaire de comprendre l'interaction entre les particules \'energ\'etiques et le plasma thermo-nucl\'eaire qui constitue l'environnement des r\'eactions de fusion, afin de la contr\^oler. La th\`ese que nous pr\'esentons ici s'inscrit dans cet effort. Le coeur du travail men\'e est l'\'etude d'un type d'instabilit\'e, le Beta Alfv\'en Eigenmode (BAE), que peuvent exciter les particules \'energ\'etiques, et dont on peut craindre qu'il d\'egrade fortement non seulement le confinement des particules \'energ\'etiques mais aussi le confinement du plasma dans sa globalit\'e. Dans un premier temps, nous nous attacherons \`a d\'ecrire les caract\'eristiques de ce mode et nous d\'eriverons sa relation de dispersion ainsi que sa structure. Dans une seconde partie, nous effectuerons l'\'etude de la stabilit\'e lin\'eaire de ce mode en pr\'esence de particules \'energ\'etiques. Cette \'etude nous a permis de d\'efinir un crit\`ere analytique rendant compte de la capacit\'e des particules \'energ\'etiques \`a exciter le BAE. Ce crit\`ere sera discut\'e et confront\'e aux r\'esultats d'exp\'eriences men\'ees durant la th\`ese. Cette \'etude lin\'eaire pr\'esentant cependant quelques limites, il nous est apparu important de nous poser la question de la possibilit\'e d'une modification de la stabilit\'e du BAE li\'ee \`a l'utilisation d'une description non-lin\'eaire. Nous sugg\'ererons dans cette pr\'esentation un processus, v\'erifi\'e analytiquement et num\'eriquement, dont peut r\'esulter l'existence d'\'etats m\'eta-stables pour le BAE.


\abstract



\vspace*{5cm}
The goal of magnetic fusion research is to extract the power released by fusion reactions and carried by the product of these reactions, liberated at energies of the order of a few MeV. The feasibility of fusion energy production relies on our ability to confine these energetic particles, while keeping the thermonuclear plasma in safe operating conditions. For that purpose, it is necessary to understand and find ways to control the interaction between energetic particles and the thermonuclear plasma. Reaching these two goals is the general motivation for the work conducted during the PhD. More specifically, our focus is on one type of instability, the Beta Alfv\'en Eigenmode (BAE), which can be driven by energetic particles and impact on the confinement of both energetic and thermal particles. In this work, we study the characteristics of BAEs analytically and derive its dispersion relation and structure. Next, we analyze the linear stability of the mode in the presence of energetic particles. First, a purely linear description is used, which makes possible to get an analytical linear criterion for BAE destabilization in the presence of energetic particles. This criterion is compared with experiments conducted in the Tore-Supra tokamak during the PhD. Secondly, because the linear analysis reveals some features of the BAE stability which are subject to a strong nonlinear modification, the question is raised of the possibility of a sub-critical activity of the mode. We propose a simple scenario which makes possible the existence of meta-stable modes, verified analytically and numerically. Such a scenario is found to be relevant to the physics and scales characterizing BAEs.

\acknowledgements




I am indebted to several people I had the chance to meet during my time as a PhD
student, for their advice and for their help, 
for their knowledge and inexhaustible curiosity, 
and for their support and chearful presence.
\\ \\
{\bf Starting with English speaking people...}

I want to thank Andrei Smolyakov and Fulvio Zonca for taking part
to my PhD committee, for commenting on my work during the whole lengthtime 
of my PhD, for their continuous support and for their 
incredibly complete, effective and fast e-mails.

I also want to thank Taik-Soo Hahm, Simon Pinches, Herbert Berk and Boris Breizman,
for welcoming me in their respective labs, for taking the time to introduce me into 
their work and for broadening my overall physical understanding.
\\
\\
{\bf Je poursuivrai en fran\c cais...}

... pour remercier Alain Becoulet d'avoir accept\'e ma candidature de th\`ese 
et Tuong Hoang d'\^etre toujours disponible, pr\'esent et 
attentif pour nous autres, th\'esards. Merci \'egalement \`a Val\'erie Icard et
St\'ephanie Villechavrolle d'avoir toujours r\'epondu avec le sourire \`a 
l'ensemble de mes questions et requ\^etes, bien que parfois maladroites. 

Je souhaite \'egalement adresser mes remerciements \`a l'ensemble des membres 
francophones du jury, Laurent Villard, Patrick Mora, Hinrich L\"utjens et Pascale
Hennequin, pour leur lecture attentive de mon manuscrit, pour leurs questions et
leurs remarques pertinentes.\\

J'en arrive maintenant \`a ces personnes qui ont rythm\'e le quotidien de ma
th\`ese et sans lesquelles elle n'aurait pu aboutir, en commen\c cant par mon 
directeur de th\`ese, Xavier Garbet, que tout simplement je n'aurais pu r\^ever
meilleur. Je lui suis tr\`es reconnaissante pour tout ce qu'il m'a appris 
avec le soucis constant de me me faire comprendre les m\'ecanismes de la fusion
au del\`a des fronti\`eres strictes de mon sujet, et en faisant preuve d'une 
clart\'e dont je ne peux que faire l'\'eloge; 
pour les conseils pertinents et vari\'es 
et pour les id\'ees toujours nouvelles qu'il a su 
me donner dans mon travail de th\`ese; 
mais aussi pour cette libert\'e d'action et d'expression dont j'ai toujours eu 
le sentiment de pouvoir faire usage avec lui, quelque que soit le sujet abord\'e;
et m\^eme un peu pour son humour. Je remercie \'egalement mon co-directeur de 
th\`ese, Roland Sabot, pour m'avoir aid\'ee \`a faire le lien etre th\'eorie et 
exp\'erience, ce qui m'a sembl\'e passionnant.

Je salue avec la chaleur toutes les personnes du groupe ``MHD+rapides'', 
auquel je souhaite longue vie, et je veux plus particuli\`erement remercier ces 
personnes avec lesquelles j'ai pu travailler sur cette th\'ematique: 
Joan Decker qui, il est vrai, m'a longtemps intimid\'ee 
mais avec lequel les discussions sont toujours pleines d'int\'er\^et;
Patrick Maget dont je ne peux qu'appr\'ecier le calme et l'esprit critique; 
Guido Huysmans et Lars-Goran Eriksson dont l'aide et les 
conseils m'ont \'et\'e extr\^emement pr\'ecieux; 
Marc Goniche et son enthousiasme toujours pr\'esent;
Virginie Grandgirard, dont j'ai pu profiter de l'efficacit\'e l\'egendaire.

Merci \ \'egalement aux diverses personnes qui incarnent pour moi  
l'atmosph\`ere chaleureuse et sympathique du CEA et qui ont \'et\'e
pour moi de tr\`es bon conseil en diverses occasions: 
Marina Becoulet, Chantal Passeron, Clarisse Bourdelle, Gloria Falchetto,
Yanick Sarazin, Philippe Ghendrih, R\'emi Dumont et Maurizio Ottaviani; 
ainsi qu'\`a diverses 
personnes sans l'aide desquelles r\'ealiser des exp\'eriences sur 
Tore-Supra aurait \'et\'e impossible, tout particuli\`erement Philippe 
Moreau.\\

Les choses auraient \'et\'e bien plus moroses sans la pr\'esence de cette 
troupe de ``non-permanents'' rencontr\'es durant la th\`ese: 
une autre fille (ouf !) vraiment super sympathique, Ga\"elle Chevet...
dans un milieu assez masculin, mais j'avoue on ne peut plus amical;
des coll\`egues de bureau, qui ont su me supporter, \'egayer l'{\it insoutenable}
labeur de la th\`ese,  all\'eger les p\'eriodes de stress ou les petites
irritations, Guilhem Dif-Pradalier au charme ravageur qui s'est 
r\'ev\'el\'e un v\'eritable ami, et Stanislas Pamela, la personne
la plus facile \`a vivre que je connaisse... en particulier, lorsque souffle 
le vent;
des ain\'es que j'ai tr\`es longtemps consid\'er\'es comme des mod\`eles, 
l'imp\'etueux Patrick Tamain
et l'irr\'eprochable Eric Nardon;
un compagnon de route plein d'humour et de d\'etachement Alessandro Casati
que j'esp\`ere ne pas avoir laiss\'e trop sceptique quant \`a mes sujets de
discussion;
des cadets qui ont toute mon affection, 
Antoine Merle avec lequel j'esp\`ere une exploration fructueuse du plus beau sujet de th\`ese du monde,
et 
David Zarzoso que je remercie pour la relation de confiance qui s'est \'etablie entre nous.

Merci \`a l'amiti\'e de Kees de Meijeire, 
\`a la sagesse de Maxime Lesur,  
aux remises en question avec Alessandro Macor,
aux longues discussions et gentils d\'esaccords avec Fr\'ed\'eric Schwander,
aux chambrages de Thomas Gerbaud, 
au r\'ealisme de Roland Bellessa,
\`a la s\'er\'enit\'e de Diego Molina,
\`a la g\'en\'erosit\'e de Daniel Villegas,
au caract\`ere accueillant de Nicolas Fedorczak,
\`a la sympathie de Shaodong Song,
au sourire et propos souriants de Emile Van-Der-Plas,
\`a l'honn\^etet\'e de \"Ozg\"ur G\"urcan,
\`a la simplicit\'e et la sinc\'erit\'e imm\'ediate des rapports qui peuvent 
s'\'etablir avec Zwinglio Guimaraes,
\`a l'\`a-propos de J\'er\'emie Abiteboul,
aux d\'ebats avec Francois-Xavier Duthoit.

Merci enfin \`a celle qui m'a permis d'\'echapper \`a ce monde de scientifiques
et qui aura \'et\'e ma confidente au quotidien, Charlotte Largeron. 
Je regretterai les ap\'eros et les tisanes, et toujours un peu de l'avoir trop
souvent r\'eveill\'ee.\\

Il me reste \`a remercier ceux qui sont pour moi des soutiens constants
depuis bien longtemps. Merci \`a mes parents et \`a mon fr\`ere de m'avoir 
appuy\'ee dans mes choix en toutes circonstances, et tout simplement de prendre 
soin de moi. Merci \`a St\'ephane et \`a sa fen\^etre {\it gmail} toujours 
ouverte pour adoucir mon quotidien.


\tableofcontents
\listoffigures
\listoftables
\dedication[]






\chapter{Introduction}
\pagenumbering{arabic}\setcounter{page}{1}
\label{chapter_Introduction}

\section{Energy from fusion}
\begin{figure}[ht!]
\begin{minipage}[t]{.491\linewidth}
\begin{center}
\includegraphics[width=\linewidth]{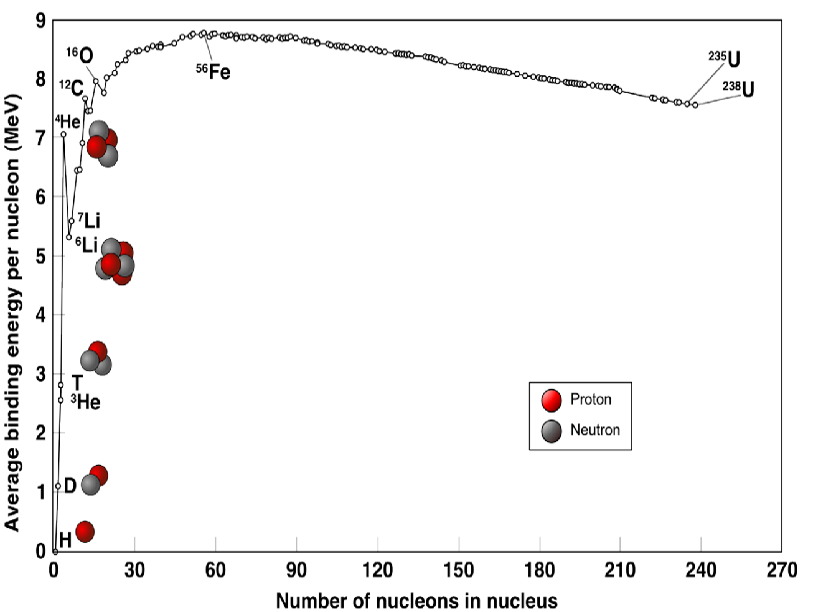}
\caption{\label{fig_FusionEnergy}
\footnotesize The binding energy curve.}
\vspace*{0.2cm}
\end{center}
\end{minipage}
\hfill
\begin{minipage}[t]{0.507\linewidth}
\begin{center}
\includegraphics[width=\linewidth]{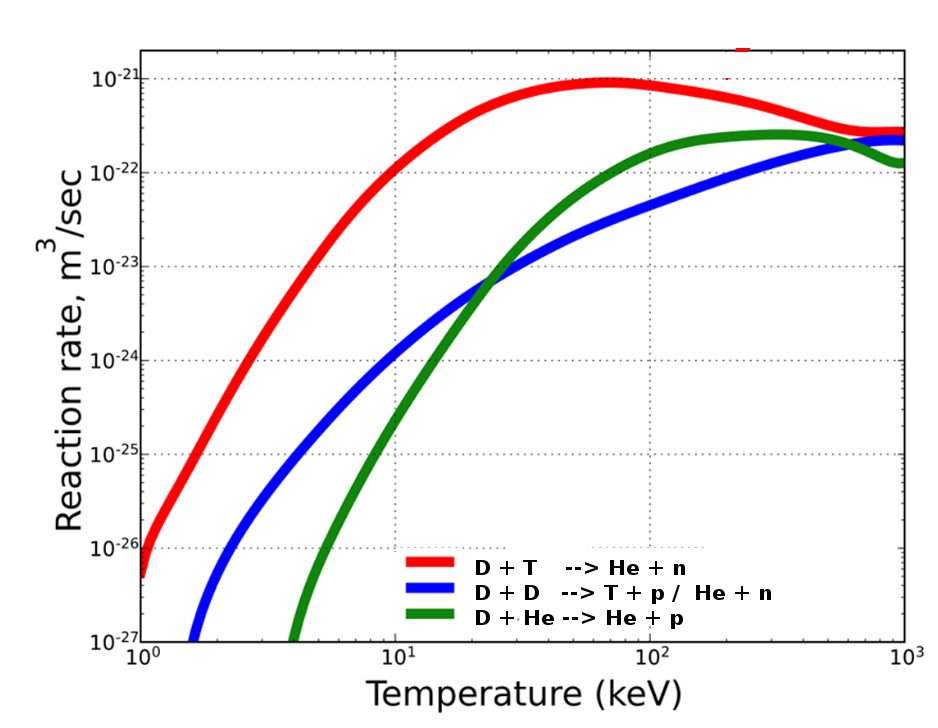}
\caption{\label{fig_DTreaction}
\footnotesize Cross-section of the Deuterium-Tritium reaction compared to 
other fusion reactions.}
\end{center}
\end{minipage}
\end{figure}

The binding energy curve, reproduced 
in Fig.~\ref{fig_FusionEnergy},
shows that the fusion of two light nuclei can lead to a more strongly 
bounded state, and consequently release energy. 
The goal of civilian fusion research is to find ways to
produce and extract this energy. \\

Currently, the most studied fusion reaction is the fusion of
two hydrogen isotopes, a 
{\bf deuterium}  nucleus (D) and a {\bf tritium} nucleus (T),
leading to the creation of a
 helium atom (also called  {\bf alpha particle}) 
and a {\bf neutron}, which carry the liberated energy
\begin{equation}
{}^2_1\text{D} + {}^3_1\text{T}
\longrightarrow {}^4_2\text{He } (3.56 \text{ MeV}) + \text{n }
(14.03 \text{ MeV}).
\end{equation}
Other fusion reactions exist which can liberate energy, but the 
one described above has  the highest cross-section 
for the energies which can be accessed in laboratories, 
as illustrated in Fig.~\ref{fig_DTreaction}. 
However, as can be seen in this figure, even for the relatively
favorable D-T reaction, fusion cross-sections are neglibly  small
below a few keV because of the necesssity to overcome Coulomb
repulsion. For this reason, an efficient production of energy from fusion
 requires to bring reactants to energies
between {\bf 10 keV to 30 keV}. \\

Due to several limitations of non-thermal methods \cite{Rider_95}, 
thermal approaches have been preferred to reach these energies, where
reactants are kept close to thermal equilibrium and heated to temperatures
exceeding several keVs. At these temperatures, equivalent to about $10^8$~K, 
matter is in the {\bf plasma state}, that is, fully ionized.
Consequently, the corresponding background where fusion reactions can
take place is  referred to as a {\bf hot thermonuclear plasma}.

Because a hot plasma tends to expand and may eventually get quenched
on the surrounding walls, and in order to enhance the 
probability of fusion reactions, 
it is necessary to {\bf confine} the plasma to a sufficiently high 
density.
Two techniques have been developed for this purpose.
\begin{itemize}
\item {\bf Inertial confinement} is based on the heating and 
  compression of pellets of a solid mixture of deuterium and tritium 
  by large laser beams or accelerated particle beams.
\item {\bf Magnetic confinement} relies on the use of large magnetic
  fields. When charged particles are plunged into a strong magnetic 
  field, their perpendicular motion with respect to this field
  is a rotational motion around the field lines, 
  or {\bf gyromotion},  which averages out to the lower order.
  Hence, confinement is achieved because particles are
  constrained to follow the field lines. Several magnetic devices have been
  studied, either with closed or open field lines, but it may be shown 
  that a closed 
  system designed to ensure a confinement in all three spacial 
  dimensions is necessarily {\bf topologically equivalent to a 
  torus}. Several torus-like designs, 
  such as spheromaks \cite{Jarboe_94} or stellerators \cite{Lyon_90} 
  are currently under study, but the configuration assessed to be the most 
  promising at the moment, is the {\bf tokamak} (from the Russian
  {\it Toro\"idalna\"ia kamera s magnitnymi katushkami} or toric magnetic 
  chamber).  A schematic of a tokamak is given in
  Fig.~\ref{fig_Tokamak}. 
\begin{figure}[ht!]
\begin{center}
\begin{minipage}{\linewidth}
\begin{center}
\includegraphics[width=0.65\linewidth]{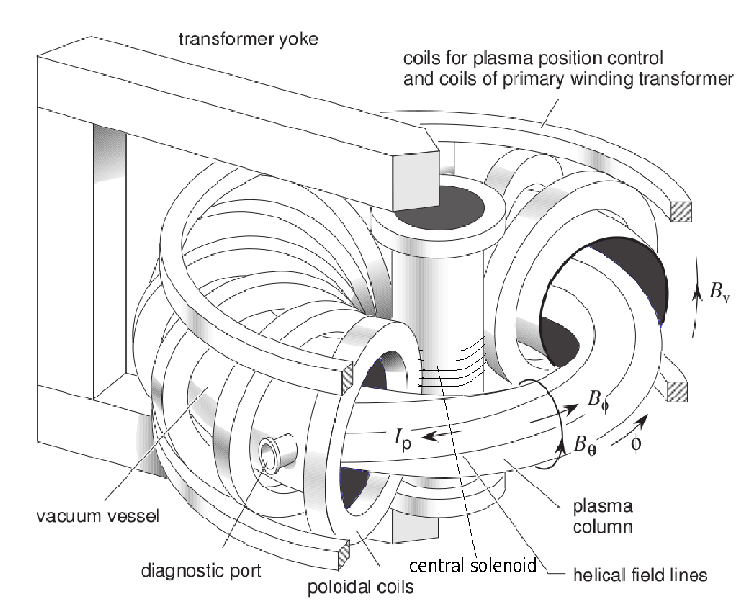}
\caption{\label{fig_Tokamak}
\footnotesize The tokamak and its confining magnetic set-up.}
\end{center}
\hfill
\end{minipage}
\end{center}
\end{figure}
As can be seen in this figure, the plasma of a tokamak 
is confined in an axisymmetric chamber, using a helical 
magnetic field, generated both by vertical coils
and by the existence of a plasma current.\\
The tokamak geometry is the framework of the presented thesis.
\end{itemize}

\section[Controlling the dynamics of energetic particles]
{The necessity to control the dynamics of energetic particles}
To allow for sufficient fusion reactions to take place, 
the plasma power  balance or {\bf Lawson balance} needs to be reached, 
that is
\begin{equation}
\text{Losses} = P_{\text{losses}} + \frac{W_{\text{plasma}}}{\tau_E} = 
 f_{\alpha} P_{\text{fusion}} + P_{\text{external}} 
= \text{Gains}
\end{equation}
where $P_{\text{losses}}$ refers to unavoidable energy losses such as 
radiations, which are independent from the plasma transport properties. 
$W_{\text{plasma}}$ is the plasma energy content and $\m{\tau}_E$ is the 
so-called  {\bf confinement time} which is a measure of the confinement 
quality, such that $W_{\text{plasma}}/\tau_E$ returns the power dissipated 
because of diffusion and convection processes.
$P_{\text{fusion}}$ is the power liberated by fusion
reactions, and $f_\alpha$
{\bf is the fraction of fusion power which remains in the plasma}. 
Finally, $P_{\text{external}}$ corresponds to the external heating
power needed to reach the balance. Frequently, one refers to this
external power input using another quantity, the 
{\bf amplification factor $\bf Q$},
\begin{equation}
Q=\frac{P_{\text{fusion}}}{P_{\text{external}}}
\end{equation}
which is certainly a quantity one may want to optimize. 
\\

For this optimization, most fusion research aims at increasing the
confinement time $\tau_E$. However, $f_\alpha$ may also be a critical
parameter. Traditionally, $f_\alpha$ is estimated to be equal to 1/5.
Indeed, neutrons, being neutral particles, have no chance to
interact with the plasma or to be confined by the magnetic field.
Consequently, this factor of 1/5 corresponds to the fraction of fusion
energy carried by the alpha particles, and assumes that all this
energy gets deposited on the bulk plasma. Several experimental observations
however have shown that  suprathermal particles, also referred to as 
{\bf energetic or fast particles},
between 70 keV and a few MeV can be expelled from the confinement 
chamber without depositing their energy on the bulk plasma. 
In the DIII-D tokamak, 
the heating of ions to about 75keV using beam injection 
resulted in the expulsion of over 50 \% of beam ions. In the TFTR tokamak
(Tokamak Fusion Test Reactor \cite{Duong_93}), ions heated by high 
frequency waves were expelled at high energy and even bored a hole in
a vacuum port \cite{White_95}. 
Recalling that alpha particles are liberated with
energies of a few MeV,  these observations suggest that 
{\bf the fraction of fusion energy carried by the alpha particles 
may not be retained in the plasma, due the possible expulsion of 
alpha particles}. 
$f_\alpha$ could be much
lower than 1/5, which would be detrimental for a reactor efficiency.
Moreover, {\bf a dangerous dynamics of the suprathermal particles is
  put to light by these experiments, 
  which may strongly damage the chamber walls}.
Hence the necessity to {\bf control the dynamics of suprathermal 
particles}, and in particular of alpha particles, with the following goals:
\begin{itemize}
\item {\bf Confining the energy of the fusion produced alpha
  particles}
\item {\bf Confining the suprathermal fuel ions and their energy.}
  External heating sources may bring fuel ions to suprathermal
  energies, but this process should not lead to their expulsion.
\item {\bf Slowing down suprathermal particles before they reach the
  plasma boundary}, in order to avoid materials damaging.
\item {\bf Avoiding a strong dilution of the fuel plasma by the alpha
  particles}. Even though it is of interest to keep the energy carried
  by the the alpha particles  inside the thermonuclear plasma,
  the presence of non-reacting particles   degrades the 
  plasma efficiency because of the related  dilution. 
  Ideally, one wants to get rid of the  alpha particles once
  they  have deposited their energy on the fuel plasma.\\
\end{itemize}

Present day experimental tokamaks mainly operate with pure deuterium plasmas,
to avoid the hardships linked to the handling of radioactive tritium. 
Consequently, in all present day experiments, the power balance is 
achieved using external power, as the major heating source.

An objective of fusion research is of course to reduce the amount of external 
power to be provided (or in other words, to increase $Q$), an ideal situation
being of course to build a self-consistent device,
that is to reach the so-called {\bf ignition}  ($Q = +\infty$).
A first step in this direction will be
achieved by the next generation fusion tokamak {\bf ITER} ({\it iter}
meaning the way in  latin), designed to be built around 2018 
 and intended to demonstrate the feasibility  of fusion energy 
production. As can be seen in Tab.~\ref{tab_ITER} 
\begin{table}[h!]
\begin{center}
\begin{tabular}{lccc}
& Tore Supra & JET & ITER\\
\hline\hline
Major radius $R_0$ (m)           &2.4&3.0&6.2\\
Minor radius $r_0$ (m)           &0.7&1.2&2.0\\
Plasma volume $V_p$ (m$^3$)      &25 &155&830\\
Maximum Plasma current $I_p$ (MA)&1.7&5-7&15\\
Toroidal field on axis $B_T$ (T) &3.8&3.4&5.3\\
Gas                              &D-D&D-D/D-T&D-D/D-T\\
Fusion power                     &$\sim$kW&$50$kW/$10$MW&$500$MW\\
Amplification factor $Q$         &$\ll1$&$\sim 1$& $>10$\\
\hline\hline
\end{tabular}\\
\caption{\footnotesize\label{tab_ITER}
Tore Supra, JET and ITER parameters}
\end{center}
\end{table}
where we compared ITER objectives with current tokamaks, the 
French tokamak Tore-Supra and the Joint European Torus (JET),  
ITER will represent a quantitavive jump, from the point of view of 
scaling and performance, but it will also a qualitative 
jump, with a real attempt to extract energy from fusion reactions.
Whereas $Q$ has not exceeded 1 in current devices,
one of {\bf ITER's goal is to achieve a $Q$ value of 10, that is, 
for $f_\alpha = 1/5$, to heat the plasma with twice as much fusion
power as external power}. 
Consequently,  ITER will be a major step in the 
understanding of the dynamics of alpha particles and suprathermal 
particles in general, for several reasons.
First, suprathermal particles confinement will be studied in a
confining chamber of realistic size.
Secondly, first studies of a non-localized, self-organized heating source
(the fusion energy carried by the alpha particles) will be  possible.
Then, interactions of alpha particles with the plasma collective 
behaviors will take place. And finally, 
experiments aiming at controlling the alpha particles
dynamics will be possible.

In the way towards the understanding and control 
of ITER alpha particles dynamics,
it is necessary to find models describing the interactions of 
suprathermal particles with the thermonuclear plasma. 

\section{Foretaste of the modelling difficulties}
As explained in the previous section, an ideal situation would
be to achieve the {\bf deposition of the alpha particles energy} on the 
fuel plasma and to {\bf tranport the slown down
particles out of the plasma chamber}. 
These two processes may occur via {\bf interactions with the main plasma}.

\begin{figure}[ht!]
\begin{center}
\begin{minipage}{\linewidth}
\begin{center}
\includegraphics[width=\linewidth]{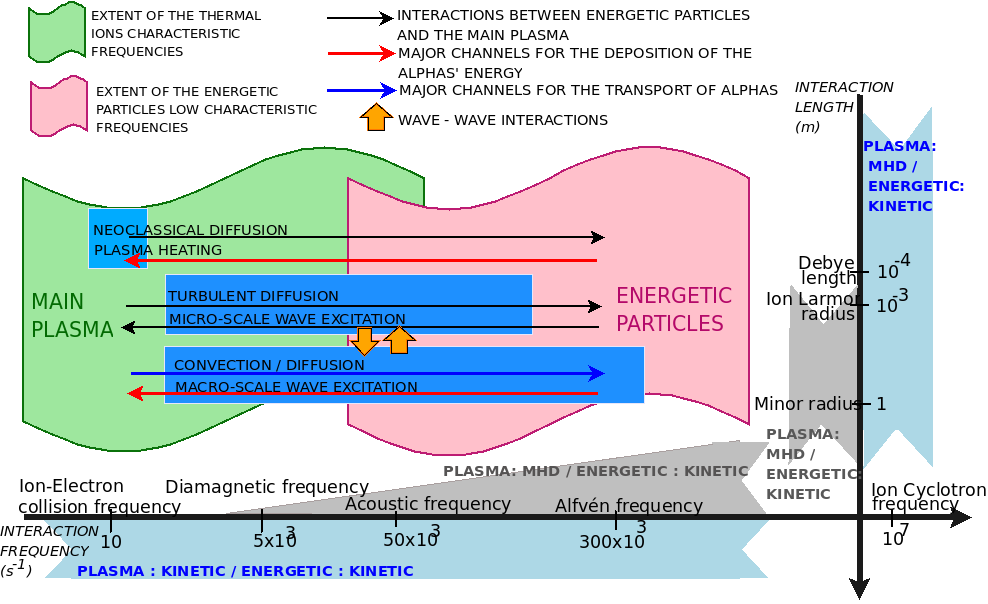}
\caption{\label{fig_TimeAndLengthScales}
\footnotesize Time and length scales involved in the interactions 
between alpha particles and the main plasma.}
\end{center}
\end{minipage}
\end{center}
\end{figure}

A favorable situation would be to have a time scale separation 
between the energy deposition and the transport of alpha particles, in 
order to ensure the particles do not reach the plasma boundary
before they have delivered their energy. Unfortunately, the situation
is not such. On the contrary, both the interactions at the origin of the 
energy deposition and  at the origin of transport are spread over 
various time scales as well as various length scales. An attempt to 
summarize the types of interactions taking place
between alpha particles and the main plasma is given in 
Fig.~\ref{fig_TimeAndLengthScales}.

Mainly, one may distinguish between
\begin{itemize}

\item short scale, individual Coulomb interactions, also simply 
called {\bf collisions}, taking place inside a sphere of radius
the  {\bf Debye length} of about $10^{-4}$m.

Collisions allow some heating of the plasma electrons,
and they induce a so-called {\bf neoclassical transport}. 
However, this transport occuring with very low frequency and  random
directions is not expected to be dominant for suprathermal particles
because of their high energies.

\item micro-scale collective behaviors, taking place via the 
excitation of {\bf micro-scale waves} 
with eigenfrequencies ranging from a few kHz to a hundred of kHz, and
a broad radial spectrum ranging from about $10^{-4}$m to a few 
millimeters. 

Micro-scale waves are at the origin of the main thermal transport
process, {\bf turbulent diffusion}. However, they are  not expected 
to be such important for suprathermal particles. Indeed, the typical
{\bf gyro-radius} of the suprathermal particles, 
that is, the characteristic radius of their gyromotion, 
is larger than a few millimeters. 
Consequently, the effects of micro-scale fluctuations is averaged out 
during one gyromotion of a suprathermal particle. 
For ITER, turbulent diffusion
of suprathermal alphas has been estimated to be 20 times smaller than 
the one of thermal alphas \cite{Angioni_08}.
Inversely, alpha particles are not expected to induce a significative
micro-scale activity (at least not directly), as long as their density 
remains limited.
Hence, micro-scale waves may not be the most important channel for the
alpha energy deposition.

\item macro-scale collective behaviors, taking place via the
excitation of {\bf macro-scale waves}, with typical scale lengths of
the order of several centimers to one meter, and typical 
eigenfrequencies ranging from a few kHz to 300 KHz.\\
Due to their broad extent, these waves can interact with suprathermal 
particles. Moreover their eigenfrequencies correspond to the typical 
eigenfrequencies of the suprathermal particles, 
allowing for {\bf resonances} and wave amplication to take place.
Moreover, because of their broad extent, they can convect particles 
very rapidly on large distances.
{\bf Hence, macro-scale waves can be expected to be
 the main channel in the interaction of
energetic particles with the plasma.}
\end{itemize}

Different plasma models are associated to the various time and length 
scales. 

{\bf Kinetic theories} describe the plasma collective motion 
in the 6 dimensional phase space using a distribution
function $F({\bf x}, {\bf v})$, which assumes a certain 
space or time averaging of the isolated charged particles 
\cite{NicholsonBook}
and of their associated discontinuous Coulomb potential. 
In quasineutral plasmas, the relevant averaging length is 
the {\bf Debye length}, which represents the necessary distance for 
the Coulomb potential of an indidual charged particle to be screened  
by particles of the opposite charge. This way, individual behaviors
get averaged. 
Nevertheless, the {\it cumulative} effects of collisions may
still be retained. 
The main equation of kinetic theories is the {\bf Boltzmann equation} 
with collision operator $\mathcal{C}$ and source $\mathcal{Q}$,
\begin{equation}
\frac{dF}{dt} = \mathcal{Q} + \mathcal{C}\cdot F
\label{eq_Boltzmann}
\end{equation}
also called {\bf Vlasov equation} when $\mathcal{C}\cdot F = 0, \mathcal{Q}=0$. 
Eq.~\ref{eq_Boltzmann} can describe all time and length scales relevant 
to magnetized fusion plasmas phenomena of interest.

{\bf Fluid theories} are based on the use of moments of the Boltzmann 
equation, that is, space dependant 
quantities of the form $\ c\int d{\bf v} {\bf v}^nf$
with $c$ a constant, $n\in \mathds{N}$ (density, velocity...) 
and  avoid the heavy calculation of the full velocity distribution.
One major fluid description is the {\bf Magneto-HydroDynamic (MHD)
model} which describes the thermal plasma (electron and fuel ions) 
as a single fluid, simply with the use of density, velocity and
pressure. 
MHD is appropriate for the description of {\bf macroscopic interactions}, 
and for relatively {\bf fast waves compared to the characteristic 
eigenfrequencies of the described populations}. 
The latter condition is called {\bf hydrodynamic approximation}.
It is often not so relevant to the electron dynamics but one may show 
that MHD still makes sense with minor error when the opposite limit is
verified for the electrons.
An important implication of the hydrodynamic approximation 
is  that resonances may not occur between the  described population 
and the waves. In the absence of resonances, 
the velocity distribution of particles cannot be broken, and the choice
to describe only 
a few moments of distribution functions gets fully  justified.\\

As explained earlier, macro-scale instabilities are the most direct
interaction channel between energetic particles and the thermal
plasma, and their characteristic interaction length is 
appropriate for MHD.  Unfortunately, the hydrodynamic approximation 
is not verified for suprathermal particles, which followingly need 
kinetic modelling. For oscillations of the order of or below the 
acoustic frequency, MHD also becomes incomplete for the representation 
of the thermal plasma.
Consequently, modelling the interaction of energetic particles with 
the thermal plasma requires a {\bf mixing of MHD and kinetic theory} 
for modes oscillating around the Alfv\'en frequency, and {\bf purely 
kinetic theory for lower frequencies}. 
For both populations, {\bf nonlinear effects} may play a role, and
such effects can account for the transport of fast particles. 
Finally, the study may also get 
complicated by the possibility of {\bf wave coupling 
between macro and micro waves}, which allow micro-waves to be 
an indirect channel in the transport and energy deposition of
energetic particles.

\section{Thesis outline}
The topic of the PhD work presented here is the study of one type of
macro-scale mode, known to interact strongly with energetic particles:
the {\bf Beta Alfv\'en Eigenmode (BAE)}.

BAEs are waves oscillating in the acoustic frequency range. This makes
their study particularly challenging and important for the efficiency of
burning plasmas.
Indeed, in this frequency range, waves can resonate with both suprathermal 
particles and  thermal ions.
Consequently, {\bf they may be  a particularly good channel for a direct 
energy transfer from the suprathermal particles to the fuel ions}. 
Besides, the acoustic frequency is located at a cross-point between 
MHD studies and turbulent transport physics, which suggests {\bf that 
acoustic waves could draw a link between  the dynamics of energetic 
particles and the transport of the thermal plasma particles}.
 
During the lengthtime of the PhD work, we derived 
{\bf the BAE dispersion relation} in a way which 
provides a direct description of the {\bf mode structure} and using a 
{\bf formalism which connects kinetic theory and MHD theory continuously}. 
Next, we analyzed the stability of BAEs in the presence of energetic ions. 
In a first attempt, a purely linear description was used, making 
possible the obtention of an {\bf analytical linear criterion for BAE 
destabilization in the presence of energetic particles}, and the latter 
criterion was {\bf compared with experiments, conducted in the Tore-Supra 
tokamak} during the time of PhD work. 
Finally, because the linear analysis revealed some features of the BAE
stability which can be subject to a strong nonlinear modification, 
the question was raised of the possibility of a subcritical activity of the 
mode. In this direction, {\bf a simple model was developed which gives some
hints into the existence of metastable modes}.\\

Beyond the particular features of BAEs, general questions have been tackled, 
related to the dynamics of modes in the presence of energetic particles 
and from a broader perspective, in the presence of a resonant drive.
The present thesis does not intend  to forget this general framework.
For this reason, we start our presentation with a review of the main 
physics involved in the interaction of energetic particles with 
macro/meso-scale waves, in {\it Chapter 2}, in order to define the points 
of relevance to be discussed in the analysis of BAEs:  the role of geometry for 
the stability of waves with a {\it finite} frequency, the time scales involved, 
the required behaviors to retain in the modelling.
In this chapter, fundamental concepts and notation of tokamak physics will 
be introduced.

The next chapters are more focused on the description and analysis of BAEs 
developed during the PhD work.
{\it Chapter 3} explains the general ``kinetic-MHD''variational model 
used  for the modelling of BAEs.
In {\it Chapter 4}, we derive the BAE dispersion relation and structure, using 
the previously defined framework. 
In {\it Chapter 5}, the results of our theoretical and experimental analysis 
of the BAE linear stability is presented.
Finally, {\it Chapter 6} offers some directions for the nonlinear description
of BAEs, and attempts to provide some preliminary response to the question of
a possible nonlinear modification of the BAE stability properties.

\begin{savequote}[20pc]
\sffamily
People talk fundamentals and superlatives and then make some changes 
of detail.    
\qauthor{Oliver Wendell Holmes Jr. (1841 - 1935)}
\end{savequote}






\chapter[Fundamentals]
{Fundamental concepts underlying the interaction of 
collective modes with energetic particles}
\label{chapter_Fundamentals}
We start this thesis with a basic  review of the fundamental concepts 
underlying the interaction of energetic particles with a tokamak 
thermonuclear plasma.

The first part of this chapter is dedicated to the description of 
particle trajectories in a tokamak
equilibrium, in a form which is particularly appropriate to understand
particle resonant behaviors. This description will provide first
insight into the tricky issue of energetic particles
confinement.

Next, the interaction of suprathermal particles with the main plasma 
itself will be considered from a  theoretical point of view, with an 
attempt to highlight and make accessible the most important results 
achieved in the modelling of this interaction, as well as the
formalisms involved.

Finally, the experimental opportunities offered by the tokamak 
Tore-Supra with regards to this study will be depicted.


\section{Magnetic configuration and particle trajectories}
\label{section_MagneticConfiguration}

\subsection{Magnetic configuration}
\subsubsection{General features}
\begin{figure}[ht!]
\begin{center}
\begin{minipage}{0.9\linewidth}
\begin{center}
\includegraphics[width=0.9\linewidth]{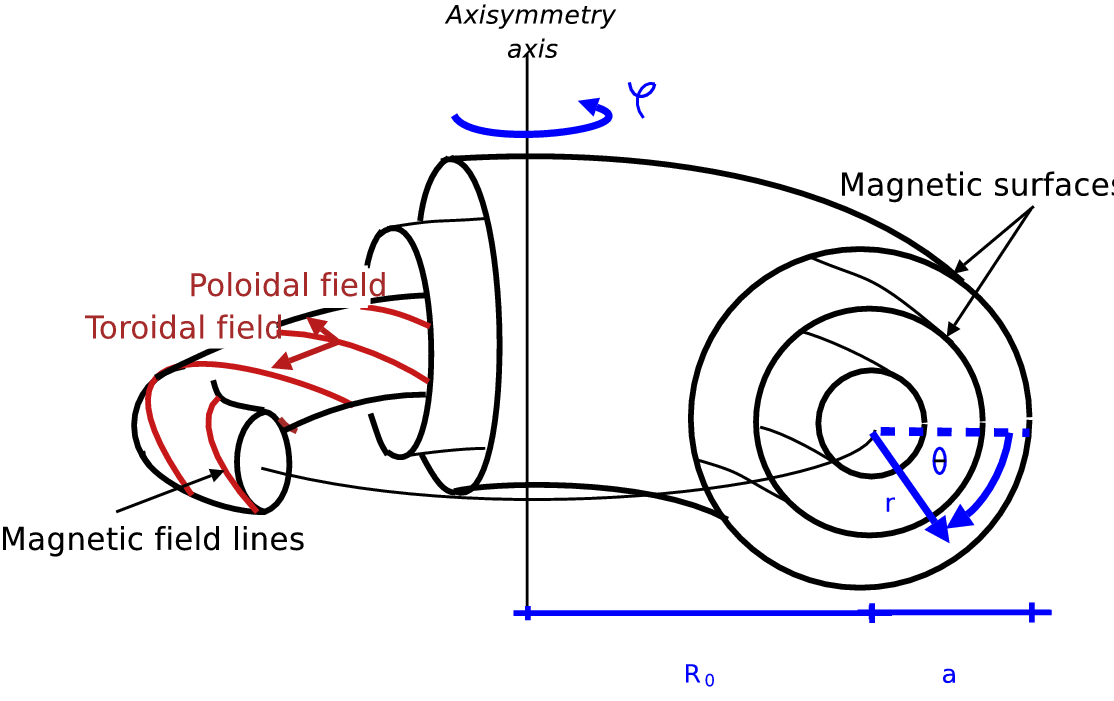}
\caption{\label{fig_MagneticConfiguration}
\footnotesize The tokamak core magnetic configuration 
and its traditional toroidal coordinates.}
\end{center}
\end{minipage}
\end{center}
\end{figure}

As  already overseen in the thesis introduction, 
a tokamak thermonuclear plasma lies in an axisymmetric torus-like chamber, 
where charged particles are confined by a strong magnetic field. When the 
plasma cross-sections are approximately circular, its dimensions can be 
characterized by a {\bf major radius} $R_0$ representing the plasma center
and a {\bf minor radius} 
$a$ represented in Fig.~\ref{fig_MagneticConfiguration}, 
but additional parameters such as elongation or ellipticity, not
considered in this thesis, may be necessary to account for more 
complex cross-sections. 
The ratio of these two lengths is called the 
{\bf inverse aspect ratio},
\begin{equation}
\epsilon = \frac{a}{R_0}
\end{equation}
and it is a measure of the importance of the effects related to the 
torus-like shape of the tokamak, compared to a purely cylindrical device. 
$\epsilon$ is a traditionnally considered to be a 
small parameter ($\sim 0.3$ in Tore-Supra).\\

At equilibrium (referred to with the notation $ _\eq$), 
the tokamak magnetic configuration consists of well-defined embedded 
flux surfaces tangent to helical field lines, as illustrated
in Fig.~\ref{fig_MagneticConfiguration}. 
The helicity of the field lines is necessary for a stable three 
dimensional plasma confinement, and it means that the magnetic field
has  components both in the axisymmetry direction, 
and in the meridian cross-sections. More explicitely,
for any equilibrium magnetic field ${\bf B}_\eq$ with the
topological shape of a torus, there exists a
set of coordinates ($-\Psi$, $\theta$, $ \varphi$) and a
function $q(\Psi)$ such that 
\begin{equation}
{\bf B}_\eq = \nabla \Psi \times\nabla (\varphi-q(\Psi)\theta),
\label{eq_BDefinition}
\end{equation}
where $\Psi$ is flux surface label (i.e, a function which is constant
on magnetic surfaces) and can be understood as a radial coordinate, 
$\varphi$ is an angle which surrounds the device main vertical axis
and defines the so-called {\bf toroidal} direction, 
and $\theta$ is an angle which wraps the confinement chamber and
defines the {\bf poloidal} direction. 
One should also note that the word {\bf toroidal} often also 
refers to the effects related to the torus configuration, 
in opposition to those related to an open cylindrical configuration.
The field curvature towards the axisymmetry axis is an example
of such toroidal effects.

Using this representation and vocabulary, the two characteristic 
components of the magnetic field can now be 
more accurately  defined as a toroidal field  
${\bf B}_{\text{T}\eq}  =- q(\Psi)\nabla \Psi\times\nabla \theta$,
and a poloidal field ${\bf B}_{\text{P}\eq} =  \nabla \Psi \times\nabla
\varphi $, which is  typically small compared to ${\bf B}_{\text T\eq}$, 
${\bf B}_{\text P\eq} \sim \epsilon {\bf B}_{\text T\eq}$.
In usual tokamak devices, the toroidal field is generated by external
fields and the poloidal field  by the generation of an equilibrium 
plasma current flows in the toroidal direction.\\

More precisely, it can be shown that in Eq.~\ref{eq_BDefinition},
$\Psi$ and $q$ are uniquely defined \cite{HazeltineBook}. 
For a given flux surface, $-\Psi$
\footnote{The minus sign is a traditional convention made to allow a
fast identification of $\Psi$ with the magnetic vector potential
which is in the axisymmetry direction.}
is the poloidal flux crossing the ribbon-like surface 
$\mathcal{S}_\theta$ streched between the magnetic axis and the 
flux surface for a fixed value of $\theta$ (normalized to $2\pi$),
\begin{equation}
-\Psi = \frac{1}{2\pi}\int_{\mathcal{S}_\theta} ds \ {\bf B}_\eq
\cdot \frac{\nabla \theta}{|\nabla \theta|},\\
(ds>0).
\end{equation}
If we now similarly define the toroidal magnetic flux going though
 $\mathcal{S}_\varphi$, a surface of constant $\varphi$ bounded by a
 given flux surface,
\begin{equation}
\Phi = \frac{1}{2\pi}\int_{\mathcal{S}_\varphi} ds \ {\bf B}_\eq
\cdot \frac{\nabla \varphi}{|\nabla \varphi|}, 
\end{equation}
$q$ can be shown to be a flux surface label  defined by
 \begin{equation}
q \equiv -\frac{d \Phi}{d\Psi} = \frac{{\bf B}_\eq\cdot\nabla\varphi}
{{\bf B}_\eq \cdot\nabla\theta}.
\end{equation}

$q$ is an important parameter for tokamak studies, called the 
{\bf safety factor}. It is a measure of the magnetic field helicity, 
whose absolute value 
precisely corresponds to the number of toroidal turns that a field
line does while performing a single poloidal turn.
For a plasma current flowing in the direction of the toroidal magnetic
field, 
$(-\Psi, \theta,\varphi)$ is a right-handed coordinate system, and 
$q$ is positive.

Note that $\Phi$ and $q$, being flux surface labels, may be chosen as
alternate {\it radial} coordinates. 
More commonly, the {\bf radial coordinate} $r=a\sqrt{\Phi/\Phi(a)}$ 
is considered, since it can be shown to have an approximate 
regular behavior, close to  the intuitive idea of a radius.

Using $\bf r$, we can define a commonly used quantity, the {\bf shear}
which is simply the normalized derivative of $q$,
\begin{equation}
s = \frac{r}{q}\frac{dq}{dr}.
\label{eq_ShearDefinition}
\end{equation}\\

Finally, we note that in a tokamak configuration, 
axisymmetry provides a simplified 
expression for the field.
Indeed, $\varphi$ may be chosen to be the axisymmetry angle, such 
that the coordinate system $(-\Psi, \theta, \varphi)$ becomes
partially orthogonal. Hence, ${\bf B}_\eq$ can be written
\begin{eqnarray}
{\bf B}_\eq = \nabla\Psi\times\nabla\varphi + I\nabla\varphi,
\label{eq_BAxisymmetricDefinition}
\end{eqnarray}
with $I$ a flux label
$I(\Psi)= R^2  {\bf B}_\eq\cdot\nabla\varphi$, $R$ the 
distance to the axisymmetry axis \cite{HazeltineBook}.

\subsubsection{Simplified large aspect ratio equilibrium}
\label{sssection_LargeAspectRatioEquilibrium}
The calculation of the flux surfaces arrangement involves the 
resolution of  the so-called  Grad-Shafranov equations 
\cite{BiskampBook}. 
For a circular set-up and with a large aspect ratio assumption 
$\epsilon^{-1}\gg 1$, the solution of the Grad-Shafranov equation 
to the second order in $\epsilon$
is a set of nested circular flux surfaces radially shifted in the 
torus outward direction. In a poloidal cross-section, 
the corresponding flux surfaces characteristic equations can be 
expressed in a simple way using the radius $R$ and the vertical
direction $Z$ by
\begin{equation}
R =  R_0 + r\cos\theta - \Delta(r),\quad
Z =  r\sin\theta,
\label{eq_GradShafranovshift}
\end{equation}
where $\Delta$ is called the Grad-Shafranov shift. 
For simplicity, most subsequent computation make use of this
approximation, sometimes with the additional assumption that 
$\Delta = 0$.

Applying Amp\`ere's law, the toroidal field is found to be
inversely proportional to $R$,
\begin{equation}
B_{\text{T}\eq} = \frac{B_0R_0}{R},\\
I(\Psi) = B_{\text{T}\eq}R=B_0R_0
\label{eq_BTnonuniformity}
\end{equation}
with $B_0$, the central field (taken at $R=R_0$ and $r=0$).
In the large aspect ratio limit with $\Delta=0$, it follows
that
\begin{equation}
B_\eq \approx B_0(1-\frac{r}{R_0}\cos\theta),\\
q = \frac{r}{R_0}\frac{B_{\text{T}\eq}}{B_{\text{P}\eq}}
\end{equation}

\subsection{Particle trajectories at equilibrium}
\label{ssection_ParticleTrajectories}
Let us now describe the motion of a charged particle in the 
equilibrium electromagnetic field. As an addition to the magnetic 
field  ${\bf B}_\eq$ described above, there may  also be an 
equilibrium electric field 
${\bf E}_\eq=-\nabla\phi_\eq-\partial_t{\bf A}_\text{solenoid}$, 
mainly induced by the magnetic flux generated in the tokamak
central solenoid (see Fig.~\ref{fig_Tokamak}) for the plasma
current generation.

From now on, we will make use  of the words {\it perpendicular} 
and {\it parallel} to describe the plasma dynamics, 
and of the corresponding notations $\perp$ and $\|$.
Unless otherwise noted, the latter adjectives should be understood
with reference to the magnetic field, and more precisely to the 
equilibrum magnetic field, when a linear analysis is carried out.

\subsubsection{Basic description of particle trajectories in 
a tokamak at equilibrium}
As explained earlier, magnetic confinement is based on the idea that 
a magnetic field can enforce particles to follow its lines. More
accurately, the perpendicular velocity of a charged particle 
immersed in a constant magnetic field is transformed into a 
rotational velocity around the field lines, whereas its parallel 
velocity is kept unperturbed. Hence its net parallel velocity. 
In a tokamak, the picture is more complicated because of the 
non-uniformity of the magnetic field and the presence of an electric 
field.
Nevertheless, to the lower approximation, a charged species of a 
tokamak rotates around the magnetic field lines with a 
{\bf gyrofrequency} $\m{\Omega}_{c}$ and  gyration or {\bf Larmor
  radius} $\rho_{\perp }$ given by
\begin{eqnarray}
\Omega_{c} = \frac{e B_\eq }{m}, \quad \rho_{\perp } 
= \frac{mv_{\perp }}{eB_\eq}
\equiv \frac{v_{\perp}}{v_{t}}\rho
\end{eqnarray}
where $e$, $m$, $v_\perp$, $v_{t}$ are respectively the charge, the
mass, the perpendicular and the thermal velocity of the species $s$.
$\rho$ is more frequently used and called the {\bf thermal
gyroradius}, or simply gyroradius.

More precisely, when the magnetic field is almost uniform
at the scale of the particle motion, ie.
\begin{equation}
\rho\frac{\nabla B_\eq }{B_\eq} 
\sim \frac{\rho}{L_{p}}\equiv \rho^* \ll 1 
 \  ( L_p \text{ is a typical equilibrium plasma scale}),
\label{eq_rhostar}
\end{equation}
which is typically the case in a tokamak 
(for electrons $\rho^*_e  < 10^{-4}$, 
for thermal deuterium ions $\rho^*_i \sim 10^{-3}$ 
and for suprathermal ions $\rho^*_h \sim 10^{-2}$), 
the motion of a charged particle {\bf x}, 
can be divided into a small scale, fast gyrating motion and 
a larger scale, slower {\bf guiding-center} motion {\bf X}, 
independent from the angle of the gyromotion (or gyroangle) $\gamma$, 
\begin{eqnarray}
{\bf x} &=& {\bf X} + \m{\rho} 
\text{, with } \m{\rho}=O(\rho^*{\bf{X}})
\label{eq_GuidingCenterDefinition}
\\
\dot{\bf x} &=& \dot{{\bf X}} + \dot{\m{\rho}}
\text{, with } \dot{\m{\rho}}\sim \dot{{\bf{X}}} .
\label{eq_GuidingCenterVelocity}
\end{eqnarray}

When expanding the charged particle motion using 
Eq.~\ref{eq_GuidingCenterDefinition}, the guiding-center motion 
is found to depend only on its position {\bf X} and of {\bf two 
motion invariants}, its {\bf magnetic moment} 
$\mu = mv_\perp^2/2B_\eq$ 
and its {\bf energy} ${\sf E} = \mu B_\eq +mv^2_\|/2 + e \phi_\eq
$,  which may be assessed at the position {\bf X} to the order
considered in Eq.~\ref{eq_GuidingCenterVelocity}.
Remark that if the invariance of {\sf E} is exact in a conservative
system at equilibrium, the invariance of $\mu$, derived  from the
expansion \ref{eq_GuidingCenterDefinition}, is only verified to the 
$0^{th}$ order in $\rho^*$, and
$\mu$ is thus often referred to as the {\bf adiabatic invariant}
\footnote{Note that $\mu$ is sometimes taken to be an invariant by 
definition, at any considered order in $\rho^*$. 
As will be clearer in the coming developments, it is
equivalent to say that the guiding-center motion is taken to be
independent of the gyroangle by definition. 
In this case, $\mu$ is not simply given by the expression 
$\mu=mv^2_\perp/2B_\eq$, and asymptotic developments exist.}.
To the expansion order considered in Eq.~\ref{eq_GuidingCenterVelocity}, 
this limit is not an issue. 

\begin{figure}[ht!]
\begin{center}
\begin{minipage}{\linewidth}
\begin{center}
\includegraphics[width=\linewidth]{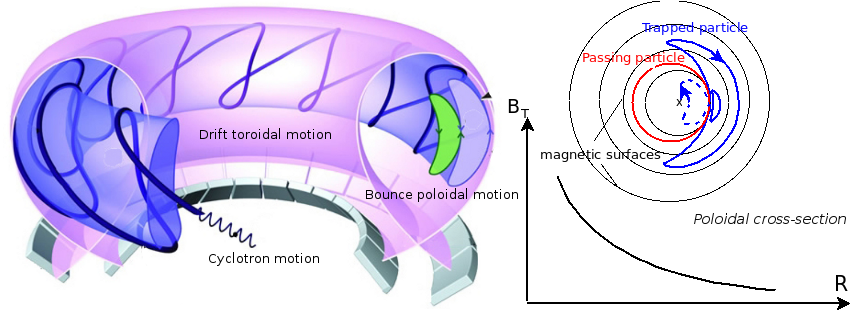}
\caption[\footnotesize Particle trajectories in a tokamak.]{
\label{fig_PassingTrapped}
\footnotesize 
Particle trajectories in a tokamak. 
The left hand side picture illustrates the three angular 
components of a trapped particle trajectory. 
The right hand side picture is a poloidal projection of the trajectories
of a passing and a trapped particle.}
\end{center}
\end{minipage}
\end{center}
\end{figure}

A direct consequence of the invariance of $\mu$ is that some 
particles remain confined in the outer part of the confinement 
chamber as illustrated in Fig.~\ref{fig_PassingTrapped}. 
The latter particles are called {\bf trapped particles} 
and their trajectories have the shape of a {\it banana} when
projected onto a poloidal cross-section.
To understand this behavior, we need to recall that the magnetic 
field of a toroidal configuration cannot be uniform, but varies
like $1/R$ (Eq.~\ref{eq_BTnonuniformity}).
Let us now consider  a particle 
characterized by the invariants {\sf E} and $\mu$ and use the
approximation that ${B}_\eq \approx B_{ \text {T} \eq}$, it
directly comes that $\mu = ({\sf E}-mv^2_\|/2)/B_\eq 
\propto R({\sf E}-mv^2_\|/2)$. 
Hence, if  the particle approximately follows a helical field line 
and comes nearer to the axisymmetry axis, $R\rightarrow 0$, 
the simultaneous  invariance of $\sf E$ and $\mu$ may enforce a
cancellation of $v_\|$, which means that the particle (or more exactly
its guiding-center) will bounce back to the outer region of the
tokamak. If $v_\|$ does not cancel, the particle is called 
{\bf passing particle}.\\

The existence of trapped particles is a first evidence that the 
motion of charged particles is not purely along the field lines. 
More explicitely, it can be shown
that the guiding-center motion is, as expected, mainly parallel but 
that it also contains a {\bf drift}. 
Explicitely,
\begin{equation}
\dot{\bf X} = v_{\|} {\bf b} +
    {\bf v}_{g} + O(\rho^{* 2} \ v_{t})
\label{eq_DriftVelocity}
\end{equation}
where  ${\bf v}_g$, called the drift  velocity, contains three
physical components related to the presence of an electric field and
to  the magnetic field non-uniformities, 
${\bf v}_g = {\bf  v}_{{\bf E}\times{\bf B}} 
+ {\bf v}_{\nabla B} + {\bf v}_{c} $ with \\
\\
\begin{tabular}{llcll}
$\bullet$&
${\bf v}_{\bf{E\times B}} $
&=&$\frac{{\bf E}_\eq\times {\bf B}_\eq}{B_\eq^2}$&
, the ${\bf{E\times B}}$  drift
\\\\  
$\bullet$& 
$ {\bf v}_{\nabla B}$&=&
$\frac{\mu}{e}{\bf b}_\eq 
\times\frac{\nabla B_\eq}{B_\eq}$&
, the grad-$B$ drift
\\\\
$\bullet$& 
$ {\bf v}_{c}$&=&$\frac{mv_\|}{eB_\eq}{\bf b}_\eq
\times \m{\kappa}$ &
, the curvature drift
\end{tabular}\\ \\
where $ {\bf b}_\eq = {\bf B}_\eq/{ B}_\eq$, 
and $\m{\kappa} = {\bf b}_\eq\cdot \nabla {\bf b}_\eq $ 
is the field local curvature. When  ${\bf E}_\eq$ is 
chosen such that the ${\bf E}\times{\bf B}$ drift is only 
first order compared to $\rho^*$, which is the case in standard 
discharges (where $E_\eq$ is mainly parallel for toroidal 
current generation), ${\bf v}_g$ is shown to be only
first order in $ \rho^*$.

Note that the above developments would still be correct with 
slowly varying fields, 
\begin{equation}
\frac{1}{\Omega_c}\frac{1}{B_\eq}\frac{dB_\eq}{dt}
\sim \rho^*.
\end{equation}
An additional drift associated with this time variation can be 
found in the litterature 
\\\\
\begin{tabular}{llcll}
$\bullet$& 
${\bf v}_{P} $ &=& 
$\frac{m}{e} {\bf b}_\eq \times \frac{d {\bf v}_g}{dt}$& ,
the polarization  drift
\end{tabular}\\\\
but it is only second order in $\rho^*$ with our conventions.

It follows from the described drifts
that the particle trajectories deviate from the field lines, 
which we represented in Fig.~\ref{fig_PassingTrapped}.

\subsubsection{Hamiltonian description of particle motion}
Even if the direct expansion of the charged particle motion 
mentioned in the previous section may lead to a good physical 
understanding of its main dynamics, it rapidly becomes a hazardous 
work, when one wants to change the coordinate system, the geometry, 
add a perturbation or  simply find higher order expressions.

For this reason, the developments to come will rather make use of the 
equivalent hamiltonian description of particle motion, which is 
particularly practical to unveil motion invariants or check the 
validity of their conservation, for example the energy conservation. 
The notions of Hamiltonian mechanics used in this thesis are 
summarized in  Appendix~\ref{appchapter_HamiltonianNotions}, and
expressed using the conventions of Ref.~\cite{Brizard_07}, 
where a Lagrangian is
mathematically a differential 1-form.\\

The motion of a charged particle immersed in a magnetic field, 
${\bf B}  = \nabla \times {\bf A}$ 
and an electric field ${\bf E} = -\nabla \phi -\partial_t{\bf A}$ in 
the six dimensional space-velocity phase space, is a Hamiltonian system, 
which can be expressed in the coordinate system 
$({\bf x}, \mathbf{p})=({\bf x}, m{\bf v} + e{\bf A})$, with
\begin{eqnarray}
\text {the {\bf Lagrangian} \ }&
\underline {
\hat{\Gamma}} 
({\bf x}, {\bf p}, t)  
&= {\bf p}\cdot {\sf d}{\bf x} - \hat{H}{\sf d}t,
\label{eq_ChargedParticleLagrangian}\\
\text{the {\bf Hamiltonian}}&
\hat{H}({\bf x}, {\bf p}, t)  
&= \frac{|\mathbf{p}-eA|^2}{2m} +e\phi.
\end{eqnarray}

A Hamiltonian system described by the Lagrangian 
$ \underline{ \Gamma} ({\bf Z}, t) = 
{\bf \Gamma}({\bf Z}, t)\cdot{\sf d} {\bf Z}  
- H({\bf Z}, t){\sf d}t$
 with ${\bf Z} =(Z^a)_{a=1...6} \in {\bf R}^6$ a phase-space
coordinate system, 
verifies Hamilton's equations:
\begin{equation}
\frac{dZ^a}{dt}=[Z^a, H]+[Z^a,
  Z^b]\frac{\partial \Gamma_b}{\partial t}
\label{eq_HamiltonsEquations}
\end{equation}
where $[,]$ are called the {\bf Poisson Brackets}. 

The Poisson Brackets correspond to a bilinear antisymmetric 
function depending of the Lagrangian components 
${\bf \Gamma}({\bf Z},t)$. They are  correctly definined in 
Appendix~\ref{appchapter_HamiltonianNotions}.
Nethertheless, only two simple situations will be tackled in 
the thesis:
\begin{itemize}
\item When the values of the Poisson Brackets are known in a 
  given coordinate system 
  ${\bf Z}\in {\bf R}$, then for any function
  $f({\bf Z},t)$ and $g({\bf Z},t)$
\begin{equation}
  [f, g] = 
  \frac{\partial{f}}{\partial Z^a}[Z^a,Z^b]
  \frac{\partial{g}}{\partial Z^b}.
\end{equation}
\item When the coordinate system is {\bf canonical}, 
  that is, when the Lagrangian is of the form 
  $ \underline{ \Gamma} ({\bf Z}, t) 
  = Z_1dZ_4 + Z_2dZ_5 + Z_3 dZ_6 - H({\bf Z}, t){\sf d}t$,  
  the expression of the Poisson Bracket is
  \begin{equation}
    [,] = \partial_{\mbox{\boldmath$\mathcal{X}$}}
    \partial_{\mbox{\boldmath$\mathcal{P}$}} 
    -\partial_{\mbox{\boldmath$\mathcal{P}$}}
    \partial_{\mbox{\boldmath$\mathcal{X}$}}
    \label{eq_CanonicalVariables}
  \end{equation}
  where 
  $\m{\mathcal{X}} = (Z_1, Z_2, Z_3)$,  and  $\m{\mathcal{P}}$
  gathers the so-called momenta, $\m{\mathcal{P}} 
  = (Z_4, Z_5, Z_6)$.
\end{itemize}
The Lagrangrian describing the charged particle motion 
\ref{eq_ChargedParticleLagrangian} is obviously expressed 
in canonical coordinates. Thus, Hamilton's equations 
\ref{eq_HamiltonsEquations} are easily computed,
\begin{eqnarray}
\frac{d{\bf x}}{dt} &=& \partial_{\bf p}\hat{H} = {\bf p}/m \\
\frac{d{\bf p}}{dt} &=& -\partial_{\bf x}\hat{H} 
= e({\bf E}+{\bf v}\times{\bf B}) + e\frac{d {\bf A}}{dt}
\end{eqnarray}
which shows that the Hamilton's equations are nothing but the usual
Lorentz force balance. Nevertheless, we now have a powerful formalism
to make coordinate transformations.

\ssssection{Coordinate transformation in Hamiltonian systems
{\rm \&} Application to the guiding-center transformation}
Hamilton's principle 
(see Appendix \ref{appchapter_HamiltonianNotions})
implies that the physics is conserved in a coordinate transformation, 
${\bf Z}\rightarrow {\bf Z'}$, if there exists a total derivative
${\sf d}S$ \cite{GoldsteinBook},
\begin{equation}
\underline{\Gamma}' ({\bf Z'}, t) 
=\underline{\Gamma} ({\bf Z}, t)+{\sf d}S
\label{eq_LagrangianTransformation}
\end{equation}
Littlejohn \cite{Littlejohn_83} made use of this principle 
to derive the equations 
of motion for the guiding center in an equilibrium field 
${\bf B}_\eq$. Starting from the Lagrangian 
$\underline{\hat{\Gamma}}$ of 
Eq.~\ref{eq_ChargedParticleLagrangian} 
and after successive modifications
of the form \ref{eq_LagrangianTransformation}, 
he found to the first order in $\rho^*$
the guiding-center relevant Lagrangian in the form
\begin{eqnarray}
({\bf x}, {\bf p}) &\rightarrow&({\bf X}, \mu, v_\|, \gamma) \\
\nonumber\underline{\hat{\Gamma}} 
&\rightarrow& \underline{\Gamma}_{\text{gc}} 
= {\bf A}_\eq^*\cdot {\sf d}{\bf X} 
+\mu {\sf d}\gamma - H_{\text{gc}}{\sf d}t\\
&&\text{ with } H_{\text{gc}} =  \frac{1}{2}mv^2_\|  
+ \mu B_\eq({\bf X})+ e\phi_\eq({\bf X}),
\label{eq_GuidingCenterLagrangian}
\end{eqnarray}
and the corresponding basis Poisson Brackets,
\begin{equation}
\begin{array}{lcllcl}
\left[X_i, X_j\right] &=& 
- (\nabla X_i \times \nabla X_j)
\cdot\frac{\mathbf{b}_\eq}{{eB^*_{\eq \| }}},&
\left[ \mathbf{X}, mv_{\|}\right] &=&
\frac{{\bf B}^*_\eq}{B^*_{\eq\|}}\\
\left[\mu, X\right]&=& 0,  &\left[\mu, v_{\|}\right] &=& 0
\end{array}
\label{eq_LittlejohnPoissonBrackets}
\end{equation}
where the $(^*)$ used by this author may be understood as a first
order correction to the traditional fields:
 ${\bf A}^*_\eq =  {\bf A}_\eq 
+ (mv_\|/e) {\bf b}_\eq$, 
${\bf B}^*_\eq = \nabla \times {\bf A}^*_\eq$,
$B^*_{\eq\|} = {\bf b}_\eq\cdot {\bf B}^*_\eq$.\\\\

As required, this formulation leads to the same dynamics as the one 
described in the previous sub\-section. 
Indeed, expanding formulas~\ref{eq_LittlejohnPoissonBrackets} to recover 
the usual fields and keeping only first order corrections, it comes
\footnote{Note a useful formula:
${\bf b}\times\nabla\times {\bf b} = -{\bf b}\cdot\nabla {\bf b}$.
}

\begin{equation}
\left[X_i, X_j\right]= - (\nabla X_i \times \nabla X_j) 
\cdot\frac{\mathbf{b}_\eq}{{eB_\eq}}, \quad
\left[ \mathbf{X}, mv_{\|}\right] =
{\mathbf{b}_\eq}-\frac{mv_{\|}}{eB_\eq}( \m{\kappa} \times 
\mathbf{b}_\eq)
\label{eq_ModifiedPoissonBrackets}
\end{equation}
which directly allows to recover the usual drifts
\begin{eqnarray}
\nonumber\frac{d {\bf X}}{dt} 
&=& [{\bf X}, H_\eq] 
= v_\|{\bf b}_\eq + {\bf v}_c  + {\bf v}_{{\bf E}\times{\bf B}}
+ {\bf v}_{\nabla B} \ .
\label{eq_GuidingCenterEquation}
\end{eqnarray}
Formulas~\ref{eq_ModifiedPoissonBrackets} are the one used in the thesis, 
for their simplicity.
However, it has to be noted that they do not conserve phase-space volumes
exactly, contrary to formulas~\ref{eq_LittlejohnPoissonBrackets}
\cite{Brizard_07}.

\subsubsection{Action-angle variables}
In the previous section, we explained that a Hamiltonian description
of particle motion could be powerful, and lead to simple insightful
motion equations when expressed in canonical variables.
However, when going from the particle variables $({\bf x}, {\bf p})$
to the guiding-center variables $({\bf X}, \mu, {\sf E},\gamma)$, 
canonicity is lost. 
Fortunately, in the tokamak geometry, it is possible to display
a {\bf canonical system of variables which is consistent with the 
decoupling of the gyromotion and guiding-center motion}.
Moreover, if follows from the tokamak periodicity in $\theta$ and 
$\varphi$ that the particle motion is quasiperiodic at equilibrium, 
and the chosen system of coordinates can be taken to be a system of 
{\bf action-angle variables} $(\m{\alpha}, \m{J})$.
Action-angle variables are a particular type of canonical variables
appropriate for periodic systems where the ``spatial'' variables are 
angles and the momenta (or actions) are motion invariants, that is, 
\begin{equation}
\dot{\bf J}  = -\frac{\partial H_\eq}{\partial \m{\alpha}} = 0, \quad
\dot{\m{\alpha}} = \frac{\partial H_\eq}{\partial \bf J}
= \m{\Omega}_\eq({\bf J})
\end{equation}
Hence, this description does not only provide canonicity but also 
physical motion invariants, to which  $\mu$ belongs.
Moreover, the characteristic eigenfrequencies of the periodic
particle motion can be directly derived, and the question of the 
time decoupling of the different periodic motion  directly assessed.
In particular, for the understanding of the resonances between waves 
and energetic  particles, it is necessary to know these 
eigenfrequencies.

A derivation of the set of  action-angle variables used in this 
thesis is provided in Appendix~\ref{appsection_ActionAngleVariables},
which closely follows Refs.~\cite{WhiteBook, Kaufman_72, GarbetHDR}.
The motion is found to be divided into three angular periodic motions
\begin{equation}
\m{\alpha} = (\alpha_1,\alpha_2,\alpha_3) 
           = \m{\Omega} \ t 
           = \m{\alpha}_0 \ \
           + \m{\Omega}\int^{\theta}_0\frac{d\theta}{\dot{\theta}}           
\end{equation}
(where $\m{\alpha}_0$ stands for the initial phase-space position)
with invariant eigenfrequencies
\begin{equation}
\begin{array}{lclcl}
\Omega_{1} &=& \Omega_b
\oint\frac{d\theta}{2\pi}\frac{1}{\dot{\theta}}\ \dot{\gamma}
&\approx& 
\Omega_b\oint\frac{d\theta}{2\pi}\frac{1}{\dot{\theta}}
\frac{eB_\eq}{m} 
\vspace{0.3cm}\\
\Omega_{2} &\equiv& \Omega_b 
= 2\pi\left(\oint\frac{1}{\dot{\theta}}\right)^{-1} 
&\approx&
2\pi\left(\oint\frac{1}{{\bf b}_\eq\cdot\nabla\theta\ v_\|}\right)^{-1}
\vspace{0.3cm}\\  
\Omega_{3} &=& 
\Omega_b\oint\frac{d\theta}{2\pi}\frac{1}{\dot{\theta}}\ \dot{\varphi}
&\approx& \Omega_b\oint\frac{d\theta}{2\pi}\frac{1}{\dot{\theta}}\
{\bf v}_D\cdot\left[-q'(\bar{\Psi})\theta\nabla\Psi
+ \nabla(\varphi-q(\bar{\Psi})\theta)    \right]
\vspace{0.3cm} \\
&&&& + \  \delta_\text{passing}\ q(\bar{\Psi})\Omega_b
\end{array}
\label{eq_GeneralEigenfrequencies}
\end{equation}
where the first angular motion is found to be related to the 
gyromotion $\gamma$, the second  to the poloidal motion 
described by $\theta$ also called the {\bf bounce motion}, 
and the third, called {\bf precessional drift}, to  the particle 
drift in the toroidal direction. 
The three angular motions can be clearly identified in the 3-D 
picture of Fig.~\ref{fig_PassingTrapped}.
The {\it bounce integral} 
$(\Omega_b/2\pi)\oint(d\theta/\dot{\theta}) ...$, 
present in the eigenfrequencies expression, allows to remove the 
fields $\theta$-dependence.
For passing particles, $ \oint = \oint_0^{2\pi}$, whereas for 
trapped particles oscillating between the $\theta$-angles 
$[-\theta_0,\theta_0]$,  
$ \oint = (1/2)\int_{-\theta_0}^{-\theta_0}$ (the full closed banana)
$ \approx \int_{-\theta_0}^{\theta_0}$.

The corresponding invariant momenta are
\begin{equation}
\begin{array}{lclcl}
J_1  &=& \frac{m}{e}\mu \vspace{0.3cm}\\
J_2  &=& 
\oint \frac{d\theta}{2\pi}\frac{B_\theta}{B_\eq} mv_\| 
+  e\oint\frac{d\theta}{2\pi}\Phi 
&\approx& \oint \frac{d\theta}{2\pi}\frac{r^2}{qR_0}\ mv_\| 
          + \delta_\text{passing} \ e\Phi(J_3) \vspace{0.3cm}\\
J_3 &=& e\Psi +  \frac{I(\Psi)}{B_\eq} mv_{\|}
    &\approx& e\Psi + Rmv_\|
\end{array}
\label{eq_GeneralInvariants}
\end{equation}
where $B_\theta$ is the covariant component of the field along $\theta$.

$J_1$ is nothing  but the gyromotion adiabatic invariant, 
Similarly, $J_2$ is an adiabatic invariant and
$J_3$ is an exact invariant related to the equilibrium 
axisymmetry in the toroidal direction, also simply 
called  {\bf toroidal momentum}.

We also provide in Eqs.~\ref{eq_GeneralEigenfrequencies} and 
Eqs.~\ref{eq_GeneralInvariants} some more approximate expressions 
of the invariants and motion characteristic frequencies, which are 
easier to use and interpret (This will be useful later on.).
To make these approximations, we simply kept the lower order expressions 
in $\rho^*$ and $\epsilon$.
Explicitely, we made use of the simplified expressions of the 
geometry given in section \ref{sssection_LargeAspectRatioEquilibrium} 
and we neglected the drift motion when the parallel velocity had a finite 
contribution.
This way a discontinuity appears between passing and trapped particles 
($\delta_\text{passing}=1$ for passing particles and 0 for trapped
particles). 
The drift motion disappears in $\Omega_2$ and hence, the flux $\Phi$ 
becomes an invariant in the expression of $J_2$ for passing particles.
The drift motion simply remains in the expressions of the toroidal 
drift, where we expanded the radial drift around a reference flux 
surface $\bar{\Psi}$. 

Fully explicit expression of the equilibrium motion eigenfrequencies 
Eq.~\ref{eq_GeneralEigenfrequencies} can even be found within these 
approximations. 
The calculation of the normalized bounce and drift frequencies  
$ \bar{\Omega}_b$ and $\bar{\Omega}_d$ such that
\begin{equation}
\Omega_2 = \Omega_b 
= \pm\frac{1}{qR_0}\sqrt{\frac{2{\sf E}}{m}}\bar{\Omega}_b, \quad
\Omega_3 = \Omega_d + \delta_\text{passing}q(r)\Omega_b
=\frac{q(r)}{r}\frac{E}{eB_0 R_0}\bar{\Omega}_d
+\delta_\text{passing}q(r)\Omega_b
\label{eq_NormalizedBounceDrift}
\end{equation}
is given in Appendix~\ref{appsection_KbKd} for the simplified geometry
described in Eq.~\ref{sssection_LargeAspectRatioEquilibrium}.
We also give in this Appendix some more general expressions taken 
from Ref.~\cite{Zonca_07} which contain effects of the Grad-Shafranov
shift. The latter expressions are the one used  in the thesis.\\

We now have a powerful description of the particle trajectories
which, as explained earlier, is particularly appropriate
for the computation of resonant behaviors. 

Indeed, using the expression of the motion eigenfrequencies 
\ref{eq_GeneralEigenfrequencies}, one may 
directly catch the frequency range of the modes which may resonate
and followingly exchange energy with the particles. As explained
earlier $\Omega_1\gg\Omega_2, \Omega_3$. Since $\Omega_3$ simply 
contains the lower order drift motion, we often get an additional
time scale separation $\Omega_d <\Omega_b$. However, since 
$\Omega_b \propto \sqrt{\sf E}$,
$\Omega_d \propto {\sf E}$, this separation is not strong for larger
energies (orders of magnitude corresponding to our specific parameters
of study will be provided later on in this thesis).

Moreover, we will see later on, that the non-conservation of the
equilibrium invariants in the presence of a wave is the source of 
resonant wave-particle energy transfers: a computation of these 
transfers is consequently cleaner in the action-angle space.

Nevertheless, one should note that 
the action-angle variables may not allways be the 
most intuitive coordinates since they imply a mixing of space and 
velocity coordinates. For this reason, the energy invariant {\sf E} 
is sometimes chosen instead of $J_2$. 
When doing this, another notation will be made use of, ie.
$(\mu = J_1, {\sf E}, P_\varphi=J_3)$, in order to make clear 
that {\sf E} not ${\bf J}_2$ should be kept constant when deriving 
by $\mu$ or $P_\varphi$ (The $P_\varphi$ notation
simply result from  the remark that $J_3$ is nothing but the toroidal 
momentum, Appendix \ref{appsection_ActionAngleVariables}.). 
Besides, noticing that 
$e\Psi/mRv_\| \sim (1/\rho^*)(\epsilon^2) (r/a)^3$, and taking 
$\rho^*\sim 10^{-3}, \epsilon\sim0.3$ 
consistently with typical Tore-Supra parameters, it comes that far
enough from the magnetic axis, $e\Psi$ dominates in the expression 
of $J_3$. Hence, it is often more convenient (and possible) to 
interpret $J_3$ as a purely radial coordinate.

\subsubsection{Application to energetic particle trajectories}
We compared in Tab.~\ref{tab_ThermalHotCharacteristics} some features of the 
thermal and suprathermal particles' trajectories corresponding to a typical
Tore-Supra discharge (where the suprathermal ions are assumed to be heated by 
{\it Ion Cyclotron Resonant Heating in the minority heating 
scheme}) and to ITER parameters 
(where suprathermal particles are taken to be fusion born alpha particles). 
First, characteristic frequencies 
(corresponding to very trapped particles, 
$\bar{\Omega}_b\sim 1$, $\bar{\Omega}_d\sim 1$) are provided and
put in parallel with the Alfv\'enic and acoustic frequency ranges.
Next, typical gyration widths associated with the motion 
invariants are given:
the Larmor radius and the banana width of trapped particles 
(considered here for marginally trapped particles, 
which display the largest width).
\begin{table}[h!]
\begin{center}
\begin{tabular}{lrrrr}
&D$^+$ in TS 
&H$^+$ in TS 
&D$^+$ in ITER 
&$\alpha$ in ITER\\
&(${\sf E}= 5$ keV) & (${\sf E}= 300$ keV)
&(${\sf E}= 15$ keV)& (${\sf E}= 3.56$ MeV)\\
\hline\hline
Acoustic range      &$\sim 32$ kHz & &$\sim 21$  kHz\\
Shear Alfv\'en range&$\sim 500$ kHz& &$\sim 190$ kHz\\
$\Omega_c/2\pi$     &28  MHz &58  MHz &40 MHz  &13 MHz\\
$\Omega_b/2\pi$     &46  kHz &500 kHz &31 kHz  &270 kHz\\
$\Omega_d/2\pi$     &0.4 kHz &24  kHz &0.12 kHz&28 kHz\\
Larmor radius       &2.5 mm  &1.5  cm &3.0 mm  &9.0 cm\\
Banana width        &8.  mm  &4.5  cm &1.0 cm  &30 cm\\
\hline\hline
\end{tabular}\\
\caption
[\footnotesize
Features of the thermal and energetic ion trajectories in 
Tore-Supra and ITER]
{\footnotesize\label{tab_ThermalHotCharacteristics}
Features of the thermal and energetic ion trajectories 
in Tore-Supra and ITER, 
calculated with the values of Tab.~\ref{tab_ITER}, 
a normalized radius of 0.3, a density of $5e19$ m$^{-3}$ and $q=1.0$.}
\end{center}
\end{table}

Tab.~\ref{tab_ThermalHotCharacteristics} shows that the time scale
separation between the gyromotion and guiding-center motion is fully
verified for both energetic and thermal particles, but that the 
scale separation of the bounce and drift motions becomes less clear
when going to higher energies. 
Besides, it confirms the idea developed in the thesis introduction 
that resonances are possible with the acoustic frequency range for 
both thermal and suprathermal particles, which makes this frequency 
range particularly important for the understanding of wave-particle 
resonances. In particular, resonances with macro-scale structures 
are expected to be more relevant since micro-scale perturbation 
($\sim 1$ mm) fall below the typical Larmor radius of energetic 
particles.
\parpic{
\includegraphics[width=0.45\linewidth]{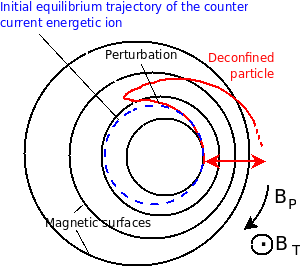}
}
Finally, the calculation of the banana width of barely trapped
particles  gives a first insight in the particles confinement, 
because it is a measure of the deviation from the field lines. 
As can be seen in the table, the banana width of barely 
trapped energetic particles is a non-negligible fraction of 
devices' minor radius 
($5$ cm$/ 70$cm in Tore-Supra, $0.3$m/$2$m in ITER).
We may already expect possible losses of confinement when a passing
particle is slightly perturbed and gets slightly trapped
(see the sche\-matic on the side). Let us now  analyze how 
such a perturbation may occur.
\vspace{1cm}

\section{Theoretical review of energetic particle driven 
modes}
\label{chapter_TheoreticalReview}
We now review the main features of the interaction between energetic 
particles and the thermal plasma.
It was explained earlier that energetic particles were expected to 
interact preferably with macro-scale waves, which are traditionally
described in the MHD formalism. After explaining some basics of MHD 
and MHD modes, we tackle the problem of the modes stability in the 
presence of an energetic particle drive. We display the particular geometric
effects which stabilize these modes and lead to a classification between 
{\bf gap modes} and {\bf energetic  particle modes (EPMs)}. 
Finally, various theories developed to explain 
the nonlinear evolution of these modes and the correlated transport 
of energetic particles will be depicted.

\subsection{Basic Magneto-HydroDynamic waves}
\label{ssection_BasicMHD}
\subsubsection{MHD equations, MHD energy principle}
\label{sssection_MHDEquations}
MHD is the traditional frame for the description of macroscopic 
instabilities ($>1$ cm in typical tokamak conditions).
In the standard ideal MHD formalism, a quasi-neutral plasma of ions
and electrons is described as a single fluid using three momenta,
its mass density $\rho_M$, its velocity ${\bf V}$ and its scalar 
(isotropic) pressure $P$, and one vector field, 
the magnetic field ${\bf B}$. Ideal MHD equations are
\begin{eqnarray}
\frac{d \rho_{\text{\sc m}}}{d t} + 
\rho_{\text{\sc m}} \nabla\cdot{\bf V} &=& 0 \\
\rho_{\text{\sc m}}\frac{d{\bf  V}}{dt} 
+ \nabla P &=& {\bf J}\times{\bf B}
\label{eq_MHDVelocity}\\
\frac{dP}{dt}  + \Gamma P\nabla \cdot {\bf V} &=& 0
\label{eq_MHDPressureClosure}\\
{\bf E} + {\bf V}\times {\bf B} &=& 0
\label{eq_MHDOhmsLaw}\\
\nabla\times{\bf E}  &=& -\frac{\partial {\bf B}}{\partial t} 
\label{eq_MHDMagFlux}\\
\nabla\times{\bf B}  &=& \mu_0{\bf J} \\
\nabla \cdot {\bf B} &=& 0.
\end{eqnarray}
with $d/dt \equiv \partial_t + {\bf V}\cdot \nabla$.

The first three equations are the evolution equations 
of the three fluid moments, and Eq.~\ref{eq_MHDPressureClosure} is 
the MHD pressure closure, characterized by the adiabatic 
compressibility coefficient $\Gamma$, equal to $5/3$ for an 
isotropic pressure.
Eq.~\ref{eq_MHDOhmsLaw} is the ideal Ohm's law which fully determines 
the electric field ${\bf E}$ as a function of ${\bf B}$. 
The last three equations result from the low frequency limit of 
Maxwell equations (no displacement current considered), and they 
define the plasma current  $\bf J$ completely as a function of 
$\bf B$. In particular $\mu_0$ is the vacuum permeability.

Though not completely useful ($\bf E$ could be replaced directly 
in Eq.~\ref{eq_MHDMagFlux}), Ohm's law contains important features 
of the MHD formalism. First, it directly returns  that 
$ E_\| = 0$,
which means that charge separation is not considered to take 
place in the parallel direction. This condition will be refered 
to as the {\bf MHD condition} in the remainder of the text.
Secondly, after crossing with $ {\bf B}$, it implies
that ${\bf V}_\perp = {\bf E}\times{\bf B}/B^2$, showing that
the MHD perpendicular velocity reduces to the fluid 
${\bf E}\times{\bf B}$ drift, which simply results from the
particles ${\bf E}\times{\bf B}$ drift (generalized here to 
non-equilibrium fields).
The dominating character of the ${\bf E}\times{\bf B}$ drift is a 
major assumption of MHD, which is consistent with the hydrodynamic 
limit $\omega \rightarrow \infty$ \cite{HazeltineBook}.  
Indeed, when fast fluctuations are considered 
$\partial_t\rightarrow+\infty$, 
${\bf E} = -\nabla \phi - \partial_t{\bf A}\rightarrow\infty$, 
with {\bf A} the vector potential (${\bf B} =  \nabla\times{\bf A}$).
Hence drifts involving {\bf E} becomes dominant compared to 
other effects involving spatial gradients only.\\

Let us now determine the linear normal modes 
(with eigenfrequency $\omega$) which may develop in an MHD 
plasma. For this,  consider a fluctuation of eigenfrequency 
$\omega$ around a known equilibrium state, 
\begin{equation}
\begin{array}{ll}
\rho_\text{\sc m} 
= \rho_{\text{\sc m} \eq} 
+ 2\re\left(\rho_{\text{{\sc m}}\omega}e^{-i\omega t}\right),& 
\quad {\bf V} = 2\re\left({\bf V}_\omega e^{-i\omega t}\right), 
\vspace{0.2cm}\\
P=P_\eq+2\re\left(P_\omega e^{-i\omega t}\right),&
\quad {\bf B}={\bf B}_\eq 
+ 2\re\left({\bf B}_\omega e^{-i\omega t}\right).
\end{array}
\label{eq_MHDLinearExpansion}
\end{equation}

A standard procedure to analyze these modes is to make use of 
a so-called {\bf MHD energy principle}. Linearizing the MHD 
equations, and defining the MHD displacement $ \m{\xi}$ such that 
$\dot{\m{\xi}}  = {\bf V}$
($-i\omega{\m{\xi}_\omega}  = {\bf V}_\omega$), 
the linearized MHD equations can be put in the form 
$\omega^2\rho_\text{\sc m\eq}\ \m{\xi}_\omega
= \mathcal{F}(\m{\xi_\omega})$,
with $\mathcal{F}$ a self-adjoint operator 
(for the scalar product, 
$({\bf u},{\bf v})\rightarrow
\int d{\bf x}^3{\bf u}^*{\bf v}$). 
Multiplying this 
equation by $\m{\xi}_\omega^*$ and integrating over all space, 
it comes
\begin{eqnarray}
\begin{split}
\delta I \equiv\omega^2\int d^3{\bf x}
\frac{\rho_{\text{\sc m}}}{2}|\xi_\omega|^2
&=  \frac{1}{2}\int d^3{\bf x} \
  \frac{|{\bf B}_{\omega\perp}|^2}{\mu_0}
+ \frac{1}{\mu_0}\left|B_{\omega\|}
- \m{\xi}_{\omega\perp}
  \cdot \nabla P_\eq \frac{\mu_0}{B_\eq}\right|^2
+ \Gamma P_\eq|\nabla\cdot\m{\xi}_\omega|^2\\
& \quad
- J_{\eq\|}({\bf \xi}_{\omega\perp}\times{\bf b}_\eq)
  \cdot{\bf B}_\omega
- 2 (\m{\xi}_\omega\cdot\nabla P_\eq)
(\m{\bf \kappa}\cdot\m{\xi}_\omega^*)\\
& = \frac{1}{2}\int d^3{\bf x} \m{\xi}_\omega^*\ \mathcal{F}(\m{\xi_\omega})
\ \equiv\  \delta W_\text{MHD} \\
\end{split}
\label{eq_dWMHD}
\end{eqnarray}
where $\delta I$ is the plasma kinetic energy, 
$\delta W_\text{MHD}$ is called the {\bf MHD potential energy}.
Due to the self-adjointness of $\mathcal{F}$, $\omega^2$ and 
$\delta W_\text{MHD}$ are directly shown to be real,
and a given mode may either be purely oscillating and stable 
($\omega \in {\bf R}, \ \delta W_\text{MHD}>0$)
or purely growing/damped 
($\omega \in i{\bf R}, \ \delta W_\text{MHD}<0$).
With this energy formulation, 
one may directly determine the stability of the given equilibrium 
under a given perturbation $\m{\xi}_\omega$.

Simple observation of Eq.~\ref{eq_dWMHD}  shows what may be
the destabilizing features of a given equilibrium. The first three 
terms of $\delta W_\text{MHD}$ are positive and hence, stabilizing.
In  particular, compressibility and magnetic 
field tension (included in ${\bf B}_{\omega\perp}$) act as restoring
stabilizing forces.
The two destabilizing mechanisms are included in the last two terms:
the first one being related to the existence of an equilibrium 
current leads to so-called {\bf current driven }, 
or {\bf kink instabilities}
and the second one depending on the respective sign of the pressure 
gradient compared to the field line curvature is the so-called 
{\bf pressure driven}, or {\bf interchange instabilities}.

\subsubsection{MHD waves in a uniform plasma}
First insight into the nature of stable linear MHD waves can be 
obtained assuming a uniform magnetized equilibrium, 
${\bf B}_\eq({\bf x}) = {\bf B}_0$, 
$P_\eq({\bf x}) = P_0$ ...

From the velocity equation, it clearly appears that the fluid 
inertia (the left hand side) has to balance pressure effects as well 
as the magnetic field tension, which enters ${\bf j}\times {\bf B}$.
In linear analysis, two characteristic velocities appear, which 
account for this  behavior,
\begin{equation}
\begin{array}{llcl}
\text{the Alfv\'en  velocity } v_A, 
& v^2_A =
& \frac{B^2_0}{\mu_0\rho_0} \vspace{0.2cm}\\
\text{and the sound  velocity } c_s,    
& c^2_s =
& \frac{\Gamma P_0}{\rho_0},
\end{array}
\label{eq_VaCsVelocities}
\end{equation}
which respectively provide a measure for the restoring force induced
by magnetic pressure, and the one resulting from compressibility 
(the kinetic pressure).
In magnetized fusion plasmas, $c_s$ is smaller that $v_A$. 
The ratio of the plasma pressure to the magnetic pressure
\begin{equation}
\beta \equiv \frac{2\mu_0 P_0}{B_0^2}
      = \frac{2}{\Gamma}\frac{c^2_s}{v_A^2}
\end{equation}
is a third important small parameter in magnetized fusion plasmas 
($\sim 10^{-2}$ in Tore-Supra, where the typical values of the 
pressure and field are taken at the plasma center).

More precisely, linearizing the MHD equations to the first order 
in a perturbation of the form $\propto  e^{i({\bf k.x}-\omega t)}$, 
three types of waves are found,
\begin{itemize}
\item the incompressible {\bf shear Alfv\'en wave}, 
of dispersion relation
\begin{equation}
\omega^2 = k_\|^2v_A^2
\end{equation}
This is a transverse wave, propagating in the parallel direction, 
where the perturbed  magnetic field and perturbed velocity 
(parallel to each other)  are perpendicular to 
the equilibrium magnetic field. Note that here $\omega^2$ is 
reminiscent from the mode inertia 
(as in Eq.~\ref{eq_dWMHD}), and $k^2_\|v^2_A$ from the field line 
bending restoring force.

\item 
the {\bf fast} (+ sign in Eq.~\ref{eq_MagnetoSound}) 
and the {\bf slow}   (- sign in Eq.~\ref{eq_MagnetoSound}) 
{\bf magnetosonic waves}
\begin{equation}
\omega^2 = \frac{k^2 (v_A^2 + c_s^2)}{2}
\left( 1 \pm \sqrt{1-
 \frac{4k^2_\|}{k^2}\frac{v_A^2c_s^2}{(v_A^2+c_s^2)^2}} \right)
\quad\left\{ 
\begin{array}{lll}
k^2v^2_A\\ \\
k_\|^2c_s^2 
\end{array}\\
\right.
\text{ if } v_A\gg c_s
\label{eq_MagnetoSound}
\end{equation}
which couple kinetic and magnetic pressures. 
\end{itemize}

In a tokamak configuration, the equilibrium parallel length 
scale is of the order of $R_0$ (or $qR_0$), whereas the typical 
equilibrium perpendicular is of the order of the minor radius $a$. 
Hence, for a large aspect ratio $1/\epsilon$, 
the tokamak geometry ``naturally'' leads to longer parallel 
scales, $ k_\|\ll k_\perp$. 
When considering a perturbation, scale lengths may change 
significantly.
However, interesting modes tend to be localized where $k_\|=0$. 
In pure MHD plasmas, it simply comes from the fact that the 
magnetic field tension is proportional to $\propto k_\|v_A$ 
(it will be derived in a clean way in the course of the thesis, 
but the shear Alfv\'en wave dispersion already gives some indication 
of this) and stabilizing. 
Consequently, 
MHD instabilities are more easily excited where this tension 
cancels. When suprathermal particles are added to the picture, 
stable MHD modes may be driven unstable by energetic particles. 
Nevertheless, so-called resonant surfaces (verifying $k_\|=0$) 
remain of interest because they can allow for a mode localization.
Hence, it is often relevant to consider modes with $k_\|\ll k$.
With the additional assumption of a low-$\beta$ plasma, a 
time separation appears between the three MHD waves,
\begin{equation}
k_\|c_s\ll k_\|v_A\ll k v_A.
\end{equation}

Considering the energetic ions typical eigenfrequencies, 
it can be shown that resonances may occur between energetic
ions and modes of the acoustic (or sound) and shear Alf\'ven 
frequency ranges.
This was illustrated in Tab.~\ref{tab_ThermalHotCharacteristics},
where we compared the typical eigenfrequencies
of suprathermal trapped particles with a typical value of the
acoustic frequency range ($c_s/R_0$) and a typical value
of the shear Alfv\'en frequency range ($v_A/R_0$).

\subsection{Alfv\'en spectrum in a sheared plasma}
\label{ssection_AlfvenSpectrum}
\subsubsection{Sheared plasmas and phase mixing}
When moving to a non-uniform plasma, the situation is more 
complicated. The three MHD waves described above get fully 
coupled, and non-uniformity leads to non-coherent local 
behaviors at the origin of dispersion and damping.\\

\parpic{
\includegraphics[width=0.3\linewidth]{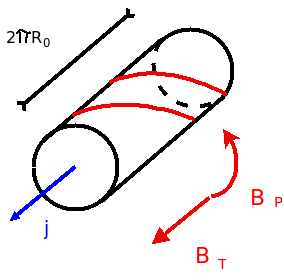}
}
To picture this phenomenon, we  make use of a cylindrical 
plasma of radius  $a$ and length $2\pi R_0$,
which can be seen as a tokamak plasma of null inverse aspect ratio,
$\epsilon=0$. 
Besides, we assume the plasma density to be uniform and the 
existence of a purely toroidal plasma current, localized
at the cylinder center (for example with a current density
$j(r)\propto (1-r/a  )^2$).
With $\epsilon=0$, the equilibrium toroidal field 
(along the cylinder axis) is approximately uniform, 
${\bf B}_{\text{T}\eq} ({\bf x}) = B_0$, whereas the poloidal
field is non-uniform (or {\it sheared}), 
${\bf B}_{\text{P}}(r)$.

This way, we obtain a cylindrical plasma with a sheared magnetic
field, but one should note that the physics developed below 
can also be recovered with a more simple slab geometry and 
a sheared density (see Ref.~\cite{Hasegawa_74}).
The case developed here is simply more convenient for later 
developments to the tokamak geometry.
To mimic the tokamak conditions, we also define an equivalent 
toroidal angle $\varphi$, such that $R_0\varphi$ holds as a 
coordinate for the toroidal direction and
we assume that
${ B}_{\text{P}\eq}/{B}_{\text{T}\eq} \sim \epsilon \ll 1$.

Let us  now consider a linear normal mode. In a cylindrical
configuration (characterized by an invariance in 
the direction of the cylindrical axis and in the 
$\theta$ direction), it is possible to take it of the form
$\propto e^{i({\sf m}\theta + {\sf n }\varphi)}$ with 
${\sf (n,m)}$ two integers. For this perturbation, it comes
\begin{equation}
i k_\| = 
{\bf b}_\eq\cdot\nabla \varphi
\left({\sf n} + \frac{\sf m}{q(r)}\right) 
\approx \frac{1}{R_0}
\left({\sf n} + \frac{\sf m}{q(r)}\right) 
\label{eq_kparallel}
\end{equation}
where we recall that ${\bf b}_\eq = {\bf B}_\eq/B_\eq$.
The safety factor $q$ ($q(r) = (r/R_0)(B_\text{T}/B_\text{P})$
in the cylindrical limit) is a non-constant function of $r$
(we avoid the odd case where ${\bf B}_\text{P}\propto r$).
Hence, $k_\|$ is also a non-constant function of $r$, which means in 
particular that pure Alfv\'en waves $\omega =\pm k_\|(r)v_A$
cannot have a global structure.
On the contrary, this equality seems to suggest that a 
{\bf continuous frequency spectrum} 
of shear Alfv\'en waves exist, where each frequency is 
associated to a different radius (and an infinitely 
localized eigenfunction).
In a global analysis however, this is not possible,
which means that a wave cannot be a pure shear Alfv\'en 
wave. On the contrary, radial continuity enforces a mixing
of the different polarizations involved in the
three MHD waves described above.\\

To obtain a better understanding of the physics involved, we 
can expand the MHD equations using a perturbation of the form 
Eq.~\ref{eq_MHDLinearExpansion}.
Under the assumption that the considered perturbation verifies
$k_\|\ll k_\perp$ and using that $\beta$ is a small parameter in 
magnetized fusion plasma, we already explained that the three
classical MHD waves are characterized by a time scale 
separation. 
When only the first assumption is made, it is possible to 
focus on the lower frequency modes only, independently from the
coupling to the fast magnetosonic branch. Such a decoupling
leads to the so-called {\bf shear Alfv\'en law}
\cite{HazeltineBook}, which is an equation on the fluid 
vorticity, $\nabla\times{\bf V}$.
Using  the smallness of $\beta$ 
(to cancel sound wave effects) and $\epsilon$ 
(to simplify the differential operators), 
it is possible to express the shear Alfv\'en law using one
unknown field only, the electric potential $\phi_\omega$. 
For general geometry,
the shear Alfv\'en law reduces to
\begin{eqnarray}
\begin{split}
&0=\nabla\cdot\left[\frac{\omega^2}{v^2_A}
\nabla_\perp\phi_\omega\right]
+B_\eq\nabla_\|
\left\{\frac{1}{B^2_\eq}\nabla\cdot
\left[B_\eq^2\nabla_\perp
\left(\frac{1}{B_\eq}\nabla_\|\phi_\omega\right)
\right]\right\}+\\
&{\bf B}_\eq\times\nabla
\left(\frac{1}{B^2_\eq}\nabla_\|\phi_\omega\right)
\cdot \nabla\left(\frac{J_{\eq\|}}{B_\eq}\right)
+\frac{2\mu_0}{B^2_\eq} {\bf B}_\eq\times\m{\kappa}_\eq
\cdot\nabla_\perp
\left[ \left(\frac{{\bf b}_\eq}{B_\eq}\times\nabla P_\eq
\cdot\nabla_\perp\phi_\omega\right)\right] 
\end{split}
\label{eq_AdaptedShearAlfvenLaw}
\end{eqnarray}
where we wrote $\nabla_\|$ = ${\bf b}_\eq\cdot\nabla$, and 
generalized the definition of the Alfv\'en velocity to space 
dependent densities and magnetic fields, $v_A(r)$.
A similar result will be derived later on. At the moment, we
can already identify the main physical components involved:
inertia ($ \propto \omega^2$), field line bending force
($\propto v_A^2\nabla_\|^2$),  kink   ($\propto J_{\eq\|}$)
and interchange ($\propto\m{\kappa}_\text{\eq}\cdot\nabla P_\eq$).
\\

Let us apply this equation to a cylindrical configuration, 
and perturbed quantities of the form 
$ X_\omega^{\sf m}e^{i{\sf m}\theta + i{\sf n}\varphi}$.
The cylindrical geometry enables 
$\m{\kappa} = \kappa(r) \nabla r $, and the various
$\sf (m, n)$ Fourier components do not couple. 
Projection of Eq.~\ref{eq_AdaptedShearAlfvenLaw} onto
the $\sf (m, n)$ harmonic returns
\begin{eqnarray}
-\frac{d}{d r}
\left\{ r^3\left[\frac{\omega^2}{v_A^2} 
- (k^{{\sf m}}_{\|})^2\right]
\frac{d}{d r} \left(\frac{\phi^{\sf m}_\omega}{r}\right) \right\} 
&=&\omega^2 r^2\frac{\phi^{\sf m}_\omega}{r}
\frac{d}{dr}\left(\frac{1}{v_A^2}\right)\\
\nonumber&&+ (1-{\sf m}^2)
\left(\frac{\omega^2}{v_A^2} - (k^{\sf m}_{\|})^2\right)
\phi^{\sf m}_\omega
- {\sf m}^2\frac{2\mu_0}{B^2_\eq}
\frac{dP_\eq}{dr}\kappa_r\phi_\omega^{\sf m}.
\label{eq_AdaptedShearAlfvenLawCylinder}
\end{eqnarray}
This equation suggests the possibility of a smooth eigenfunction 
far from the {\bf  Alfv\'en resonance} condition, 
$\omega^2 = k^{\sf m}_\| {}^2v_A^2$, 
but it shows that the existence of a surface of radius $r_A$ 
verifying the Alfv\'en resonance leads to a singularity in the 
mode structure.

Formally, the existence of an Alfv\'en resonance at $r=r_A$
can be treated using a perturbative two scale 
analysis  separating the smooth and discontinuous structure 
variations of the mode \cite{Hasegawa_74}.
Using temporarily  the notation 
$\bar{E}^{\sf m}_{\omega} = \phi^{\sf m}_\omega/r$, the two scale 
analysis leads to the decomposition of the form
$ \bar{E}_\omega^{\sf m} = \bar{E}^{\sf m}_{\omega 0}(r_0)
+ \bar{E}^{\sf m}_{\omega 1}(r_0,r_1)$ where 
$\bar{E}^{\sf m}_{\omega 1}\ll \bar{E}^{\sf m}_{\omega 2}$, 
$r_0\sim r$ is a slow variable similar to equilibrium quantities, 
and $r_1 \sim r-r_A$ is a fast variable, 
$\partial_{r_0}\ll\partial_{r_1} $. 
To the lower order, 
\begin{eqnarray}
\frac{d}{dr_1} \left[r^3\left(\frac{\omega^2}{v_A^2}
-(k^{\sf m}_{\|})^2\right)
\frac{d\bar{E}^{\sf m}_{\omega 1} }{dr_1} \right]=0 
&\Rightarrow& \frac{d \bar{E}^{\sf m}_{\omega 1}}{dr} 
= \frac{C(r)}
{\left.\frac{d}{dr}\left(\frac{\omega^2}{v_A^2}
-(k^{\sf m}_{\|})^2\right)\right|_{r_A}(r-r_A)}
\label{eq_LogDiscontinuity1}
\\
&\Rightarrow& \bar{E}^{\sf m}_{\omega1} \propto C(r)\ln |r-r_A|
\label{eq_LogDiscontinuity2}
\end{eqnarray}
with C a slowly varying continuous function. 

A {\bf logarithmic singularity} appears close to the {\bf Alfv\'en 
resonant surface} $r_A$, where $\omega^2 = k_\|^2v_A^2$. 
Whereas the mode structure involves various polarizations away
from $r_A$ (and in particular a radial dispersion), singularity  
can be understood as the excitation of a  ``pure'' shear Alfv\'en 
wave at the Alfv\'en resonant surface.
Since a shear Alfv\'en wave propagates in the parallel
direction only (hence on a given radial surface), information
gets accumulated at $r_A$, leading to the logarithmic singularity 
\cite{Vlad_99}. In particular, the singularity can be shown to be 
associated to a {\bf discontinuity of the Poynting flux}
\cite{Hasegawa_74, Hasegawa_76, Vaclavik_91}.
Another way to understand the logarithmic behavior and the correlated 
development of {\bf small scale structures} is to separate the plasma 
into various radial plasma shells characterized by its own eigenfrequency 
$\pm k_\|(r)v_A(r)$.
Because each shell tends to respond better to a signal corresponding
to its eigenfrequency, radial coherence is lost. This is called
{\bf phase-mixing} \cite{Grad_69}.
As a consequence of this lack of coherence, these shear Alfv\'en-like
waves happen to be very localized.
Hence, non-localized solutions are expected in Fourier space 
(ie: for the Fourier transform of $ \bar{E}_\omega^{\sf m}(r)$), which 
in analogy with quantum mechanics are to be associated with a 
{\bf continuous frequency spectrum} of solutions. The idea developed 
above is now justified.

\begin{figure}[ht!]
\begin{center}
\begin{minipage}{\linewidth}
\begin{center}
\includegraphics[width=0.45\linewidth]{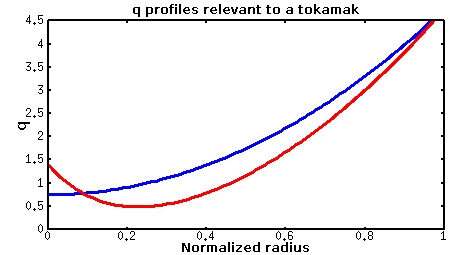}
\includegraphics[width=0.45\linewidth]{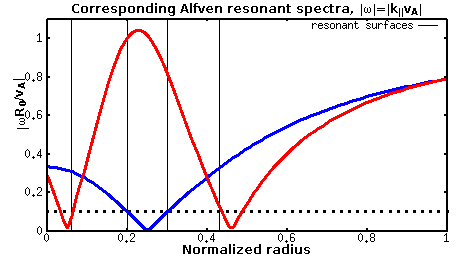}
\caption[\footnotesize Typical sheared q profiles and corresponding 
resonant spectra.]
{\label{fig_SAWSpectra}
\footnotesize 
Typical sheared $q$ profiles of a tokamak and corresponding 
Alfv\'en resonant spectra for modes with a single poloidal 
and toroidal component, $({\sf m}=-1, {\sf n}=1)$. 
As an example, the resonant surfaces for an acoustic mode
($\omega R_0/v_A \sim \beta \sim 0.01$) are indicated, and show 
where such a mode should display a singularity.
}
\end{center}
\end{minipage}
\end{center}
\end{figure}

Note that depending on the geometry, the surface where the Shear 
Alfv\'en law Eq.~\ref{eq_AdaptedShearAlfvenLaw} presents a singularity, 
or {\it resonance}, may have a somewhat more complicated 
dispersion relation than $\omega = \pm k_\|v_A$. 
In the following, such surfaces will be
called {\bf Alfv\'en  resonant surfaces}, and the spectrum of frequencies 
which can be solution of the corresponding dispersion relation, the 
{\bf Alfv\'en continuous (resonant) spectrum}. Finally, the modes 
oscillating with a frequency of the continuous spectrum will be 
refered to as {\bf continuum modes}.
The latter notions are illustrated in Fig.~\ref{fig_SAWSpectra}, where we
applied them to standard sheared q-profiles.\\

An important consequence of the singularity is that it implies 
a  damping of the considered waves, called {\bf continuum damping}.
Recalling that MHD is only supposed to acccount for relatively
large structures, this damping can physically be understood as a 
loss of information towards the small scales.
We can already infer that moving to a kinetic description 
may eventually allow for some of these modes to live undamped (at least 
at the lowest order), the so-called {\bf kinetic Alfv\'en waves}.

Interestingly however, it was shown that if an MHD mode is damped by 
continuum damping, and if the kinetic modelling of the wave displays
a "strong" (kinetic) damping mechanism,
the damping rate calculated by MHD does not differ from the one 
calculated with the kinetic formalism. 
In other words, the damping rate does not depend on the detailed 
process or on the details of the mode structure \cite{Hasegawa_76}.

The verification of the existence of a damping and the computation of 
the resulting damping rate can be done using the interpretation of the 
MHD equations as an {\bf energy relation},
in a similar fashion as for the derivation of the MHD energy principle. 
Multiplying Eq.~\ref{eq_AdaptedShearAlfvenLawCylinder} by 
$\bar{E}_\omega^{\sf m}$ and integrating over space under the assumption 
that there exists one single Alfv\'en resonance at $r_A$, it comes
\begin{eqnarray}
\nonumber - \delta W(\omega,\bar{E}^{\sf m}_\omega) &=&
 {\mathcal P}\int rdr\left\{
\left[r^2\left(\frac{d\bar{E}^{\sf m}_\omega}{dr}\right)^2
+({\sf m}^2-1)\bar{E}^{\sf m \ 2}_\omega\right]
\left(\frac{\omega^2}{v_A^2}-(k^{\sf m}_{\|})^2\right)\right.\\ 
\nonumber &&\left.-\omega^2r^2\bar{E}^{\sf m \ 2}_\omega\frac{d}{dr}
\left(\frac{1}{v_A^2}\right)
+{\sf m}^2\frac{2\mu_0}{B^2_\eq}\frac{dP_\eq}{dr}
\kappa_r\bar{E}^{\sf m\ 2}_\omega\right\}\\
&=& \left.r^3\left[\frac{\omega^2}{v_A^2}-(k^{\sf m}_{\|})^2\right]
\left(\frac{d}{dr}\bar{E}_\omega^{\sf m}\right)\bar{E}^{\sf m}_\omega
\right|^{r^+_A}_{r^-_A}
= C(r)\bar{E}^{\sf m}_\omega|^{r_A^+}_{r^-_A}
\label{eq_ContinuousModeJump}
\end{eqnarray}
where the energy function used here can be easily identified as
$\delta W(\omega, \bar{E}^{\sf m}_\omega) 
= - \delta I + \delta W_\text{MHD}$, 
and the last equation comes from Eq.~\ref{eq_LogDiscontinuity1}.
At the lowest order, the right hand side small scale
singularity is usually assumed negligible compared to the expected 
large mode structure, and the main mode eigenfunction is found as a 
solution of
$-\omega_0^2\delta W(\omega_0, \bar{E}^{\sf m}_{\omega 0})=0$.
Expanding Eq.~\ref{eq_ContinuousModeJump} to the next order, we can
obtain the effect of the singularity on this eigenfunction.
In particular, with the expansion
$\omega = \omega_0 + \delta \omega_r+i\gamma$, the damping rate is found
to verify
\begin{equation}
-\frac{\partial (\delta W)}{\partial\omega_r}
(\omega_0, \bar{E}^{\sf m}_{\omega 0})\ i\gamma
= \frac{ -i\pi {\rm \sgn(\omega_0)} \ C^2(r_A)}
{\left|\frac{d}{dr}
\left(\frac{\omega_0^2}{v_A^2}-(k^{\sf m}_{\|})^2\right)(r_A)\right|}.
\label{eq_ContinuumDamping}
\end{equation}
Eq.~\ref{eq_ContinuumDamping} can be seen as an {\bf energy balance}, where 
$\partial_rW$ is to be interpreted as the a measure of the {\bf wave energy}
of the low order solution of Eq.~\ref{eq_ContinuousModeJump}
\footnote{We will come later to these ideas in subsection 
\ref{ssection_PositiveEnergyWaveDensity}}, 
and the right hand side represents an energy well.
Assuming in a first approximation that 
$\delta I\propto \omega^2$ and that $\delta W_\text{MHD}$ has no dependance 
in $\omega$ (it may not be true in general since the structure of 
$\phi_\omega$ can depend on $\omega$),
$-\partial_\omega\delta W(\omega_0, \bar{E}^{\sf m}_{\omega0})>0$.
Hence, $\gamma<0$, the singularity implies a {\bf damping}, as announced. 
This absorption of the wave energy suggested in the seventies that the 
coupling to the shear Alfv\'en spectrum, could be an interesting way to 
{\bf heat the plasma by a direct injection of Alfv\'en type waves.}\\

Note that we can obtain a better idea of the damping time scales,
coming back to the resolution of Eq.~\ref{eq_LogDiscontinuity1} 
which may be rewritten
\begin{equation}
\left(\partial_{tt} +(k^{\sf m}_{\|})^2v_A^2\right)
\partial_r\bar{E}^{\sf m}
=C(r, t) 
\label{eq_ShearAlfvenSolutions}
\end{equation}
close to the continuum. Eq.~\ref{eq_ShearAlfvenSolutions} 
admits solutions of the form \ref{eq_ShearAlfvenSolutions}
$\partial_r\bar{E}^{\sf m}_\omega=C(r)\exp(\pm  i k_\|v_A t)$ 
\cite{Hasegawa_74}. 
Consequently, when $t\rightarrow +\infty$, $\partial_r \rightarrow \pm
i\partial_r(k^{\sf m}_\|v_A)\ t$ and diverges, 
reflecting the fact that the mode is characterized by a singularity. 
Finally
\begin{equation}
\bar{E}^{\sf m}_\omega = \pm\frac{C(r)}{i\partial_r(k_\|v_A)\ t}\ \
e^{\pm i k_\|v_A t},
\end{equation}
and the time behavior of the damping is observed to {\bf be proportional 
to $\bf 1/t$}.

\subsubsection{Gap Modes and  Energetic Particle Modes (EPM) in toroidal
geometry}
From the previous section, we know that a mode is damped where Alfv\'en 
resonances occur.
For example, let us consider a mode with  (${{\sf n}=1, {\sf m}=-1}$). 
We already explained that a mode often develops close to a surface 
where $k_\|= {\bf b}_\eq\cdot\nabla\varphi({\sf n}+{\sf m}/q)= 0$, 
that is, where $q$ is rational, and in this case equal to one. 
Take  for example $\omega$ to be an acoustic frequency,
Alfv\'en resonance occurs for 
$|R_0k_\|| = |\omega/v_A| \sim \beta $ ($\sim 0.01$  in Tore-Supra).
Expanding $R_0k_\|$ linearly around the $q=1$ surface
\begin{equation}
|R_0k_\|| \approx |R_0k'_\|||\delta x|
          =  |{\sf m}/q^2||q'||\delta x|,
\label{eq_KparExpansion}
\end{equation} 
and considering a
standard monotonic Tore-Supra $q$ profile, two Alfv\'en resonances
appear on both sides of the rational surface, with a typical 
distance to it $|\delta x| \sim 5$mm. This characteristic length is 
small for a macroscopic mode, and compared to Tore-Supra minor radius 
$0.7$ m.

{\bf In a toroidal geometry,  various poloidal components }
(various {\sf m} numbers in the above decomposition) 
{\bf are coupled together}, because the system is  not fully invariant 
in the poloidal direction ($\epsilon \neq 0$). 
Due to this coupling which implies more global structures, continuum 
damping may be thought to be even more stabilizing than in cylindrical 
geometry. However, it is not the case because of the existence of 
{\bf gaps in the shear Alfv\'en spectrum}, where a mode can live undamped.

From now on and until the end of this thesis, to avoid any ambiguity 
in the description of toroidal geometry and to allow for toroidicity induced 
poloidal coupling to be explicit, the quantities $\rho_s$ and $v_A$ need to 
be understood as independent from the poloidal angle $\theta$ 
($\rho_s(r), v_A(r)$). In other words, the central field $B_0$ is used when 
assessing these quantities.\\

\begin{figure}[ht!]
\begin{center}
\begin{minipage}{\linewidth}
\begin{center}
\includegraphics[width=0.8\linewidth]{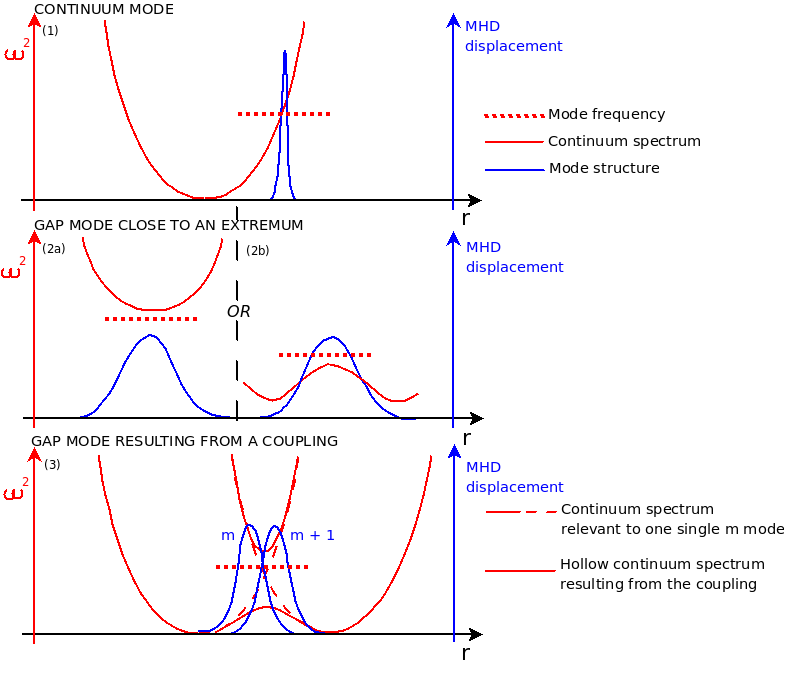}
\caption{\label{fig_SAWSpectra_withstruc}
\footnotesize 
Continuum modes and gap modes.}
\end{center}
\end{minipage}
\end{center}
\end{figure}

We may distiguish two types of conditions for a gap to occur 
\cite{Heidbrink_08}. 
Typical representations corresponding to the various types of modes 
described in this section are given in Fig.~\ref{fig_SAWSpectra_withstruc}.
\begin{itemize}
\item {\bf Particular equilibrium properties naturally result in 
    gaps in the Alfv\'en spectrum 
    \cite{Appert_82, Appert_84,Appert_85,Berk_01}.}
  
  Looking at the formula $\omega^2 = k^{\sf m\  2}_\|v_A^2$ for the 
  localization of Alfv\'en resonances,   
  it is obvious that some profiles are more favorable for
  the existence of modes with a large extent.
  A non-monotonic $q$ profile \cite{Berk_01} and/or a non monotonic
  density profile may lead to finite minima of $k^{{\sf m} 2}_\|v_A^2$. 
  Below these minima, a mode with a frequency $\omega$ 
  can have a rather large extent. An example of such 
  modes are the {\bf Reversed Shear Alfv\'en Eigenmodes} (RSAEs)
  \cite{Berk_01, Breizman_03}.\\
  Besides, some physical phenomena neglected 
  in Eq.~\ref{eq_AdaptedShearAlfvenLaw} (such as compressibility), 
  can change the frequency localization of the Alfv\'en 
  resonances. 
  For example, when compressibility is taken into account
  the characteristic equation of the resonant spectrum has the form 
  $\omega^2 = k_\|^{{\sf m}\ 2}v_A^2+ \omega_0^2$  
  (in other words, the resonant spectrum is not purely of the shear
  Alfv\'en type but involves terms related to sound waves), and 
  {\bf Beta Alfv\'en Eigenmodes} with a eigenfrequency below 
  $|\omega_0|$ are not damped.

\item {\bf Wave-wave coupling in toroidal geometry can allow for the 
    existence of standing shear Alfv\'en waves}.
  
  We have explained earlier that the singularity induced by an Alfv\'en 
  resonance at $r_A$ could be associated to the purely parallel phase 
  velocity of shear Alfv\'en waves. 
  Hence the idea that parallel transport and singularity vanishes 
  if two shear Alfv\'en waves are coupled together to form a standing 
  wave. This behavior may be visualized as the apparition of a {\it gap} 
  in the shear Alfv\'en wave spectrum, as illustrated in 
  Fig.~\ref{fig_SAWSpectra_withstruc}.
  
  In toroidal geometry, the coupling of different poloidal 
  harmonics directly leads to such types of gaps, called 
  {\bf toroidicity induced gaps}, and associated with 
  {\bf toroidal Alfv\'en Eigenmodes (TAE)} \cite{Kieras_82, Cheng_86}.
\end{itemize}
The creation of the toroidicity induced gap appears directly when 
applying Eq.~\ref{eq_AdaptedShearAlfvenLaw} to a toroidal geometry.
In this equation, the divergence operator depends on 
$R = R_{0} (1+(r/R_{0})\cos\theta)$ and hence on $\theta$, which
leads to poloidal coupling.
Accordingly, poloidal coupling is found to involve terms proportional 
to the inverse aspect ratio, and the Right Hand Side (R.H.S.) 
of Eq.~\ref{eq_AdaptedShearAlfvenLaw} can be put in the form 
\cite{Berk_91}
\begin{eqnarray}
-\frac{d}{dr}
\left[r^3\left(\frac{\omega^2}{v_A^2}-(k^{\sf m}_{\|})^2\right)
\frac{d}{dr}\left(\frac{\phi^{\sf m}_\omega}{r}\right)\right]
-\frac{d}{dr}
\left[
r^3\bar{\epsilon} (r)\frac{\omega^2}{v_A^2}
\frac{d}{dr}
\left(\frac{\phi_\omega^{\sf m+1}+\phi_\omega^{\sf m-1}}{r} 
\right)\right]=R.H.S.\quad
\label{eq_AdaptedShearAlfvenLawToroidal}
\end{eqnarray}
where $\bar{\epsilon}$ is of the order of and proportional to 
$\epsilon$.

Focusing on two neighboring poloidal components only, 
and making use of the previous notation $\bar{E}^{\sf m}_\omega$, 
this is again
\begin{eqnarray}
-\frac{d}{dr}r^3\underline{\underline{D}}\frac{d}{dr}
\left(\begin{array}{ll}\bar{E}^{\sf m+1}_\omega\\
\bar{E}^{\sf m}_\omega\end{array}\right) = R.H.S., { \rm with }\ 
\underline{\underline D}=
\left(
\begin{array}{ll}
\omega^2/v_A^2-(k^{\sf m }_\|)^2&\bar{\epsilon}(\omega^2/v_A^2) \\
\bar{\epsilon}(\omega^2/v_A^2) & \omega^2/v_A^2-(k^{\sf m+1 }_\|)^2 
\end{array}
\right)\
\end{eqnarray}

\begin{itemize}
\item[]
Far from the shear Alfv\'en resonances $\omega^2/v_A^2-(k^{\sf m/m+1}_\|)^2$,
the terms of the diagonal dominate ($\bar{\epsilon}\ll 1 $), and 
the situation is close  to the one of a cylinder. 

The situation is different when both resonances are verified. 
The double cancellation of the resonant terms allows for both poloidal 
components  to have similar, strong weights, and off-diagonal terms 
representing the poloidal coupling become of major importance. 
Consequently, the resonant spectrum (which leads to a singularity), 
of equation $\det(\underline{\underline D})\-=0$, does not cancel 
where both shear Alfv\'en waves couple. On the contrary, solving  
$\det(\underline{\underline D})\-=0$  displays a {\it gap}, 
represented in Fig.~\ref{fig_SAWSpectra_withstruc}.
\end{itemize}

Thanks to gaps, we now know that continuum damping may be avoided 
on relatively large radial extents. There remains
to show that {\bf radially localized modes} can live in these 
{\bf frequency gaps}, which implies two necessary conditions on the 
frequency and on the existence of a {\it defect} able to localize 
a mode \cite{Chen_07}.

Such a study is classically carried out in Fourier space, where the
Fourier transform of the radial 
dispersion equations (Eq.~\ref{eq_AdaptedShearAlfvenLawCylinder}, 
\ref{eq_AdaptedShearAlfvenLawToroidal}), 
takes the form of a Schr\"odinger equation, and 
regular (ie, localized in radial space) 
localized (ie, non discontinous in radial space) solutions are
looked for. In analogy with quantum mechanics
localization in Fourier space, leads to a 
{\bf discrete frequency spectrum}, in opposition to the 
shear Alfv\'en continuous spectrum.

Considering the above first type of gaps for example, it can be 
shown that localization is possible close to a point $r_0$,
where $k^{\sf m}_\|$ is well approximated by 
$ (k_\|^{\sf m})^2(r) 
=  (k_{\|}^{\sf m}(r_0))^2 + (1/2)(k^2_\|)''(r-r_0)^2$, 
and in the presence of a defect. 
Such a situation occurs at the minimum of the  $q$ profile where 
 $k_\|'(r_0)=0$ (leading to RSAEs) 
or close to a resonant surface where $k^{\sf m}_\|(r_0) = 0$.
For RSAEs, the mode localization at the minimum of a $q$-profile 
leads to the so-called {\it Alfv\'en cascades}, that is to modes 
with a shifting localization and a shifting frequency, which 
follows the time variations  of the $q$ profile.
\begin{figure}[ht!]
\begin{minipage}{\linewidth}
\begin{center}
\includegraphics[width=0.45\linewidth]{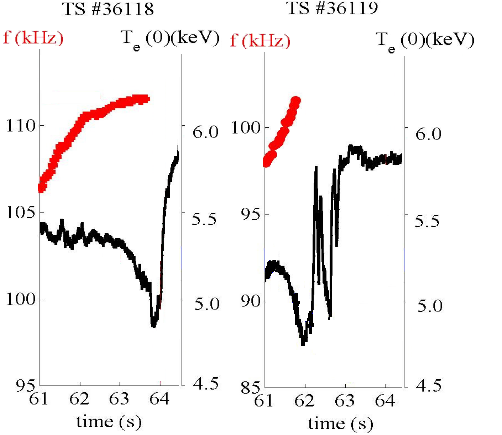}
\caption[\footnotesize
Alfv\'en Cascades in Tore-Supra.]
{\label{fig_TS_CascadeAlfven}
\footnotesize 
Frequency spectrum of Reversed Shear Alfv\'en Eigenmodes
(RSAEs) identified in Tore-Supra \cite{Goniche_08}. 
Frequency cascading is associated to both density fluctuations and 
oscillations of the q-profile.}
\end{center}
\end{minipage}
\end{figure}

Assuming for simplicity that ${\sf m}\gg 1$, and accordingly 
that equilibrium variations ($\sim r$) are negligible compared to 
radial variations of $k_\|$, and using the normalized radius,
$\bar{r} = (r-r_0)(|{\sf m}|s_0/r_0) $ where $s_0$ is a characteristic
value of the shear (defined in Eq.~\ref{eq_ShearDefinition}, 
and simply taken to be $s_0 = 1$ for when it cancels at the minimum 
of the $q$ profile),
Eq.~\ref{eq_AdaptedShearAlfvenLawCylinder} takes the form
\begin{equation}
\frac{d}{d \bar{r}}\left\{ \left[ \bar{r}^2 -\Lambda^2  \right]
\frac{d}{d \bar{r}} \phi^{\sf m}_\omega\right\} = 
\frac{1}{s_0^2}[\bar{r}^2-\Lambda^2]
\phi^{\sf m}_\omega - \mathcal{G}\phi^{\sf m}_\omega
\label{eq_RSAE}
\end{equation}
where 
$\Lambda^2 = 
(\omega^2/v_A^2-k^{\sf m}_{\|}(r_0)^2)
\times 2({\sf m}^2s^2_0/r^2_0)/(k^2_\|)''$, 
contains the information relative to the eigenfrequency.

Now taking the Fourier transform of $\phi^{\sf m}_\omega(\bar{r})$, 
ie,  
$
\phi(\vartheta) = \int_{-\infty}^{+\infty}  d\bar{r}
\, \phi_\omega^m(\bar{r})\  e^{-i  \bar{r}\vartheta}
$
and using the change of variable, 
$\Psi(\vartheta) = \sqrt{1 + s_0^2\vartheta^2}\phi(\vartheta)$, 
it comes
\begin{equation}
-\frac{\partial^2\Psi}{\partial\vartheta^2} + 
\left(-\Lambda^2
+ \frac{s_0^2}{(1+s_0^2\vartheta^2)^2}
-\frac{s_0^2\mathcal{G}}{(1+s_0^2\vartheta^2)} 
\right)\Psi=0
\label{eq_SchrodingerWell}
\end{equation}
The condition for a mode to be in the gap (ie, to avoid singularity 
in Eq.~\ref{eq_RSAE}) is $\Lambda^2 < 0$, which means that the mode
eigenfrequency is in the outside region of the resonant 
spectrum curve (see Fig.~\ref{fig_SAWSpectra_withstruc}).
The condition for a 
mode localization depends on $\mathcal{G}$ 
and it is simply the necessary condition to create a localizing potential, 
in the Schr\"odinger like equation \ref{eq_SchrodingerWell}. 
As was shown in Ref.~\cite{Manhajan_94,Berk_01,Breizman_03}, various
physical components can enter $\mathcal{G}$, and 
make possible the existence of localized modes in the frequency gaps:
{\bf these {defects} include equilibrium current \cite{Manhajan_94}, energetic 
particles \cite{Berk_01} or toroidal effects} \cite{Breizman_03}...

Similarly, the dispersion relation of TAEs \cite{Cheng_85, Chen_94} 
reads
\begin{equation}
-\frac{\partial^2 \Psi}{\partial\vartheta^2}
-\Lambda^2(1+\bar{\epsilon}cos\vartheta)\Psi
+\frac{s_0^2}{(1+s_0^2\vartheta^2)^2}\Psi=0
\end{equation}
where $s_0$ corresponds to the non-zero shear taken at the radial 
intersection of the two coupled shear Alfv\'en wave.
Whereas $\bar{\epsilon}$ creates a gap, the finite shear $s_0$
is a defect which prevents a purely sinusoidal non-localized mode
\cite{Zonca_92}.\\

Finally, we have shown that global modes can develop in a toroidal plasma, 
which are not damped by continuum damping at the lowest order. 
As we will explain in the next section, 
these modes can be driven unstable by energetic particles.
Besides, one should not forget these modes, which are damped in a 
pure MHD description, since the latter 
can be driven unstable 
when energetic particles are present. 
New modes can also result from the presence of energetic particles, 
such as {\it Energetic Geodesic Acoustic Modes} \cite{Fu_08}.

Following the MHD results presented so far, a separation is 
traditionnally made between:
\begin{itemize}
\item {\bf Gap Modes} which live in a gap of the resonant Alfv\'en 
spectrum, and are not damped at the lowest order, so that only a 
small drive can excite them. 
\item {\bf Energetic Particle Modes (EPMs)} which are damped or do 
not exist without a fair amount of energetic particles. 
EPM excitation is characterized by a relatively large threshold.
\end{itemize}

{\it A priori}, gap modes may be thought to be deleterious and 
dangerous modes.
Indeed, the fact that they are only weakly damped makes them
particularly easy to drive, but not very good  
candidates for transferring the energy of energetic particles 
to the thermal plasma. 
Moreover, their global structure can be particularly 
dangerous for the confinement of energetic particles, whose 
trajectories may be heavily modified by the mode fluctuations.

However, one should not forget that additional damping processes
are missing in the above picture, which have an effect on the 
stability of gap modes. 
Example of such missing dampings are higher order 
{\bf continuum damping at the plasma boundary} \cite{Zonca_92}, 
{\bf radiative damping} 
which occurs close to the resonant Alfv\'en spectrum, 
kinetic damping resulting from resonances with
the thermal plasma, such as 
{\bf ion Landau damping}      \cite{Zonca_96} or 
{\bf electron Landau damping} \cite{Betti_92}.
In particular, it is not possible to neglect ion Landau damping for 
the description of the Beta Alfv\'en Eigenmodes, although these modes
are considered to be gap modes.

Moreover, it may not be fully meaningful to focus on gap modes only, 
if the population of energetic particles and the corresponding drive
is {\it de facto} non  negligible.

\subsubsection{Non-ideal effects}
Before dealing with the interaction of the modes described above with
energetic particles, let us simply make a few comments on the limits 
of ideal MHD used so far, for the description of the 
thermal plasma.

For this, let us recall the fundamental assumptions of MHD, given
in the thesis introduction,
\begin{itemize}
\item the MHD time-scale validity: for low collisionality plasmas, MHD 
depends on the MHD hydrodynamic approximation, according to which 
$\omega$ is larger than the characteristic thermal plasma frequencies.
\item the MHD spatial-scale validity: MHD only makes possible the 
description of macro-scale structures which are much larger than the 
thermal ion Larmor radius.
\item the MHD closure: the plasma description is reduced to three 
moments in MHD, two scalars: the pressure and the density, and one 
vector field, the velocity. This relies in particular on the 
assumption that the plasma velocity distributions are always 
Maxwellians.
\end{itemize} 

Two limits of ideal MHD directly follow the previous discussion, 
and call for kinetic modelling.

First, {\bf continuum damping} which takes place in the  Alfv\'en 
continous spectrum, can be seen as {\bf conversion of macro-scale MHD waves to 
micro-scale kinetic waves} (when looking at the modes as propagative).
This conversion does not necessarily enforce the use of kinetic theory 
because MHD damping rates are equal to kinetic damping rates, 
when kinetic waves are fully dissipated \cite{Hasegawa_76}. 
However, non-dissipated kinetic waves  can exist inside the Alfv\'en 
continuous spectrum  which do not present the same
stability properties as gap or continuum modes \cite{Lauber_03}.
Kinetic Toroidal Alfv\'en waves (KTAEs) are an example of such kinetic
waves.
More generally, kinetic effects such as {\bf micro-scale effects, 
also called Finite Larmor Radius (FLR) effects} (that is, 
effects of the order of the few ion Larmor radii) 
enforce the existence a potential
well in Eq.~\ref{eq_SchrodingerWell}. It follows a discretization
of the Alfv\'en continuous spectrum \cite{Zonca_98}
and the possibility for non-discontinuous radial modes to exist, 
as long as kinetic damping mechanisms are not too strong.

Secondly, {\bf wave-particle resonances} are not appropriately modelled
using MHD, because of the {\bf hydrodynamic limit} and the consistent use 
of fixed Maxwellian shapes for the velocity distributions.
Even if MHD is often reasonable outside of the hydrodynamic region, 
this extension may only be true when resonances are negligible.
When continuum damping is null and kinetic resonant damping mechanisms
become the major damping processes \cite{Zonca_96,Chavdarovski_09} 
(ion Landau damping for example)
kinetic modelling is desirable.

Finally, some physics is missing in MHD. 
We explained in section \ref{sssection_MHDEquations} 
that the {\bf ideal MHD flow was purely a result of the 
${\bf E}\times{\bf B}$ drift}, consistently with the hydrodynamic 
approximation. 
In particular the so-called diamagnetic flow (or velocity) which is 
related to pressure gradients is not taken into account. 
This assumption breaks when going to lower frequencies 
(see Fig.~\ref{fig_TimeAndLengthScales}), and additional modes with
a finite eigenfrequency come into place (such as {\bf kinks  
oscillating with around the so-called diamagnetic frequency} in the 
{\bf diamagnetic gap}).
Similarly, MHD lacks charge separation in the parallel direction 
$E_\|=0$, and this may become an issue  when electron inertia is
non negligible.

\subsection{Energetic particle drive}
\label{ssection_EnergeticParticleDrive}
Let us move to the interaction of the modes described above 
with energetic particles.
{\it A  priori}, the idea that energetic particles can transfer energy 
to the thermal plasma modes is not obvious. On the contrary, first 
theories rather predicted that fast particles would stabilize MHD modes 
\cite{Rosenbluth_83}.
Because suprathermal are characterized by large eigenfrequencies, 
the idea was that they {\bf could adapt to the mode fluctuations rapidly, 
and hence mitigate them}. 
However, first experiments designed to stabilize MHD kink modes with fast
particles led to the excitation of unexpected new instabilities, the so-called 
{\bf fishbone modes}.

What was later found for the latter kink modes is that they could indeed be
stabilized by energetic particles, as long as their oscillation frequency 
verified $\omega\ll\Omega_{h,3}$ (which in general also implies 
$\omega \ll \Omega_{h,2},\Omega_{h1}$, where $h$ stands for {\it hot} particles), 
and this was indeed interpretated as a mitigation effect induced by the fast 
response of the energetic particles \cite{Porcelli_91}.
However, the situation is strongly modified when resonances can take place
$\omega \sim \Omega_{h,1},\Omega_{h,2},\Omega_{h,3}$.
Nowadays, it is well known that a bounded frequency window exists, 
where fast particles are stabilizing for the kink instability 
\cite{White_89, Porcelli_91}, but that both resonant damping and resonant drive
can take place above this window.\\

A first intuition of the interaction of energetic particles with MHD modes can 
be obtained by the simple addition of a scalar pressure gradient corresponding 
to the hot particles in Eq.~\ref{eq_MHDVelocity}.
When moving to the energy formulation, multiplication by 
$\m{\xi}^*_\omega$  returns
\begin{equation}
  -\m{\xi^*}_\omega \cdot \nabla P_{h\omega}
= -\m{\xi}^*_\omega \cdot {\bf J}_{h\omega}\times {\bf B}_\eq 
= -i\omega\ {\bf J}_{h\omega} \cdot {\bf V}_\omega^*\times{\bf B}_\eq
=  i\omega\ {\bf J}_{h\omega}\cdot{\bf E}^*_\omega
\label{eq_EnergeticParticlesWork}
\end{equation}
where  we neglected the fast ion velocity (assumption of closeness
to the velocity balance) in the first equality,
and made use of Ohm's law for the thermal plasma flow in the last one.

From Eq.~\ref{eq_EnergeticParticlesWork}, it appears that 
a {\bf work} on the fast ion population has to be added in  the 
energy relation Eq.~\ref{eq_dWMHD} \cite{Porcelli_91}, 
and this work can lead to a 
{\bf resonant wave-particle energy transfer}.
This additional work implies a new term in the MHD
energy relation, which is traditionally written with the notations
\begin{equation}
\delta { W}
= -\delta{I}+\delta{W}_{f}+\delta {W}_k=0
\label{eq_NNFishboneDispRel}
\end{equation}
where $\delta I$ represents inertia  as in Eq.~\ref{eq_dWMHD}, 
$\delta{W}_{f}$ is a generalized MHD energy potential 
(supposed to contain all {\it fluid} behaviors), and 
$\delta W_k$ represents  the contribution of the energetic 
particles (which can only be modelled using {\it kinetic} 
theory). Eq.~\ref{eq_NNFishboneDispRel} has been given the name of
{\bf fishbone-like dispersion relation} \cite{Chen_84,Biglari_91}, 
and it can be shown to be of very general use in the study of the 
interaction of energetic particles with the thermal plasma, even when the 
plasma is given a kinetic description \cite{Zonca_EPS07}. Even if the 
form of Eq.~\ref{eq_NNFishboneDispRel} may seem  very intuitive, 
one should note that it implicitely contains an important feature
of the physics at stake: the existence of two 
radial scales as in Eq.~\ref{eq_LogDiscontinuity1}, which allows for 
a separation between localized inertial dynamics and longer scales 
associated with fluid  behaviors and energetic particles.

There now remains to understand the effect of $\delta W_{k}$.
Depending on the time scale involved, it can 
be either stabilizing or destabilizing.
\begin{itemize}
\item If the addition of a hot pressure were to be understood with 
traditional MHD, that is in the hydrodynamic approximation
$\omega \gg \Omega_{h,1},\Omega_{h,2},\Omega_{h,3}$, 
one would expect a destabilizing effect. 
Indeed, we explained earlier that interchange instabilities were 
sensitive to the relative direction of the equilibrium field curvature 
compared to the one of the pressure gradient, and it may be shown
that this feature favors instabilities in the outside part of the 
torus, that is in the low field side half part of the torus.
Hence, energetic trapped particles which mainly evolve in the 
outside part of the torus (and are the dominant energetic particles 
under some heating processes like Ion Cyclotron Frequency Heating, 
ICRH)  are destabilizing. However, the hydrodynamic approximation 
is not relevant.
\item When $\omega \ll \Omega_{h,1},\Omega_{h,2},\Omega_{h,3}$, 
we already explained that stabilization could be expected.
\item Now, $\omega \sim \Omega_{h,1},\Omega_{h,2},\Omega_{h,3}$ is a 
particular case where  particles behave {\bf resonantly} and 
can exchange energy with the mode. 
This response is called {\bf resonant response} 
(whereas the two previous responses are referred to as 
{\bf reactive responses}), and it may be either stabilizing or 
destabilizing.
\end{itemize}
Hence the importance of time scales... \\

Resonances are of particular interest, because they allow for  
{\bf secular energy exchanges between particles and waves}.
Such energy transfers come from the {\bf non-conservation of the equilibrium
motion invariants} of resonant particles.
{\bf Their magnitude and sign linearly depend on the gradients of the involved 
particle distribution function}.
More precisely, if the distribution function of a hot particle population
(but it is also the case for any other population) is $F_{h\eq}$, and resonance 
occurs for $\omega = \sum_i{\sf n}_i\Omega_{h,i}$, the gradients of interest can 
be shown (and in fact {\it will} be shown in chapter 
\ref{chapter_LinearStability}) to be the gradients 
${\sf n}_i\partial_{J_i}F_{h\eq}$.
Because the type of instabilities we are looking at are much slower than
$\Omega_{h, 1}$, it is reasonable to consider ${\sf n}_1=0$.
Hence two types of mechanisms are usually distinguished for resonant drive or
damping: energy induced transfers (with some subdivisions such as anisotropy
induced excitation \cite{Berk_05}, or {\bf Landau damping} \cite{ONeil_65} 
due to the negative energy slope of a thermal velocity distribution)
and radial gradient induced energy transfers, related to the non-conservation
of the third invariant.

An interesting point concerns the numbers ${\sf n}_i$ which are necessary
for the resonant condition $\omega = \sum_i {n}_i\Omega_{h,i}$ to be met.
{\it A priori}, the ${\sf n}_i$'s may span the whole  range of integers, which 
somehow questions the possibility to avoid resonances. 
The question is all the more striking that the magnitude of the energy transfers 
is proportional to the ${\sf n}_i$'s, which suggests resonances may be more
efficient when $\omega$ is well separated from the particle characteristic
frequencies...
In fact, this is not so because the ${\sf n}_i$'s cannot be taken independently
from the wave structure it resonates with. It is possible to show that
resonant drive by energetic particles is most efficient for medium 
${\sf n}_i\sim$ 5-10 \cite{Fu_92}.

\subsection{Nonlinear behaviors}
\label{ssection_NonlinearPresentation}

We now know that interaction between energetic particles and 
some modes of the thermal plasma are possible. 
In particular, we put forward that a purely kinetic behavior, 
{\bf wave-particle resonance},
can allow for an energy transfer from the energetic particles 
to the waves, which is of interest for the confinement of the 
alpha particles energy in burning plasmas.

Nevertheless, only a  nonlinear analysis can tell us whether
a wave can be sustained and act as an energy channel, or
whether it is damped nonlinearly. Only a nonlinear analysis can 
tell us whether a mode will transport energetic particles and imply 
dangerous losses.

Classically, two kinds of  nonlinearities are considered.
\begin{itemize}
\item {\bf MHD nonlinearities} concern the thermal plasma and the
mode structure and frequency nonlinear evolution.
\item {\bf Kinetic nonlinearities} concern the kinetic energetic 
response, related to its resonant behavior.
\end{itemize}
\subsubsection{MHD nonlinearities}
The fully nonlinear treatment of the MHD equations leads to the 
coupling of various perturbed fluid moments and fields, 
and can result  in higher order poloidal components. 
This mechanism can be refered 
to as {\bf mode-mode coupling}, 
and was analyzed for the TAE and for one type of EPM, called the 
{\bf precessional fishbone mode} 
(a purely energetic mode, contrary to the previously mentionned 
{\bf diamagnetic fishbones}).

For TAEs, mode-mode coupling (considered with a linear response of 
the energetic particles) has been shown to lead to a shift of the mode 
frequency out of the gap region, and hence to an increased damping 
and to a saturation \cite{Zonca_95}. 
For the precessionnal fishbones, mode-mode coupling leads to the 
generation of an ${\sf m}=0$ plasma rotation and  magnetic
field, and displays a more ambiguous role of the nonlinearities. 
Far above the mode instability threshold, mode-mode coupling 
leads to saturation. However, close to threshold, 
MHD nonlinearities have been found to produce en explosive growth 
leading to a finite time divergence  \cite{Odblom_02}. 

The latter behavior suggests a dominant role of MHD nonlinearities, 
under certain conditions. However, kinetic nonlinear  theories
are considered more successful at the moment, 
because number of observations could be related to their 
predictions.

\subsubsection{Nonlinear saturation of gap modes 
via kinetic nonlinearities}

\ssssection{Nonlinear trapping}
A successful explanation of the role of kinetic nonlinearities 
in the evolution of fast particle driven {\it gap modes} is 
based on the theory of {\bf nonlinear particle trapping}.
When a mode has a finite amplitude, energetic resonant particles 
can get trapped inside its structure (in six dimensional phase space)
and their motion invariants are no longer conserved.
If particles bounce inside the mode structure much faster than
the mode growth, their trajectories are fully nonlinear and it 
is possible to solve for the energetic particle evolution and
for the mode growth separately.

For a gap mode, the calculation of the mode growth can be done close to
threshold, with a perturbative treatment of the energetic population.
To the lower order, the linear real gap mode can is a solution of 
Eq.~\ref{eq_NNFishboneDispRel} and verifies $\delta W_{L}(\omega_0)=0$, 
$\omega_0$ real.

If the mode linear structure can be assumed unchanged nonlinearly 
(MHD nonlinearities are neglected) but the resonant population of energetic
particles is treated nonlinearly, expansion of the latter equation 
to the first order classically leads to an energy-like relation of the form
\cite{BerkBreizman_90}
\begin{equation}
(\partial_\omega \delta W_{L}) \ \delta \omega
= -\delta W_{k,NL} -i\gamma_d \partial_\omega \delta W_L,
\label{eq_NLdW}
\end{equation}
where the subscripts $L$ and $NL$ respectively stand for {\it linear} 
and {\it nonlinear}. 
$\delta W_{k,NL}$ represents the energetic particle kinetic response treated 
nonlinearly, and $\gamma_d$ accounts for first order background damping
mechanisms, simply modelled with a constant rate in most studies of the 
nonlinear saturation of fast particle driven modes via nonlinear trapping.

Following the nonlinear trapping theory, Boltzmann equation 
Eq.~\ref{eq_Boltzmann} for the energetic particle distribution $F_h$ can be put 
in the form 
\cite{BerkBreizman_99}
\begin{equation}
\frac{d\tilde{F}_h}{dt} - {\rm C}\cdot\tilde{F}_h
= - \frac{\partial F_{h\eq}(\omega_0+\re(\delta\omega))}{\partial t}  
+ \left(\mathcal{Q}+ C \cdot F_h\right)
\end{equation}
where $\tilde{F}_h = F_h-F_{h\eq}(\omega_0+\re(\delta \omega))$, and 
$F_{h\eq}$ stands for the value of the equilibrium distribution function
where resonance occurs with $\omega_0+\delta\re\omega$ (a nonlinear shift
of the frequency is allowed).

\begin{figure}[ht!]
\begin{center}
\begin{minipage}{\linewidth}
\begin{center}
\includegraphics[width=0.9\linewidth]{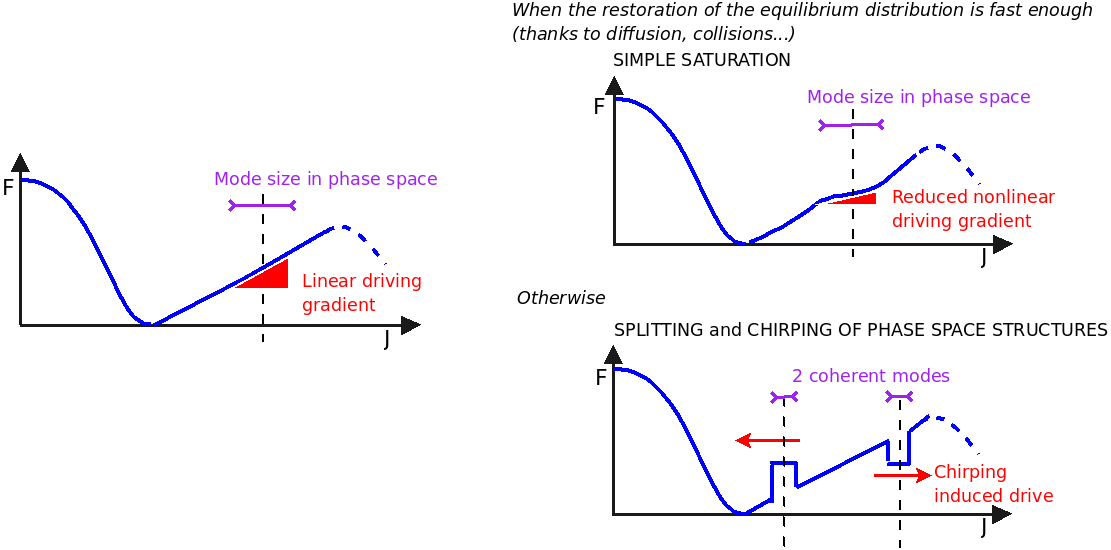}
\caption{\label{fig_BOTSaturation}
\footnotesize Two saturation regimes predicted for gap modes close
to marginal stability: simple saturation and frequency chirping.}
\end{center}
\end{minipage}
\end{center}
\end{figure}

For a mode to be driven, the $\gamma_d$ related damping needs to be 
overcome by the energetic particle drive.
Nonlinearly, particle trapping occurs and may be understood as the 
mixing of trajectories of neighboring particles. It results in
a flattening of the gradients at the origin of the drive and eventually
cancels the drive (see Fig.~\ref{fig_BOTSaturation}).
Nethertheless, saturation of the mode amplitude can take place if some 
residual drive is conserved  to balance $\gamma_d$
(and allow  for $\im(\delta\omega) = 0$ in Eq.~\ref{eq_NLdW}).
Two mechanisms have been identified to allow for a residual  drive. 
When {\bf collisions} ($C\cdot\tilde{F}_h$ in Eq.~\ref{eq_NLdW}) 
are strong enough to reduce 
trapping, some driving slope is conserved. Otherwise, phase space
structures resulting from trapping can move, ({\bf or chirp}), in 
phase space ($\partial_tF_{h\eq}$ in Eq.~\ref{eq_NLdW}). This chirping is 
to be associated to a mode frequency chirping and delivers some energy 
to the mode. 
These two saturation mechanisms are illustrated in 
Fig.~\ref{fig_BOTSaturation}, and they have been found to account for 
several experiments \cite{Berk_05,Pinches_06}.

\ssssection{Energetic particle transport}
When a particle is nonlinearly trapped, breaking of one or several
of its initial motion invariants, can lead to the situation pictured 
below Tab.~\ref{tab_ThermalHotCharacteristics}, where a barely passing 
particle is scattered into a trapped particle 
(in the usual sense of section \ref{tab_ThermalHotCharacteristics}) 
and is no longer confined \cite{Pinches_06, Hsu_92}.
If the phase space structure, where homogeneization occurs,
extends in the radial direction, a particle can get 
{\bf convected} radially.
However, saturation due to nonlinear trapping has been shown to occur at 
low levels (meaning that the structure remains small)
, 
and a strong radial transport can only occur if several modes are 
close by and their structures overlap
.
For example, an overlaping of the TAE and fishbone structures
has been postulated in JET to explain the observation of energetic 
ion losses \cite{Sharapov_00}. 
However, according to simulations \cite{Sigmar_92,Candy_97},
a strong transport may only occur when a lot of modes are involved
such that {\bf stochastization} and hence diffusion
\cite{Berk_95} take place.

\subsubsection{Strongly nonlinear kinetic evolution of Energetic Particle Modes}
When dealing with EPMs, the perturbative treatment of energetic 
particles presented in the previous  section is no longer possible, 
and the mode characteristics (its structure and frequency) are 
strongly dependent on the energetic particle population itself. 
The latter feature prevents the use of a time scale separation between
the mode and the fast particle evolutions, and leads to the 
resolution of time dependent equations.
A possible form for the mode evolution has been suggested in 
Ref.~\cite{Zonca_05}
\begin{equation}
\partial_\omega\delta W_L \ \partial_t \mathcal{A} =  
-\delta W_{k,NL}(t)\mathcal{A}-\Delta\mathcal{A} 
\label{eq_NLdWt}
\end{equation}
where $\mathcal{A}(r,t)$ stands for the mode amplitude, 
$\Delta$ is a dispersive operator. 
Analogy with Eq.~\ref{eq_NLdW} is clear, but now the amplitude is
made explicit in order to treat its time dependence on the same 
footing as the time evolution of $\delta W_{k,NL}(t)$. An additional
feature of this model  is that $\mathcal{A}$ is now given a radial
dependence, which allows for a radial dispersion of the mode structure.
Because the traditional variational formulation assumes that integration
over has already been carried out, the $r$ dependence should be 
understood as a larger radial scale that the one corresponding to the 
mode size.

Since the time evolution of  $\delta W_{k,NL}$ is now needed, a full evolution
equation needs to be solved to get the energetic particle response, in
agreement with Eq.~\ref{eq_NLdW}.
Because the time evolution of the mode is supposed to be as fast as the 
energetic particle evolution, nonlinear trapping does not have time to take 
place, such that the particle trajectories are {\bf decorrelated} from the 
wave and the fast particle distribution can be treated with a formalism 
which has some analogy with a {\it quasi-linear} treatment.
This calculation was carried out in \cite{Zonca_07b} and returns a time 
evolution equation of the energetic particle population, which has the form
\begin{equation}
\partial_t\left[\partial_t
\left(\hat{\delta W}_{k,NL}\right)
\mathcal{A}^2\right]
\propto \mathcal{A}^4 \\ 
\text{with } \hat{\delta W}_{k,NL}\propto\frac{1}{\mathcal{A}^2}\delta W_{k,NL}
\label{eq_Balistic}
\end{equation}
From Eq.~\ref{eq_Balistic}, it appears that the typical nonlinear time 
verifies $ \tau_{NL} \propto \mathcal{A}^{-1}$. 
The consequence is that nonlinear transport can be {\it balistic} 
(that is, faster than the traditional diffusive time, 
$\tau_{NL}\propto\mathcal{A}^{-2}$).\\

This analysis predicts a transition to a stronger (balistic)
transport of energetic particles. In particular, it makes possible
a strong and fast particle tranport corresponding to a simultaneous 
radial shift of 
both the energetic population and the mode structure, that may be seen as 
the radial propagation of  an unstable front. 
This prediction is in agreement with the strong transport observed during
the so-called ALEs (Abrupt Large amplitude Events) in the tokamak JT-60U.  


\section[Experimental and modelling tools in Tore-Supra]
{Experimental and modelling tools for the study of 
energetic particle  modes in Tore-Supra}
Let us take a (short!) break from theory to present the experimental
device,  which was made use of during this work to proofcheck the 
theoretical predictions.

\subsection{Tore-Supra}
\label{ssection_ToreSupra}
\parpic{
\includegraphics[width=0.4\linewidth]{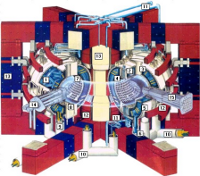}
}
The tokamak Tore-Supra is a relatively large tokamak
with a circular cross section,  whose 
dimensions are described in Tab.~\ref{tab_ITER}.
Its equilibrium configuration is well approximated by the 
embedded set  of circular flux tubes with a shift, given in 
Eq.~\ref{eq_GradShafranovshift}.\\

Tore-Supra discharges make use of deuterium  only. 
The most basic tunable parameters are the toroidal 
magnetic field, the toroidal induction current and the plasma 
average density. 

In standard Tore-Supra discharges, 
{\bf the field} $B_\text{T}$ belongs to 
$[2.2 {\rm T}, 3.8 {\rm T}]$ and can be tuned with precision 
($\pm 10^{-3}{\rm T}$), but standard operation points 
$B_\text{T} = 2.2, 2.8, 3.2, 3.8T$ are used 
(in particular for the tuning of the external heating sources).

{\bf The toroidal plasma current} at the origin of the 
poloidal field, can be created using two mechasnisms. An {\bf 
induction current} results from a varying vertical magnetic 
flux generated by the central solenoid 
(see Fig.~\ref{fig_Tokamak}). Induction is the easiest
and most tunable mean to generate current, and it is in particular 
necessary to start  the plasma. Standard Tore-Supra discharges can
be tuned with a induction current between  0.6MA and 1MA.
An additional current resulting from the injection
of   {\bf High Frequency waves}  in the plasma can be added to 
sustain the initial plasma current. 
This current source is usually more localized than induction, and 
allows for a control of the current radial profile.

Finally, the central density can be adjusted and belongs to the 
interval $ [4\times10^{19} {\rm m}^{-3}, 7\times 10^{19} {\rm m}^{-3}]$ in
standard discharges.\\

A direct {\bf heating} of the plasma occurs due to Joule's law and 
following the existence of a plasma current. 
Besides, injection of high frequency waves into the plasma
can also improve the heating:
\begin{itemize}
\item {\bf Ion Cyclotron Radio-frequency Heating (ICRH)} makes use
of the ion cycloton resonance $\Omega_c$ to transfer energy to the 
plasma ions.
However, requirements for the coupling of the waves with the plasma 
make easier the radio-frequency heating of an intermediate ion  
species, which deposits its energy on the plasma  in 
a second step. In Tore-Supra, this intermediate species or 
{\bf minority species} (representing $2$ to $9$\% of the plasma ions) 
is  usually taken to be hydrogen ions.
Under ICRH heating, a tail of energetic hydrogen ions is formed
and determining how they can transfer their energy to the main plasma
is similar to the problem of transfering the alpha particles energy.
Typical energy coupled to the plasma (deuterium + hydrogen) can 
reach 8MW in Tore-Supra.
\item {\bf Electron Cyclotron Radio-frequency Heating (ECRH) }
makes use of the electron cyclotron resonance, and allows for a very
localized and tunable heating of the plasma. In Tore-Supra, less
that 1MW of ECRH can be coupled to the plasma.
\item {\bf Lower Hybrid (LH) waves} make use of the Landau resonance
to transfer energy to the plasma. Their main use is to induce
the plasma current, but they can also heat the plasma.
In Tore-Supra, a maximum of 3 MW of LH can be reached at the moment.
\end{itemize}

\subsection{Diagnostics}
\label{ssection_Diagnostics}
\subsubsection{Overview}
Various diagnostics are set up in Tore-Supra, which apply either
for the measure of equilibrium parameters or for the measure of 
fluctuations.\\

Various equilibrium parameters can be measured. We give in the following 
a few examples of such diagnostics which are relevant to our topic.
The density profile is accessed using Thompson scattering, interferometry 
or microwave reflectometry, with a typical error of $10 \%$. 
The electron temperature 
profile is obtained using Electron Cyclotron Emission measurements, 
with an uncertainty below 5\%.
The measure of the minority fraction can be performed thanks to an 
analysis of the edge neutrals, but it is characterized by a strong error 
of about $50\%$.
Finally, neutral Beam Injection  allows for the measure of the ion 
temperature at one point of the plasma, but the diagnostic could not 
be used during our experiments.\\

Diagnostics with a high frequency acquisition (until 1MHz), 
 can be used for the analysis of fluctuating MHD modes 
(typically, BAEs have a frequency of about 50 kHz, TAEs of about
200 kHz). 

Tore-Supra core {\bf microwave reflectometer} measures density 
fluctuations with eigenfrequencies below 300 kHz 
(its aquisition frequency is 1MHz) 
and with scale lengths longer than 1cm.
The main advantages of this diagnostic are its high sensitivity
(Relative density fluctuations, normalized to the main plasma density 
can be found as low as $\delta n/n\sim 10^{-4}$.), and the possibility
to change the target radial location in one given discharge.
Disadvantages come from some uncertainty on the targetted radial 
location and a relatively important noise.
In the following, we provide some more details about this diagnostic,
which was the main tool of the analysis performed in the thesis.

{\bf Electron Cyclotron Emission (ECE)} makes possible the measure 
of electron temperature fluctuations, with scale lengths longer than 
1cm, and with a good precision on the targetted radial location 
($<$ 1cm). The current set-up in Tore-Supra allows  two kinds of 
measurements.
{\bf Fast ECE}  allows to catch fluctuations below 30 kHz,
with a fast radial scan of the tokamak core. {\bf ECE correlation}
is used for higher frequencies (below 300 kHz), but simply allows to 
target one plasma radial location.

{\bf Soft X-Rays diagnostics} measure the plasma emissivity which
is a complex function of electron temperature, and of the densities
of electrons and various ion species. Fluctuations below 100 kHz can be 
accessed. The main advantage of this diagnostic is that it is charaterized 
by several lines of sights (coming from the top and from the side of the 
machine) which scan a whole plasma cross-section simultaneously. 
A disadvantage is its lower sensitivity.

A main disadvantage of Tore-Supra diagnostics for the study of MHD 
fluctuations is the low performance of the Mirnov coils measurements 
(which do not detect fluctuations above a few KHz) due to protective 
tiles which surrounds the plasma chamber, and cut the magnetic signal. 
A detrimental consequence for MHD studies is the  impossibility
to access the toroidal numbers of the mode with frequencies above a 
few KHz.

\subsubsection{Microwave reflectometry}
A microwave reflectometer makes use of the waves propagation properties 
in a plasma, and in particular of the notion of {\bf cut-off layer}.
The latter properties can be described using a so-called  
{\bf wave index} $N$, 
which depends on the plasma (and in particular on its density) 
and on the wave characteristics. 
A cut-off layer is a layer where $N^2=0$, and it can 
be shown that a wave propagating in a region where $N^2>0$ which 
encounters a cut-off layer, gets reflected.

The idea of reflectometry is to inject waves in a plasma at well chosen
frequencies which allow for them to be reflected, and then to measure the 
phase shift $ \Phi_\mathcal{R}$ between the injected and the reflected wave.
If the injected wave has a frequency $\omega_\mathcal{R}$, it directly 
comes that
\begin{equation}
\Phi_\mathcal{R} 
= 2 \frac{\omega_\mathcal{R}}{c}\int^{R_c}_{R_\mathcal{R}} N(R,\omega_\mathcal{R})dR
-\frac{\pi}{2},
\label{eq_PhaseDifference}
\end{equation}
where $[R_\mathcal{R},R_c]$ is simply  the distance between the reflectometer 
($\mathcal{R}$) and the cut-off layer.\\

A first use of the reflectometer is the determination of the equilibrium 
density profile. From the knowledge of $N(R,\omega_\mathcal{R})$ in 
Eq.~\ref{eq_PhaseDifference}, one can deduce the distance 
$[R_\mathcal{R},R_c]$
and link it to the cut-off density. When modifying $\omega_\mathcal{R}$,
the full equilibrium density profile can be reconstructed.

At fixed $\omega_\mathcal{R}$, that is for one targetted cut-off layer,
the analysis of high frequency fluctuations can be linked to density 
fluctuations using  Eq.~\ref{eq_PhaseDifference}.
Assuming that the mode scale length is greater than the free space 
wavelength of the injected wave (typically $>1$cm), the cut-off layer 
is strongly enhanced by plasma fluctuations near this layer. 
Thus, in traditional configurations, 
the difference in phase shift (compared to the phase shift associated to 
the equilibrium)
can be related to density fluctuations using
\cite{Mazzucato_98}
\begin{equation}
\delta \Phi_\mathcal{R} = M_\mathcal{R}\sqrt{\pi}\ \frac{\omega_\mathcal{R}}{c}
\left(\frac{1}{k_r\nabla N^2}\right)^{1/2}
\frac{\delta n_e}{n_{e\eq}}
\end{equation}
where $M_\mathcal{R}=n_e\partial N^2/\partial n_e$, and $\nabla N^2$ depend 
on the injected wave and  need be assessed at the cut-off. 
$n_e$ is the electron density, and $k_r$ stands for the mode radial mode 
number (the calculation of Ref.~\cite{Mazzucato_98} is done for a plane
wave $\propto e^{ik_rx}$ ). Once the density fluctuations of a a mode has
been extracted, $\delta n_e$, the MHD displacement can be obtained from 
$\delta n_e$ using the density transport equation. In its simplest form, 
it returns
\begin{equation}
\xi =\delta n_e/\ \frac{dn_e}{dr}.
\label{eq_ReflectoMHDDisplacement}
\end{equation}\\

The reflectometer used in this thesis sends waves in the tokamak 
equatorial plane (the horizontal plane cutting the torus chamber in two).
It is configured to work with the higher cut-off of the so-called 
{\bf X-mode}. 
This means that the injected wave  refraction index verifies
\begin{equation}
N_X^2 = 1 - \frac{\omega^2_{pe}
\left(1-\omega^2_{pe}/\omega_\mathcal{R}^2\right)}
{\omega_\mathcal{R}^2-\omega^2_{pe}-\Omega^2_{ce}},
\end{equation}
and the injected wave frequencies are chosen to match the higher cut-off,
of equation
\begin{equation}
\omega_\mathcal{R} = \frac{1}{2}
\left(\sqrt{\Omega^2_{ce}+ 4\omega_{pe}^2}+\Omega_{ce}\right).
\end{equation}
where $\Omega_{ce}$ is the electron cyclotron frequency and radially 
depends on the fiels, 
$\omega_{pe}$ is the electron plasma frequency and depends on the 
density, $\omega^2_{pe} = n_ee^2/(m_e\epsilon_0)$ ($m_e$ the electron 
mass, $\epsilon_0$ the dielectric permittivity). 
In particular $M_\mathcal{R}\approx 2$.

\begin{figure}[ht!]
\begin{center}
\begin{minipage}{\linewidth}
\begin{center}
\includegraphics[width=0.65\linewidth]{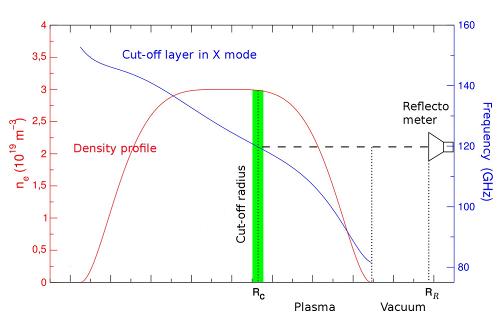}
\caption{\label{fig_CutOff}
\footnotesize
Schematic of the reflectometry set-up, and cut-off layer in X-mode.}
\end{center}
\end{minipage}
\end{center}
\end{figure}

The cross-section of the cut-off is represented in 
Fig.~\ref{fig_CutOff}, 
where it appears that the plasma can be scaned from the low field side
(the outward part of the torus) to a third of the high field side, when
the reflectometer frequency, $\omega_\mathcal{R}$ , 
is moved from 100 GHz to 140 GHz. In fact the radial extent which can be 
scaned by reflectometry depends on $B_{\eq \text{T}}$
(\ $B_{\eq \text{T}}=3.8$ T in Fig.~\ref{fig_CutOff}) and 
does not allow to catch the core region for low fields. 
For these reasons, core reflectometry requires at relatively large 
field $B_{\eq\text{T}} > 2.8 $T.

For the study  of fluctuations, we already mentionned an additional
limitation of the reflectometry set-up to fluctuations of 
eigenfrequencies below 300 KHz 
(due to the reflectometry acquisition frequency, 1MHz). 
A final limitation is due to storage capacities.
Two choices can be made:
\begin{itemize}
\item In {\it mode 1}, $\omega_\mathcal{R}$ can be increased by steps, 
such that a radial scan of the  plasma can be done.
In this case, steps cannot be longer 15 ms.
\item In {\it mode 2}, a single radial point is targetted (more 
precisely, 3 points a later developments of the diagnostic), but a 
longer acquisition time is possible (until 0.5s).
\end{itemize}


\subsection{CRONOS, PION}
\label{ssection_CRONOSPION}
For the analysis of experiments, two codes have been used in this 
work.

{\bf CRONOS} is a transport code which integrates various modules aiming at 
including MHD equilibrium requirements, the effects of multiple species in 
the plasma, and various heating sources \cite{Basiuk_03}.
It can be used either for the interpretation of experiments, or in a 
predictive way. In the interpretative mode, it  reconstructs the plasma 
equibrium  (and returns for example such quantities as the main ion 
or the electron temperature profiles $T_i$, $T_e$ or the $q$ profile), 
based on the experiments characteristics and on various 
diagnostic measurements. 

To obtain a precise effect of ICRH, and in particular to obtain the 
ICRH fast ion distribution function, we made use of a Monte-Carlo code 
called {\bf PION} \cite{Eriksson_93}, which takes as input various 
equilibrium parameters and returns the deposited ICRH power deposited
on each species and the resulting deuterium and hydrogen distribution 
functions.

\subsection{Experiments design}
\label{ssection_ExpDesign}
Any experiment involves a huge number of free parameters.
We gave a sample of these parameters in 
section~\ref{ssection_ToreSupra}.
One may be aware of some limitations on the range of parameters that
can be reached.

In the study of fast particle driven modes, it is interesting
to decouple equilibrium parameters (such as the field, the main plasma
temperatures...), from the fast particle population. 
For that purpose, varying the source of suprathermal particles 
is of interest, but of course, one may check it is done independently 
from other parameters.
In the experiments conducted in this work, the fast particle 
population is created using ICRH. As will be explained below, ICRH power
coupling is dependent on equilibrium parameters as well.\\

We listed in the following limitations on the experiments parameters.
\begin{itemize}
\item {\bf General limitations}
  
  Besides traditional technical limitations, 
  such as the maximum current in the toroidal coils,
  or the maximum flux (which limits the discharge time lengths),  
  the sustainability of a plasma depends on MHD stability, 
  as we know.
  One strong limit on the density is the Greenwald density \cite{Troyon_84},
  $\beta_N \equiv (\beta/100)*(aB_0)/I_p(\text{MA}) \leq 3 \%$.
\item {\bf Power limitations}

  Power injection of heating waves in the plasma is of course limited
  by the capacitites of the antennas generating the waves.
  Besides, the resonant absorption of ion cyclotron or LH waves depends on 
  plasma parameters. In particular, ICRH requires a relatively high
  density, and a window of minority fraction. This limits the possiblity to
  decouple the choice of these parameters.
  Finally, the expulsion of fast ions  explained in the thesis introduction
  is particularly risky for the antennas, and has been found to induce
  detrimental localized heat loads.
\item {\bf Diagnostic requirements}\\ 
  Finally, the parameter extent is limited by the diagnostic requirement.
  We already explained that the use of core reflectometry is only possible
  for relatively large toroidal field, $B_0>3.2$ T.
\end{itemize}
\section{Summary}
{
\it
In this chapter, we presented some general physics of relevance for the 
understanding of the motion of energetic particles, and we reviewed important
physics involved in the resonant excitation of waves with a finite frequency
in tokamak, by energetic particles. Finally, we introduced the experimental
device, Tore-Supra, where we conducted experiments to study the latter physics.

Important ideas can be retained:
\begin{itemize}
\item In the absence of perturbations, the motion of a particle in a tokamak
  can be divided into {\bf three periodic motions}, in general characterized 
  by a time-scale separation, and appropriately described in a set of
  action-angle angles, which does not exactly match the natural geometric
  tokamak coordinates.
 
  In the presence of a wave, particles whose periodic motion match its 
  oscillation frequency are {\bf resonant and can exchange energy 
  with the wave. This allows energetic particles to drive macro- to 
  meso- scale modes}.

\item 
  The natural formalism for the study of such  macro- to meso- scale modes in a 
  plasma is the MHD formalism and ordering, but MHD has some limitations.
  In particular, small scales cannot be reached by MHD, neither resonant
  phenomena.

\item In a general sheared geometry, modes ocillating with a finite frequency
  are damped via phase-mixing (a process called  {\bf continuum damping}) and 
  hence stabilized. However, depending on geometry, some modes, the {\bf 
    gap modes}, can avoid the latter damping. 
  As a consequence, they can be very unstable and grow, with only a small 
  amount of  resonant drive by the energetic particles. 
  For this reason, gap modes are the most dangerous modes, although other 
  {\bf energetic particle modes}
  can also be driven in the presence of a strong amount of energetic particle 
  drive.

\item Nonlinear descriptions are necessary to determine the effects of these
  instabilities on the plasma and in particular of the transport of energetic 
  particles. Depending on the regime (the involved time scales and the amount
  of drive), {\bf different regimes of saturation and of particle transport 
    can be predicted}.

\item Experiments can be conducted in the Tore-Supra tokamak using various
  heating devices to create a population of fast particles. Diagnostics 
  exist to catch some features of the modes that develop.
\end{itemize}
}
\begin{savequote}[20pc]
\sffamily
Equation (1.2-9) is a second order, nonlinear, vector, 
differential equation which has defied solution in its 
present form. It is here therefore we depart from the 
realities of nature to make some simplifying assumptions...    
\qauthor{Bate, Mueller, White, 1971, {\it Fundamentals of Astrodynamics}}
\end{savequote}




\chapter[From linear gyrokinetic theory to linear Magneto-Hydrodynamics]
{From the Gyrokinetic to the Magneto- HydroDynamic linear 
description of instabilities}
\label{chapter_GyrokineticToMHD}

We now describe the formalism and plasma model used in this thesis 
for the study of collective instabilities driven by energetic particles.

The theoretical review of the previous chapter indicated that MHD provides
a good understanding of the main physics at stake in the interaction 
between macro-scale modes  and energetic particles. 
However, we also indicated some limits of MHD in section 
\ref{ssection_AlfvenSpectrum}.

For the study of Beta Alfv\'en Eigenmodes, of interest in this 
thesis, kinetic effects cannot be overlooked. The main reason 
is the expected important role of resonances with the thermal 
plasma, which we mentionned in the thesis introduction \cite{Zonca_96}.
Besides, experiments display acoustic fluctuation frequencies 
very close to the MHD Alfv\'en continous spectrum, previously 
defined. This suggests that conversion to kinetic modes and 
small radial scales cannot be fully neglected.\\

Consequently, {\bf a kinetic plasma model} is used in this thesis, 
for both the thermal plasma and  the energetic population.
The electric and magnetic field evolution equations, are taken into 
account using an equivalent {\bf variational formulation} of Maxwell 
equations, in the same fashion as in Ref.~\cite{Edery_92}. 
Compared to Ref.~\cite{Edery_92}, we bring some clarification of the 
coordinate systems used for the kinetic modelling, and the formulation 
is put in a form which can be directly related to the Shear 
Alfv\'en law, 
Eq.~\ref{eq_AdaptedShearAlfvenLaw}.

An advantage of a variational formulation is to provide an 
energy principle which 
directly extends the MHD kinetic energy principle to kinetic 
theory. When used as a dispersion relation, an energylike 
relation such as the MHD energy equation \ref{eq_dWMHD} provides 
a nice simple starting point to analyze the stability of a mode.
For example, such simple conclusions on the stabilizing effect of 
various physical phenomena drawn ealier 
(for example the stabilizing effect of bended field line,
via magnetic tension), can be generalized.
However, one should note that moving to kinetic theory greatly 
modifies the form of the ideal MHD energy equation.
In MHD, it takes the form
$(\omega^2/2)\int d^3{\bf x}\ \rho_{\text{\sc m}}|\xi_\omega|^2 
= (1/2)\int d^3{\bf x}\xi_\omega^*\mathcal{F}(\xi_\omega)
= \delta W_\text{MHD}$, with $\mathcal{F}$ a hermitian operator, and 
$\delta W_\text{MHD} (\xi_\omega, \xi_\omega^*)$ a functional which
does not explicitely  depend on $\omega$. 
With kinetic effects included, the
$\omega$ independence of the right hand side disappears in general.
Worse, the dispersion relation $\mathcal{F}$ is no longer hermitian
and the dispersion relation becomes fully complex.

A more technical advantage of the use of an integrated form of 
the Maxwell equations, is the possibility to switch easily from 
one coordinate system to another by a simple change of variables.
In particular, a rigorous derivation of resonances with the  
particles eigenfrequencies (which is best understood in the 
action-angle variables) can be combined with a more geometric 
picture provided by the usual non-canonical guiding-center 
coordinates (which separates velocity and space coordinates).\\

We start this presentation by a general description of the 
variational principle, 
with an attempt to provide some insight in the physical meaning 
involved when dealing with stability issues. Next, we explain
the plasma kinetic model used in this work, the so-called  
{\bf gyrokinetic model}, expanded here in the two coordinate system
of interest to us, the action-angle variables 
and the non-canonical guiding-center variables introduced in the 
previous chapter.
Finally, we combine the two models to display a gyrokinetic energy 
relation, and show how the MHD limit, 
leading to the traditional MHD energy equation, can be recovered.

\section{Variational dispersion relation and instabilities}

\subsection{Variational formalism and energylike relation}
\label{ssection_EnergylikeRelation}
For the   instabilities of interest 
in this thesis, the electromagnetic fields can be described using 
the {\bf low frequency Maxwell equations}
\begin{eqnarray}
\nabla \cdot  {\bf E} &=& -\frac{\rho}{\epsilon_0}
\label{eq_Poisson} \\
\nabla \times {\bf B} &=& \mu_0{\bf j}
\label{eq_Ampere} \\
\nabla \cdot  {\bf B}      &=& 0\\
\nabla \times {\bf E} &=& -\frac{\partial {\bf B}}{\partial t}
\end{eqnarray}
where the  displacement current $\mu_0\epsilon_0\partial_t{\bf E}$
is neglected.
The charged particles effects come into 
$\rho   ({\bf x}, t) = \sum_s \rho_s ({\bf x}, t)$ 
and   ${\bf j}({\bf x}, t) = \sum_s {\bf j}_s ({\bf x}, t)$  
which are the total charge density and total current density, 
summed over all the species $s$.
From  now on, the effects or properties of each species will be 
specified whenever there may be an ambiguity, with the convention that 
$s$ refers to any species, $s=i$ when the focus is on the thermal main
ion, $s=e$ when it is on electrons and $s=h$ for the hot species.

Equivalently, it is possible to use the electric and magnetic 
vector potentials
\begin{eqnarray}
{\bf E} &=& -\nabla \phi - \frac{\partial  {\bf A}}{\partial  t}\\
{\bf B} &=& \nabla \times {\bf A}
\end{eqnarray}
and  a variational principle to solve Poisson and Amp\`ere 
Eqs.~\ref{eq_Poisson}, \ref{eq_Ampere}.
Under the assumption that the studied domain is surrounded by an 
ideal conductor (such that there is no surface term in 
Eq.~\ref{eq_FullElectromagLagrangian})
and similarly as for the hamiltonian formulation  of the particles 
motion, it is easy to see that 
Poisson and Amp\`ere equations are equivalent to the 
extremalization of the electromagnetic action
$\int dt \  \mathcal{L}$ under  variations of the potentials 
$\mathbf{A}$  and $\phi$ (for fixed $\rho$ and $\bf j$), 
where $\mathcal{L}$ is the electromagnetic Lagrangian defined by
\begin{equation}
  \mathcal{L}(\mathbf{A}, \phi)=\int  d^{3}\textbf{x}
  \left( \frac{\epsilon_0\mathbf{E}^2}{2}
          - \frac{\mathbf{B}^2}{2\mu_0}\right)
     + \int  d^{3}\textbf{x}
  \left(\mathbf{j}\cdot\mathbf{A}-\rho \phi \right).
  \label{eq_FullElectromagLagrangian}
\end{equation}
$\mathcal{L}$ clearly contains information on the fields
energies $\propto {\bf E}^2$ and $ {\bf B}^2$, whereas the second
term of Eq.~\ref{eq_FullElectromagLagrangian} can be interpreted as
the interaction of the fields with the particles.
If the term $\propto \bf E^2$ representing  the electric field energy
is neglected,  extremalization by $\phi$ returns $\rho = 0$, that 
is {\bf electroneutrality}. 
Since only collective perturbations are of interest to us
(that is with wavelengths longer than the Debye length), 
the assumption of electroneutrality is made in the following.
\\

For the study of electroneutral coherent perturbations, ie. for 
perturbed quantities of the  form 
$X=X_{\omega}e^{-i\omega t}+c.c.$, a simpler 
variational form is the extremalization of the reduced 
Lagrangian
\cite{Edery_92}
\begin{eqnarray}
  \mathcal{L_{\omega}}=-\int d^{3}\textbf{x}
  \frac{\mathbf{B}_{\omega}^{\dagger} \mathbf{B}_{\omega} }{\mu_0}
     + \sum_{s}  \mathcal{L}_{s\omega}
\text{\quad with \ } \mathcal{L}_{s\omega}=\int d^{3}\textbf{x}
\left({\bf j}_{s\omega}\cdot\mathbf{A}_{\omega}^{\dagger}-\rho_{s\omega}
\phi_{\omega}^{\dagger} \right)
\label{eq_OmElectromagLagrangian}
\end{eqnarray}
under variations of the virtual potentials $\phi_\omega^{\dagger}$ and 
${\bf A}_\omega^{\dagger}$, where $\mathcal{L}_{s\omega}$ represents again 
the interaction between the fields and the species $s$, 
and the virtual fields are chosen to cancel at the plasma boundary.
Taking the $\omega$ component of Eqs.~\ref{eq_Poisson} 
and \ref{eq_Ampere}, and multiplying them 
respectively by $\phi_\omega^{\dagger}$ and ${\bf A}^{\dagger}_\omega$,
it comes that $\mathcal{L}_\omega ({\bf A}_\omega, \phi_\omega,
{\bf A}^\dagger_\omega, \phi^{\dagger}_\omega)=0$ 
for any virtual fields $\phi_\omega^{\dagger}$, ${\bf A}^{\dagger}_\omega$, 
once the physical fields have been found.
Inversely,  when Eq.~\ref{eq_OmElectromagLagrangian} is extremalized 
by these virtual fields, the $\omega$ component of Poisson and Ampere
equations are directly recovered.

This variational principle is simpler than the one corresponding to 
Eq.~\ref{eq_FullElectromagLagrangian} because the real physical
fields are made completely independent from the virtual fields.
In particular, using a self-consistent model for the particle fields, 
that is
${\rho}_{s\omega}(\phi_\omega, {\bf A}_\omega)$,
${\bf j}_{s\omega}(\phi_\omega, {\bf A}_\omega)$ 
is not in contradiction with
the variational principle, whereas the previous procedure (where 
extremalization is with regards to $\phi$ and $\bf A$) 
requires the use a more {\bf complete electromagnetic Lagrangian} 
including the particle energy. 
Of course, there is a cost to this simplicity. 
First, $\mathcal{L}_\omega$ is linear in $\phi_\omega^\dagger$, 
${\bf A}^{\dagger}_\omega$, hence the idea of extremalization does not 
provide strong physical insight (it would have been different if
extremalization would have correspond to minimization for example).
Secondly, the physical interpretation of $\mathcal{L}_\omega$ is more 
ambiguous than the complete real electromagnetic Lagrangian, in
particular when resonance come into play \cite{Berk_88}.
Nevertheless an {\bf energylike relation} can still be 
displayed.\\

Since $\mathcal{L}_\omega ({\bf A}_\omega, \phi_\omega,
{\bf A}^\dagger_\omega, \phi^{\dagger}_\omega)=0$  is verified for any virtual 
field once the physical fields ${\bf A}_\omega$, $\phi_\omega$ have been 
determined, it is in particular true for $\phi_\omega^*$ and 
${\bf A}_\omega^{*}$, if the latter cancel at the plasma boundary.
Assuming an ideal surrounding conductor however, the latter verification
is not necessary, and it directly comes from the integration of Maxwell
equation that $\mathcal{L}_\omega ({\bf A}_\omega, \phi_\omega,
{\bf A}^*_\omega, \phi^{*}_\omega)=0$. Hence, we obtain
an {\bf energylike relation} similar to the one provided by MHD 
(built with the use of $\xi_\omega^*$).
It is particularly interesting to recover such familiar quantities
as the perturbed magnetic field energy 
$\int d^3 {\bf x}|{\bf B}_\omega|^2/\mu_0$, which are closely related to 
the physical quantities.

However, the {\it linear} MHD energy principle which allows
an easy determination of stability based on the sign of the different  
terms involved in the energylike relation cannot be extended in general.
We explained that the linearized MHD equations could be put in the  form 
$\omega^2\rho_\text{\sc m\eq}\ \m{\xi}_\omega
= \mathcal{F}(\m{\xi_\omega})$, with $\mathcal{F}$ a hermitian operator.
When trying the same with the fields of interest in here, 
that is using a self-consistent linear model for the particle fields, 
$\rho_{s\omega}(\phi_\omega, {\bf A}_\omega)$,
${\bf j}_{s\omega}(\phi_\omega, {\bf A}_\omega)$, 
the electromagnetic fields equations can be 
put in the form $\mathcal{F}(\omega,\phi_\omega, {\bf A}_\omega)=0$, 
but there is not garanty that $ \mathcal{F}$ be hermitian, and  the 
energylike relation and corresponding eigenvalues and 
eigenfunctions are in general fully complex.
\footnote{Note that the use of the so called adjoint 
fields \cite{Berk_88}, that is of the solution of the 
adjoint operator $\mathcal{F}^{\dagger}$, instead of the conjugate 
fields provides another energy-like relation, from which one may 
derive some properties of the eigenvalues $\omega$. 
When $\mathcal{F}$ is hermitian, these adjoint fields are the same 
as the conjugates.}

It is important to note that the variational procedure described above 
in Eq.~\ref{eq_OmElectromagLagrangian}, does not a priori assume any 
linearization, even one single $\omega$ was considered.
Plasma nonlinearities certainly imply a mixing of various $\omega$. 
However, Maxwell equations return  the electromagnetic 
fields as a linear function of the particle fields. 
Hence, for any field  $X(t)$, 
$X_\omega$ can be understood as the Fourier transform of X 
($X(t) = \int X_\omega e^{-i\omega t}d\omega$) 
and the $\omega$ component of Maxwell and Poisson equation can be 
extracted. 
Hence, the energy-like dispersion relation 
$\mathcal{L}_\omega(\phi_\omega, {\bf A}_\omega,
\phi^*_\omega, {\bf A}^*_\omega)$
can be used in the nonlinear regime.

\subsection{Physical interpretation of the energylike dispersion relation}
\label{ssection_PositiveEnergyWaveDensity}
 Let us now try to have some insight in the physical meaning of the 
energylike relation.

Focusing on the particle part of the Lagrangian, the 
following relation comes out
\begin{eqnarray}
\nonumber
  - \int d^3{\bf x}{\bf j}_{s\omega}\cdot {\bf E}^*_\omega &=& 
  \int d^3{\bf x}{\bf j}_{s\omega}\cdot 
  \left(\nabla \phi^*_\omega +i \omega {\bf A}^*_\omega\right)
= i \omega \int d^3{\bf x}
\left( {\bf j}_{s\omega}\cdot {\bf A}^*_\omega
- \rho_{s\omega}\phi_\omega^* \right) \\
&=& i\omega\mathcal{L}_{s\omega}
\label{eq_ComplexImpedance}
\end{eqnarray}
where we made use of charge conservation 
$-i\omega \rho_{s\omega} + \nabla \cdot {\bf j}_{s\omega}=0$ in the second 
equality. Thus, the particle Lagrangian $\mathcal{L}_{s\omega}$ is found 
to be directly related to the {\bf work} done by the electric field on 
the particles, from which wave-particle energy transfers follow.
More precisely, considering  the {\it real} total current and fields
to have one single $\omega$ perturbed component,
${\bf j}_s={\bf j}_{s\eq}+({\bf j}_{s\omega}e^{-i\omega t}+cc.)$,
$\rho_s=\rho_{s\eq}+(\rho_{s\omega}e^{-i\omega t}+cc.)$,
and assuming $\omega$ to be real,
the {\bf net power transfer from the particles to the wave}
(after averaging of the sinusoidal periodic variations
on one period $T = 2\pi/\omega$) is found to be of the form
\begin{equation}
-\frac{1}{T}\int_{0}^T \int d^3\mathbf{x} \mathbf{j_{s}}\cdot \mathbf{E} 
= -\int d^3\mathbf{x} {\bf j}_{s\eq}\cdot {\bf E}_{s\eq} 
- 2\omega\im(\mathcal{L}_{s\omega}).
\label{eq_JE}
\end{equation}
Since $-\int d^3\mathbf{x} {\bf j}_{s\eq}\cdot {\bf E}_{s\eq}$,
is simply the energy transfer from the particles to the equilibrium
fields (if there is any), 
the remaining term, $2\omega\im(\mathcal{L}_{s\omega})$ appears like the 
power transfer from the wave to the particles of species $s$.

What about the real part of $\mathcal{L}_{s\omega}$?
By analogy with a circuit component, the left hand side of 
Eq.~\ref{eq_ComplexImpedance} can be  interpreted as
the product of a complex current $\mathcal{I}$ and of a complex 
electric potential jump \ $\mathcal{U}$ considered in a certain plasma 
volume. Writing $\mathcal{Z}$, the impedance characterizing the plasma
response (that is $\mathcal{U}=\mathcal{Z}\mathcal{I}$), 
the left hand side of Eq.~\ref{eq_ComplexImpedance} can be
rewritten $-\mathcal{I}\mathcal{U}^* = -\mathcal{Z}|\mathcal{I}|^2$.
In circuit theory, the impedance is usually put in the form 
$\mathcal{Z} = \mathcal{Z}_R + i \mathcal{Z}_I$, where 
$\mathcal{Z}_R$ and $\mathcal{Z}_I$ are respectively the electric
component's {\it resistance} and {\it reactance}.
Taking the imaginary part of Eq.~\ref{eq_ComplexImpedance}
(still with $\omega$ real), it appears that 
$\re(\mathcal{L}_{s\omega} )$ corresponds to the reactance of the circuit
component \cite{Berk_88}. For this reason, $\re(\mathcal{L}_{s\omega})$ 
is called the species {\bf reactive response}, or {\bf reactive energy}.
It corresponds to the reactive response qualitatively described in 
section \ref{ssection_EnergeticParticleDrive}.

Finally, we find two types of responses. One response, 
the reactive response, behaves likes a reactance. Hence, it is not 
expected to modify the system energy, but energy may be ``stored'' or 
delivered. This response typically leads to exchanges between 
the equilibrium potential energy and the energy of waves.
The other response is linked to energy transfers between waves and 
particles, and contains  such effects  as resonant excitation or 
damping, or dissipative effects such as collisions. 
When generalized to the Lagrangian \ref{eq_OmElectromagLagrangian} 
(not simply to the species $s$ Lagrangian $\mathcal{L}_{s\omega}$), 
$\im(\mathcal{L}_{\omega})$ can also contain continuum damping. 
This is obvious when looking at equation \ref{eq_ContinuumDamping},
if we admit that the Lagrangian $\mathcal{L}_\omega$ is not much 
different 
$\delta W$ given earlier (The link between the two will be made clearer
in the following.).
By analogy with circuit components, $\im (\mathcal{L}_\omega)$ 
could be associated with {\it dissipation} (understood with a positive 
or negative sign).
However, as noted in Ref.~\cite{Brizard_92}, the idea of dissipation is 
confusing. Indeed, $\im(\mathcal{L}_{\omega})$
contains resonant wave-particle energy transfers, which can be described by
Vlasov equation (Eq.~\ref{eq_Boltzmann} with 
$\mathcal{C}\cdot F = 0, \mathcal{Q}=0$). 
As will be explained in the following, Vlasov equation has a hamiltonian 
formulation, and hence should conserve energy. 
Thus, the idea of dissipation may be a bit misleading, 
and it is safer to simply say that $\im(\mathcal{L}_\omega)$ contains
resonances and dissipation (note that a complicated formalism has been
developed to separate the two \cite{Brizard_94}).
Nevertheless, because resonances and dissipation have similar behaviors
(in particular, discontinuities introduced by resonances can be solved
assuming a small order dissipation), it is often easier for physical 
interpretation to separate resonances from the remaining plasma energy and
look at them as external energy sources. This makes some sense when resonances
are simply low order \cite{Berk_88}.\\

How can instabilities arise in such a system?
Because we assumed the system to be closed (no surface terms in the 
Lagrangian), dissipation, sources, energy storage, are to be associated
with the growth or damping of the plasma kinetic energy. In other words,
$\omega$ can have an imaginary part, and {\bf instabilities} are simply 
roots of the energylike dispersion relation with a positive imaginary part 
$\im(\omega)>0$. 

When dissipation or resonances are negligible in the system, the type of 
instabilities which can arise come from the real part of the Lagrangian and 
they are called {\bf reactive instabilities} \cite{LashmoreDavies_07}. 
In this case, it is demonstrated in Ref.~\cite{Hasegawa_68} that the 
existence of a reactive instability is possible only in the presence of a 
{\bf reactively active} species, that is, a population with a conductivity 
$\sigma_s$ such that  such that $\partial_{\omega_r}(\im\ \sigma_s)>0$
($\omega_r = \re{\omega}$) for some real $\omega_r$ 
\footnote{
Note that the theorem used here, Foster reactance theorem, simply applies
when there are no root on the real axis... which excludes a major candidate,
MHD.}. 
By definition, conductivity verifies $ {\bf j}_{s}=\sigma_s{\bf E}$, 
such that Eq.~\ref{eq_JE} becomes. 
$-\int d^3{\bf x}\sigma_s|{\bf E}_{\omega}|^2 
= i\omega\mathcal{L}_{s\omega}$. 
It seems reasonable to expand the definition of a reactively active species
in the following way,
\begin{equation}
\partial_{\omega_r}
\left(\omega_r\re(\mathcal{L}_{s\omega})\right) <0,
\ \text{for at least some real } \omega = \omega_r. 
\label{eq_ReactivelyActive}
\end{equation}
According to Ref.~\cite{Hasegawa_68},  because for such instabilities, no 
external source is feeding the system, energy conservation requires the 
instability to develop with a {\bf negative energy}.

The idea of negative energy  becomes clearer when dissipation or 
resonances (seen as energy sources) are added to the picture. 
Assume the energylike relation to be almost real (ie, with a lower order 
real expression $\mathcal{L}_{\omega\eq}$) with an approximate real root 
$\omega_{r0}$ ($\mathcal{L}_{\omega\eq}(\omega_{r0})=0$), and 
add a first order dissipation $\mathcal{L}_{\omega(1)}$ 
with $\im\mathcal{L}_{\omega(1)}\neq 0$. Expansion close to 
$\omega_{r0}$ ($\omega = \omega_{r0} + \delta\omega_r + i\gamma $) returns
\begin{equation}
[\omega_{r0}\partial_{\omega_r}\re\mathcal{L}_{\omega(0)}]   \gamma
= \partial_{\omega_r}(\omega_r\re{\mathcal{L}_{\omega\eq}}) \gamma
=-\omega_{r0}\im\mathcal{L}_{(1)}(\omega_{r0}).
\label{eq_JEDampsOrDrives}
\end{equation}
The term in brackets in the first term is traditionally called the 
{\bf wave energy density}, whereas the third term is easily recognized
from Eq.~\ref{eq_JE} to correspond to a {\it source} of energy.
Intuitively, one may expect an instability to grow when energy is fed
into the system. This is the case for {\it positive energy waves}
$[\omega_{r0}\partial_{\omega_r}\re\mathcal{L}_{\omega(0)}] >0 $, but 
the existence of reactively active components verifying
\ref{eq_ReactivelyActive} or of reactive instabilities 
\cite{LashmoreDavies_07}, suggests the possibility of waves with a negative
energy density... which are damped when energy is provided to the 
system, as can be seen in Eq.~\ref{eq_JEDampsOrDrives}\cite{LashmoreDavies_05}.
Note that we already faced the difficulty to determine the sign of the wave
energy in Eq.~\ref{eq_ContinuumDamping}.

\subsubsection{Application to the fishbone-like dispersion relation, and 
to the special case of gap modes with a dominant {\sf m} component}

As will be made clearer later on, it is possible to link the Lagrangian
to the more traditional $\delta W$ introduced in the previous chapter
in a simple way
\begin{equation}
\mathcal{L}_\omega = -2\delta W
\end{equation}
and hence to apply the remarks of the previous paragraph to the 
fishbone-like dispersion relation Eq.~\ref{eq_NNFishboneDispRel},
$\delta W = -\delta I +\delta W_f + \delta W_k =0$.

From the previous section, the interpretation of the fishbone-like 
dispersion relation is simpler when a well defined eigenmode 
(with a well defined stable structure and real frequency) 
can be determined for the non-resonant part of the dispersion relation, 
and when resonant excitation of the mode (or any other kind of 
drive or dissipation) can be treated perturbatively, like an external 
source.
In particular, this description is legitimate for a gap mode close 
to {\bf marginal stability}, that is, when only a small amount of 
resonant energetic particles interact with the mode.

The derivations carried out in this thesis make use of these two 
assumptions, and treat in particular the resonant excitation and damping 
as perturbative. 
Justifications or at least determination of validity domains for the 
latter assumptions will be provided. However, one should keep in mind 
that this approach breaks down if:
\begin{itemize}
\item the solution of the non-resonant dispersion relation is associated 
to a strong damping (for example for EPMs, defined in section 
\ref{ssection_AlfvenSpectrum}) 
or it is reactively growing.
In this case, marginal stability may only be reached for a sufficiently
important resonant ``source''. And there is no garanty that even close 
to marginality a well defined mode can exist (depending on the strength 
characteristic time scales characterizing the drive and the damping 
mechanisms).
\item the resonant energetic population is de facto non-perturbative. 
In this case, the eigenmode structure may be modified and it depends on
the localization of the resonant drive for example.
\end{itemize}
In both cases, the wave cannot be separated from the resonant 
population and a wave energy density cannot be defined in a simple way.\\

Let us focus more precisely on gap modes with a dominant poloidal 
component and considered with the description of subsection
\ref{ssection_AlfvenSpectrum}.
For such modes, the various terms involved in the fishbone-like
dispersion relation can be given explicitely and we can try to see how 
our interpretation of the energylike dispersion applies.

In subsection \ref{ssection_AlfvenSpectrum}, we saw that the radial 
structure of the modes could be described in Fourier space by a
Schr\"odinger-like equation Eq.~\ref{eq_RSAE},
\begin{equation}
-\frac{\partial^2\Psi}{\partial\vartheta^2} + 
\left(-\Lambda^2
+ \frac{s_0^2}{(1+s_0^2\vartheta^2)^2}
-\frac{s_0^2\mathcal{G}}{(1+s_0^2\vartheta^2)} 
\right)\Psi=0.
\label{eq_SchrodingerWell_bis}
\end{equation}
Here, we can recall that $\Lambda^2$ is related to inertia,
$\Lambda^2 = (\omega^2/v_A^2-k^{\sf m}_{\|}(r_0)^2)
\times2({\sf m}^2s^2_0/r^2_0)/(k^2_\|)''$, and needs to verify $\Lambda^2<0$
by definition of a gap mode.
The other terms can be related to the fluid and kinetic potential energy,
involved in the fishbone-like dispersion relation. 
Indeed, when multiplying Eq.~\ref{eq_SchrodingerWell_bis} by $\Psi^*$, we 
can rebuild an energylike formula, which is directly proportional to the
traditional MHD-like energy functional by Fourier transform properties
(Parceval formula).
The sign of the proportionality constant follows from the remark that
$\omega^2$ (representing inertia) is multiplied by $(k_\|^2)''$ in the 
definition of $\Lambda^2$. It comes
\begin{equation}
\delta W_f + \delta W_k 
= C \sigma_{\hspace{-0.08cm} \pm}
\left\{
\int^{+\infty}_{0} d\vartheta \left(|\partial_\vartheta\Psi|^2 
+ \frac{s_0^2|\Psi|^2}{(1+s_0^2\vartheta^2)^2}
- \frac{s_0^2\Psi^*\mathcal{G}\Psi}{1+s_0^2\vartheta^2}
\right)
\right\}.
\label{eq_dWfanddWtheta}
\end{equation}
with $C$ some positive constant, and $\sigma_{\hspace{-0.08cm} \pm} = \sgn(k_\|^2)''$.

If $\Lambda^2 \ll 1$ and $\mathcal{G} \leq 1$, $\delta I$ can also be
accessed.
In this case, 
Eq.~\ref{eq_SchrodingerWell_bis} can be solved independently for the 
inertia dependent terms and for the fluid terms, separating a  ``large 
$\vartheta$'' and a ``small $\vartheta$'' region, as was done in
Refs.~\cite{Tsai_93, Zonca_96}.
At large $\theta$ (or small {\it radial} scales), the fluid dependent
terms are negligible, and the solution of Eq.~\ref{eq_SchrodingerWell} is 
a wave equation which has simple solutions of the form 
$\Psi_\text{ext} = \Psi_0 \exp(i\Lambda\vartheta)$, with $\Lambda$ one
of the (complex) roots $\Lambda^2$.
When $\vartheta$ goes to 1, inertia becomes negligible in 
Eq.~\ref{eq_SchrodingerWell}. The large $\vartheta$  solution can 
be connected to the fluid (low $\vartheta$ region) 
solution $\Psi_\text{in}$, using a matching of the slopes
(see Fig.~\ref{fig_MHDSchrodingerWell})
\begin{eqnarray}
\nonumber
i\Lambda 
&=&\frac{\Psi_\text{ext}\partial_\vartheta\Psi_\text{ext}|_{0+}}{\Psi_0^2}  
=\frac{\Psi_\text{in}\partial_\vartheta\Psi_\text{in}|_{+\infty}}{\Psi_0^2}
= \frac{1}{\Psi_0^2}\int^{+\infty}_{0} d\vartheta 
\left(|\partial_\vartheta\Psi|^2 
+ \frac{s_0^2|\Psi|^2}{(1+s_0^2\vartheta^2)^2}
- \frac{s_0^2\Psi^*\mathcal{G}\Psi}{1+s_0^2\vartheta^2}
\right)
\label{eq_MatchingBallooning}\\
&=&\left[\frac{1}{\Psi_0^2C\sigma_{\hspace{-0.08cm} \pm}} \right](\delta W_f + \delta W_k ), 
\text{ or equivalently }\   \delta I = \Psi_0^2 C\sigma_{\hspace{-0.08cm} \pm}
 i\Lambda
\end{eqnarray}
where the slope $\partial_\vartheta\Psi_\text{in}$  directly comes from an
integration by part of Eq.~\ref{eq_SchrodingerWell_bis}.
We now have the expression of $\delta I$! \\

Nevertheless, the story is not over, because we did not define what was
the correct root to choose for $\Lambda$.
In the following, we compare three criteria which can allow for such
a determination, or more precisely, {\bf we interpret two criteria which
can be found in the litterature, with the ideas of 
negative/positive energy waves} developed in the previous paragraph:
\begin{itemize}
\item 
{\bf The localization condition,
which we presented as an existence condition for a gap mode to live in 
the previous chapter}.

We explained that the traditional form of the existence criterion 
\cite{Berk_01}, \cite{Breizman_05} was a condition on $\mathcal{G}$ 
(explicitely of the form $\mathcal{G} > \mathcal{G}_0$) designed to 
create a sufficiently localizing Schr\"odinger well. 
The  latter requirement can be reformulated as the necessity for the 
slope calculated in Eq.~\ref{eq_MatchingBallooning} to be negative, 
as illustrated in Fig.~\ref{fig_MHDSchrodingerWell}, or
\begin{equation}
\im\Lambda>0.
\end{equation}
\begin{figure}[ht!]
\begin{minipage}{\linewidth}
\begin{center}
\includegraphics[width=0.65\linewidth]{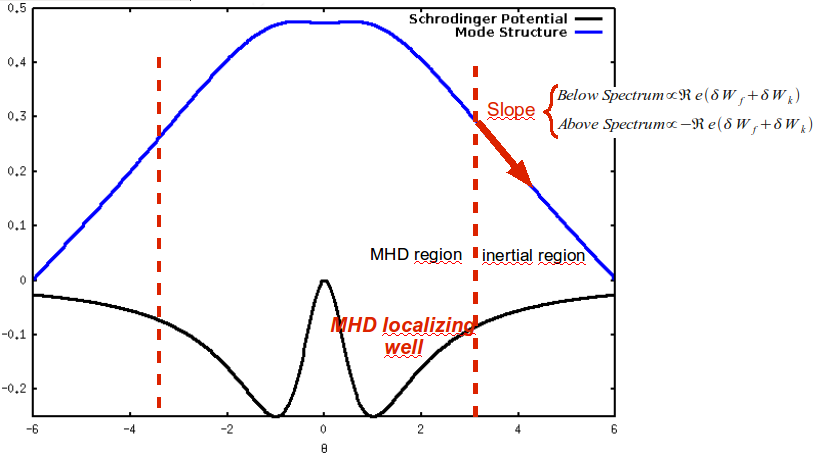}
\caption[\footnotesize
Localizing potential and mode structure in Fourier
space, relevant to a gap mode with a dominant poloidal mode number.]
{\label{fig_MHDSchrodingerWell}
\footnotesize 
Schematic of a Fourier space localizing Schr\"odinger-like well 
and of a mode structure, relevant to a gap mode with a dominant 
poloidal mode number.
The localization condition results from the necessity to overcome 
the central repulsive bump. For an inertia of the form of
Eq.~\ref{eq_MatchingBallooning}, the localization condition requires
negative slope at the frontier between the inertial and MHD regions.
}
\end{center}
\end{minipage}
\end{figure}
Note that the condition of localization in $\vartheta$-space only 
makes sense for a gap mode $\gamma \geq 0$. If the wave is damped by continuum
damping, such a condition is not relevant (the mode is peacked in radial 
space)
and it may be necessary to expand it. We will not discuss negative
$\gamma$ below.
\item {\bf The outgoing wave boundary condition} \cite{Zonca_99}.\\
This condition applies for equations of the form 
Eq.~\ref{eq_SchrodingerWell_bis} which have been shown to be quite
general, when a two-scale separation between inertia and fluid
terms applies \cite{Zonca_07}.
It reads
\begin{equation}
\partial_{\omega_r}\re\Lambda>0.
\label{eq_ZoncaCriterion}
\end{equation}
Recalling that in $\vartheta$-space, the wave-like solution 
of \ref{eq_SchrodingerWell_bis} at large $\vartheta$ is proportional to 
$\propto \exp{i\Lambda\vartheta}$, the requirement \ref{eq_ZoncaCriterion}
can be seen as the necessity for the wave to propagate towards 
the small radial scales, that is for its group velocity (in 
$\vartheta$-space) bring information to larger $\vartheta$.
\item {\bf The positivity of the wave energy density.}

If the main $\omega$-dependence is carried by inertia, the positivity
of the wave energy density close to a point $\omega_r$ 
(Eq.~\ref{eq_JEDampsOrDrives}) reads
\begin{equation}
- \sigma_{\hspace{-0.08cm} \pm}\omega_r\partial_{\omega_r}\im\Lambda > 0.
\end{equation}
Such a situation makes sense in the fishbone-like dispersion relation if
the fast particles are simply perturbative and can be looked at as an 
external source, and $\delta W_f$ has no $\omega$-dependence (as it is the 
case for $\delta W_\text{MHD}$).
\end{itemize}
Let us compare the three criteria.
For simplicity, we use the reduced notations,
$\Lambda^2 = \sigma_{\hspace{-0.08cm} \pm}(\omega^2-\omega_0^2)$, 
$\delta I = \sigma_{\hspace{-0.08cm} \pm} i\Lambda$
where $\omega_0$ a constant. 
(The positive constants have no role in the following).
Writing $\omega = \omega_r + i\gamma$, it comes
$\Lambda_r^2 - \Lambda_i^2 = \sigma_{\hspace{-0.08cm} \pm}
(\omega_r^2 -\gamma^2-\omega_0^2)$,
$\Lambda_r\Lambda_i = \sigma_{\hspace{-0.08cm} \pm}\omega_r\gamma$,
and it follows
\begin{eqnarray}
\partial_{\omega_r} \re\Lambda 
&=& \gamma\im\Lambda\frac{\sigma_{\hspace{-0.08cm} \pm} }{|\Lambda|^2}
\left(1 + \frac{\sigma_{\hspace{-0.08cm} \pm}\omega_r^2}{(\im\Lambda)^2}
\right),\\
- \sigma_{\hspace{-0.08cm} \pm} \omega_r\partial_{\omega_r}\im\Lambda 
&=& \im\Lambda\frac{\omega_r^2}{|\Lambda|^2}
\left(1 - \frac{\sigma_{\hspace{-0.08cm} \pm}(\re\Lambda)^2}{\omega_r^2}
\right).
\end{eqnarray}
For gap modes close to threshold, the fishbone like dispersion relation
is almost real $\Lambda_r\rightarrow0$, $\gamma \geq 0$, and we see that
{\bf the three criteria are equivalent for} modes which will be of interest 
to us in the following: modes oscillating with a frequency
below the continuum spectrum
(picture (2a) in Fig.~\ref{fig_SAWSpectra_withstruc}) which verify 
$\sigma_{\hspace{-0.08cm} \pm}>0$. 
In other words, {\bf the localization and outgoing wave conditions simply 
select positive energy waves}. For the case of gap modes oscillating
above the continuum spectrum 
(picture (2b) in Fig.~\ref{fig_SAWSpectra_withstruc}), the situation
is more ambiguous for the outgoing wave condition.\\

Let us finally see the consequences of the above criteria. When doing the
assymptotic matching and using for example the localization constraint, 
it follows from the sign of $\im\Lambda$
\begin{eqnarray}
\re(\delta W_f + \delta W_k) > 0, & \text{above the continuum spectrum} & \\
\re(\delta W_f + \delta W_k) < 0, & \text{below the continuum spectrum} &
\label{eq_CausalityConstraints}
\end{eqnarray}
In the presence of an external drive, new instabilities can arise outside
of the MHD unstable region. Similarly as for MHD instabilities 
$\delta W_\text{MHD}$, they are characterized by a band of energy given by
Eqs.~\ref{eq_CausalityConstraints}. We will explain the signification of 
Eqs.~\ref{eq_CausalityConstraints} more precisely for BAEs.

\section{Gyrokinetic description}
\label{section_GyrokineticDescription}
The variational method presented above is simply a equivalent form of 
Maxwell equations. For self-consistent calculations, we now need to 
define a model for the plasma.
In chapter \ref{chapter_Fundamentals}, it was explained that kinetic
modelling was necessary for energetic particles, and could also be 
desirable for the main plasma. 
For the special case of Beta Alfv\'en Eigenmodes, 
several reasons call for a kinetic modelling of all the plasma species
and will be listed in the next chapter. 

In the following, we derive the so-called {\bf linear gyrokinetic 
equation},  used in the latter analysis. 
Compared to traditional derivations \cite{Catto_81, Frieman_82, Antonsen_80, 
Brizard_07},
the following one aims at providing a formulation, 
which {\bf simultaneously applies to the 
non-canonical guiding center coordinates and to the action-angle coordinates} 
introduced in chapter \ref{chapter_Fundamentals}. 
Such a formulation legitimates the subsequent simultaneous use of the two 
sets of coordinates in the variational formulas in section 
\ref{section_GyrokineticEnergy}, which is useful for a clean derivation 
of resonances.

\subsection{Basics of gyrokinetic theory}
\label{ssection_BasicsOfGyrokinetics}
We indicated in the thesis introduction, chapter~\ref{chapter_Introduction},
that the collective behavior of a plasma species $s$ could be 
described using a six dimensional distribution function 
$F_s({\bf x},{\bf v})$, following Boltzmann equation \ref{eq_Boltzmann}. 
The so-called {\bf drift-kinetic} and {\bf gyrokinetic theories} are 
simplifications of the Boltzmann equation, which take advantage of the time 
scale separation between collective fluctuations and the fast gyromotion of 
a particle in a strongly magnetized plasma, to reduce the description to a 
5 dimensional problem.  Whereas drift-kinetic theories apply for smooth 
 gradients compared to the species typical gyroradius $k_\perp\rho_s\ll1$ 
(typically equilibrium characteristic gradients when $s=i$), 
{\bf gyrokinetic theories are used for the modelling of perturbations with 
$\m{k_\perp\rho_s \sim 1}$}. 
The latter ordering (when considered with respect to the thermal ions) 
is relevant to micro-scale perturbations, which characterize 
microturbulence. 
But gyrokinetic theories are also needed
when perturbations are of the size of a few ion Larmor radii only, 
a scale which characterize the singularities of MHD modes.

The possibility to reduce the problem dimensionality is easy to understand
when thinking of the separation between the slow {\bf guiding-center} 
motion (renamed {\bf gyrocenter} motion  when a perturbation is present, 
because of the necessity to re-define the slow scale motion and the motion
adiabatic invariant, in this case) and the fast gyromotion of a particle.
If perturbations are slow compared
to the fast gyromotion, it is {\it intuitively} possible to {\it time}-average 
this motion
\footnote{Note that this notion of {\it time-averaging} is rejected by  
the modern gyrokinetic theories, because it lets think that some information 
is lost when moving to the gyrokinetic frame. As will be explained in the 
next pages, the information relative to the fast gyro-angle can be kept in a 
coordinate transformation.}.
Hence, the gyroangle becomes a non-necessary variable, and the 
dimensionality gets reduced. Because a particle gyromotion (for given 
fixed gradients) is not too hard to compute, the {\it time}-averaging 
procedure can be directly linked to a {\it space}-averaging procedure 
along the gyromotion of typical size $\rho_s$, called {\bf gyro-average}. 
The latter space-averaging procedure should not let think however that 
only perturbations with  scales larger that $\rho_s$ can be modelled. 
Gyrokinetic theories are really designed to allow for small amplitude
perturbations with $k_\perp\rho_s \geq 1$
(It makes sense for perturbation lengths which are fractions of $\rho_s$).

From the latter description, we already see that it will be necessary to 
choose an appropriate system of coordinate to describe the guiding-center
(or gyrocenter) motion. And we provided two of them in subsection
\ref{ssection_ParticleTrajectories}, the non-canonical guiding center
variables ({\bf X},$\mu$,{\sf E},$\gamma$), 
and the action-angle variables (${\m{\alpha},{\bf J}}$).

\subsubsection{Assumptions}
Gyrokinetic theory is valid in the presence of a perturbation, and can 
be used for the modelling of instabilities. However, it has some 
requirements on both the equilibrium and perturbation structure.

The main requirement is that the {\it unperturbed} 
plasma be strongly magnetized, that is the condition of Eq.~\ref{eq_rhostar},
\begin{equation}
\rho_s\frac{\nabla B_\eq }{B_\eq}\equiv \rho_s^* \ll 1 
\end{equation}
which ensures the possibility to define a slow guiding-center motion and a
fast gyromotion.
Traditional gyrokinetic theories also consider the equilibrium electric 
field to be reduced 
\begin{eqnarray}
\frac{E_{\eq\perp, \ \|}}{v_{ts} B}  =  O(\rho^*)\ 
\text{such that in particular }
  \frac{v_{{\bf E}\times{\bf B}}}{v_{ts}}  = O(\rho^*_s)  
\end{eqnarray}
The latter limitation on the equilibrium ${\bf E}\times{\bf B}$ drift
is referred to as the {\bf drift-kinetic ordering}\cite{HazeltineBook}, 
and it is opposed to the so-called {\bf MHD ordering}, 
${v_{{\bf E}\times{\bf B}}}\sim{v_{ts}}$ (this designation will be made 
clearer in section \ref{section_MHDLimit}).
The drift-kinetic ordering is usually valid in the plasma core, but may
break at the edge where strong ${\bf E}\times{\bf B}$ flows are 
possible.\\

Gyrokinetic theories enable the description of perturbations with a
limited amplitude (which cannot break the overall confinement), and a
relatively limited frequency $\omega$ compared to the species 
gyrofrequency, $\Omega_{s,c}$. 
Explicitely, the following small parameters are used
\begin{equation}
\frac{\omega}{\Omega_{s,c}} \sim \epsilon_\omega \ll 1,\\
\frac{\mathcal{X}_\omega}{\mathcal{X}_\eq} = \epsilon_\delta\ \ll 1,
\end{equation}
where $\mathcal{X}$ stands for any charateristic quantity.
However, as explained before, the mode length scale is allowed to be as 
small as the species gyroradius, but in the perpendicular direction only
\begin{equation}
\rho_s k_\perp = \epsilon_\perp \quad\text{ arbitrary},\\
\frac{k_\|}{k_\perp} = \epsilon_\| \ll 1
\end{equation}
This restriction on $k_\|$ comes from the fact that perturbations
with a large $k_\|$ are strongly stabilized by the magnetic field tension
and consequently not very dangerous.

We now know all the small parameters involved in (linear and nonlinear) 
gyrokinetic theory \cite{Brizard_07}.
An ordering which is relevant to core thermal ions is the following
$\epsilon_\delta\sim\epsilon_\omega\sim\epsilon_\|\sim\rho^*$.
For our derivations, we will assume 
$\rho^*_e \ll \rho_i < \rho_h \sim \epsilon_\delta\sim\epsilon_\omega
\sim\epsilon_\|$, such that gyrokinetic theory applies to 
the three considered populations \cite{Frieman_82}. \\


For our derivations, we will make use of the {\it linear} gyrokinetic 
equation, and simply retain first order terms in the perturbation 
($\epsilon_\delta$). For a clean separation of the equilibrium and 
perturbed distribution function, $F_s = F_{s\eq} + f_{s\omega}$, the 
equilibrium distribution function needs to verify the (reasonable) 
orderings
\begin{equation}
\mu \partial_\mu F_{s\eq}|_{{\bf X},\mu,\gamma } 
\sim {\sf E}\partial_{\sf E} F_{s\eq}{}|_{{\bf X},\mu,\gamma} 
\sim P_\varphi \partial_{P_\varphi} F_{s\eq}|_{{\bf X},\mu,\gamma } 
= O(1).
\end{equation}
For simplicity, we go a little bit further and assume the equilbrium 
distribution to be only slighty anisotro\-pic, such that 
$\mu \partial_\mu F_{s\eq}|_{{\bf X},\mu,\gamma } = O(\epsilon_\delta)$.

Besides, for simplicity, no equilibrium electric field, no collisions and 
no sources will be considered in our derivation (A gyrokinetic treatment
of collisions is a delicate task \cite{Garbet_09}.).

\subsubsection{The gyro-average operator}
The {\bf gyro-average operator} is the space-averaging operator which 
allows to get rid of the details of the gyromotion, to retain only the 
slower motion relevant to the guiding-center (or gyrocenter) dynamics,
\begin{equation}
J_0\cdot =\frac{1}{2\pi} \int^{2\pi}_0 d\gamma{|_{{\bf X}, {\sf E}, \mu}} 
\ ... 
= (1/2\pi)\int^{2\pi}_0 d{\alpha_1}{|_{\alpha_2, \alpha_3, {\bf J}}} 
\ ...\ \  .
\end{equation}
\footnote{Note that the equivalence of the two fast-angle averaging 
  operators is not so trivial {\it a priori}, because two different fast 
  angle systems could be defined for two different coordinate systems. 
  Because the action-angle defined in this case, 
  have been designed to match the traditional non-canonical guiding-center 
  variables (in particular, $J_3$ depends on $(\Psi, E, \mu)$) equivalence
  is verified.}
A convenient form of this operator is given by its Fourier expression. 
Consider a function $\mathcal{F} ({\bf x})$ of  Fourier transform 
$\mathcal{F}({\bf k})$, and separate the particle gyromotion
${\bf x}= {\bf X}+\m\rho$, with $\m\rho$ the gyromotion of the form
$\m{\rho}=\rho_\perp(\cos\gamma\ {\bf e}_{\perp 1}+\sin\gamma\ 
{\bf e}_{\perp 2})$,
with ${\bf e}_{\perp 1}$ and ${\bf e}_{\perp 2}$  two orthogonal unit vectors 
perpendicular  to the magnetic field in {\bf X}.
It comes
\begin{eqnarray}
\nonumber J_0 \cdot \mathcal{F} \ ({\bf X}) 
&=& \int^{2\pi}_0 d\gamma{|_{{\bf X}, {\sf E}, \mu}} \mathcal{F} ({\bf x}) 
= \int_{-\infty}^{+\infty} d{\bf k} \mathcal{F}({\bf k}) 
\int^{2\pi}_0 \frac{d\gamma{|_{{\bf X}, {\sf E}, \mu}}}{2\pi}
e^{-i {\bf k}\cdot{\bf X}
   - i {\bf k}\cdot\mbox{\boldmath $\rho$}(\gamma)} \\
&=& \int d{\bf k} \mathcal{F}({\bf k}) \ e^{-i {\bf k}\cdot{\bf
    X}}J_0 (k_{\perp}\rho_{\perp})   
\end{eqnarray}
where  $J_0()$ in the final term is the usual notation for the 
$0^\text{th}$ order Bessel function. 
Consequently, the gyro-average operator is found to be equivalent to a 
simple multiplication by a Bessel function in Fourier space (which 
justifies its notation $J_0$ !). \\

\begin{figure}[ht!]
\begin{minipage}{\linewidth}
\begin{center}
\includegraphics[width=0.5\linewidth]{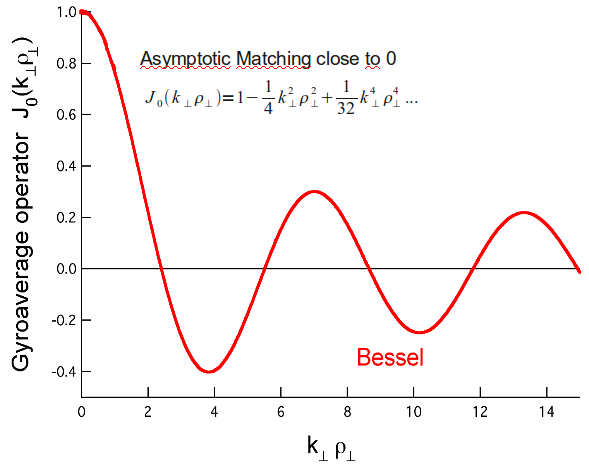}
\caption{\label{fig_Gyroaverage}
\footnotesize $0^{th}$ order Bessel function $J_0$.}
\end{center}
\end{minipage}
\end{figure}

We represented the $J_0$ Bessel function in Fig.~\ref{fig_Gyroaverage}.
Immediate physical insight can be obtained from this picture. 
When $k_\perp\rho_s\ll 1$, $J_0\rightarrow 1$, the gyroradius of the 
particle is so small that it does not see any perturbation, and the 
gyro-averaging is nothing but the identity operator. 
When on the contrary, $k_\perp\rho_s\gg 1$, $J_0\rightarrow 0$, the particle
encounters several important variations of the fluctuation levels during 
its gyromotion. Thus, their averaged effect is null.
Taking for example a perturbation of a few millimeters (a few thermal ion 
Larmor radii), the orderings $\rho_i\sim 1$mm, $\rho_e\sim 0.1$ mm,
$\rho_h\sim 10$mm (energetic ions) show that the electron gyromotion should 
not modify the electron response to the mode, energetic ions will not be 
sensitive to the perturbation. For thermal ions, the effect of the 
gyromotion is more ambiguous, and requires a high order expansion of the 
gyro-average operator.

\subsubsection{Derivations of the gyrokinetic equation}
The gyrokinetic equation is {\it an} equation on a 
{\bf 5D distribution function}, 
expressed in a set of {\bf equilibrium based coordinates} where one fast 
angle representing the fast gyromotion is not present.
From this definition, no unicity is expected. Hence, a derivation of the
gyrokinetic equation requires to define the basis coordinate system, as 
well as the relation between the analyzed 5D equation and the real 6D 
distribution function.

Two different approaches have been developed to display such a 
gyrokinetic equation.
And their equivalence is not absolutely trivial  {\it a priori}.

\begin{itemize}
\item {\bf In traditional gyrokinetic theory}, Vlasov equation,
  $d_tF_s = 0$, is expressed in the equilibrium coordinates, for example
  in the guiding-center coordinates, 
\begin{equation}
\frac{dF_s}{dt} =
\frac{\partial F_s}{\partial t} + \frac{d{\bf X}}{dt}\cdot
\frac{\partial F_s}{\partial \bf{X}} + \frac{d\mu}{dt}\cdot
\frac{\partial F_s}{\partial \mu} +\frac{d{\sf E}}{dt}\cdot
\frac{\partial F_s}{\partial \sf E} + \frac{d \gamma}{dt}\cdot
\frac{\partial F_s}{\partial \gamma} = 0
\label{eq_TraditionalGyrokinetic}
\end{equation}
and an equation on $J_0\cdot F_s$ is derived based on a direct 
gyro-average of this equation \cite{Catto_81, Frieman_82, HazeltineBook}.
The advantage of this method is that it offers a direct understanding
of the orderings at stake from a single look at 
Eq.~\ref{eq_TraditionalGyrokinetic}
(the range of $\partial_t \sim \omega, \text{or} \ \Omega$ for 
example, the order of the drifts in $\dot{\bf X}$...).
However, this derivation is difficult because the $\gamma$-dependence
needs to be handled at the same time in $F_s $ and in the total derivatives, 
which makes averaging a tough work, especially for nonlinear gyrokinetic
theories.
\item {\bf Modern gyrokinetic theory} \cite{Brizard_07, Hahm_88}
  expands the idea underlying the definition of the equilibrium 
  guiding-center coordinates, ${\bf Z}_\text{gc}$, to the perturbed case. 
  Making use of the techniques of the Hamiltonian formalism, and in 
  particular, of the Lie transformations,   
  they define the equations of motion of a {\it virtual} particle, the 
  {\bf gyrocenter} ${\bf Z}_\text{gy}$,   whose trajectory is independent 
  from the fast gyroangle.
  Writing $\mathcal{T}$ the transformation
  $\mathcal{T}: {\bf Z}_\text{gc} \rightarrow {\bf Z}_\text{gy}$, the 
  expression of Vlasov equation in the guiding-center variables is 
  derived using
  \begin{eqnarray}
    &&\frac{d F_{s, \text{gy}}}{dt}=0 \text{, where "}d_t 
    \text{" is independent of }\gamma \\
    &&\text{and \ } F_{s,gc} ({\bf Z}_\text{gc}) 
    = F_{s,gy} (\mathcal{T}\cdot{\bf Z}_\text{gc}) 
  \end{eqnarray}
  With this method, all the fast angle dependence is put in 
  one single term, $\mathcal{T}$.
  Moreover, the use of a Hamitonian formalism ensures the 
  conservation of energy and phase-space volumes.
\end{itemize}

\subsection{Formulation of the linear gyrokinetic equation in a coordinate
independent way}
\label{ssection_GyrokineticDerivation}
We now derive {\bf the  linear gyrokinetic equation} in 
a coordinate-independent form, {\bf valid for a simultaneous use in the 
canonical action-angle coordinates 
$\left(\mbox{\boldmath $\alpha$}, \mathbf{J}\right)$
and in the non-canonical guiding-center coordinates 
({\bf X}, $\mu, {\sf E}, \gamma$)}.

We already explained that the action-angle variables were the 
appropriate coordinates to compute the particle eigenfrequencies, 
and as a consequence resonant behaviors. In linear theory, they 
are also a very convenient calculation tool because they allow for 
an explicit inversion of the Vlasov full time derivative (which, will 
greatly simplify our calculations). On the other hand,
the non-canonical variables are closer to the fluid moments and to the 
geometric features of modes.
For these reasons, we want to make sure that the {\it perturbed} 
fast-angle distribution function we will later introduce in the 
Lagrangian \ref{eq_OmElectromagLagrangian} makes sense in both coordinate
systems.
The reason why this question is not fully trivial is that a perturbed
distribution has {\it a priori} no reason to remain the same in a
coordinate transformation 
$ f_{\omega, {\bf Z}}({\bf Z}) \neq f_{\omega, {\bf Z'}}({\bf Z'})$.

The following derivation follows the same averaging method as 
traditional gyrokinetic theories, but will not fully overlook the lessons
of modern theories. For purely linear derivations, it is indeed
not needed to compute the full gyrocenter motion to concentrate all the
gyroangle information.  Only a {\it partial} coordinate transformation
is necessary (and makes things much simpler than the traditional linear
derivations \cite{HazeltineBook}), 
which will be made in the following. Moreover, this {\it small} step will 
be observed to give the two coordinate systems of interest the same 
behavior. To make this small step, some classical formulas of Hamiltonian 
transformations are required, which can be found in 
Appendix~\ref{appchapter_HamiltonianNotions}.\\

In any coordinate system ${\bf Z} = (Z^a)_{a=1...6}$, Vlasov equation for 
a species $s$ can be written in a hamiltonian form,
\begin{equation}
0 = \frac{dF_s}{dt} =\frac{\partial{F_s}}{\partial{t}}
+\frac{d Z^a}{dt}\frac{\partial{F}}{\partial{Z^a}}
=\frac{\partial{F_s}}{\partial{t}} - [H_s, F_s]
- \frac{\partial \m{\Gamma}_s}{\partial t}\cdot[{\bf Z},F_s]
\label{eq_HamiltonianVlasov}
\end{equation}
where we simply used  Hamilton's equations 
Eq.~\ref{eq_HamiltonsEquations}, and the particle motion is described 
by the Hamiltonian $H_s$ and the Lagrangian
$\underline{\m{\Gamma}}_s = \m{\Gamma}_s\cdot{\sf d}{\bf Z} - H_s{\sf d}t$.

At equilibrium, $\partial_t=0$,  
$H_s=H_{s\eq} = \frac{1}{2}mv^2_{\|} + \mu B_{\eq}$ 
and by definition of the action-angle variables,  $H_{s\eq}$ 
is only a function of the three invariants $\bf J$. 
Expressed in the action-angle variables, Vlasov equation reads
\begin{equation}
[H_{s\eq}, F_{s\eq}] 
= -\partial_{\bf J}H_{s\eq}\cdot\partial_{\m{\alpha}}F_{s\eq}= 0
\end{equation}
Since the particle eigenfrequencies 
$\partial_{\bf J}H_{s\eq} = \m{\Omega}_{s\eq}$ are in general not null,
$F_{s\eq}$ can also be taken to be a function of the {\bf J} invariants
only, or equivalently of the invariants ($\mu$, $\sf E$, $P_\varphi$).

Consider a perturbation, 
$H = H_{ s\eq}  + h_{s}$, \mbox{\boldmath $\Gamma$}$_s$
= \mbox{\boldmath $\Gamma$}$_{s\eq}$ $+\m{\gamma}_s$,
$F_s = F_{s\eq} + f_{s}$ (with $h_s$, $\gamma_s$ and $f_s$ first order 
quantities $\sim \epsilon_\delta$).
For general, not necessarily canonical, coordinate systems, a 
perturbation can modify both the Hamiltonian and the Poisson bracket 
structure given by \mbox{\boldmath $\Gamma$}$_s$. 
The guiding-center/gyrocenter coordinate transformation consists in 
removing the fast-angle dependence of these two quantities, 
in order to display fast-angle free Hamilton's equations.
In the nonlinear regime, this is necessary because of the complexity 
which results from the simultaneous mixing of three fast angle 
dependences in Eq.~\ref{eq_HamiltonianVlasov} 
(Due to the perturbation, fast-angle dependence may appear in $H_s$, in 
the Poisson brackets [,], and in $F_s$.).
In the linear approximation however, second order perturbations are 
neglected, and only two fast angle dependences may get mixed together. 
Thus, it looks reasonable to use a simpler method. In the following, 
we make a transformation (${\bf Z} \rightarrow {\bf \bar Z}$) removing 
the perturbation from the Poisson brackets (only), in order to concentrate
the fast angle dependence of the $d_t$ operator in the Hamiltonian, 
and hence, ease the linear expansion.

From Appendix \ref{appsection_CoordinateTransformations}, 
we know that this can be done using appropriate transformation generating 
functions (and a null gauge $S = 0$, for the notation of this Appendix). 
Equilibrium quantities are not modified by the transformation whereas
the perturbed Hamiltonian becomes 
(Eq.~\ref{eq_TransformedHamiltonian})
\begin{equation}
\bar{h}_s = h_s-({\mbox{\boldmath $\gamma$}}_s
- \bar{\mbox{\boldmath $\gamma$}}_s)\cdot \dot{\bf Z} 
\end{equation}
where we enforce $\bar{{\mbox{\boldmath $\gamma$}}_s} = 0$ to remove the 
perturbation from the Poisson bracket.
In this new coordinate system, the perturbation (hence the fast angle 
dependence)
disappears from the brackets  such that the Vlasov equation can be linearized 
easily. To the first order in $\epsilon_\delta$, we obtain
\begin{equation}
 \partial_t{ \bar{f}_s }- \left[H_{s\eq}, \bar{f}_s\right]
=  \left[\bar{h}_{s}, F_{s\eq} \right]
\label{eq_FirstOrderVlasov}
\end{equation}
where [,] are the {\it unperturbed} Poisson brackets.\\

We define $\bar{G}_s = \bar{f}_s - \partial_{\sf E}F_{s \eq}\bar{h}_s$. 
From Eq.~\ref{eq_FirstOrderVlasov}, $\bar{G}_s$ verifies
\begin{equation}
\partial_t \bar G_s - [H_{s \eq}, \bar G_s]
= - \partial_t\bar{h}_s\partial_{\sf E}F_{s\eq} 
+ [\bar{h}_s , \mu]\partial_{\mu} F_{s\eq}
+ [\bar{h}_s , P_\varphi]\partial_{P_\varphi} F_{s\eq}
\label{eq_G}
\end{equation}
and has interesting characterics.
\begin{itemize}
\item $\bar{G}_s$ can be related to the initial distribution function,
using Eq.~(\ref{eq_TransformedDistribution}). It comes
$ \bar f_s = f_s -(h_s -  \bar{h}_s) \partial_{\sf E}F_{s\eq}$  such that
$ f_s = \partial_{\sf E}F_{s \eq}h_{s} + \bar{G}_s$.
\item Eq.~(\ref{eq_G}) does not depend on the coordinate system. 
On a one hand, the perturbed Hamiltonian $\bar{h}_s$ of a canonical system is 
equal to $h_s =  e_s(\phi- {\bf v}\cdot{\bf A})$ (with $\phi$
and {\bf A} perturbed potentials), and the suggested coordinate 
transformation ${\bf Z} \rightarrow {\bf \bar Z}$ is simply  identity, 
since the bracket structure is by definition independent on the 
perturbation.
In particular, $ \bar{h}_s = h_s = e_s(\phi - {\bf v}\cdot{\bf A})$ 
for action-angle variables.
On another hand, we displayed in Eq.~\ref{eq_GuidingCenterLagrangian} the 
particle Lagrangian for the non-canonical guiding-center variables. 
In fact, Littlejohn calculation of the guiding-center Lagrangian is valid 
in the presence of perturbed fields as well, such that a perturbation may
enter both the Lagrangian symplectic structure $\m{\Gamma}_s$, 
$\m{\gamma}_s\cdot {\bf \dot Z} = {\bf A} \cdot {\bf v}$, and the Hamiltonian
$h_s = e\phi_s$
\footnote{In Eq.~\ref{eq_GuidingCenterLagrangian}, $B_\eq$ shall not be 
expanded to include perturbed quantities because it is simply a result of 
the definition of $\mu$.}. Thus, 
$\bar{h}_s = e_s(\phi-{\bf v}\cdot{\bf A})$ in the non-canonical coordinates 
as well. 

Since the values taken by Poisson Brackets are not changed when the 
coordinate system is changed, the equality of the effective hamiltonians, 
shows that Eq.~\ref{eq_G} is independent from the chosen set of guiding 
center coordinates.
Thus, it should also be true for its solution $\bar G_s$.
\end{itemize}
Following the second point, one can choose to apply Eq.~\ref{eq_G} to the 
action-angle variables, to evaluate the orderings at stake. 
Using the orderings of section \ref{ssection_BasicsOfGyrokinetics}, 
it clearly comes that the terms
of the left hand side are respectively of the order of 
$ \epsilon_\delta\omega$, \
$ \epsilon_\delta {\bf \Omega}_{s\eq}\cdot\partial_{\m{\alpha}}$,
and the terms of the right hand side of the order of
$ \epsilon_\delta\omega$, \
$ \epsilon_\delta\rho_s^*\Omega_{s,1}\partial_{\alpha_1}$, \
$ \epsilon_\delta\Omega_{s,3}\partial_{\alpha_1}$. 
As a consequence, the fast angle dependent part of $\bar{G}_{s}$ 
($\partial_{\alpha_1}\neq 0$) may only be low order in 
$\rho^*_s\leq\epsilon_\delta$. And to the first order in $\epsilon_\delta$, it
is only needed to retain the gyro-averaged part of $\bar{G}_s$. The problem
is reduced to the equations
\begin{eqnarray}
f_s &=& \partial_{\sf E}F_{s \eq}h_{s} + J_0\cdot\bar{G}_s \quad \text{with}
\label{eq_AdiabaticNonadiabatic}
\\
\partial_t &(J_0\cdot \bar G_s)& - \ [H_{s\eq}, J_0\cdot \bar G_s] = -
\partial_t (J_0\cdot \bar{h}_s)\partial_{\sf E}F_{s\eq} +
[J_0\cdot \bar{h}_s ,P_\varphi]\partial_{P_\varphi} F_{s\eq}.
\label{eq_GyrokineticG}
\end{eqnarray}
Eq.~\ref{eq_GyrokineticG} can be called the linear gyrokinetic equation,
and it is related to the perturbed distribution function by 
Eq.~\ref{eq_AdiabaticNonadiabatic}. As can be seen in the latter equation,
a fast angle contribution has been retained in the expression of the 
perturbed distribution function ($\propto h_s$). 
The latter contribution is called the {\bf adiabatic term}, whereas 
$\bar{G}_s$ is the {\bf non-adiabatic term}. As shown by this
derivation, $\bar{G}_s$ can be used for our two reference coordinate 
systems, but from Eq.~\ref{eq_AdiabaticNonadiabatic}, it clearly
appears that it is not the case for $f_s$($h_s$ being different in the
two sets of coordinates).

\subsubsection{Application to  action-angle variables}
In action-angle variables, a linear perturbation can be Fourier 
expanded. Taking the perturbation to one single frequency $\omega$, we
can write
\begin{equation}
  \ h_{s\omega} = \sum_{\mathbf{n}} h_{s, \mathbf{n}\omega}\left(\mathbf{J}\right) 
  e^{ i\mathbf{n}\cdot\mbox{\boldmath $\alpha$}} + c.c. \ ,   \\
  f_{s\omega} = \sum_{\mathbf{n}} f_{s,\mathbf{n}\omega}\left(\mathbf{J}\right)
  e^{ i\mathbf{n}\cdot\mbox{\boldmath $\alpha$}} + c.c. \ .
\end{equation}
Eq.~(\ref{eq_GyrokineticG}) returns for each triplet 
$\mathbf{n} = ({\sf n}_1, {\sf n}_2, {\sf n}_3)$,
\begin{equation}
  (J_0 \cdot \bar G)_{s, \mathbf{n} \omega}
  = \delta_{\mathbf{n}=\mathbf{n^*}}\ \ 
  \frac{\omega \partial_{\sf E} F_{s \eq} + (F_{s\eq}/T_{s\eq})\ 
    {\sf n}_3 \Omega_{* s} }
  {\omega - {\bf n} \cdot \mathbf{\Omega}_{s}} \ h_{s,\mathbf{n}\omega}
\label{eq_GyrokineticActionAngle}
\end{equation}
where ${\Omega}_{*s}$ is the {\bf diamagnetic frequency}
${\Omega}_{*s} = T_{s\eq} \partial_{P_\varphi}\ln(F_{s\eq})
=T_{s\eq} \partial_{J_3|{\sf E}}\ln(F_{s\eq})$ and contains effects related 
to equilibrium pressure gradient (For later use, we also use the
practical notation ${\bf \Omega}_{*s}\equiv (0,0,{\Omega}_{*s})$.).
$T_{s\eq}$ is the species $s$ equilibrium temperature. In order to remain 
general and to considere non-Maxwellian distribution functions, we 
generalize the definition of the temperature writing
\begin{equation}
-\frac{1}{T_{s\eq}} = \partial_{\sf E} \ln F_{s\eq}
\label{eq_TemperatureDefinition}
\end{equation}
where $T_{s\eq}$ may be a function of the three equilibrium motion 
invariants. 
Finally, $\mathbf{n^*}$ refers to any triplet with ${\sf n}_1=0$.
The condition ${\bf n} = {\bf n}^*$  results from the 
gyro-average, which corresponds to the removal of any component 
with ${\sf n}_1\neq 0$ in the action-angle Fourier space.

\subsubsection{Application to the guiding-center non-canonical variables}
In order to compute Eq.~\ref{eq_GyrokineticG} in the non-canonical variables 
$({\bf X}, {\sf E}, \mu, \gamma )$, we need to use the unperturbed 
fundamental Poisson brackets given in Eq.~\ref{eq_ModifiedPoissonBrackets}. 
It directly comes
\begin{equation}
  \left(\partial_{t} +  v_{\|}\nabla_{\|} +
    \textbf{v}_{gs}\cdot\nabla\right)J_0\cdot \bar G_s=
\left(-T_{s\eq}\partial_{\sf E}F_{s \eq} \partial_t +
  F_{s\eq}\mathbf{v}_{* s}\cdot\nabla\right)
\frac{J_0    \cdot \bar{h}_s}{T_{s\eq}}
\end{equation}
Here, 
\begin{equation}
    \textbf{v}_{* s}=  \frac{T_{s\eq}}{e_s B_{\eq}}
    \left(\textbf{b}_{\eq}
    \times\frac{\nabla F_{s\eq}}{F_{s\eq}}\right)  \\
  \label{diamagneticdrift}
\end{equation}
is the velocity counter-part of the previously defined diamagnetic 
frequency, and it is simply called the {\bf diamagnetic velocity} 
($i{\bf n}\cdot\m{\Omega}_{*s} \equiv {\bf v}_{*s}\cdot \nabla $).
${\bf v}_{gs}$ is naturally the usual drift derived from the 
equilibrium Hamiltonian Eq.~\ref{eq_DriftVelocity}, with a null 
${\bf E}\times{\bf B}$-drift (no perpendicular electric field is
considered for simplicity).

\section{A gyrokinetic energy functional for the study of shear Alfv\'en
  waves}
\label{section_GyrokineticEnergy}
Now that we have determined the plasma response, it can be introduced in the 
Lagrangian \ref{eq_OmElectromagLagrangian}.
In the following section, we expand the electromagnetic Lagrangian with the 
gyrokinetic plasma response, in order to study modes of the shear Alfv\'en 
type.
More precisely, we focus on modes with eigenfrequencies 
$\omega \leq v_A/qR$ ($1/qR$ being the natural $k_\|$ scale length), but no
lower bound (in a sheared plasma $k_\|$ {\it a priori} varies radially and may 
cancel).

Considering the time scale separation $k_\|v_A \ll k_\perp v_A$, it means in 
particular that our developments to come will not take into account fast 
magnetosonic waves. 
Nethertheless, we already said that the lower shear Alfv\'en frequency range
was more relevant to the energetic particle eigenfrequencies.\\

For simplicity, we also assume the various species equilibrium distribution
to be even in $v_\|$, such that there cannot be any equilibrium current 
in particular.
Recalling that the equilibrium distribution functions need to be functions
of motion invariants, $F_\eq(\mu, {\sf E}, P_\varphi)$, this property is verified
if we make the approximation $ P_\varphi \approx e\psi$.

\subsection{Conventions for the study of waves of the shear Alfv\'en type}
From now on, perturbations will be considered with a single frequency 
$\omega$.
It can be shown that for frequencies which are small compared to the one
of compressional Alfv\'en waves, the perturbed parallel magnetic field 
verifies
\begin{equation}
B_{\omega\|} = -\mu_0 \frac{\nabla P_\eq \times {\bf b}_\eq}{B_\eq^2}
\cdot \frac{\nabla_\perp \phi_\omega}{i\omega},
\label{eq_PerpendicularPressureBalance}
\end{equation}
where $P_\eq$ is the total plasma pressure. An easy way to understand this
equality is to consider the perpendicular pressure balance in 
Eq.~\ref{eq_MHDVelocity} (under the assumption that the left hand side
inertia does not adjust fast enough)
along with the incompressible pressure state equation, or to cancel the 
second term of the MHD principle \ref{eq_dWMHD}, which can be seen as the 
oscillating energy of compressional Alfv\'en waves (whereas the first and
the third terms are to be associated respectively with the shear Alfv\'en
and sound waves).

Using Eq.~\ref{eq_PerpendicularPressureBalance}, a nice simplification 
appears. The problem is found to be equivalent to a simpler problem with a
reduced perturbed  hamiltonian 
$\bar{h}_s=e_s(\phi_\omega -  v_\| A_{\omega\|})$ 
(with no perpendicular magnetic potential), 
under the condition that the drift velocity be replaced by
\begin{equation}
    \mathbf{v}_{gs}= \frac{1}{e_sB_{\eq}}
    \left(m_s v_{\|}^{2}+\mu B_{\eq}\right)
    \left(\mathbf{b}_{\eq}
    \times\frac{\nabla B_{\eq}}{B_{\eq}}\right)
\label{eq_SimplifiedDrift}
\end{equation}
or in other words, to take the curvature drift proportional to the 
$\nabla B$ drift.

This condition is often understood as a low $\beta$ approximation, because 
it leads to a simultaneous disappearing of the perturbed parallel field 
$B_{\omega\|}$ and of the difference between the two drifts in the equations,
(and the two are proportional to $\beta$, as can be seen for example for 
$B_{\omega\|} \propto \mu_0P_\eq/B_0^2$ in 
Eq.~\ref{eq_PerpendicularPressureBalance}). 
But this cancellation is better understood as a compensation of the two 
effects. 
A more rigorous derivation of Eqs.~\ref{eq_PerpendicularPressureBalance} and
\ref{eq_SimplifiedDrift} is given in 
Ref.~\cite{Edery_92}
.\\

The {\bf problem is now reduced to  two scalar fields}, the electric 
potential and the  parallel magnetic potential. 
For the study of MHD like modes, it is interesting to make use of two 
different scalar fields ($\psi, \mathcal{E}$), defined by \cite{Chen_91} 
\begin{equation}
\nabla_{\|} \psi = -\partial_t A_{\|}\\
\text{and} \\ \mathcal{E} = \psi - \phi \ .
\label{eq_PsiEDefinition}
\end{equation}
Obviously, $\psi$ is directly related to $A_\|$, whereas $\mathcal{E}$ can 
be linked to the parallel electric field. Indeed 
$ E_{\|} = - \mathbf{b}_{\eq}\cdot \nabla \phi - \partial_t A_{\|} =  
\mathbf{b}_{\eq}\cdot \nabla \mathcal{E} = 0$, such that the 
{\bf ideal MHD constraint} $E_\|=0$ can be directly  inferred from 
$\mathcal{E}$.

Note that  the definitions of $\psi$ and $\mathcal{E}$ are somehow ambiguous
when $\nabla_\| = 0$. Considering one tokamak  mode of toroidal mode number 
{\sf n} and one of its poloidal component of number {\sf m} 
($\propto \psi_\omega^{\sf m}(r)\exp (i( {\sf m}\theta + {\sf n}\varphi)$),
$k_\|$ is again 
$\nabla_\| = ik_\| = {\bf b}_\eq\cdot\nabla\varphi({\sf n} + {\sf m}/q)$
(Eq.~\ref{eq_kparallel}). Because $q$ is sheared, $\nabla_\|$ may only 
be null on some surfaces unless ${\sf n} = {\sf m} = 0$. 
Consequently the derivations to come are only valid for $({\sf n,m})
\neq(0,0)$,
but they can also apply to the electrostatic limit (taking $\psi=0$).\\

\subsection{Lagrangian reduction}
\label{ssection_LagrangianReduction}
\subsubsection{Expression of the field and and particle terms}
The Lagrangian field component can be calculted in the reference fields
$(\psi_\omega, \mathcal{E}_\omega)$, 
\begin{eqnarray}
\nonumber
{\bf b}_\eq\cdot\nabla\times {\bf B}_\omega 
&=& {\bf b}_\eq\cdot\nabla\times\nabla\times (A_{\omega\|}{\bf b}_\eq)
= \nabla_\| \left(B_\eq\nabla_\|\left(\frac{A_{\omega\|}}{B_\eq}\right)\right)
- {\bf b}_{\eq}\cdot\nabla^2(A_{\omega\|}\ {\bf b}_{\eq})\\
& \approx &-\nabla^2_\perp A_{\omega\|} 
= - \frac{1}{i\omega}\nabla_\perp^2\nabla_\|A_{\omega\|}
\end{eqnarray}
where the approximation is valid for ${\bf B}_\eq\approx {\bf B}_\text{T\eq}$.

The Lagrangian particle terms, $\mathcal{L}_{s\omega}$, require to express
the charges and currents using Vlasov equation, in the gyrokinetic 
approximation. 
This can be done easily in the action-angle canonical variables, and 
it returns 
\begin{equation}
   \mathcal{L}_{s\omega}=\int d^{3}\textbf{x}
   \frac{n_{s\eq}e_s^2}{T_{s\eq}}\left|\phi_{\omega} \right|^2
    - \sum_{\mathbf{n} =\mathbf{n}^{\ast}}
   \int d\Gamma \frac{F_{s\eq}e_s^2}{T_{s\eq}}
   \frac{\omega-\mathbf{n} \cdot \mathbf{\Omega}_{\ast s}}
   {\omega-\mathbf{n} \cdot \mathbf{\Omega}_s}
   \left|J_0 \cdot \left( \phi-v_{\|}   A_{\|}\right)_{\mathbf{n}\omega} \right|^2.
  \label{eq_ParticleLagrangian}
\end{equation}
We recover the particle adiabatic response in the first term, whereas the
second term is obviously resonant.

\begin{description}
\footnotesize
\item\quad\quad\ $\blacksquare-$\\
To derive Eq.~\ref{eq_ParticleLagrangian}, we can for example express the 
perturbed particle velocity in the canonical system
$({\bf x},{\bf p} = m_s{\bf v}+e_s{\bf A})$
where the expression of the perturbed velocity is easily found to be
$\textbf{v}_{s\omega}
= (-e_s/m_s)\textbf{A}_{ \omega}
= (-e_s/m_s)A_{ \omega \|} \textbf{b}_{\eq}$
to derive the perturbed linear destributions of charges and currents,
\begin{equation}
\begin{array}{lcl}
\mathbf{j}_{s \omega} &=&e_s \left(\int d^3\textbf{p}f_{s\omega}\textbf{v}_s
+ F_{s\eq} (\frac{-e_s}{m_s}\textbf{A}_{\omega}) \right) \\
\rho_{s \omega} &=&e_s\int d^3\textbf{p}f_{s\omega}.
\label{eq_jrho}
\end{array}
\end{equation} 
The perturbed particle Lagrangian for the species $s$  becomes
\begin{equation}
\begin{array}{lcl}
   \mathcal{L}_{s\omega}
&=& e_s\int d\Gamma f_{s\omega}\textbf{v}_{s}.\textbf{A}_{\omega}
-\frac{e_s^2}{m_s}\int d\Gamma F_{s\eq}|
\textbf{A}_{\omega}|^2-\int d\Gamma f_{s\omega}e_s\phi_{s \omega} 
\end{array}
\end{equation}
where $d\Gamma=d^{3}\mathbf{x}$ is a phase space volume element.

We can now switch to another canonical set of variables, 
$(\m{\alpha}, $\m{J}$)$, such that 
$d\Gamma=d^{3}\mbox{\boldmath {$\alpha$}} d^{3}\mathbf{J}$,
and use the linear gyrokinetic approximation
$ f_{s\omega} = \partial_EF_{s, eq}h_{s\omega} + J_0\cdot\bar{G}_{s\omega}$,
Eq.\ref{eq_GyrokineticActionAngle} to find
\begin{equation}
\begin{split}
 \mathcal{L}_{s\omega} =&
 -\frac{e_s^2}{m_s}\int d\Gamma F_{s\eq}|A_{\omega\|}|^2 -e_s\int d\Gamma J_0
\cdot \bar G_{s\omega }(\phi_{ \omega}-v_{ \|}A_{\omega\|})  \\
 &+  e_s^2\int d\Gamma \frac{F_{s\eq}}{T_{s\eq}}\left(|v_{ \|}A_{\omega \|}|^2
+ |\phi_{\omega}|^2
 - v_{\|}(\phi_{\omega \|}^*A_{\omega\|}+
\phi_{\omega \|}A_{\omega\|}^*)\right)\\
=&\int d^{3}\textbf{x} \frac{n_{s\eq}e_s^2}{T_{s\eq}}
   \left|\phi_{\omega} \right|^2
    -e_s \int d\Gamma  \left(J_0 \cdot \bar G_{s\omega}\right) \left[J_0 \cdot
\left( \phi_{\omega}-v_{\|} A_{\omega\|}\right)\right]^{\ast}
\label{eq_ParticleLagrangianExpansion}
\end{split}
\end{equation}
Note that the last equality is valid for an equilibrium distribution which is 
even in $v_{\|}$ (Note that it is possible to relax this assumption to include 
current driven modes, overlooked in this analysis \cite{Edery_92}).
\hfill $-\blacksquare$\\
\end{description}
Eq.~(\ref{eq_ParticleLagrangian}) is the usual expression for studying the
stability of drift waves. To study electromagnetic modes, it is convenient to 
reformulate the resonant term using the fields $\psi$ and $\mathcal{E}$ which 
can be compared to the MHD orderings. 
We made the expansion using the more tractable action-angle Fourier expansion,
but with an attempt to introduce elements of the traditional guiding-center
coordinates, in order to recover some traditional fluid and kinetic results, in
the following parts. This expansion is explained in smaller characters below.

\begin{description}
\footnotesize
\item\quad\quad\ $\blacksquare-$\\
To carry out  the expansion of Eq.~(\ref{eq_ParticleLagrangian}), we 
constantly  made  use of the coordinate independence of the operators involved
in the linear gyrokinetic equation  to switch from action-angle variables
to the non-canonical guiding-center coordinates. 
Expansion of the perturbed Hamiltonian
\begin{equation}
\begin{split}
  \frac{1}{e_s} J_0\cdot \bar{h}_{s\omega}
 = & J_0\cdot (\phi_{\omega} - v_{\|}A_{\omega\|})
 = J_0\cdot (\phi_{\omega}+ \frac{v_{\|}\nabla_{\|}\psi_{\omega}}{-i\omega})\\
 =& J_0\cdot \left[ \frac{-i\omega + v_{\|}\nabla_{\|} + \mathbf{v}_{gs}
\cdot \nabla }{-i\omega} \psi_{\omega} - \frac{\mathbf{v}_{gs}
\cdot \nabla}{-i\omega}\psi_{\omega} - \mathcal{E}_{\omega} \right]\\
 =& J_0\cdot \left[ \frac{ \partial_t - [H_{s\eq}, \cdot]}{-i\omega} \psi_{\omega}
   - \frac{\mathbf{v}_{gs}\cdot \nabla}{-i\omega}\psi_{\omega}
   - \mathcal{E}_{\omega} \right]\\
\left(\frac{1}{e_s} J_0\cdot \bar{h}_{s\omega}\right)_{\mathbf{n}= \mathbf{n^*}}
 =&  \left[ \frac{-i\omega+i\mathbf{n} \cdot \mathbf{\Omega}_s}{-i\omega}
\psi_{\omega\mathbf{n}}\right]
 - \left[\frac{\mathbf{v}_{gs}
\cdot \nabla}{-i\omega}\psi_{\omega}
+ \mathcal{E}_{\omega} \right]_{\mathbf{n}}
\end{split}
\end{equation}
leads to
\begin{equation}
\begin{split}
   \mathcal{L}_{s\omega}
     = +&\int d^{3}\textbf{x} \frac{n_{s\eq}e_s^2}{T_{s\eq}}
        \left|\phi_{\omega} \right|^2
       - \sum_{\mathbf{n}^{\ast}}
        \int d\Gamma \frac{F_{s\eq}e_s^2}{T_{s\eq}}
	\left(1+\frac{ i\mathbf{n} \cdot \mathbf{\Omega}_{\ast s}}
	{-i\omega}\right)\left(1-\frac{ i\mathbf{n} \cdot \mathbf{\Omega}_{ s}}
	{-i\omega}\right)|\psi_{\omega\mathbf{n}}|^2\\
       +&\sum_{\mathbf{n}^{\ast}}  \int d\Gamma \frac{F_{s\eq}e_s^2}{T_{s\eq}}
        \left(1+\frac{ i\mathbf{n} \cdot \mathbf{\Omega}_{\ast s}}{-i\omega}\right)
	\left( \psi_{\omega\mathbf{n}} \mathcal{E}_{\omega\mathbf{n}}^{\ast}
	+\psi^{\ast}_{\omega\mathbf{n}} \mathcal{E}_{\omega\mathbf{n}}\right)\\
	-& \sum_{\mathbf{n} =\mathbf{n}^{\ast}} \int d\Gamma
        \frac{F_{s\eq}e_s^2}{T_{s\eq}}
	\frac{\omega-\mathbf{n} \cdot \mathbf{\Omega}_{\ast}}
        {\omega-\mathbf{n} \cdot \mathbf{\Omega}_s}
        \left|\left(\frac{ \mathbf{v}_{gs} \cdot \nabla\psi_{\omega}}{-i\omega}
       + \mathcal{E}_{\omega}
        \right)_{\mathbf{n}} \right|^2.
\end{split}
\end{equation}
where we suppressed any integration term even in the parallel velocity
$v_{\|}$. 
Rewriting  $ ( \mathcal{E}_{\omega\mathbf{n}}^\ast\psi_{\omega\mathbf{n}}
+ \mathcal{E}_{\omega\mathbf{n}}\psi_{\omega\mathbf{n}}^\ast)
= |\mathcal{E}_{\omega\mathbf{n}}|^2+|\psi_{\omega\mathbf{n}}|^2
-|\phi_{\omega\mathbf{n}}|^2$,
and using electroneutrality:
\begin{equation}
 \sum_s\int d^3\mathbf{p} (F_{s\eq}e_s^2/T_{s\eq})\mathbf{v}_{\ast s}
= \mathbf{b}_{\eq}\times \nabla\left(\sum_se_s\int  d^3\mathbf{p} F_{s\eq}\right) =0
\end{equation}
and moving to the non-canonical guiding-center variables,
we finally get an MHD-like energy principle \ref{eq_KineticdW}
\hfill $-\blacksquare$\\
\end{description}

We can finally  put the Lagrangian in a form which unavoidably reminds us 
of the MHD energy functional.
\begin{equation}
\begin{split}
    \mathcal{L}_{\omega}= -&\int d^{3} \textbf{x} \frac{1}{\mu_0}
    \left|\frac{\nabla_{\bot}\nabla_{\|}\psi_{\omega}}{-i\omega}\right|^2
    +\sum_{s}  \int d   \Gamma \frac{F_{s\eq}e_s^2}{T_{s\eq}}
    \left|\mathcal{E}_{\omega} \right|^2 \\
    +& \sum_{s} \int d\Gamma \frac{F_{s\eq}e_s^2}{T_{s\eq}}
     \left(  1+\frac{\mathbf{v}_{\ast s} \cdot \nabla}{-i\omega} \right)
    \left[ (1 - J_0\otimes J_0) (\left|\phi_{\omega}\right|^2
     - \left|\mathcal{E}_{\omega}\right|^2) \right] \\
    + & \sum_{s} \int d\Gamma \frac{F_{s\eq}e_s^2}{T_{s\eq}}
    \left(J_0 \cdot \psi_{\omega}\right)^{\ast}
    \left( \frac{ \mathbf{v}_{gs} \cdot\nabla}{-i\omega}\right)
    \left(J_0 \cdot \psi_{\omega}\right) \\
    +& \sum_{s} \int d\Gamma \frac{F_{s\eq}e_s^2}{T_{s\eq}}
    \left(\frac{\mathbf{v}_{\ast s} \cdot
    \nabla\left(J_0 \cdot\psi_{\omega}\right)}{-i\omega}\right)^{\ast}
    \left(\frac{ \mathbf{v}_{gs}\cdot \nabla
    \left(J_0 \cdot\psi_{\omega}\right)}{-i\omega}\right) \\
    -& \sum_{s} \sum_{\mathbf{n}=\mathbf{n}^{\ast}} \int d\Gamma
    \frac{F_{s\eq}e_s^2}{T_{s\eq}}
    \frac{\omega -
    \mathbf{n} \cdot \mathbf{\Omega}_{\ast}}
    {\omega-\mathbf{n} \cdot \mathbf{\Omega}_s}
    \left|J_0 \cdot \left(
    \frac{ \mathbf{v}_{gs} \cdot \nabla\psi_{\omega}}{-i\omega}
    + \mathcal{E}_{\omega}
    \right)_{\mathbf{n}} \right|^2
\end{split}
\label{eq_KineticdW}
\end{equation}
In Eq.~\ref{eq_KineticdW}, the notation used for the non-resonant part
needs some clarification:
the form $\hat{ L}(fg)$, with $f$ and $g$ two scalar fields should be understood
as the effect of a hermitian operator
$\hat{ L}(fg)= \hat{ L}(gf) = f (\hat{ L}g)$, and the $\otimes$ symbol
stands for a bilinear function such that 
$ (A\otimes B)(fg)=(Af)\times(Bg)$ for $f$, $g$ two scalar fields.

In this functional, we can recognize the first term to be the magnetic 
tension  or {\bf magnetic field line bending term}. 
The next three contributions are to be associated with {\bf inertia} and
include  {\bf  polarization (with Finite Larmor Radius corrections), and  
diamagnetic effects, related  to ${\bf v}_{*s}$}. 
The fifth term is the MHD-like {\bf interchange drive},
and the last term which is not present in MHD, describes the wave-particle 
resonant interaction.

\subsubsection{Application to a thermal plasma}
We now apply the kinetic energy functional \ref{eq_KineticdW} to a 
purely thermal plasma with one single ion species $i$ (and no energetic 
particles).
In other words, the electron and ion thermal equilibrium distribution
functions are assumed to be Maxwellians with respect to energy, ie:
\begin{equation}
  F_{s\eq}= \frac{n_{s\eq}\left( P_\varphi \right)}
{\left[ 2\pi m_s T_{s\eq}\left(P_\varphi \right)\right]^{3/2}}
\exp\left\{-\frac{E}{T_{s\eq}\left(P_\varphi \right)}\right\}.
\label{eq_Maxwellian}
\end{equation}
where $n_{s\eq}$ is the species density and $T_{s\eq}$ is the species 
temperature.
$T_{s\eq}$ is in agreement with the definition of the temperature given 
earlier \ref{eq_TemperatureDefinition}, but it now  depends on $P_\varphi$ only.

Moreover, we assume that $P_\varphi = e_s\Psi + R_0m_sv_\| \approx e_s\Psi$ such 
that $P_\varphi$ is close to a {\it  radial} coordinate 
(such that we can write the density and the temperature to be functions
of $r$ only, $n_s(r)$ and $T_s(r)$).
Doing this, the particles radial drift away from their reference magnetic
surface is neglected.
Because this approximation is more problematic for trapped 
particles, it is often called the {\bf thin banana width approximation}.
Using the conservation of $P_\varphi$, 
$\delta P_\varphi = 0 = $
$ (e_sB_0)((r/R_0)\delta r + \rho_s(\delta v_\|/v_{ts}))$, 
this approximation is observed to scale like $ (R_0/r)\rho_s$ for 
trapped particles and may not be valid at the center (for  ions). 
However, it is usually a  good approximation for passing particles 
(with a reduced $\delta v_\|$).
With the latter approximation, the Maxwellian Eq.~\ref{eq_Maxwellian} 
is similar to the more natural idea of thermal equilibrium (ie, a velocity 
Maxwellian at a given position, as is used in MHD for example). However,  one 
should not forget that in general, the equilibrium distribution has to be taken a 
function of the invariants, and that $P_\varphi$ does not rigously correspond to 
a position
.

Finally, the ion and electron populations can be combined in the Lagrangian
under  a few reasonable assumptions. If there is no additional population,
electroneutrality at equilibrium implies  $n_{i\eq}=n_{e\eq}\equiv n$.
Besides, we assume that 
$|\nabla T_{i \eq}|/T_{i \eq} \approx | \nabla T_{e \eq}|/T_{e\eq}$, and use the 
notation $e = e_i = -e_e$.\\

We now focus on modes with typical wavelengths going from a few ion Larmor 
radii to the mesocale $ k_\perp a \leq 0.1$. More precisely, we assume
the following orderings to be verified, 
$ \nabla n/n \sim  |\nabla T_{\eq}|/T_{\eq} \ll k_{\perp} < 1/\rho_i$, 
such that equilibrium variations can be removed at the scale of the perturbation
of interest. 
Combining all the assumptions, it comes 
\begin{equation}
\begin{split}
    \mathcal{L}_{\omega}= -&\int d^{3} \textbf{x} \frac{1}{\mu_0}
    \left|\frac{\nabla_{\bot}\nabla_{\|}\psi_{\omega}}{-i\omega}\right|^2
    +  \int d^3\mathbf{x} \frac{n e^2}{T_{i\eq}}
    \left(1+\frac{1}{\tau_e}\right)
    \left|\mathcal{E}_{\omega} \right|^2 \\
    +&  \int d^3 \mathbf{x} \frac{ne^2}{T_{i\eq}}
     \left<\left(  1+\frac{\mathbf{v}_{\ast i} \cdot \nabla}{-i\omega} \right)
     (1 - J_0\otimes J_0) \right>(\left|\phi_{\omega}\right|^2
    -\left|\mathcal{E}_{\omega}\right|^2) \\
    - &  \int d^3 \mathbf{x} \frac{ne^2}{T_{i\eq}}
    \left<\left( \frac{ \mathbf{v}_{gi} \cdot\nabla}{-i\omega}\right)
    (1 - J_0\otimes J_0)\right>\left| \psi_{\omega}\right|^2 \\
    +&  \int  d^3 \mathbf{x}\frac{ne^2}{T_{i\eq}}
    \left<\left(\frac{\mathbf{v}_{\ast i} \cdot \nabla}{-i\omega}\right)
    \left(\frac{ \mathbf{v}_{gi} \cdot\nabla }{-i\omega}\right)
    (\tau_e+J_0\otimes J_0)\right>|\psi_{\omega}|^2 \\
    -& \sum_{s} \sum_{\mathbf{n}=\mathbf{n}^{\ast}} \int d\Gamma
    \frac{F_{s\eq}e_s^2}{T_{s\eq}}
    \frac{\omega-\mathbf{n} \cdot \mathbf{\Omega}_{\ast s}}
    {\omega-\mathbf{n} \cdot \mathbf{\Omega}_s}
    \left|J_0 \cdot \left(
    \frac{ \mathbf{v}_{gs} \cdot \nabla\psi_{\omega}}{-i\omega}
    + \mathcal{E}_{\omega}
    \right)_{\mathbf{n}} \right|^2
\label{eq_ThermalKineticdW}
\end{split}
\end{equation}
where $\left<...\right> = (1 / n_{i\eq})\int d^3 \mathbf{p} F_{i\eq} ...$,
$\tau_e=T_{e\eq}/T_{i\eq}$. In deriving Eq.~\ref{eq_ThermalKineticdW},  the 
drift-kinetic equation for the electrons, $J(k_\perp\rho_e) = 1$, has been used
and the $J_0$ notation of Eq.~\ref{eq_ThermalKineticdW} should be understood 
as the thermal ion relevant gyro-average operator.
The identity Eq.~\ref{eq_ThermalKineticdW} will be the starting point 
of our analysis  of BAEs.\\

Before, analyzing Eq.~\ref{eq_ThermalKineticdW}, it is relevant to check its
validity. Several non-variational gyrokinetic formulations exist, which have
been developed for the study of shear Alfv\'en type waves
\cite{Chen_91,Chu_92,Zonca_96}. One should  normally recover similar equations 
when applying the variational principle, and extremalizing $\mathcal{L}_\omega$
under variations of $\phi^*_\omega$ and  $A^*_{\omega\|}$, or almost
equivalently under variations of 
$\psi_{\omega}^\ast$ and $ \mathcal {E}_\omega^\ast$ 
\footnote{The word {\it almost} here is simply used to recall the indetermination
  of $\psi_{\omega}^\ast$ where $k_\| = 0$, 
  which results from its definition. 
  As will be clearer in Eq.~\ref{eq_InertialEigenEquation}, this may have an 
  impact.}.

And indeed, one can recover the equations of Ref.~\cite{Zonca_98}. 
Differentiation according to $\mathcal{E}_\omega^\ast$ at fixed $\psi_\omega^\ast$
returns a {\bf modified electroneutrality equation}
\begin{equation}
   - \left(1+\frac{1}{\tau_e}\right) \mathcal{E}_{\omega}
   + \left<\left(  1+\frac{\mathbf{v}_{\ast i} \cdot \nabla}{-i\omega}\right)
     (1 - J_0^2) \right>\psi_{\omega}
   =- \sum_{s} \frac{T_{s\eq}}{ne_s} \left<J_0 \cdot K_s\right>
\label{eq_Electroneutrality}
\end{equation}
with $K_s$ the solution of {\bf {\it a} linear gyrokinetic equation}
\begin{equation}
  \left(\partial_{t} +  v_{\|}\nabla_{\|} +
  \textbf{v}_{gs} \cdot \nabla \right) K_s=
  \frac{F_{s\eq}e_s}{T_{s\eq}} \left(\partial_{t} +
  \textbf{v}_{\ast s}\cdot\nabla\right) \left[J_0\cdot
  \left( \frac{\mathbf{v}_{gs} \cdot\nabla\psi_\omega}{-i\omega}
+ \mathcal{E}_\omega\right)\right].
  \label{eq_GyroZonca}
\end{equation}
Differentiation according to $\psi_\omega^\ast$ at fixed
$ \mathcal{E}_\omega^\ast $ returns the {\bf vorticity equation}
\begin{equation}
\begin{split}
  -\frac{v_A^2}{\omega^2}\nabla_{\|}(\rho_i^2\nabla_{\bot}^2)
    \nabla_{\|}\psi_\omega
  + \left<\left(  1+\frac{\mathbf{v}_{\ast i} \cdot \nabla}{-i\omega} \right)
    (1 - J_0^2) \right>\phi_{\omega}
  - \left< \left( \frac{ \mathbf{v}_{gi} \cdot\nabla}{-i\omega}\right)
    (1 - J_0^2) \right>\psi_{\omega} \\
  -\frac{\rho_i^2v_A^2}{\omega^2} \mathbf{b}_{\eq}\times \nabla
  \left< \beta \right>
  \cdot \nabla_{\perp} \left(\mathbf{b}_{\eq}
    \times \kappa\cdot \nabla_{\perp} \psi_{\omega}\right)
  = \sum_s  \frac{T_{s\eq}}{ne_s}\left<\left(\frac{\mathbf{v}_{gs}\cdot
        \nabla}
   {-i\omega}\right) J_0\cdot K_s \right>
\label{eq_Vorticity}
\end{split}
\end{equation}
where in $v_A^2 = B_{0}^2 /\mu_0 n m_i$ 
(the field is taken at the plasma center), again 
$\beta = 2\mu_0n (T_{i\eq}+T_{e\eq})/B_{0}^2$ takes into account electrons 
and ions. 
Eqs.~\ref{eq_Electroneutrality}, \ref{eq_GyroZonca} and \ref{eq_Vorticity} 
are in agreement with the equation of Ref.~\cite{Zonca_98}.

Note that we can recognize  in the vorticity equations some terms of the 
Shear Alfv\'en Law  \ref{eq_AdaptedShearAlfvenLaw}, now expanded to include
diamagnetic, FLR and resonant behaviors.

\section{Link with MHD}
\label{section_MHDLimit}
The {\bf MHD ordering} is traditionally recognized to be appropriate for a
large {\bf total} ${\bf E}\times{\bf B}$-drift velocity
\begin{equation}
\frac{{\bf E}\times {\bf B}}{B^2}\sim v_{ti}
\end{equation}
and it is for this reason opposed to the  {\bf drift-kinetic ordering}
\begin{equation}
\frac{{\bf E}\times {\bf B}}{B^2}\sim \rho^*_iv_{ti}
\end{equation}
which we made use of to derive the linear gyrokinetic  equation 
\cite{HazeltineBook}.
We already explained in the first paragraph of section \ref{ssection_BasicMHD}
that a strong ${\bf E}\times{\bf B}$ flow was to be associated with the 
requirement of fast dynamics. The consequence of this ordering is that several
terms considered to be low order compared to the ${\bf E}\times{\bf  B}$
related effects (diamagnetic terms, $E_\|$) are not present in MHD,
and this appears both when studying equilibrium properties and fluctuations.
In particular, in the presence of a coherent fluctuation of frequency 
$\omega$, the previously defined MHD displacement can be directly related
to the electromagnetic fields 
fluctuations, via the ${\bf E}\times{\bf B}$ fluctuations only
(constrained  by the ideal Ohm's law),
\begin{equation}
\m{\xi}_\omega =\frac{{\bf B}_\eq\times{\nabla\phi_\omega}}{-i\omega}.
\label{eq_XiPhi}
\end{equation}

Despite this ordering difference, we can recover the 
{\bf low $\m\beta$ limit of the MHD energy equation} \ref{eq_dWMHD}, 
when cancelling the parallel electric field ($\mathcal{E}_\omega =0$) and 
keeping enough terms to retain the $ {\bf E\times B}$ drift only.
In our functional, ${\bf E}\times{\bf B}$ fluctuations correspond to  
second order terms  in $k_{\perp}\rho_i$, and neglecting lower order terms
kills any diamagnetic effects other than the interchange drive. 
This possibility to recover the MHD limit is to be related to the fact 
that our linear expansion of the MHD equations assumed a null equilibrium 
electric field and was consequently not in contradiction with the 
drift-kinetic ordering. 

A clean way to proceed is to use the {\bf hydrodynamic approximation} and 
to adopt the limit $\omega \rightarrow +\infty $.
This ordering increases the relative weight of the electromagnetic potential
term $|\mathcal{E}_\omega|^2$ in the functional and
consequently implies its cancellation while extremalizing by
$\mathcal{E}_\omega^*$. It also cancels diamagnetic effects
since  $\omega_*/\omega = ({\bf V}_{*i} \cdot {\bf k})/\omega\rightarrow 0$.
Besides,  the hydrodynamic ordering $\omega \gg k_{\|}v_{te}\gg k_{\|}v_{ti}$ 
removes any resonance.
Finally, to the second order in $\rho_ik_{\perp}$, the Lagrangian
given in Eq.~\ref{eq_ThermalKineticdW} reduces to
\begin{eqnarray}
  \nonumber
  \mathcal{L}_{\omega}= &-&\int d^{3} \textbf{x} \frac{1}{\mu_0}
  \left|\frac{\nabla_{\bot}\nabla_{\|}\psi_{\omega}}{-i\omega}
  \right|^2
  + \int d^3 \mathbf{x} \frac{ne^2}{T_{i\eq}}
  \rho_i^2 \left|\nabla_{ \perp }\psi_{\omega}\right|^2 \\
  \nonumber
  &+& \int  d^3 \mathbf{x}\frac{ne^2}{T_{i\eq}}(1+\tau_e)
  \left<\left(\frac{\mathbf{v}_{\ast i} \cdot \nabla}{-i\omega}\right)
  \left(\frac{ \mathbf{v}_{gi} \cdot\nabla }{-i\omega}\right)
  \right>|\psi_{\omega}|^2 \\
  \nonumber
  &-&  \int  d^{3} \textbf{x} \frac{1}{B^2_{0}}
  \left| \frac{ {\bf{b}}_{\eq}\times \kappa \cdot \nabla\psi_{\omega}}{-i\omega}
  \right|^2 \left[\sum_{s} T_{s\eq}\int d^{3} \textbf{p}
  F_{s\eq} \left(\frac{m_sv_{\|}^2 + \mu_sB_{\eq}}{T_{s\eq}}\right)^2  \right]\\
\end{eqnarray}
or again
\begin{eqnarray}
  \nonumber
  \mathcal{L}_\omega
  = &-&\int d^{3} \textbf{x} \frac{1}{\mu_0}
  \left|\frac{\nabla_{\bot}\nabla_{\|}\psi_{\omega}}{-i\omega}\right|^2
  + \int d^3 \mathbf{x} \frac{1}{\mu_0v_A^2}
  \left| {\bf b}_{\eq}\times \nabla_{\perp} \psi_{\omega}\right|^2  \\
  &+& 2 \int  d^3 \mathbf{x}\frac{1}{B^2_{0}}\left(\frac{{\bf b}_{\eq}
      \times \nabla P
      \cdot\nabla_\perp \psi_\omega}{-i\omega}\right)
  \left(\frac{  {\bf b}_{\eq}\times \kappa \cdot\nabla_\perp
      \psi_\omega}{-i\omega}\right)^*
  \label{eq_RecovereddWMHD} \\
  \nonumber
  &-&  \int  d^{3} \textbf{x} \frac{ c^2_s}{\mu_0v_A^2}
  \left|2 \frac{ {\bf{b}}_{\eq} \times
      \kappa \cdot \nabla\psi_{\omega}}{-i\omega}
  \right|^2
\end{eqnarray}
where $\left\langle J_0^2\right\rangle(.)\equiv \Gamma_0\left(.\right)$
has been approximated by its asymptotic limit close to 0, to the order 
$ k_\perp^2\rho_s^2$.
$c_s^2=\Gamma\left(T_{e\eq}+T_{i\eq}\right)/m_i$ is the
{\bf MHD sound speed}, with $\Gamma$  the adiabatic compression 
index (similar  to Eq.~\ref{eq_VaCsVelocities}). 
Note that our kinetic derivation gives
$\Gamma = 7/4 $, which is different from the value of $5/3$
found in the usual 3-D adiabatic pressure equation of state.
This difference is the result of the temperature anisotropy
between the parallel and perpendicular directions implied by the form
 of ${\bf v}_{gs}$ we considered. However, if we now neglect anisotropy,
the usual value of $\Gamma$ can be shown to be recovered.

The Lagrangian of Eq.~(\ref{eq_RecovereddWMHD}) is directly 
recognized as
\begin{equation}
  \mathcal{L}_{\omega}=-2\delta W_\text{MHD} + \omega^2\int d^{3}\mathbf{x}
  nm_i \left|\mbox{\boldmath $\xi$}_{\perp} \right|^2
\label{eq_LagrangianAnddWMHD}
\end{equation}
where we deduced $\m{\xi}$ from Eq.~\ref{eq_XiPhi} and $\delta W_\text{MHD}$ is 
a reduced form of the MHD energy functional \ref{eq_dWMHD}, similar to the 
one developed in Ref.~\cite{Chu_92} and called the Slow Sound Approximation.
One easily identifies the first three terms of 
Eq.~\ref{eq_RecovereddWMHD}
as the field line bending tension, kinetic energy, and interchange.
The $B_\|$ stabilization of Eq.~\ref{eq_dWMHD} is missing due to the 
low $\beta$ approximation.
Finally, using Eqs.~\ref{eq_XiPhi}, it comes 
\begin{equation}
\nabla\cdot \m{\xi} = \nabla\cdot \m{\xi}_\perp 
= \frac{1}{B_\eq^2}\left({\bf B}_\eq\times\nabla
\times{\bf B}_\eq-\nabla B_\eq^2\right)
\cdot\m{\xi}_\perp= 2\m\kappa\cdot\m{\xi}
\label{eq_DivKappa}
\end{equation}
and the last term is directly recognized as the compressibility related
term of Eq.~\ref{eq_dWMHD}. It is interesting to note that in this approach,
{\bf compressibility is intimately related to geometry} via curvature, 
$\m{\kappa}$.\\

It is worth noting a difference between the direct derivation of 
$\delta W_\text{MHD}$ starting from the MHD equations and the kinetic 
derivation provided here. 
This difference explains that the MHD energy functional can
only be recovered with an approximation,  the Slow Sound Approximation (SSA) 
of Ref.~\cite{Chu_92}, corresponding to $\omega \gg c_s/R$ and similar to the 
hydrodynamic limit $\omega \gg {\bf n}\cdot \m{\Omega}_s$ assumed in here.

In MHD, the removal of the SSA displays a resonant term which is reminiscent 
of the wave-particle kinetic resonances, and behaves in a similar fashion 
as the Alfv\'en resonant continum. This term is associated to a so-called 
{\bf Slow Sound Continuum} corresponding to $\omega^2 = k_\|^2c_s^2$. It 
can be recovered from kinetic resonances of the form 
$ 0 = \omega - {\bf n}\cdot \m{\Omega}_i = \omega - k_\|v_\|$
(relevant to passing ions, when $\omega\sim k_\|c_s$),
with the {\it formal} substitution 
\begin{eqnarray}
\mathcal{ R}_{kinetic}
&=&\frac{1}{m_i}\Gamma \left[T_{e\eq} + T_{i\eq}\frac{1}{4\Gamma}
 \left\langle \left(\frac{ m_i v_{\|}^{2}+\mu B_{\eq}}{T_i} \right)^2
\left( \frac{\omega}{\omega - k_\|v_\| }\right)
\right\rangle\right]
\label{eq_RKinetic}\\
\longrightarrow  \mathcal{R}_{fluid}&=&\quad
c_s^2 \frac{\omega^2}{\omega^2 - k^2_\| c_s^2}
\label{eq_RFluid}
\end{eqnarray}
in the resonant term of Eq.~\ref{eq_ThermalKineticdW} (where we simply 
assumed $\mathcal{E} = 0$, considered all ion resonances to be of 
the form $\omega - k_\|v_\|$, and neglected electron resonances 
$\omega \ll {\bf n}\cdot\Omega$ to derive the first equation above).
With this substitution, the fluid and kinetic approaches can be made
consistent, and the form of Eq.~(6) of Ref.~\cite{Chu_92} and
Eqs.(2) and (3) of Ref.~\cite{Gorelenkov_07} can be recovered. 
Nevertheless, one should note that such a substitution makes sense in the 
limit $k_\|v_\|/\omega \longrightarrow 0$ only, also associated to 
a {\bf high-q} limit ($k_\| \propto 1/q$), that is, close to the
SSA. 
In this limit, resonances do not occur but 
{\it transit corrections} may result from Eqs.\ref{eq_RKinetic} and 
\ref{eq_RFluid}, which are similar in fluid and kinetic theories,
\begin{equation}
\mathcal{R}_{kinetic}\approx c_s^2 \left(1
+ \frac{6}{7(1+\tau_e)^2}\frac{k^2_\|c_s^2}{\omega^2}\right)
\approx  \mathcal{R}_{fluid} \approx c_s^2
\left( 1 + \frac{k^2_\|c_s^2}{\omega^2}\right).
\label{eq_ComparedTransitCorrections}
\end{equation}
Otherwise, when $\omega^2$ approaches $k^2_\|c_s^2$, the MHD Slow Sound 
Continuum arises. But the latter is not fully consistent with 
collisionless kinetic theory, and it is consequently safer to directly start
with the more precise kinetic formalism.\\

The MHD functional \ref{eq_RecovereddWMHD}, with the transit corrections
\ref{eq_ComparedTransitCorrections} is relevant to study shear Alfv\'en waves 
or $n=0$ electrostatic modes in the MHD-SAA limit.
This system has been studied extensively \cite{Chu_92,Huysmans_95},
and it shows how finite compressibility modifies the Alfv\'en spectrum
and forces the occurence of a low frequency gap  where the BAE can exist,
even when the slow sound continuum is not taken into account \cite{Chu_92}.
In the next chapter, we study the equivalent kinetic functional to 
investigate tokamak acoustic modes and we underline the role of kinetic effects.


\section{Summary}
{
\it
In this chapter, we presented the kinetic formulation, later used in the thesis
for the study of Beta Alfv\'en Eigenmodes.

\begin{itemize}
\item Using a variational principle, we explained that Maxwell equations
  could be put in the form of a complex {\bf energy balance}, 
  which allows to compute particle-wave energy 
  transfers, and to define the notion of wave-energy density, and positive
  energy waves. 

  We saw in particular, that waves with a {\bf positive energy density} could 
  be driven {\bf unstable}, when the sign of particle to wave energy resonant
  energy transfers was positive.

  Next, in a effort to understand this idea of positive energy density, 
  we compared it with some requirements
  of gap modes, classically used in the litterature \cite{Chen_94, Zonca_99}: 
  the idea that gap modes should be {\bf localized in the radial Fourier space} 
  (that is to say not discontinuous in the "normal" radial space, by definition)
  or the idea that {\bf energy propagation should be towards the small scales}
  \cite{Zonca_99}. Our conclusion was that {\bf all three requirements were 
  equivalent for gap-like modes}, with a dominant {\sf m} component. 
  
\item Next, we derived the kinetic response of plasma particles to an 
  electromagnetic fluctuation, using the traditional gyrokinetic framework.
  In this work, we made use of some properties of Hamiltonian transformation
  to make a derivation of the linear gyrokinetic equation in a {\bf form
  which simultaneously applies for action-angle variables and for the 
  non-canonical guiding center variables}, which is appropriate to deal
  with resonances.
  
\item We finally combined this gyrokinetic response with the variational
  formulation of Max\-well equations, to display an {\bf energylike relation,
    which extends the traditional MHD energy equation}.
  
  In this work, we chose to write this energylike relation in a form which
  directly applies to MHD-like fluctuations. 
  This allowed us to {\bf recover the classical equations used for the
  study of shear-Alfv\'en type waves \cite{Zonca_98}, 
  as well as the MHD functional as a limit}.
\end{itemize}
}


\begin{savequote}[20pc]
\sffamily
What is the "Beta-induced Alfv\'en Eigenmode?"
\qauthor{Heidbrink, Phys. Plasmas 6, 1147 (1999)}
\end{savequote}




\chapter{The Beta Alfv\'en Eigenmode}
\label{chapter_BAEdescription}
The question raised by W. W. Heidbrink in a paper of 1993, reveals the 
ambiguity surrounding the nature of the Beta Alfv\'en Eigenmode.
The latter ambiguity reflects the complexity of the acoustic frequency 
band where resonances are possible with both thermal and suprathermal 
particles (see Fig.~\ref{fig_TimeAndLengthScales})
and where various physical effects (compressibility, effects related to 
the previously defined diamagnetic terms, sound waves) can enter into play.

The main goal of this chapter is to provide a kinetic description of the 
mode, 
with an attempt to clarify the physics involved in its dispersion 
relation and structure, and in particular the role of some kinetic effects
which are missing in MHD, such as Finite Larmor Radius (FLR) effects.
A related objective is to compare BAEs with other modes traditionally 
recognized to oscillate in the acoustic frequency range, to allow for an 
unambiguous identification of BAEs in experiments.
Based on this analysis, an interpretation of some acoustic modes observed 
in the Tore-Supra tokamak will be given.

The chapter is organized as follows.
An introduction presents some basic observations of the mode and the 
particular features of our description. Sections 
\ref{section_ApproximationsForBAEDerivation}, 
\ref{section_BAEInertialLayer} and 
\ref{section_BAEDispRelAndStructure}
contain the derivation of the mode frequency and structure, 
and respectively present the approximations, the local inertial physics 
and the global structure and dispersion relation of Beta Alfv\'en 
Eigenmodes.
Section~\ref{section_BAEGAMDegeneracy} compares BAEs with well known 
modes involved in transport studies, the so-called Geodesic Acoustic 
Modes (GAMs). Finally, Section~\ref{section_IdentificationToreSupraModes}
presents the acoustic modes observed in Tore-Supra and offers an 
interpretation for their identification.\\

From now on and until the end of the thesis, we get rid of the
subscript $ _\eq$ to refer to the equilibrium, whenever there is no
possible confusion. In particular, $T_i, T_e, v_{ti}, \tau_e, n_h$ will
have to be understood as equilibrium quantities.

\section{Introduction}

\subsection{A variety of modes in the acoustic frequency range}
Recently, quasi-coherent modes have been observed in Tore-Supra in the
{\bf acoustic frequency range \cite{Udintsev_06,Sabot_06}, that is with a 
frequency of the order of $v_{ti}/R_0\sim c_s/R_0$}. Their identification
requires to understand the types of modes that can develop in this range.\\

{\bf BAEs} are traditionally recognized to be 
{\bf electromagnetic fluctuations} {\bf with  finite {(\sf m, n)}} 
mode numbers, oscillating with an acoustic frequency in a
{\bf gap of the Alfv\'en resonant spectrum induced by compressibility}
(The last term of Eq.~\ref{eq_RecovereddWMHD}, possibly with the correction 
Eq.~\ref{eq_ComparedTransitCorrections}). The name comes from the dual
physics which is associated with the mode existence: 
the presence of an {\it Alfv\'en} resonant spectrum (symbolized by the 
frequency $\omega_A \equiv v_A/qR \approx v_A/R_0$) and {\it compressibility} 
which is at the origin of the mode {\it acoustic} frequency 
($\omega_{BAE}\sim v_{ti}/R_0$). Indeed, taking the ratio of these two 
characteristic frequencies, {\it beta} is recovered : 
$ 4v^2_{ti}/ v^2_A = \beta$.

First observations of BAEs were performed in the TFTR and DIIID tokamaks 
\cite{Turnbull_93} in the presence of energetic ions heated by Neutral 
Beam Injection, and the modes were found to have a strong deleterious 
impact on the confinement of the energetic population \cite{Heidbrink_93}.
Later, BAEs, likely driven by finite amplitude magnetic islands,
were also identified in the tokamaks FTU 
\cite{Buratti_05, Annibaldi_07} and TEXTOR \cite{Zimmerman_05}.
Finally, in JET, a high frequency branch of the fishbone modes was
observed to oscillate with typical BAE frequencies
\cite{Nabais_05,Zonca_09}, and may be associated to the 
same compressible physics as the BAE.\\

Nevertheless, other modes have been predicted and/or observed in the 
acoustic frequency range.

An important example is {\bf Geodesic Acoustic Modes (GAMs)}. GAMs are 
well-established modes in turbulent transport studies, 
which are thought to strongly interact with turbulence 
\cite{Hallatschek_01,Miyato_04,Scott_05,Itoh_05,Naulin_05,Sugama_06,
Angelino_06}. They are usually taken to be acoustic oscillations with
{\bf a main $({\sf n}=0, {\sf m}=0)$ component} and related to the same
physics as BAEs: {\bf geometry related compressibility}.
Because the existence of GAMs is strongly related to turbulent dynamics, 
and because electrostatic fluctuations dominate turbulence, they
are usually  thought to be well described in the {\bf electrostatic limit} 
(that is, with negligible magnetic fluctuations).
Some experimental oscillations, obtained with non-magnetic diagnostics,
have been identified as GAMs: in  DIII-D using Beam Emission Spectroscopy
\cite{McKee_03}, in JIPT-IIU \cite{Hamada_05},  JFT-2M \cite{Ido_06} and T-10
\cite{Melnikov_06} using heavy ion beam probes, in Asdex-Upgrade by Doppler
reflectometry \cite{Conway_05} and in the Heliac H1 using Langmuir probes
\cite{Shats_06}. 
Nevertheless, more recent experiments also reported the existence of 
$({\sf n}=0, {\sf m}=0)$ magnetic fluctuations with an acoustic frequency,
and the latter were also refered to as GAMs \cite{Boswell_06}.

Other modes, BAAEs (Beta Alfv\'en Acoustic Eigenmodes), 
electromagnetic modes with lower frequencies than BAEs and related to the 
shear of the tokamak equilibrium \cite{Gorelenkov_07, Gorelenkov_09}, 
modes related to diamagnetic effects
(ITG/Ion Temperature Gradient or  AITG/Alfv\'en Ion Temperature Gradient
\cite{Zonca_98, Nazikian_06}), 
as well lower branches of RSAEs (the Reversed Shear Alfv\'en Eigenmodes
\cite{Breizman_05}, 
already presented in chapter \ref{chapter_Fundamentals}) are also expected 
in the acoustic frequency range.

This variety makes the study of the acoustic frequency range particularly 
challenging and controversial.

\subsection{An ambiguous interpretation of BAEs}
\begin{figure}[ht!]
\begin{center}
\begin{minipage}{\linewidth}
\begin{center}
\includegraphics[width=0.55\linewidth]{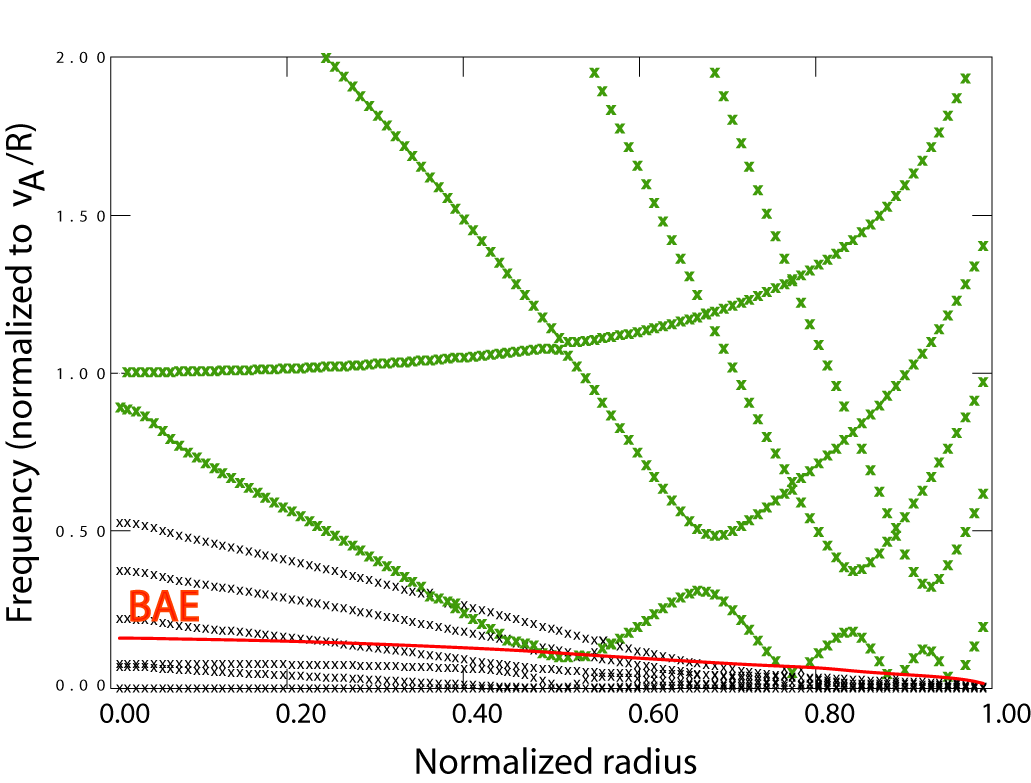}
\caption
[\footnotesize
Resonant continuum spectrum for Tore-Supra discharge \#42039.
]
{\label{fig_CS42039_wCompressibility}
\footnotesize 
Continuum spectrum calculated with CSCAS \cite{Huysmans_95}, based
on the CRONOS transport reconstructed equilibrium for shot \#42039 
($t\sim 9.5$s), whose main characteristics can be found in
Figs.~\ref{fig_Specgram42039} and \ref{fig_Psd42039}.
The spectrum is plotted for a mode with a toroidal mode number $n=1$ and 
and an assumed compressibility of 1.67.
The green curves correspond to the lines of the Alfv\'en spectrum, the black
curves to the slow continuum.
}
\end{center}
\end{minipage}
\end{center}
\end{figure}

Soon after the first observations of BAEs, the existence of a 
{\bf compressibility induced gap} in the MHD resonant Alfv\'en spectrum was 
identified \cite{Chu_92, Huysmans_95} with MHD codes (For illustration,
such a gap has been calculated in Fig.~\ref{fig_CS42039_wCompressibility}
for a Tore-Supra relevant discharge).

Analytically, when compressibility is added but the Slow Sound Continuum is 
neglected (see section \ref{section_MHDLimit}), it is easily shown that 
the local Alfv\'en resonant spectrum of a mode with a main finite 
$({\sf m, n})$ component takes the form 
\begin{equation}
\Lambda^2\equiv\frac{\omega^2 - \omega^2_{BAE}}{\omega_A^2} 
         = q^2R^2_0(k^{\sf m, n}_\|)^2 \\
\text{instead of}\\
\Lambda^2\equiv\frac{\omega^2}{\omega_A^2} = q^2R_0^2(k^{\sf m, n}_\|)^2 
\label{eq_BAEResonantSpectrum}
\end{equation}
(with $ \omega_A = v_A /qR_0$) close to the surface where $k_\|^{\sf m, n}$ 
cancels,
which can also be seen as modification of inertia (or the necessity to define
a {\bf generalized inertia} $\Lambda^2$).
Moreover, $\omega_{BAE}$ is an acoustic frequency 
(of the order of the experimental observations) which slightly varies
depending on the chosen MHD-like model, 
\begin{equation}
\begin{array}{rcll}
  \omega^2_{BAE} &=& 2c^2_s/R_0^2   
  & \text{- in the MHD-SAA.}\\
  \omega^2_{BAE} &=& (v^2_{ti}/R^2_0)(7/2 + 2\tau_e) & 
  \text{- when ions only are in the hydrodynamic limit}\\
  &&&\text{and anisotropy is considered,  $\Gamma = 7/2$,}
\end{array}
\label{eq_SeveralOmBAE}
\end{equation}
such that gap modes are possible below $\omega_{BAE}$ ($\omega<\omega_{BAE}$). 
Next, the existence of an extremum of the resonant spectrum where  
$(k^{\sf m, n}_\|)^2$ 
cancels, makes reasonable the existence of a discrete mode in this gap 
(with the same arguments as those given in  the derivation of 
Eq.~\ref{eq_SchrodingerWell}), and the interpretation of BAEs as gap modes
close to an extremum pictured in Fig.~\ref{fig_SAWSpectra_withstruc} 
(schematic 2.b).\\

This description is interesting because it shows that compressibility, 
which is usually thought to be {\it stabilizing} (it is a positive 
contribution to $\delta W_\text{MHD}$~\ref{eq_dWMHD}) can have a destabilizing 
impact on MHD stable modes.
However, this picture is somehow limited and has been subject to 
controversy.
\begin{itemize}
\item A first reason is that neglecting the Slow Sound Continuum 
  in the MHD formalism only makes sense if the mode frequency verifies 
  $\omega \gg k_\|c_s$ for any relevant poloidal component of the mode. 
  In particular, for the mode sidebands (which are absolutely necessary for 
  computating compressibility),  $k^{\sf m \pm 1}_\| \approx \pm 1/qR_0$, where 
  $ k^{\sf m, n}_\| = 0$. Hence,  it is not clear that for an acoustic mode, 
  the approximation that the Slow Sound Continuum has no effect can apply.
  
  For this reason, an alternate interpretation of the gap represented in 
  Fig.~\ref{fig_CS42039_wCompressibility} has been given,  attributing this 
  gap to a {\it coupling between a sound wave of poloidal number 
  ${\sf m}\pm 1$
  and an Alfv\'en wave of poloidal number $\sf m$}, in the same 
  fashion as for the TAE gap \cite{Huysmans_95} 
  (the coupling of two Alfv\'en waves). This interpretation is reasonable in 
  MHD and indeed, it was shown that the MHD Slow Sound Continuum 
  impacts on the shape of the compressibility induced gap.
  
  However, we explained in section \ref{section_MHDLimit}, 
  that a kinetic description was more reasonable for describing the resonant 
  regime $\omega\sim c_s/qR_0 \sim v_{ti}/qR_0$ \cite{Turnbull_93, Zonca_96}. 
  For this reason, the MHD picture of a coupling between a sound wave and an 
  Alfv\'en wave can be misleading.

  First, a resonant condition of the form $\omega - k^{\sf m, n}_\|(r)v_\| = 0$ 
  (a certain curve in the $(r, v_\|)$ space) is generally less localizing
  radially than a resonant condition of the form 
  $\omega^2 - (k^{\sf m,n}_\|(r))^2c_s^2=0$ 
  (a line in $(r,v_\|)$ phase space, parallel to $r=0)$), even if it is true
  that for a thermal population, the first resonant condition 
  is usually more active where $ v_\| \sim v_{ts}$, which also tends to enforce
  some radial localization.

  Secondly, $c_s \ll v_A$ leads to a cross 
  point of the Alfv\'en and Slow Sound Continua which is very close to the 
  resonant surface, such that the mode is expected to have a 
  dominating ${\sf m}$ component rather than two poloidal components of equal 
  weight (like it is the case for TAEs).

\item A second reason is that several experiments \cite{Boswell_06, Nguyen_09}
  return mode frequencies which are close to the values of $\omega_{BAE}$ 
  given by the formulas \ref{eq_SeveralOmBAE}, that is close to 
  the continuum spectrum. It is in particular the case for the modes observed
  in Tore-Supra, as will be shown later on. We already explained that in this
  case, small scales (FLR) 
  which may only be accessed with kinetic theory, become important 
  \cite{Zonca_98}.

  More generally, for a given generalized inertia $\Lambda^2$, closeness to
  the resonant spectrum can be associated to the closeness to the roots
  of the $\Lambda^2 = 0$. For this reason, the roots of this generalized 
  inertia have been given a special name, {\bf accumulation points}
  \cite{Chen_94}. 
  With a resonant spectrum of the form
  $\Lambda^2(\omega)= q^2R^2_0(k^{\sf m, n}_\|)^2$, it is clear that accumulation
  points separate the resonant spectrum $\Lambda^2 >0 $
  from the frequency where discrete gap modes can take place $\Lambda^2<0$
  (see again Fig.~\ref{fig_SAWSpectra_withstruc}).
  In this sense, the generalized inertia defined in here is fully consistent
  with the $\Lambda^2$ defined in Eq.~\ref{eq_SchrodingerWell}.

\item Finally, this picture does not take into account some physics
  which can be relevant to the acoustic frequency range, and which is not
  included in MHD. For example, it is interesting {\it a priori} to keep 
  $E_\|$
  because it  contributes to resonances 
  (see the numerator of the resonant term in Eq.~\ref{eq_ThermalKineticdW}) 
  and hence wave-particle energy transfers.
  Secondly, such modes as ITG/AITG may only be found when diamagnetic terms 
  are included \cite{Zonca_99}.  
\end{itemize}

The multiplicity of mode names given in the previous subsection results
from the diverse physics involved, but in a lot of parameter ranges, these
physical features need to be considered simultaneously.
Moreover, as will be clearer in the following 
(and in particular in Eq.~\ref{eq_GammaExpansion})
the addition of kinetic effects, such as high order FLR, imply a degeneracy
removal of the MHD modes \cite{Zonca_98}.
It follows that from a kinetic point of view, an infinite spectrum usually needs
to be substituted to the picture of well defined separated discrete modes 
(though in general, {\bf accumulation points} may be displayed).
For this reason, more recent analysis favor a more global description of 
the mode spectrum rather than a separate description of each possible mode
\cite{Chavdarovski_09}.

\subsection{Objectives of the performed derivation}
In the first part of this work, we will calculate the BAE dispersion relation 
using the {\bf variational gyrokinetic energy principle 
of Eq.~\ref{eq_ThermalKineticdW}}, with an attempt to catch some missing 
kinetic features of the MHD model which are relevant to modes on Tore-Supra, 
and to provide a description of the mode structure.

First, the use of Eq.~\ref{eq_ThermalKineticdW} will make possible a 
separation between the electron and ion responses, which is important when 
dealing with frequencies of the order of the thermal particles equilibrium 
characteristic frequencies.

Next, to derive its geometry, the {\bf mode will be expanded into its various 
poloidal Fourier components $\sf m$, and FLR effects will be calculated to a 
high order},
relevant to fluctuations which are close to  $\omega_{BAE}$, 
that is, close to the resonant spectrum.
Earlier MHD \cite{Chu_92} and  kinetic \cite{Zonca_96, Zonca_98} derivations 
of BAEs were carried out in the {\bf ballooning formalism}, 
briefly explained in Appendix \ref{appchapter_BallooningRepresentation}. 
In the work presented here, 
the Fourier expansion in the poloidal components (only) 
will return the coupling between poloidal sidebands which is intrinsically 
related to compressibility and makes possible an analysis of the relative
weight of the sidebands involved. 
Moreover, the mode radial structure will be calcultated directly in a possibly
more intuitive way.

Nevertheless, the following calculation will not provide the complete 
calculation 
of the Slow Sound Continuum related resonance, and {\bf transit corrections 
only} (with the correct {\it kinetic} coefficients though) will be provided, 
which are relevant to the high-$q$ limit.
The full computation of the Slow Sound Continuum related resonances 
(ie, of resonances with the passing ions, see section 
\ref{section_MHDLimit}) has been done  in  Ref.~\cite{Zonca_96}, 
and the additional effect of resonances with trapped ions can also be found
in the literature \cite{Chavdarovski_09}.
Besides, diamagnetic effects (negligible for typical Tore-Supra parameters)
will not be taken into account.

With these methods and approximations, the results of 
Eq.~\ref{eq_SeveralOmBAE} will be recovered and extended.\\

In a second part, a comparison of BAEs with electrostatic GAMs will be 
performed, and the degeneracy of BAEs and GAMs, developed in 
Ref.~\cite{Zonca_08}  analyzed in the light of our formalism.

\section{Approximations  of the derivation}
\label{section_ApproximationsForBAEDerivation}

In the following, we study BAEs in a simplified {\bf circular equilibrium} 
using a poloidal and toroidal Fourier component expansion. \\

Concerning equilibrium, we take advantage of the smallness of $\beta$,
$r/R_0$ and $L_p/R_0$
(Recall that $L_p$ a typpical equilibrium plasma poloidal scale at the mode 
location) 
considering $r/R_0\sim\beta$ and $ L_p/R_0 >\sim \beta^{1/4}$.
The different choices made for $L_p/R_0$ and $r/R_0$ (which one could have 
choosen with the similar ordering $\epsilon \sim \sqrt{\beta}$, following the 
given numerical values provided earlier)
come from the fact that modes on Tore-Supra have been observed at the plasma 
center ($r\sim 0.2a$), where equilibrium gradients are flatter.
A consequence of  the smallness of $r/R_0$ is that the operators 
$\nabla_{\perp}$,
$\nabla_{\|}$ and $J_0$ which follows, are to be understood as independent
on $\theta$. Moreover we take $q\sim s \sim \tau_e\sim 1$.\\

For the {\it linear} description of modes, axisymmetry enables the
selection of one single $\sf n$ component whereas toroidicity
implies poloidal coupling, such that we can simply write the various 
field quantities in the form
\begin{equation}
\phi_\omega = \sum_{\sf m}\phi^{\sf m}_\omega(r)
\exp(i{\sf m}\theta + i{\sf n}\varphi).
\end{equation}
In particular, with the notation $\nabla \equiv i{\bf k}$, 
$k_r$ needs to be understood as an operator whereas 
$k_\theta = {\sf m}/r$ for the poloidal component {\sf m}. 
Next, BAEs (as well as GAMs) are considered to belong to the acoustic frequency 
range  $\omega \sim v_{ti}/R_0$ with a meso-scale structure
$1/L_p\ll{\bf k}_\perp \approx (k_r, k_{\theta}) \ll 1/\rho_i$. 

For BAEs in particular, we expand the computation close to a surface 
$r_s$ 
\footnote{Often called {\it resonant} surface, but not to be confused with
the Alfv\'en resonant surface.}  
where $k_\|^{\sf m}(r_s) = 0$ (without neglecting neighboring radial variations)
where  localization is reasonable, as explained above. Since the
{\sf m} poloidal component corresponding to the latter surface is likely 
to be dominant in this area, we refer to it as the mode {\it main poloidal
component}. From now on, we also use the notation {\sf m} to refer to this 
component
(whereas  ${\sf m} \pm 1, \pm 2$ will be used for its sidebands).
Taking $k_\theta = {\sf m}/r$, we finally consider $\rho_ik_\theta\sim \beta$
($\sf m$ from 1 to 10),
but we do not enforce the radial scale, following the remark that a two-scale
structure is expected to arise from this study (see section 
\ref{ssection_AlfvenSpectrum}).

\subsection {The energy functional in the acoustic frequency region}
\label{ssection_IntermediateFrequencyRegion}
Let us first reduce the functional Eq.~\ref{eq_ThermalKineticdW} to a simpler 
expression valid in the acoustic frequency range 
(hence, for both for BAEs, GAMs or BAAEs at this point).\\

For $\omega \sim v_{ti}/R_0$, {\bf diamagnetic effects} verify
\begin{equation}
\frac{{\bf n}\cdot\m{\Omega}_{*i}}{\omega} 
\equiv 
\frac{{\bf v}_{*i}\cdot {\bf k}}{\omega}
\sim  \frac{\rho_iv_{ti}k_\theta/L_p}{(v_{ti}/R_0)}
= (\rho_ik_\theta)\left(\frac{R_0}{L_p}\right) 
\sim< O(\beta^{3/4})
\label{eq_DiamagneticOrdering}
\end{equation}
In general, the ordering we use makes diamagnetic terms negligible in the 
acoustic frequency range (this will be clearer in the next paragraph).\\

Secondly, we know from subsection \ref{ssection_GyrokineticDerivation}
that the resonance $\omega/(\omega - {\bf n}\cdot\m{\Omega}_s)$ 
in action-angle variables, is equivalent to the  {\it operator} 
$(-i\omega)(-i\omega + v_\|\nabla_\| +{\bf v}_{gs})^{-1}$. 
For {\bf electrons} characterized by a large thermal velocity, 
$ T_{e} \sim T_{i} \Rightarrow v_{ti}/v_{te}
\sim\sqrt{m_e/m_i}\sim 0.02\sim 2\beta$.
Hence, for the natural length scale $k_\| \sim 1/qR_0$ ($q\sim 1$) and
for finite {\sf m} and {\sf n} mode numbers,
\begin{equation}
\frac{k_\|v_{te}}{\omega} \sim \frac{v_{ti}}{v_{te}}\sim \beta 
\end{equation} 
and the resonant operator is found to be dominated by the parallel component,
whose role is to make the resonance negligible. In other words, for
the scaling $k_\| \sim 1/qR_0$, electrons move faster than the mode
phase velocity such that their {\it non-adiabatic response} cancels.
Very close to the resonant surface where $k^{\sf m}_{\|}$  cancels for the main
component, the assumption of {\bf electron adiabaticity} can be challenged
(even if it remains valid for the resonant terms  which involve
poloidal sidebands...not all of them). 
For simplicity however, we assume this assumption to be valid in the whole 
extent of the studied domain, as long as the main poloidal component has a 
finite {\sf m} (hence {\sf n}) mode number.

Note however  that the latter assumption is not correct for GAMs, whose 
main component verifies $k_\|^{\sf m=0} =0$ (${\sf n}=0$) and falls into the 
opposite {\it hydrodynamic} limit. \\

With the above assumption, 
${\bf n}\cdot {\bf \Omega}_{*i} \ll \omega\sim v_{ti}/qR_0 \ll k_\|v_{te}$,
the functional reduces to
\begin{equation}
\begin{split}
    \mathcal{L}_{\omega}= -&\int d^{3} \textbf{x} \frac{1}{\mu_0}
    \left|\frac{\nabla_{\bot}\nabla_{\|}\psi_{\omega}}{-i\omega}\right|^2
    + \int d^3\mathbf{x} \frac{n e^2}{T_{i, eq}}
    \left(\left|\mathcal{E}_{\omega} \right|^2
    +\frac{1}{\tau_e}\left|\mathcal{\tilde{E}}_{\omega} \right|^2\right) \\
    +&  \int d^3 \mathbf{x} \frac{ne^2}{T_{i, eq}}\left\{
           (1 - \Gamma_0)(\left|\phi_{\omega}\right|^2
    -\left|\mathcal{E}_{\omega}\right|^2)
    - \left<\left( \frac{ \mathbf{v}_{gi} \cdot\nabla}{-i\omega}\right)
    (1 - J^2_0)\right>\left| \psi_{\omega}\right|^2\right\} \\
    -&  \sum_{\mathbf{n}=\mathbf{n}^{\ast}} \int d\Gamma
    \frac{e^2F_{i, eq} }{T_{i,eq}}
    \left( \frac{\omega}
    {\omega-\mathbf{n} \cdot \mathbf{\Omega}_s} \right)
    \left|J_0 \cdot \left(\frac{ \mathbf{v}_{gi} \cdot 
      \nabla\psi_{\omega}}{-i\omega}
    - \mathcal{E}_{\omega}
    \right)_{\mathbf{n}} \right|^2.
\end{split}
\label{eq_LagrangianIntermediateFcy}
\end{equation}
In order to make this expression  valid  for GAMs as well, we 
simply introduced the modified field 
$\tilde{\mathcal{E}}_{\omega} = \mathcal{E}_{\omega}
-\left\langle\mathcal{E}_{\omega}
\right\rangle_{\theta,\varphi}$ 
($\mathcal{E}_\omega$ = for finite ({\sf n, m}) components), 
which stands for the remaining resonant electrostatic term involved in the 
conservation of the GAM main poloidal component.

\subsection{Two scale-separation and  inertial  layer}
\label{ssection_InertialBAEOrderings}
In order to assess the importance of the terms of the functional
Eq.~\ref{eq_LagrangianIntermediateFcy}, there remains to determine the radial
scales. In the following, we simply focus on BAEs. For these modes, a special
role is played by the {\it field line bending tension} which is characterized by a
radial dependence, $k_\|^{\sf m} (r)$. Close to the resonant surface where 
$ k_\|^{\sf m}(r_s) = 0$, we can expand $k_\|$
\begin{equation}
k^{\sf m}_\| \approx k^{\sf m}_\|\ '\ x=  - \frac{1}{qR_0} k_\theta s x $,\\ 
with $x=(r-r_s)
\end{equation}
where the shear $q, s$ are assessed at $r_s$. This radial variation is at the 
origin of the shearing effect of section \ref{ssection_AlfvenSpectrum} and of 
the related apparition of two scales.

More precisely, the orderings of the various terms of the functional
\ref{eq_LagrangianIntermediateFcy} are shown in 
Tab.~\ref{tab_FunctionalOrderings},
\begin{table}[ht!]
\begin{center}
\begin{tabular}{c c c c c c}
\hline\hline
Field line & & High order Larmor
&  & Resonant terms linked \\
bending & Inertia & radius effects
& Interchange & to compressibility\\
\hline
$\frac{(\rho_i k_{\perp})^2(qR_0k_{\|})^2}{\beta} $ & $(\rho_i k_{\perp})^2$
& $(\rho_i k_{\perp})^4$
&$(\rho_i k_{\perp})(\rho_i k_{\theta})\frac{R_0}{L_p}$&
$\frac{\omega-\mathbf{n} \cdot \mathbf{\Omega}_{\ast}}
    {\omega-\mathbf{n} \cdot \mathbf{\Omega}_s}(\rho_ik_\perp)^2$\\
&& $(\rho_i k_{\perp})^2(\rho_i k_{\theta})\frac{R_0}{L_p} $& &\\
\hline\hline
\end{tabular}
\caption[\footnotesize Ordering of the various term in the acoustic Lagrangian]
{\label{tab_FunctionalOrderings}
\footnotesize Orderings of the functional terms. The Lagrangian terms
are of the order of 
$\mathcal{L} \sim \int d{\bf x} (ne^2/T_{i})$ 
times the values given in the table.}
\end{center}
\end{table}
where, as explained in Eq.~\ref{eq_SeveralOmBAE}, 
compressibility may be combined with inertia and lead  to a  
{\it generalized inertia} 
$\Lambda^2$, such  that (with the table conventions)
$ (\rho_ik_\perp)^2 \rightarrow 
(\omega_A/\omega)^2\Lambda^2 (\rho_ik_\perp)^2$.

We know from section \ref{section_MHDLimit}
that MHD corresponds to the case where FLR are simply expanded to the order
$(\rho_ik_\perp)^2$. This limited expansion is at the origin of the apparent 
{\it continuum damping} which characterizes MHD, and comes from disregarding 
small scales.
In order to model smaller scales and catch in a same picture gap modes and 
continuum modes, the relevant next order is $(\rho_ik_\perp)^4$. As expected, 
the relevance of such scales is related to closeness to the MHD resonant
spectrum or closeness to an accumulation point: indeed, when matching  the 
$(\rho_ik_\perp)^4$ order with generalized inertia, it comes
\begin{equation}
\Lambda^2\sim\beta (\rho_ik_\perp)^2 \ll 1.
\label{eq_ClosenessToAccumulationPoint}
\end{equation}
With a similar matching, now of the field line bending term with 
$(\rho_ik_\perp)^4$, the radial scale which is relevant to 4th order FLR is 
obtained
\begin{equation}
(\rho_i k_{\perp})^2(qR_0k_{\|})^2/\beta \sim (\rho_ik_\perp)^4 \\
\Rightarrow\\
\rho_i k_\perp \sim  \rho_i k_r \sim \beta^{1/4} \gg \rho_i k_\theta,
 qR_0 k^m_\|\sim \beta^{3/4}.
\end{equation}
It now clearly appears that small scales are localized in a region around
$r_s$ of typical length $1/k_r\sim \rho_i\beta^{-1/4}$ (a few millimeters), 
which is large enough to contain the Alfv\'en resonant surfaces (where 
$\Lambda^2=(qR_0k_\|)^2$ if $\Lambda^2$ is {\it positive}) where MHD predicts
a discontinuity for continuum modes.

This particular region where strong gradients are expected will be called 
{\bf inertial region} because it contains inertia. Away from this region,
that is for larger $r \sim 1/k_r$, terms involving $k_r$ 
get smaller and can no longer balance the field line  bending tension, 
such that $\rho_ik_r\rightarrow 0$. Other terms, which are less dependent
on $k_r$, come into play and can set up  the relevant radial scale, 
usually a macro scale relevant to MHD. For this reason this ``external region''
will be refered to as {\bf MHD region}, but one should note that it does no
longer contain either inertia or compressibility.
In particular a lower limit exist on 
$\rho_ik_\perp\geq\rho_ik_\theta\sim\beta$, which  limits the mode radial
extent to $1/k_r\sim 1/k_\theta \sim 0.1-1$m.\\ 

With this estimate of the mode radial scale, we can assess the validity of 
the approximations announced earlier.

To the relevant lower order of the functional terms ($\sim\beta$ following
Tab.~\ref{tab_FunctionalOrderings}), neglecting diamagnetic terms 
(understood as any term involving $\Omega_{*}$) is justified. 
Higher order diamagnetic corrections are uncessary 
because of the scaling \ref{eq_DiamagneticOrdering}, whereas interchange is only a very lower order 
term with the assumption $L_p/R_0>\beta^{1/4}$. In the MHD region, this latter
choice is possibly debatable and it may be more reasonable to consider 
$L_p/R_0\sim a/R_0\sim \beta^{1/2}$, such that interchange may be the higher
order effect to balance field line bending, at the relevant scale 
$1/k_r\sim\rho_i\beta^{-1/2}$. 

Inversely, neglecting the Slow Sound continuum is not fully consistent, but 
will be done for simplicity.
In the derivation, we will simply consider transit corrections with the 
ordering $k_\|v_\|/\omega = O(\rho_i k_\perp)$ which can simply make sense in 
a high q limit.
Note that a purely formal ordering using 1/6 fractions of $\beta$ can be 
developed, but using such an ordering does not change anything in the 
story... 

\section{Derivation of the BAE characteristic equation in the
  inertial  layer}
\label{section_BAEInertialLayer}
In this part, we focus on the inertial region only.

\subsection{Determination of the relevant  fields}
From the previous section, we can derive the relevant field poloidal 
components, using a Fourier expansion and extremalization of the functional 
\ref{eq_LagrangianIntermediateFcy}. 
Expansion to the fourth order FLR effects makes necessary a Fourier 
expansion  to the second order sidebands, 
such that 2 scalar fields ($\psi_\omega, \mathcal{E}_\omega$) for 5 poloidal
components: ${\sf m, m\pm 1,m\pm 2}$.

This calculation is done in Appendix \ref {appchapter_InertialFieldOrderings}
with the addtional assumption on the {\bf  sidebands symmetry}
\begin{equation}
\psi_\omega^{\sf m+1}=-\psi_\omega^{\sf m-1}, 
\psi_\omega^{\sf m+2}=\psi_\omega^{\sf m-2},
\mathcal{E}_\omega^{\sf m+1}=-\mathcal{E}_\omega^{\sf m-1}, 
\mathcal{E}_\omega^{\sf m+2}=\mathcal{E}_\omega^{\sf m-2}
\label{eq_SidebandsSymmetry}
\end{equation} 
This asssumption on the sidebands parity is made possible by the approximation 
$qR_0k_\|^{\sf m\pm1} = qR_0k_\|^{\sf m} \pm 1 \approx \pm 1$
which is consistent with the order of accuracy  of our calculation 
($(\rho_ik_\perp)^4\sim\beta $) and
leads to the degeneracy of the poloidal sidebands. Note however that this
degeneracy removal is also consistent with the inverse parity choice 
$ \psi^{\sf m +1 } = \psi^{\sf m-1}, \psi^{\sf m +2 } = - \psi^{\sf m-2} ...$, 
but that the latter choice does not return modes with an acoustic frequency
\cite{Smolyakov_08a}.\\

This expansion returns the following estimates,
\begin{equation}
\begin{array}{lclllcl}
\psi^{\sf m+1/m+2}_\omega &=& O ((\rho_ik_\perp)^4\psi^{\sf m}_\omega)&,& 
\mathcal{E}^{\sf m}_\omega &=& O((\rho_ik_\perp)^5\psi^{\sf m}_\omega),\\
\mathcal{E}^{\sf m+1}_\omega &=& O((\rho_ik_\perp)\psi^{\sf m}_\omega)&,& 
\mathcal{E}^{\sf m+2}_\omega &=& O((\rho_ik_\perp)^2\psi^{\sf m}_\omega)
\end{array}
\label{eq_SidebandsOrdering}
\end{equation}
We conclude that close to a resonant surface, the main poloidal component
$(\sf n,m)$ of a BAE mode is close to satisfy the MHD constraint
$\mathcal{E}^{\sf m}_{\omega}=0$, while its  ${\sf m}\pm 1,2$  satellites are 
almost electrostatic. Since the same derivation can be done at a neighboring
resonant surface, this means that the mode extension
of the vector potential is smaller than the one of the electric potential.

Moreover, to our order of accuracy (4th order in  $\rho_ik_{\perp}\sim \beta$),
it comes $\mathcal{E}_\omega^{m}=\psi_\omega^{m+1}=\psi_\omega^{m+2}=0$,
which means in particular that the problem is reduced to the determination
of three field components only $\psi_\omega^{m}, \mathcal{E}_\omega^{m+1}$ and
$ \mathcal{E}_\omega^{m+2}$, instead of 6. 
\subsection{Derivation}
\label{ReducedInertialFunctional}
We can now develop the Lagrangian \ref{eq_LagrangianIntermediateFcy} to the 
4th order in $(\rho_ik_\perp)$ using the fields 
$\psi_\omega^{\sf m}$, $\mathcal{E}_\omega^{\sf m+1}$, 
$\mathcal{E}_\omega^{\sf m+2}$ and 
determine the expression of the sideband fields as functions of the main 
poloidal vector potential $\psi_\omega^{\sf m}$.

\begin{description}
\footnotesize \item \quad \quad \ $\blacksquare- $\\
In a cylidrical equilibrium, the curvature verifies
${\bf b}_{\eq}\times\bf{\kappa}
=-\frac{1}{R_0}\left[\sin\left(\theta\right)
\mathbf{e}_{r}+\cos\left(\theta\right)\mathbf{e}_{\theta}\right]$ and implies 
a coupling of poloidal components  via the ${\bf v }_{gi}\cdot\nabla$
operator such that for any poloidal number $\sf m'$,
\begin{equation}
\left({\bf v }_{gi}\cdot\nabla \psi_{\omega}\right)^{\sf m'}
= i\sum_{\epsilon = \pm  1} \omega_{gi, \epsilon}\psi^{{\sf m'}+\epsilon}_\omega
\label{eq_CurvatureCoupling}
\end{equation}
where $\omega_{gi, \epsilon}$ is an operator defined as
$\omega_{gi, \epsilon} = 
\frac{1}{2}v_{gi}(+i\epsilon k_r
+\frac{{\sf m}'+\epsilon}{r}) \approx i\epsilon\frac{v_{gi} k_r}{2}
\equiv + i\epsilon \omega_{di}$.
$i k_r$ stands for the $\partial_r$ operator and 
$ v_{gi} =\frac{-1}{eB_{\eq}R_0}\left(m_i v_{\|}^{2}+\mu_i B_{\eq}\right)$. 

Using formula \ref{eq_CurvatureCoupling} and in the absence of transit 
corrections, Eq.~\ref{eq_LagrangianIntermediateFcy} takes the form
$\mathcal{L}_{\omega}
=\int d^{3}\mathbf{x} \frac{ne^2}{T_{i}} L_{\omega}$,
with
\begin{equation}
\begin{split}
      L_{\omega}= &- \rho_i^2\frac {v_{A}^2}{\omega^2}
      \nabla_{\|}\nabla^2_{\perp}\nabla_{\|}|\psi_\omega^{m}| ^2
      +\left(1-\Gamma_0\right) \left|\psi_\omega^{m}  \right|^2
      +  2\left(1-\Gamma_0+\tau_e^{-1}\right)
      \left[\left|\mathcal{E}_\omega^{m +1} \right|^2
	+\left|\mathcal{E}_\omega^{m+2} \right|^2\right] \\
        &- 2i K_1 \left[\psi^{m}\mathcal{E}_\omega^{m+1\ast}
	-\psi^{m*}\mathcal{E}_\omega^{m+1}
	+ \mathcal{E}_\omega^{m +2}\mathcal{E}_\omega^{m+1*}	
	- \mathcal{E}_\omega^{m +2*}\mathcal{E}_\omega^{m+1}   \right] \\
      &- 2 K_2 \left[ \left|\psi_\omega^{m} \right|^2
	+ 3\left|\mathcal{E}_\omega^{m+1}  \right|^2
	+ \psi_\omega^{m}\mathcal{E}_\omega^{m+2\ast}
	+\psi^{m*}\mathcal{E}_\omega^{m+2} \right]\\
      &- 6i K_3  \left[\psi_\omega^{m}\mathcal{E}_\omega^{m+1\ast}
	- \psi_\omega^{m\ast} \mathcal{E}_\omega^{m+1} \right]
      - 6 K_4 \left|\psi_\omega^{m} \right|^2,
\end{split}
\label{eq_BAE4thOrderLagrangian}
\end{equation}
and $K_n=\left\langle \left[\frac{\omega_{di}}{\omega}\right]^n
J_0^2\right\rangle \sim (\rho_i k_{\perp})^n $.
If  small order transit effects are added, the following additional terms 
should be added to $L_{\omega}$,
\begin{equation}
      L_{\omega\|}=  - 2 L_0
      \left(\left|\mathcal{E}_\omega^{m+1} \right|^2
      +4 \left|\mathcal{E}_\omega^{m+2} \right|^2\right)
        - 2 iL_1 \left[\psi_\omega^{m}\mathcal{E}_\omega^{m+1\ast}
	-\psi_\omega^{m*}\mathcal{E}_\omega^{m+1}   \right]
	-2L_2  \left|\psi_\omega^{m} \right|^2
\label{BAETransitTerms}
\end{equation}
where $L_n =  \langle\left(\frac{v_{\|}}{qR_0\omega}\right)^2
\left[\frac{\omega_{di}}{\omega}\right]^n J_0^2\rangle$,
where the transit operators have been assumed to be of order
$L_n \sim (\rho_i k_{\perp})^{n+2}$.

Next, extremalization with respect to $\mathcal{E}_\omega^{m+1\ast}$ and 
$\mathcal{E}_\omega^{m+2\ast}$ yields
\begin{eqnarray}
\mathcal{E}_\omega^{m+1} &=& i\tau_e\left[(K_1+L_1+ 3K_3)
  + 4\tau_eK_1K_2 -\tau_eK_1(1-\Gamma_0 - L_0 - K^2_1\tau_e)\right]
  \psi_\omega^m \\
\nonumber  &+& O\left((\rho_i k_{\perp})^4 \psi_\omega^m\right)\label{E2}\\
\mathcal{E}_\omega^{m+2} &=& -iK_1\tau_e\mathcal{E}_\omega^{m+1}
 +K_2\tau_e\psi_\omega^m + O\left((\rho_i k_{\perp})^3\psi_\omega^m\right) \label{E1}
\label{eq_PerturbedSidebandsExpression}
\end{eqnarray}
The sidebands are now easily expressed as functions of the main poloidal 
component$\psi_\omega^{\sf m}$, and the Lagrangian can simply be written
\begin{equation}
\begin{split}
      L_{\omega}= &[- \rho_i^2\frac {v_{A}^2}{\omega^2}
      \nabla_{\|}\nabla^2_{\perp}\nabla_{\|}+\left(1-\Gamma_0\right)
      -2K_2 - 6 K_4  - 2L_2 -2\tau_e\left(K_1^2 + K_2^2\right)\\
&-2\tau_e\left(2K_1L_1 +\tau_eK_1^2L_0-\tau_e (1-\Gamma_0)K_1^2
+ 5\tau_eK_1^2K_2 + 6K_1K_3 + \tau_e^2K_1^4 \right)\,]\,\,|\psi_\omega^{m}| ^2.
\end{split}
\label{eq_LagrangianInertialPart_Partial}
\end{equation}

There remains to calculate the explicit expressions of the operators
$K_n$'s and $L_n$'s.
If we make use of the notation
$b = \rho^2_ik^2_r$,  $\Omega=\frac{\omega R_0}{v_{Ti}}$, 
$\zeta=\frac{v_{\|}}{v_{Ti}}$, $h=\frac{v_{\perp}^2}{2v_{Ti}^2}$ 
in the following paragraphe only, it easily comes that
\begin{eqnarray}
  \left\langle ... \right\rangle &=&
  \int_{-\infty}^{+\infty}\frac{d\zeta}{\sqrt{2\pi}} e^{-\zeta^2/2}
  \int_{0}^{+\infty}dh e^{-h} ...\\
  \frac{\omega_{di}}{\omega}      &=&
  - \frac{\sqrt{b}}{\Omega} \frac{h+\zeta^2}{2} \\
  J_0^2 &=& 1-b h + \frac{3}{8} b^2 h^2.
\end{eqnarray}
Hence,
\begin{equation}
\begin{array}{lcllcl}
     \Gamma_0 &=& \left\langle J_0^2 \right\rangle_i = 1-b+\frac{3}{4}b^2
        &K_1 &=& - \frac{\sqrt{b}}{\Omega} \left(1-\frac{3}{2}b\right)\\
        L_0 &=& \frac{1}{q^2\Omega^2}\left(1-b+\frac{3}{4}b^2\right)
	&K_2 &=& \frac{7}{4}\frac{b}{\Omega^2}\left(1-\frac{13}{7}b\right)\\
	L_1 &=& -\frac{1}{q^2\Omega^2}
	\frac{\sqrt{b}}{\Omega}\left(2-\frac{5}{2}b\right)
        &K_3 &=& - \frac{9}{2}\frac{b\sqrt{b}}{\Omega^3} \\
	L_2 &=& \frac{1}{q^2\Omega^2}\frac{b}{\Omega^2}
	\left(\frac{23}{4}-\frac{33}{4}b\right)
        &K_4 &=& \frac{249}{16}\frac{b^2}{\Omega^4}.
\label{eq_Developments}
\end{array}
\end{equation}

Substituting these expressions into Eq.~\ref{eq_LagrangianInertialPart_Partial}
directly returns the functional Eq.~(\ref{eq_LagrangianInertialPart}).
\hfill $-\blacksquare$
\end{description}
The result is 
\begin{equation}
\mathcal{L}_{\omega}
=-\int d^{3}\mathbf{x} \frac{ne^2}{T_{i}}
\frac{\omega^2_A}{\omega^2}\psi^{\sf m*}_\omega
\left[  (qR_0)^2 \rho_i^2
      \nabla_{\|}\nabla^2_{x}\nabla_{\|}
      + \Lambda^2 \rho_i^2 \nabla^2_{x}
      + \sigma \rho_i^4\nabla_{x}^4 \right]\psi^{\sf m}_\omega
\label{eq_LagrangianInertialPart}
\end{equation} where
\begin{eqnarray}
\Lambda^2 &=& \frac{\omega^2}{\omega^2_A}
\left[ 1 - \left(\frac{v_{ti}}{\omega R_0}\right)^2\left(\frac{7}{2}
+2\tau_e\right)
-\frac{1}{q^2}\left(\frac{v_{ti}}{\omega R_0}\right)^4\left(\frac{23}{2}
+ 8\tau_e+2\tau_e^2\right)\right]
\label{eq_Coeff4thOrderLambda}\\
\nonumber \sigma &=& \frac{\omega^2}{\omega^2_A}\left[
\frac{3}{4} - \left(\frac{v_{ti}}{\omega R_0}\right)^2\left(\frac{13}{2}
	  + 6\tau_e + 2\tau_e^2\right)\right.\\
&&\left. \quad\quad\quad\quad\quad\quad\quad
  + \left(\frac{v_{ti}}{\omega R_0}\right)^4
\left( \frac{747}{8} + \frac{481}{8} + \frac{35}{2}\tau_e^2
+2\tau_e^3 \right)\right]
\label{eq_Coeff4thOrderSigma}
\end{eqnarray}
$\Lambda^2$ is obviously the fluid-like {\it generalized inertia} announced 
earlier, whereas $\sigma$ stands the high order FLR.

To determine the relevant equation for $\psi_\omega^{\sf m}$, there
remains to extremalize according to $\psi_\omega^{\sf m*}$.
Some caution is necessary when dealing with electromagnetic waves because
of the ambiguous definition of $\psi_\omega^{\sf m*}$ at the resonant surface,
which simply determines $\psi_\omega^{\sf m*}$, modulo a delta function.
Hence, the clean use of the variational principle, originally based on
variations of $\mathbf{A}_\omega^*$ rather than $\psi_\omega^*$,
leads to the eigenmode equation
\begin{equation}
\left[  (qR_0)^2 \rho_i^2
      \nabla_{\|}\nabla^2_{x}\nabla_{\|}
      + \Lambda^2 \rho_i^2 \nabla^2_{x}
      + \sigma \rho_i^4\nabla_{x}^4 \right]\psi^{\sf m}_\omega = C
\label{eq_InertialEigenEquation}
\end{equation}
where the C constant is some constant which results from this indetermination.

The form of this eigenmode equation is fully consistent with the one
found by F. Zonca {\it et~al.} in Eq.~(9) of Ref.~\cite{Zonca_98} for their
description of the BAE inertial region, where the effects of the C
constant disappears in their formalism because the inertial region
corresponds to large $\theta$ value, in the ballooning representation.
The coefficients found by these authors,
\begin{equation}
\Lambda^2 \text{ and } Q^2 = (k_{\theta}\rho_is)^2\sigma
\label{eq_QDefinition}
\end{equation}
have been verified to asymptotically match ours when
diamagnetic effects are neglected and the
real, high q limit is taken \cite{Zonca_08}.
Hence formul\ae~\ref{eq_Coeff4thOrderLambda}
and \ref{eq_Coeff4thOrderSigma} provide simple
tractable expressions, relevant to the high q limit.\\

At this point, we can already link our kinetic derivation to the fluid 
result of Eq.~\ref{eq_BAEResonantSpectrum}. In the absence of the 4th order 
term in Eq.~\ref{eq_InertialEigenEquation},
a resonant continuum mode (characterized by a discontinuity) can be excited 
if $\Lambda^2>0$, at the localization where $\Lambda^2 = (qR_0k_\|^{\sf m})^2$, 
and a gap exists for $\Lambda^2<0$.
$\Lambda^2$ given in Eq.~\ref{eq_Coeff4thOrderLambda} is of the form 
\ref{eq_BAEResonantSpectrum} with the additional small order transit 
correction $\propto 1/q^2$, and hence allows for a gap below the
accumulation point
\begin{eqnarray}
\omega^2 <\omega^2_\text{accumulation point}&\approx&\omega_{BAE}^2 
\left[1+ \frac{1}{q^2}
\frac{23/2 + 8\tau_e+2\tau_e^2}{7/2+2\tau_e}\right]\\
\text{with }\omega_{BAE}&=& 
 \left(\frac{v_{ti}}{ R_0}\right)
\sqrt{\frac{7}{2} +2\tau_e}
\label{eq_omBAE}
\end{eqnarray}

When now fourth order terms are added, 
discontinuity at the origin of continuum damping disappears, and the separation
between gap modes and continuum resonant modes looses significance.
More precisely, we can say that small scale kinetic effects offer a 
continuous description of the boundary between gap modes and continuum resonant
modes.
Note that it is the same 4th order expansion which makes possible the existence
KTAEs mentionned in subsection \ref{ssection_AlfvenSpectrum}.

\section{Dispersion relation and structure of the Beta Alfv\'en 
Eigenmode}
\label{section_BAEDispRelAndStructure}

As explained in subsection~\ref{ssection_InertialBAEOrderings}, 
Eq.~\ref{eq_InertialEigenEquation} is only valid in a narrow
{\it inertial} region around  the mode  main resonant surface,
whereas {\it incompressible MHD-like terms} are expected to play a more 
important role away from this resonant surface.

We now determine the {\bf solutions of Eq.~\ref{eq_InertialEigenEquation} 
which can match the solutions of the incompressible ideal MHD region}, usually 
constrained by boundary conditions.
For this, we use a classical matching procedure \cite{WhiteBook} 
from which we derive the {\bf mode structure in real (radial) space}.
As will be shown in the following for modes with a given parity,
the asympotic behavior of the ideal MHD region close to a resonant surface depends 
on {\it two} coefficients only
(note that the latter coefficients are not necessarily the mode amplitude and 
slope close to the  resonant surface), which reduce the degrees of freedom
offered by the inertial equation via asymptotic matching.
More precisely, matching these {\it two} coefficients makes 
possible to display a {\it univocal} relation linking the mode frequency and wave 
numbers, the mode {\bf dispersion relation}, derived below.

In the following, the focus is on the resolution of the inertial layer solution.
For this reason, only a reduced model will be used of for the ideal 
incompressible region, which overlooks the geometry difficulties related to the 
treatment of broad mode structures. Nevertheless, the latter model is sufficient 
to describe a mode asymptotic structure close to a resonant surface, 
where strong radial gradients
start to dominate (see Appendix \ref{appchapter_BallooningRepresentation}).
Also, putting aside the details  of radial profiles non uniformity
along with our assumption of {\it symmetry of the poloidal components}
Eq.~\ref{eq_SidebandsSymmetry}, 
makes possible to assume eigenmodes to have a given  (radial) parity in the
surrounding of the resonant surface, which is convenient for a clearer presentation
of the following presentation.

\subsubsection{Ideal incompressible solution}
We explained in section~\ref{ssection_InertialBAEOrderings} 
that MHD-like contributions can set up the larger scale of the mode.
When $\rho_ik_r$ gets smaller, the poloidal wave number number 
$\rho_i k_\theta \sim \beta$ in the field line bending tension, or the interchange 
drive can set up the leading order.
For this reason, we choose to keep these two terms in the ideal incompressible region.
Vorticity (Eq.~\ref{eq_Vorticity}) reduces to
\begin{equation}
\begin{split}
  \nabla_{\|}\nabla_{\bot}^2
    \nabla_{\|}\psi_\omega
  +\mathbf{b}_{\eq}\times \nabla
  \left< \beta \right>
  \cdot \nabla_{\perp} \left(\mathbf{b}_{\eq}
    \times \m{\kappa}\cdot \nabla_{\perp} \psi_{\omega}\right)
  = 0
\label{eq_VorticityIdeal}
\end{split}
\end{equation}

In order to avoid the lengthty character of a poloidal Fourier expansion, we 
solve this equation in the {\bf ballooning representation, which makes possible 
a 1-D treatment} of this equation (instead of the traditional 2D treatment in 
the $(r,\theta)$ space). The basics of the ballooning formalism are
presented in Appendix~\ref{appchapter_BallooningRepresentation}. 
Following this Appendix, when the perpendicular gradients are much larger than the 
parallel gradients (which is the case for modes with a large toroidal mode numbers 
{\sf n}, or close to regions where radial gradients are strong), it is possible to 
transfom Eq.~\ref{eq_Vorticity} into an equation which depends on one single 
variable $\vartheta  \in [-\infty, + \infty] $.
$\vartheta$ is an extension of the geometrical poloidal angle $\theta$, 
regarded as a coordinate along the magnetic field, and 
the transformed field $\psi(\vartheta)$ can be understood as the Fourier 
transform of $\psi_\omega^{\sf m} (x/k_\theta s)$, 
where {\sf m} is the main mode poloidal component we are focusing  on, and 
the quantity $s$ is assessed at the mode resonant surface.
In other words, 
$\psi(\vartheta) = \int_{-\infty}^{+\infty} |k_{\theta}s| dx
\, \psi_\omega^m(x) e^{-i k_{\theta}s x} $.

Following Appendix~\ref{appchapter_BallooningRepresentation}, the ballooning transformation
in an equilibrium with circular flux surfaces, returns
\begin{eqnarray}
\nabla_\|      &\rightarrow& \frac{1}{qR_0}\partial_\vartheta\\
-\nabla_\perp^2 &\rightarrow& k_\perp^2 = k_\theta^2 + k_r^2 
= k_\theta^2   (1+s^2\vartheta^2)\\
{\bf b}_\eq\times\m{\kappa}\cdot\nabla_\perp 
&\rightarrow& \frac{-ik_\theta}{R_0}
\left(\cos\vartheta + s\vartheta\sin\vartheta\right)
\end{eqnarray}
where again $s$ and $q$ are assessed at the localization of the {\sf m} resonant 
surface.
Finally, the ballooning representation of vorticity is
\begin{equation}
-\partial_{\vartheta}(1 + s^2\vartheta^2)\partial_{\vartheta} \psi
-\alpha (\cos\vartheta + s\vartheta \sin \vartheta)\psi = 0
\label{eq_IdealBallooning}
\end{equation}
where  $\alpha = -q^2R_0 \partial_r\beta$.

Using the additional transformation $\Psi = \sqrt{1+s^2\vartheta^2}\,\psi$, 
vorticity finally takes a simpler form
\begin{equation}
-\partial_{\vartheta\vartheta} \Psi + \frac{s^2}{(1+s^2\vartheta^2)^2} \Psi
- \frac{\alpha}{1+s^2\vartheta^2}
(\cos\vartheta + s\vartheta \sin \vartheta)\Psi=0
\label{eq_ModifiedIdealBallooning}
\end{equation}
The form of Eq.~\ref{eq_ModifiedIdealBallooning} is obviously the same as 
the one of Eq.~\ref{eq_SchrodingerWell} and it can be understood as a 
Schr\"odinger potential well. \\
 
Since, we now really want to determine the mode structure and an
univocal dispersion relation, we now go a little bit further than what was done 
in Eq.~\ref{eq_MatchingBallooning}, and display the mode {\bf asymptotic 
behavior} 
close  the mode resonant surface, which corresponds to large $\vartheta$ when 
working in the ballooning (/Fourier) space. 

We can extract the non-periodic part of the solutions of 
Eq.~\ref{eq_ModifiedIdealBallooning} at large $\vartheta$  using solutions of 
the form $\vartheta^{\delta}$.
It directly comes that the non-periodic large $\vartheta$ asymptotic behavior 
of a solution of Eq.~\ref{eq_ModifiedIdealBallooning} can be put in the form
\cite{Fitzpatrick_94},
\begin{equation}
\begin{array}{ll}
\Psi = \Psi_{-} + \Delta\Psi_+,  {\rm with \,} &
\Psi_- \sim \hat\Psi_0 (\vartheta)
|\vartheta|^{1/2-\nu} {\,\,\rm for\, large\,\,} \vartheta \\
& \Psi_+ \sim \hat\Psi_0(\vartheta)|\vartheta|^{1/2+\nu}
\end{array}
\end{equation}
with $\nu=\sqrt{1/4 - \alpha/s^2}$,
$\Delta$ a constant depending on boundary conditions,
$\hat\Psi_0 (\vartheta)=\Psi_0$ for even parity modes and
$\hat\Psi_0 (\vartheta)=\Psi_0 \sgn(\vartheta)$ for odd parity modes.

When it is small, the $\Delta$ constant has a simple meaning.
We already saw that we could access the MHD energy functional multiplying
by the Schr\"odinger-like equation by $\Psi^*$ and integrating by parts.
More precisely, we can write
\begin{equation}
\Psi^*\partial_\vartheta\Psi = \delta W^{bal}_\text{MHD}(\Psi)
\label{eq_Delta}
\end{equation}
for modes with a given parity, with $\delta W^{bal}_\text{MHD}$ the ideal MHD
energy potential relevant to the ballooning representation. 
Using that $\psi(\vartheta)$ is the Fourier transform of 
$\psi_\omega^{\sf m}(x/k_\theta s)$ along with Parceval identity, it is easy to 
see that $\delta W^{bal}_\text{MHD}$ related to the traditional MHD 
energy  potential by
\begin{equation}
\delta W^{bal}_\text{MHD} 
= \frac{|k_\theta s|}{2\pi R_0 r_s}
\frac{1}{\rho_i^2k^2_{\theta}} \frac{T_i}{ne^2}\frac{\omega^2}{\omega_A^2}
\delta W_\text{MHD} 
\label{eq_BallooningNormalization}
\end{equation}

So, to the first order, Eq.~\ref{eq_Delta} returns
\footnote{We define here a {\it normalized} MHD energy which is
  independent from the mode amplitude, $\delta \hat W_\text{MHD}$. 
  We introduce $|s|$ simply to recover the 
  normalization of Ref.~\cite{Zonca_07}}
\begin{equation} 
 \Delta = \delta W^{bal}_{MHD} (\Psi_-)/|\Psi_0|^2 
 \equiv\delta \hat W_\text{MHD} /|s|
\label{eq_AmplitudeNormalization}
\end{equation}

From now on, we assume that $\Delta$ is small
\footnote{This assumption is consistent with our earlier assumption of
  closeness to the BAE accumulation point $\Lambda\rightarrow0$ 
  because of the matching of Eq.~\ref{eq_MatchingBallooning}},
and that the same is true for $\alpha/s^2$, in agreement with the ordering of 
subsection 
\ref{ssection_InertialBAEOrderings}.
It follows  that $\nu \sim 1/2$, and $\Psi_- \sim \hat\Psi_0$ at large $\theta$.
Making such a substitution is equivalent to the ``constant-$\Psi$'' 
approximation used in Ref.~\cite{Chen_94},
and  allows a first estimate of $\delta \hat W$ of the form
\begin{equation}
\delta \hat{W}_\text{MHD} 
= \int_0^\infty d\vartheta \left[\frac{s^2}{(1+s^2\vartheta^2)^2}
-\frac{\alpha(\cos\vartheta+s\vartheta \sin\vartheta)}
{(1+s^2\vartheta^2)}\right]
\end{equation}
Finally, coming back to $\psi$, the {\it large $\vartheta$} behavior of the 
ideal incompressible MHD part is
\begin{equation}
\psi (\theta) \sim
\frac{\hat\Psi_0(\vartheta)}{s}\left(\frac{1}{|\vartheta|}+
\frac{\delta\hat W_\text{MHD}}{|s|}\right).
\label{eq_IdealLargeTheta}
\end{equation}

We can directly deduce the small $x$ behavior of the ideal solution close to the 
resonant surface, taking the inverse Fourier transform as in 
Ref.~\cite{Pegoraro_86, Fitzpatrick_94}.

For modes with the {\bf tearing parity} (odd $\psi$), the ideal small $x$ 
behavior is
\begin{equation}
\psi^{\sf m}_\omega(x) \sim i\frac{\Psi_0}{2s}\sgn(x)
\left( 1 + 
\frac{2}{\pi}\frac{\delta\hat W_\text{MHD}}{|s|}\frac{1}{|k_\theta s||x|}\right)
\label{eq_OddIdeal}
\end{equation}

For modes with the {\bf twisting parity} (even $\psi$), it comes
\begin{equation}
\psi^{\sf m}_\omega(x) \sim \frac{\Psi_0}{s}
\left(\frac{\delta\hat W_\text{MHD}}{|s|}\frac{\delta(x)}{|k_\theta s|} 
- \frac{1}{\pi}\ln|x|\right)
\label{eq_EvenIdeal}
\end{equation}

There now need  to verify if and under what condition solutions of the 
inertial layer can smoothly connect to the MHD solution.

\subsubsection{Solutions of the inertial region and asymptotic matching}

We restrict the resolution of the inertial equation 
\ref{eq_InertialEigenEquation}
to the case $\sigma \geq 0$, or equivalently  $Q^2\geq0$. 
This is a reasonable assumption for usual conditions, and frequencies
of the order of the accumulation point ($\sigma (\omega_{BAE})>0$ for 
$\tau_e>5.45$).

Expanding as usual $k_{\|}$ close to the resonant surface,  
$k_\|= - k_\theta s x/qR_0 $,
the inertial layer eigenmode equation can be written in
a simplified Weber-like form
\begin{equation}
\partial_{\bar x} \left( - {\bar x}^2 - a + \frac{1}{4}
\partial_{\bar x \bar x}
\right)
\partial_{\bar x} \psi_\omega^{\sf m} = \frac{C}{k_\theta^2s^2}=C_\infty
\label{eq_NormalizedInertialEigenEquation}
\end{equation}
where $a=-\Lambda^2/2\sqrt{Q}$ and  ${\bar x} = x/L_I$,
with $L_I$ a {\bf typical size of the inertial region},
\begin{equation}
L_I = \sqrt{2Q}/|k_\theta s|.
\label{eq_InertiaLength}
\end{equation}

For modes with the {\it tearing} parity, $C_\infty=0$.
Eq.~(\ref{eq_NormalizedInertialEigenEquation}) becomes 
$\left( - {\bar x}^2 - a + \frac{1}{4} \partial_{\bar x \bar x}\right)
\partial_{\bar x} \psi_\omega^{\sf m} = C'_\infty$. For $a>-1/2$, it has a 
unique explicit solution for $\partial_{\bar{x}}\psi_\omega^{\sf m} $ 
which cancels at infinity (in agreement with the requested
asymptotic behavior of the ideal solution)
\begin{equation}
\partial_{\bar x}\psi_\omega^{\sf m}
= - C'_\infty \int^1_0 dt
(1-t)^{-\frac{3}{4}+\frac{a}{2}}(1+t)^{-\frac{3}{4}-\frac{a}{2}}
e^{-t{\bar x}^2}
\label{eq_OddSolution}
\end{equation}
and this solution has the following large $\bar x$ asymptotic behavior,
\begin{eqnarray}
\nonumber \psi_\omega^{\sf m}(\bar x) &\sim&  \sgn(\bar x)
\left(\frac{C'_\infty}{|\bar x|}
+ \int^\infty_0 \partial_{\bar x}\psi^{\sf m}_\omega \right)\\
\nonumber &=& C'_\infty \sgn(\bar x)
\left(\frac{1}{|\bar x|} -\frac{\sqrt{\pi}}{2}\int_0^1 dt\
(1-t)^{-\frac{3}{4}+\frac{a}{2}}(1+t)^{-\frac{3}{4}-\frac{a}{2}}
t^{-\frac{1}{2}}
\right)\\
&=& C'_\infty \sgn(\bar x)
\left( \frac{1}{|\bar x|} - \frac{\pi}{2\sqrt{2}}
\frac{\Gamma(\frac{1}{4} + \frac{a}{2})}{\Gamma(\frac{3}{4}+\frac{a}{2})}
\right)
\end{eqnarray}
It can be matched to the asymptotic behaviors of the ideal region taking
the dispersion relation
\begin{equation}
-2\sqrt{Q}\frac{\Gamma(\frac{3}{4}-\frac{\Lambda^2}{4Q})}
{\Gamma(\frac{1}{4} - \frac{\Lambda^2}{4Q})}
=\frac{\delta\hat W_\text{MHD}}{|s|}.
\label{eq_BAEDispersionRelation}
\end{equation}

Similarly, an explicit form of the inertial solution can be found for the 
even parity modes. For $a>-3/2$,
the unique even solution which cancels at infinity is
\begin{equation}
\partial_{\bar x} \psi_\omega^{\sf m}=- C_\infty \bar x
\int_0^1 dt (1-t)^{-\frac{1}{4}+\frac{a}{2}}(1+t)^{-\frac{1}{4}-\frac{a}{2}}
e^{-t{\bar x}^2}
\label{EvenSolution}
\end{equation}
and we want to match this solution to the ideal MHD region.
We first note from the eigenmode equation
(\ref{eq_NormalizedInertialEigenEquation}) considered in the
large $\bar x$ region, that the dominant large $\bar x$ behavior
of $\partial_{\bar x}\psi^m_\omega$
is equal to $ -C_\infty/{\bar x}$, announcing the logarithm
behavior of $\psi_\omega^{\sf m} \sim - C_\infty \ln|\bar x|$, and leading to a first
matching condition to the ideal solution
\begin{equation}
C_\infty = \frac{\Psi_0}{\pi s}
\end{equation}
To recover the relevant coefficient corresponding to the ideal
$\delta$-function
behavior and hence obtain a second matching condition, we first notice that
the ideal behavior can be rewritten
$
\psi^{\sf m}_\omega(\bar x)+C_{\infty}\ln|\bar x|
\sim(C_\infty\pi \delta\hat W_\text{MHD} /
(|s|\sqrt{2Q})) \ \delta(x)
$. 
The corresponding matching condition is
\begin{equation}
\int_{-\infty}^{+\infty}  d\bar{x}\left( \psi_\omega^{\sf m}
+ C_\infty \ln|\bar x|\ \right)
=\frac{C_\infty \pi}{\sqrt{2Q}}\frac{ \delta\hat W_\text{MHD}}{|s|}
\end{equation}
After integration by parts and under the assumption that
$ a>1/2$, the right hand side becomes\\
$-\int_{-\infty}^{+\infty}\left({\bar x}\partial_{\bar x}
\psi_\omega^{\sf m}  + C_\infty \right) \ d\bar{x} $
\begin{eqnarray}
\nonumber &=&-C_{\infty} \int_{-\infty}^{+\infty} d\bar{x}\int_0^1 dt
(1-t)^{-\frac{5}{4} + \frac{a}{2}}
(1+t)^{-\frac{5}{4} - \frac{a}{2}}
\left[(-\frac{1}{4}+\frac{a}{2})(1+t)
+(\frac{1}{4}+\frac{a}{2})(1-t)\right]\\
&=&-C_\infty\frac{\pi}{\sqrt{2}}
\left[(-\frac{1}{4}+\frac{a}{2})\frac{\Gamma(-\frac{1}{4}+\frac{a}{2})}
{\Gamma(\frac{1}{4}+\frac{a}{2})}
+ (\frac{1}{4}+\frac{a}{2})
\frac{\Gamma(\frac{3}{4}+\frac{a}{2})}
{\Gamma(\frac{5}{4}+\frac{a}{2})} \right] =
-C_\infty \sqrt{2}\pi
\frac{\Gamma(\frac{3}{4}+\frac{a}{2})}{\Gamma(\frac{1}{4}+\frac{a}{2})}
\end{eqnarray}
This  leads to the exact same dispersion relation as found for the
tearing parity, ie Eq.~(\ref{eq_BAEDispersionRelation}). \\

\begin{figure}[ht!]
\begin{center}
\begin{minipage}[t]{\linewidth}
\begin{center}
\includegraphics[width=0.55\linewidth]{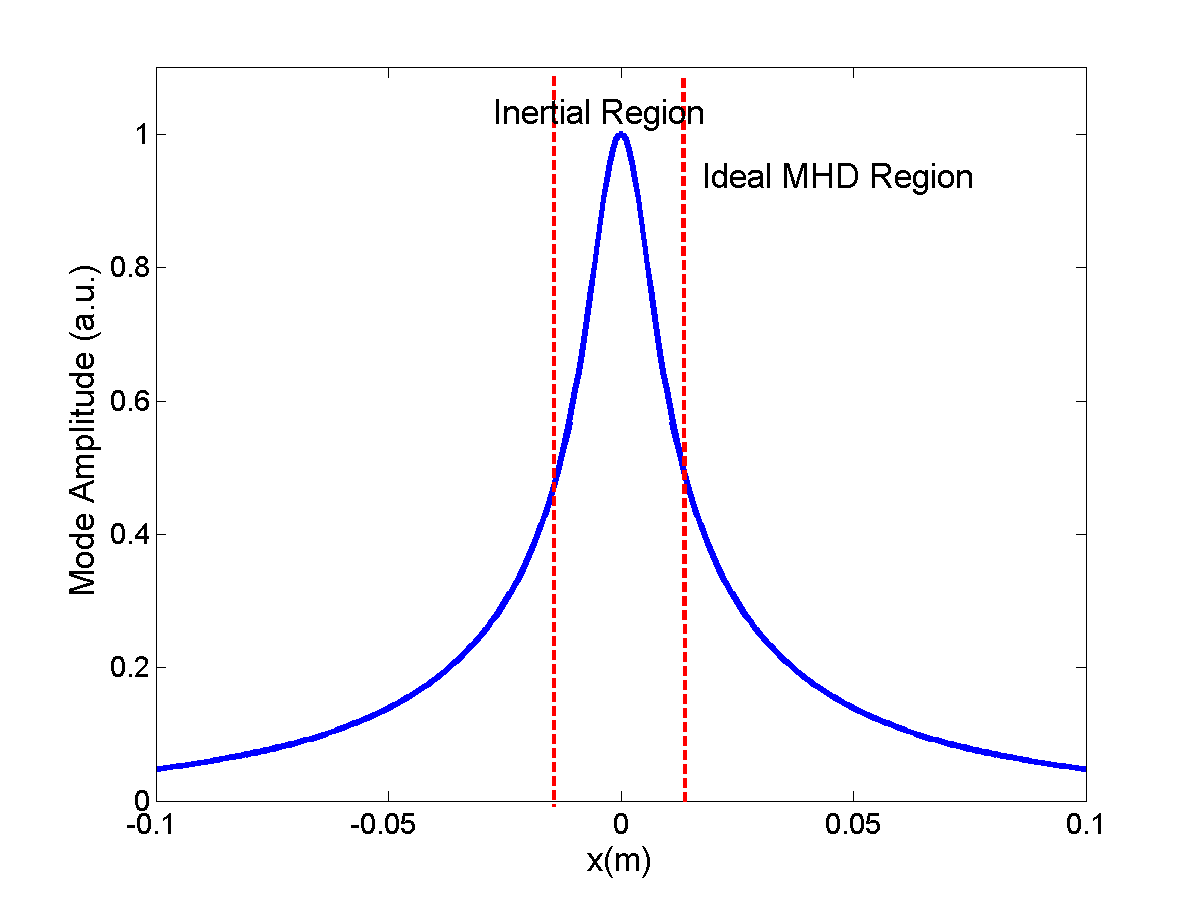}
\caption
[\footnotesize 
BAE structure in the vicinity of the BAE accumulation point, 
for the twisting parity.
]
{\label{fig_AsymptoticMatching}
\footnotesize 
Typical scales of the mode radial structure
for the  twisting parity. The curve is a numerical solution of 
the equation $\partial_{\bar x} \left( - {\bar x}^2 - a + \frac{1}{4}
\partial_{\bar x \bar x} \right) \partial_{\bar x} \psi_\omega^{\sf m}
=  (\sqrt{2|Q|} / s)^2 \  {\bar x}^2 \psi
+ (\alpha/s^2) \psi $ where the contributions of both the inertial
region and the ideal region have been kept. The parameters used are
$\beta = 2\% $, $ \alpha/s^2=0.5$ and
$\sqrt{2|Q|}/s \sim ( \rho_i k_{\theta} )^{3/4} \sim \beta^{3/4}$.}
\end{center}
\end{minipage}
\end{center}
\end{figure}

We finally arrived at the same dispersion relation found in Ref.~\cite{Zonca_98} 
in the ballooning representation, and we recovered the equivalence between even 
and odd parity modes, discovered for a large class of asymptotic problems using 
the Fourier transformation  \cite{Fitzpatrick_94}.
In addition, this derivation produces an image of the BAE radial structure  in 
real space, and shows how high radial gradients, characterizing the inertial 
region, 
can connect to lower gradients of the incompressible ideal MHD region, when the 
accumulation point is approached. An illustration of such a structure is given 
in Fig.~\ref{fig_AsymptoticMatching} for the twisting parity.
A similar two-scale structure could be displayed for the tearing parity, with a 
{\bf step like}-shape, as suggested by the {\it constant} asymptotic 
behaviors in the ideal  region in Eq.~\ref{eq_OddIdeal}.

\subsection{Discussion}
\ssssection{Structure of the Alfv\'en Spectrum}
From the form of the BAE dispersion relation Eq.~\ref{eq_BAEDispersionRelation},
we can make some comments on the role of 4th order kinetic terms on the structure
of the Alfv\'en spectrum.

First, as was pointed out in Ref.~\cite{Zonca_98}, when $\Lambda^2$ is sent to 
$+\infty$ in the BAE dispersion relation \ref{eq_BAEDispersionRelation}, the 
fluid case given in Eq.~\ref{eq_MatchingBallooning} is recovered, 
\begin{equation}
  i|s|\Lambda = \delta \hat W_\text{MHD}
\end{equation}
(and we now have a clean definition of the constants of proportionality used 
before).

When now $\Lambda^2 \rightarrow 0$, 
then $\delta \hat W_\text{MHD}\rightarrow 0$ 
and higher order FLR come into play. In particular, for a very small 
$\delta \hat W_\text{MHD}$, the dispersion relation 
Eq.~\ref{eq_BAEDispersionRelation} can be expanded and returns the solutions 
\begin{equation}
\Lambda^2 = (4l+1)Q 
+ (2/\pi)\sqrt{Q}\ [(\Gamma(l+1/2)/\Gamma(l+1))\delta\hat W_\text{MHD}/|s| \ ]\ ,
\label{eq_GammaExpansion}
\end{equation} 
with $l$ a positive integer \cite{Zonca_98}, at the origin of an infinite number 
of discrete solutions for the mode frequency $\omega$. Hence, the image of a 
separation between a discrete spectrum of {\it undamped} modes  below the 
accumulation point ($\approx \omega_{BAE}$), 
and a continuum spectrum of {\it damped} modes above this point, 
is replaced by a continuous description of the frequency spectrum, where
an  infinite number of discrete {\it undamped} solutions exist.
What we found in this calculation (for $Q^2\geq 0$) is that high order FLR 
enforce a {\it finite} (small) size for the structures, 
such that radial discontinuity (and the correlated damping) are no longer 
relevant. In other words, the mode is localized
in the Fourier $\vartheta$ space, and solution spectrum is consequently discrete.
Higher order FLR  are said to {\bf discretize the continuum}.

In particular, for $l=0$ and $\delta\hat W_\text{MHD} =0$, the solution is 
approximately $\omega^2 \approx \omega^2_{BAE} + \omega_A^2\rho_i
|k_\theta s| \sqrt{\sigma(\omega_{BAE})}$. This can also be rewritten 
$ \omega^2= \omega^2_{BAE} + k^2_\|v^2_A$  with $k_\|$ assessed 
at the typical mode extent enforced by high order FLR and given in 
Eq.~\ref{eq_InertiaLength}, such that the local coupling of the continuum shear  
Alfv\'en type wave with compressibility appears.

\ssssection{Around the causality constraint}
We explained in subsection \ref{ssection_PositiveEnergyWaveDensity} that the sign 
of a wave energy density could be a difficult issue, and often required the use 
of an additional {\it constraint} 
(a localization constraint, or an equivalent localization constraint...).
In particular, for gap modes expected to oscillate with a frequency {\it below} 
the continuous Alfv\'en spectrum and with a generalized inertia of the form 
$\Lambda^2 \propto \omega^2-\omega_0^2$, we saw that causality implied 
$\sgn (\delta W_f + \delta W_k)=\sgn(\delta \hat W_\text{MHD}) < 0$,
in agreement with the fact that $\delta \hat W_\text{MHD} $ 
was to be associated with the mode slope at the frontier between ideal and 
inertial regions, as was shown  in Fig.~\ref{fig_MHDSchrodingerWell}. 
Neglecting transit terms, the BAE inertia clearly enters this category, and 
(undamped) gap  modes  are consequently expected to verify 
$\sgn (\delta \hat{W}_\text{MHD})<0$. 

Recalling that $\delta \hat W_\text{MHD}$ defined here is incompressible
(compressibility has been moved to the generalized inertia), 
a way to understand the causality constraint, is to rewrite the full 
(normalized) MHD energy, 
$\delta W^{full}_\text{MHD} = \delta \hat W_\text{MHD} 
+ \delta W_\text{compressible}$, where we
know $\delta W_\text{compressible}>0$. 
It follows that
\begin{itemize}
\item MHD unstable modes verify $\delta W^{full}_\text{MHD}<0$      
\item MHD gap modes occur for $ 0 \leq \delta W^{full}_\text{MHD} \leq \delta 
W_\text{compressible}$
\end{itemize}
and the role of {\bf compressibility clearly appears to be stabilizing for MHD 
type instabilities, but destabilizing from the point of view of continuum 
damping}.

\begin{figure}[ht!]
\begin{center}
\begin{minipage}{\linewidth}
\begin{center}
\includegraphics[width=0.55\linewidth]{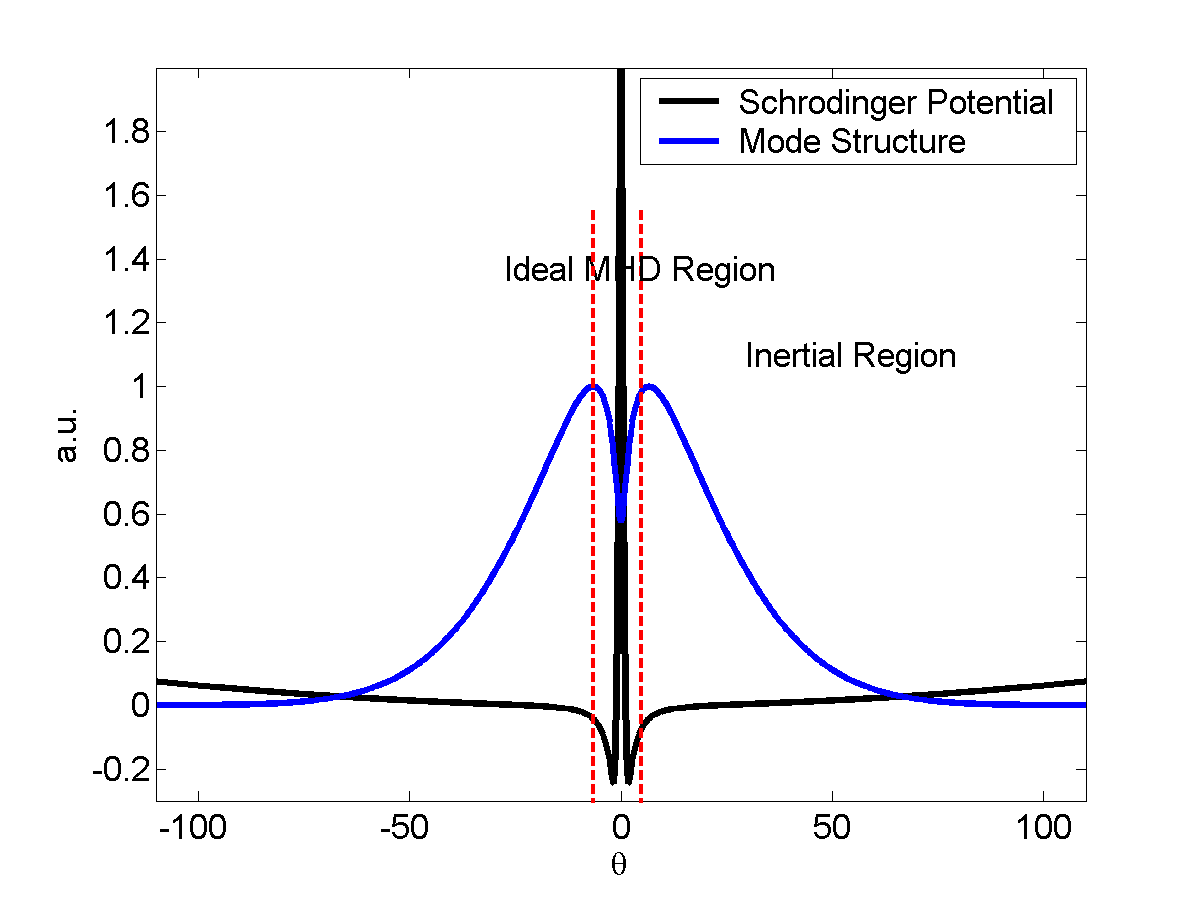}
\caption[\footnotesize
The impact of high order Finite Larmor Radius effects on the 
Fourier space potential and mode structure of the BAE.]
{\label{fig_InertialSchrodingerWell}
\footnotesize 
Schematic of the Fourier space localizing Schr\"odinger-like well and 
structure of a BAE close to the BAE accumulation point, for the twisting 
parity.

The effect of high order Finite Larmor Radius is to enforce the localization
condition in Fourier space, or in other words to set-up a finite small radial
scale.
Thanks to high order FLR, localization is possible even if the slope of the
mode at the boundary between the inertial and MHD regions is not negative.

}
\end{center}
\end{minipage}
\end{center}
\end{figure}

In this derivation however, we displayed undamped modes with a positive 
$\delta \hat{W}_\text{MHD}$ 
(the odd solutions provided above have a validity range which 
extends to $a>-3/2$ and allows for a positive $\delta \hat W_\text{MHD}$ in 
Eq.~\ref{eq_BAEDispersionRelation}. This is due to high order FLR. However, 
these effects do not completely modify the previous picture.
 $\delta \hat W_\text{MHD}$ 
can still be associated with the mode slope at the junction between the inertial 
and ideal MHD regions. However, now, inertia itself can enforce localization, 
as illustrated in  Fig.~\ref{fig_InertialSchrodingerWell}.

\section{The degeneracy of Beta Alfv\'en Eigenmodes and Geodesic Acoustic Modes}
\label{section_BAEGAMDegeneracy}
We explained in the introduction of this chapter that a reason for the 
complexity of the acoustic range is the variety of modes and mode names which 
have been attributed to acoustic oscillations.
We compare in this section the well known GAMs with BAEs, where again GAMs are 
acoustic oscillations with a main ${\sf m=0, n=0}$ component, which corresponds to
usual case treated in transport theories.
More precisely, it is possible to display an {\bf electrostatic} acoustic 
solution of our equations, with a main ${\sf m =0, n=0}$ component. Our following 
comparison will have to be understood as a comparison between {\bf electrostatic 
GAMs taken in the electrostatic limit and BAEs}. Hence, if not otherwise noted,
the word GAM will have to be understood in the electrostatic limit.\\

We show how the {\bf degeneracy of the BAE and GAM dispersion relations}, claimed 
in  Ref.~\cite{Zonca_08}, can be understood in the light of our formalism.
Even if the difference between GAMs and BAEs may be thought to be small 
(it is simply a difference in mode numbers), the idea of a degeneracy is not so
trivial, because a change in mode numbers modifies the response of the electrons 
in the wave frame. In particular, for our derivation of BAEs, we used the time 
scaling
\begin{equation}
k_\|v_{ti} \ll \omega \ll k_\|v_{te}
\end{equation}
(where $k_\|$ is understood as an operator and stands for any of the poloidal 
components), which leads to an {\it adiabatic}  electron response and a
{\it hydrodynamic}  ion response. 
Electron adiabaticity can obviously not hold for  ${\sf m=0, n=0}$. 

Such a degeneracy can have important implications.
As will be clearer in the next chapter, mode excitation by fast particles is
easier for modes with finite $(\sf n, m)$ than for $({\sf n=0,m=0})$ modes.
If the BAE/GAM degeneracy is verified, then we could expect wave-wave coupling
to lead to an excitation of GAMs, which are well known to have an impact on
turbulent transport. 

\subsection{The BAE/GAM degeneracy}

\subsubsection{Basis of the BAE/GAM degeneracy}   
We start our analysis with Eq.~\ref{eq_LagrangianIntermediateFcy}, which
applies for both GAMs and BAEs.

As follows from the preliminary comments, a major difference between BAEs
and GAMs is the electron response to the main $\sf m=0, n=0$ field component. 

For GAMs in the electrostatic limit, the electron adiabatic response, 
ie. the term proportional to 
$(1/\tau_e) |\mathcal{\tilde{E}}_\omega^{\sf m}|^2$
in Eq.~\ref{eq_LagrangianIntermediateFcy},
is balanced by the last term
of Eq.~\ref{eq_LagrangianIntermediateFcy}, when the latter is calculated for 
$\omega \gg {\bf n}^*\cdot \m{\Omega}_e$.
Consequently, the {\bf total electron response cancels}. 

Inversely, for BAEs, we assumed the electron response to be fully 
adiabatic, such that the $(1/\tau_e)|\mathcal{E}_\omega^{\sf m}|$ 
term should {\it a priori} be conserved. However, electron adiabaticity
($\omega \ll k_\|v_{te}$)
exactly means that electrons respond very {\it fast} to the perturbation
in the parallel wave frame (ie. following the wave phase velocity
in the parallel direction), and can cancel parallel charge separation 
and the correlated parallel electric field, $E_\|=0$. 
Consistently, we found from our derivations of BAEs that 
$\mathcal{E}_\omega^{\sf m}=0$, such that the electron reponse is null in
the functional Eq.~\ref{eq_LagrangianIntermediateFcy}.

So finally, the difficulty for the electron response disappears.\\

From a more formal point of view, 
we found that the BAE problem could be reduced to the determination 
of three fields, $\psi_\omega^{\sf m}, \mathcal{E}_\omega^{\sf m+1}$  and 
$\mathcal{E}_\omega^{\sf m+2}$, instead of 6. 
The same is true for electrostatic GAMs, for which it is necessary to determine 
the three components of $\phi_\omega = -\mathcal{E}_\omega$.
Using the approximations of section \ref{section_ApproximationsForBAEDerivation}  
(without any necessity to relate small parameters to $\beta$ since
the field line bending tension now cancels for the main mode), 
the development of the GAM functional exactly returns 
Eq.~\ref{eq_BAE4thOrderLagrangian} with the following substitution
\begin{equation}
\mathcal{\phi}^{\sf m+1 / m+2}_{\omega, \sf GAM}
=-\mathcal{E}^{\sf m+1 / m+2}_{\omega, \sf BAE},\\
\phi^{\sf m}_{\omega, \sf GAM}=\psi^{\sf m}_{\omega, \sf BAE},
\end{equation}
It follows that the inertial BAE Lagrangian given in 
Eq.~\ref{eq_LagrangianInertialPart} and the characteristic equation 
\ref{eq_InertialEigenEquation} hold for GAMs as well, {\bf to the  4th order in 
$\rho_i k_r$}. 

In this sense, we observe a degeneracy of the BAE and GAM modes. It does not mean
however that the two modes have the exact same eigenfunctions. 
In particular, for GAMs, $k_\|^{\sf m}=0$ in Eq.~\ref{eq_InertialEigenEquation}.
The field line bending and interchange ($k_\theta =0$) terms do not enforce 
the existence of a larger radial scale where $(\rho_ik_r)^2 $ terms become
negligible, as it is the case for BAEs. 
Hence, if is makes sense to  neglect equilibrium profiles radial 
variations (as we have done until now), there is no need to distinguish two 
regions.

Finally, we obtained that {\bf the BAE/GAMdegeneracy is verified at the local 
points where  $k^{\sf m}_{\|,\sf BAE}=0$, 
in the case where diamagnetic effects are negligible.}

\subsubsection{Limits of the BAE/GAM degeneracy}
In order to capture  the full  extent of the BAE/GAM degeneracy, we may 
come back to some of our assumptions to see which one may be considered weaker.

{\bf Diamagnetic effects} $\propto {\sf n}$ will clearly not have the same 
effects on GAMs and BAEs. However, in the range of parameters depicted here, 
their removal has been justified. Weaker assumptions may be the 
{\bf electron adiabaticity} or the {\bf GAM electrostatic feature}.
In particular, we took electrostaticity as a starting point.
Since Eq.~\ref{eq_InertialEigenEquation} has solutions for $\sf n=0, m=0$, 
the electrostatic limit is not incorrect. 
However, we did not prove that electromagnetic fluctuations are
impossible for GAMs (in the general sense).

One of the basis of the BAE/GAM degeneracy derived above is the possibility 
to reduce the problem to the description of three poloidal components only.
The cancellation of $\psi_\omega^{\sf m+1/m+2}$ is reasonable and weakly depends  
on the main mode numbers. Indeed, it comes from Eq.~\ref{eq_VorticityProjected}
of the Appendices, 
that the cancellation of an ${\sf m}_i$ ($ i\in ({\sf m, m+1, m+2})$) 
magnetic fluctuation is valid when 
$qR_0k_\|^{{\sf m}_i}/\beta$ is large .
Since $qR_0k_\|^{{\sf m}_i} \approx \pm 1 \text{ or } \pm 2$, regardless on the
main mode poloidal component, it is reasonable to take 
$\psi_\omega^{\sf m+1, m+2}=0$.

Consequently, four fields remain
\begin{equation}
\psi_\omega^{\sf m}, \mathcal{E}_\omega^{\sf m}, \mathcal{E}_\omega^{\sf m+1}, 
\mathcal{E}_\omega^{\sf m+2},
\end{equation}
and our model for BAEs considers $\mathcal{E}_\omega^{\sf m}=0$, whereas GAM
electrostaticity leads to $\psi_\omega^{\sf m}=0$. 
What we verified, is that the cancellation of one of these two components leads
to same final dispersion equation.\\

However, these cancellations can break.
\begin{itemize}
\item $\mathcal{E}_\omega^{\sf m}=0$ relies on the fast (massless) 
response of the electrons (see Eq.~\ref{eq_ElectroneutralityProjected}), to the
$\sf m=0, n=0$ fields.
This cannot hold for GAMs (in the opposite). 
For BAEs, it may also be questioned because electron
adiabaticity cannot really be considered to any order, in particular close to a 
resonant surface. {\bf Electron inertia} may lead to some corrections of the BAE
dispersion relation which do not behave the same way for (n=0,m=0) modes and for 
finite {\sf n} modes. These ideas have been developed in  
Ref.~\cite{Smolyakov_09}.
\item $\psi_\omega^{\sf m}=0$ relies on the importance of 
$qR_0k_\|^{\sf m}/\beta$ which is not verified for an $(\sf n=0,m=0)$ component.
It means that the electrostaticity of GAMs is not enforced.
\end{itemize}

Electromagnetic GAMs were found in JET experiments described in 
Ref.~\cite{Boswell_06}.
To the extent of our analysis, there is no garanty that they may have the same 
dispersion as the one we derived for BAEs.
In Ref.~\cite{Boswell_06} however, the classical MHD BAE dispersion relation was
used,  and we already said that it was in agreement with our calculation.
The reason for this agreement is simply that the authors of this article
started from an MHD formalism... 
hence with $\mathcal{E}_\omega^{\sf m}=0$ assumed {\it a priori}.

\subsection{The GAM eigenmodes}
We already discussed the locality of the BAE/GAM degeneracy. In particular, the 
GAM eigenmodes do not need to be the same as the one we described
above.
The GAM charateristic equation \ref{eq_InertialEigenEquation} can be rewritten
$  (\Lambda^2 \rho_i^2 \nabla^2_{x}
  + \sigma \rho_i^4\nabla_{x}^4)  \psi^{\sf m}_\omega = 0
$, which simply leads to a wave-like solution 
\begin{equation}
\omega^2
\approx \omega^2_{BAE} + \omega_A^2\sigma(\omega_{BAE})
\rho_i^2k_r^2
\label{GAM_dispersion_relation}
\end{equation}
characterized by the group velocity
\begin{equation}
v_{gr} = \frac{\partial \omega }{ \partial k_r}
= \frac{T_i}{eBR}
\sqrt{\frac{7}{2} + 2\tau_e} \frac{\omega^2_A}{\omega^2_{BAE}}
\sigma(\omega_{BAE})\rho_i k_r
\end{equation}
In general standard conditions (for which $\sigma(\omega_{BAE})\geq 0$), this
means an outward propagation, but it is interesting to note that 
a change of sign of $\sigma(\omega_{BAE})\geq 0$
is {\it a priori} not impossible for special conditions.

So, contrary to BAEs, GAMs are not localized to our order of approximation.
In reality, a radial shearing effect similar to the one described 
in section \ref{ssection_AlfvenSpectrum}
can take place (leading in particular to conversion to kinetic waves), 
if the radial variation of equilibrium profiles is not 
neglected. In this case, the continuous spectrum is simply $\Lambda^2 =0$, or
$\omega^2 = \omega_{BAE}^2(r)$ and similarly as for usual gap modes
with a main ${\sf m}$ poloidal component, localization is possible where
$(\omega_{BAE}^2)'=0$ \cite{Boswell_06, Zonca_08}.

\subsection{Distinguishing BAEs from GAMs}
\label{ssection_DistinguishingBAEGAM}
If we are to believe the BAE/GAM degeneracy, distinguishing them based
on their oscillation frequency may be an issue, since both modes
eigenfrequencies are expected be dominated by $\omega_{BAE}$ whereas
the distinguishing corrections $k_r$, $\delta\hat W_\text{MHD}$ are
experimentally subject to strong errorbars.

Hence, instead of considering their frequency to distinguish BAEs and
GAMs, it is more relevant to look at their {\bf structure}.

With our definitions, the main difference between BAEs and GAMs is
their toroidal mode number, but unfortunately the latter cannot be
accessed in Tore-Supra, as explained in section \ref{ssection_Diagnostics}.

The radial structure of the BAEs and GAMs also differs. 
We showed  in this paper that BAEs are radially localized, 
whereas GAMs tend to have a wave structure. 
This property of the GAMs, which we found here in the electrostatic
limit, is intrisically linked to the cancellation of the field line
bending term, and could consequently be relevant for GAMs with a
magnetic component. It makes sense as long as equilibrium profiles do
not present a localizing extremum.

Finally, we explained that the chosen symmetry Eq.~\ref{eq_SidebandsSymmetry}
was relevant to the acoustic frequency range, when the shear is limited. 
It indicates that both the
structure of electrostatic GAMs and  the {\it inertial} structure of  BAEs
are characterized by a poloidal symmetry
$\psi_{\omega} \approx \psi^{\sf m}_{\omega} 
+ 2i\psi_\omega^{\sf m+1}\sin\theta...$, and this property was also
found to be relevant to GAMs with electromagnetic fluctuations
(in Ref.~\cite{Smolyakov_08a}). For BAEs however, this symmetry does
no longer hold in the MHD ideal region, where interchange starts being
effective and implies a coupling to the $\cos\theta$ parity. 
A consequence of this difference in poloidal polarization is that the
observation of fluctuations in the torus equatorial plane (where $\theta=0$)
can only be caused by BAEs, which are sensitive to diamagnetic 
or interchange terms, in particular in the ideal MHD region.

\section{Identification of acoustic modes observed in Tore-Supra }
\label{section_IdentificationToreSupraModes}

Our previous analysis makes possible the analysis of the acoustic
oscillations measured in Tore-Supra.

\subsection{Characteristics of Tore-Supra acoustic modes}
\label{ssection_ToreSupraAcousticModes}
First acoustic fluctuations of the density \cite{Sabot_06,Sabot_09} and the 
temperature \cite{Udintsev_06} were performed in 2005.
During the PhD work, new experiments were conducted which will be reported on
in this thesis.

Tore-Supra modes have been observed in the presence of ICRH Heating in
the hydrogen minority heating scheme (see section \ref{ssection_ToreSupra}), 
in various equilibrium conditions.
Typical observation conditions were for 
a central toroidal field $B_0 \in $ [2.8, 3.8] T, 
a central density        $n_0 \in $ [4.8e19,  5.5e19]  m$^{-3}$, 
a hydrogen minority fraction  $f_{min}$ 
(fraction of the hydrogen density to the main ion density, 
deuterium) $\in$ [3, 10] \%, 
and an ICRH  power $P_{ICRH} \in$ [1.5, 5.0] MW.\\

\begin{figure}[ht!]
{\footnotesize Shot \#41925: $B_0=3.80T$, $n_0=5.04e19m^{-3}$, $I_p=0.95MA$, 
$P_{LH}=0.0MW$, $P_{ICRH}=3.9MW$, $f_{min}=4.8\%$}
\begin{center}
\begin{minipage}[b]{0.5\linewidth}
\begin{center}
\includegraphics[width=\linewidth]{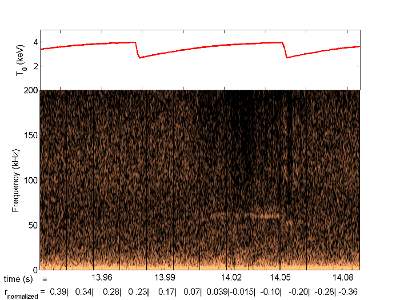}
\end{center}
\end{minipage}
\begin{minipage}[b]{0.49\linewidth}
\begin{center}
\includegraphics[width=\linewidth]{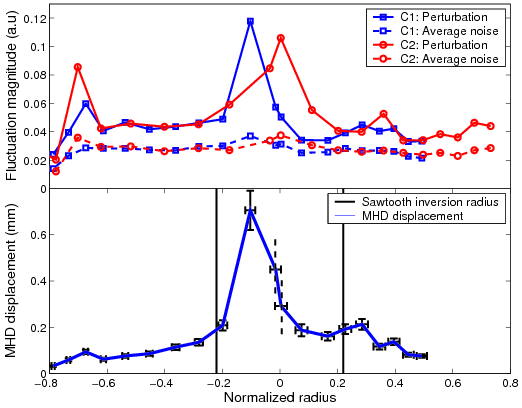}
\end{center}
\end{minipage}
\caption 
[\footnotesize
Fluctuation spectrum and radial structure of the BAE density
fluctuations in Tore-Supra for shot \#41925.
]
{
\label{fig_SpectrumAndRadialStruc41925}
\footnotesize
In the left hand side figure, the time spectrum of density oscillations has
been plotted, for the different radii successively targetted by the 
reflectometer. -/+ are used to distinguish high field side radii from low 
field side radii.\\
In the right hand side figure, the mode radial structure is obtained from the 
determination of the maximum density fluctuation level measured at each radius, 
in the acoustic frequency range.
The picked maximum acoustic fluctuations ({\it full lines}) and average ambient 
fluctuation level ({\it dotted lines})
are obtained from two reflectometry channels, C1 ({\it squares}) and C2 
({\it circles}).
The MHD displacement corresponding to the fluctuations given by C1 is derived
using Eq.~\ref{eq_ReflectoMHDDisplacement}.
Horizontal error-bars correspond to a reflectometer in X-mode. Unbounded vertical 
errorbars indicate a prohibitive  uncertainty at the plasma center.
}
\end{center}
\end{figure}

We represented in Fig.~\ref{fig_SpectrumAndRadialStruc41925} a typical 
observation performed with X-mode reflectometry, in {\it mode 2} 
(see section \ref{ssection_Diagnostics}) using 15ms windows for each radius.
We also represented in this set of figures the central temperature $T_0$ which
is subject to a series of crashes and relaxations, called 
{\bf sawtooth oscillations}. Sawtooth oscillations are well known macroscopic tokamak 
{\bf MHD oscillations} which systematically occur in Tore-Supra standard dicharges 
(in the presence of ICRH power or not) on which we
will be spending some more time in the next chapter.
As will be shown, the stability of BAEs has been observed to depend on the 
sawtooth dynamics, which calls for some caution when describing the modes.

With the parameters given above, Tore-Supra modes have been found to belong to 
the frequency range [45, 65] kHz, in agreement with the {\bf acoustic frequency} 
$\propto v_{ti}/R_0$, as illustrated in the left hand side of
 Fig.~\ref{fig_SpectrumAndRadialStruc41925}.

A typical mode structure, derived from the measure of the acoustic fluctuation
level is represented on the right of Fig.~\ref{fig_SpectrumAndRadialStruc41925},
as a function of the normalized radius $r/a$, where $a$ is the plasma radius.

In order to get rid of the role of the sawtooth dynamics (which can be expected
to play a role in the mode amplitude), two channels of the reflectometer have been
used, targetting different radii at different times.
The main features of the observed are well illustrated by the figure:
a {\bf macroscopic scale of several centimeters} (in general 1 to 15cm), 
{\bf a central localization bounded by
the $q=1$ surface} (identified in these figures by the sawtooth inversion radius)
and {\bf a relatively low amplitude} (with an estimated maximum MHD displacement of
the order of one millimeter) even for large ICRH power input 
($P_{ICRH}=3.9$MW in shot $\#41925$).
In several shots, we could find some indications that the mode fluctuations were
stronger in the torus high field side, but this was not a systematic observation.

\subsection{Identification}
\label{ssection_BAEIdentificationToreSupra}
\subsubsection{Analysis of the frequency spectrum}
\begin{figure}[ht!] 
\begin{center}
\begin{minipage}{0.48\linewidth}
\begin{center}
\includegraphics[width=0.95\linewidth]{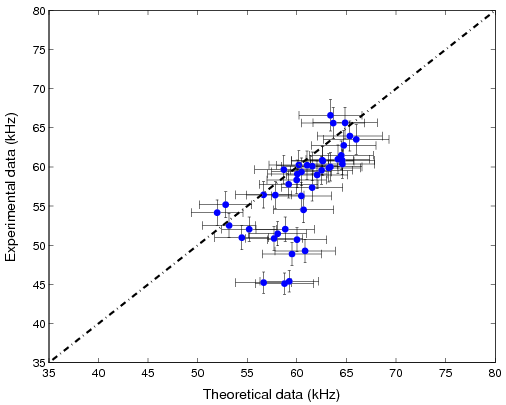}
\caption
[\footnotesize BAE frequency spectrum in Tore-Supra.]
{\label{fig_FcySpectrum}
\footnotesize 
Comparison of $\omega_{BAE}$ with the experimental frequencies 
(averaged over 10ms windows, to avoid the details of the mode 
frequency  evolution faster than the temperature relaxation time) 
between 40kHz and 80kHz and exceeding a signal/noise 
value of 3., for the full set of our experiments with 
$P_{LH}=0$. Horizontal errorbars correspond to an assumed 30\% error
on the value of $T_e/T_i$.}
\end{center}
\end{minipage}\hfill
\begin{minipage}{0.48\linewidth}
\vspace*{-1.9cm}
\begin{center}
\includegraphics[width=\linewidth]{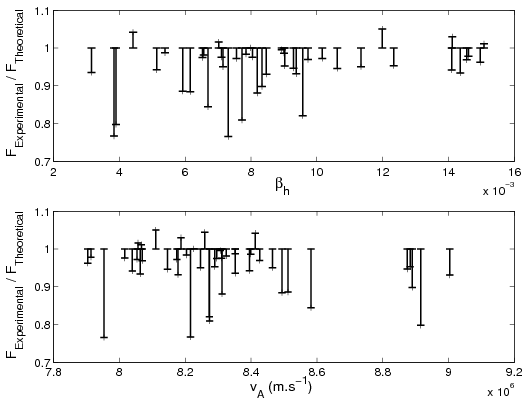}
\caption
[\footnotesize Analysis of shift between the BAE frequency spectrum in Tore-Supra 
and the theoretical value of the BAE accumulation point]
{\label{AnalyzedFcySpectrum_simple}
\footnotesize Distance of experimental points to $\omega_{BAE}$
as a function of the Alfv\'en velocity $v_A$, 
and the estimated normalized fast ion pressure $\beta_h$. }
\end{center}
\end{minipage}
\end{center}
\end{figure}

Let us now examine the {\bf mode frequency spectrum}.
Our orderings for the acoustic frequency range apply well to Tore-Supra, and 
it is consequently reasonable to expect diamagnetic effects to play a minor 
role. 
Hence, it makes sense to postulate that the measured fluctuations are of the
BAE/GAM type. To make this verification, we plotted in 
Fig.~\ref{fig_FcySpectrum} the experimental frequency spectrum as a function of the 
dominant contribution to the BAE/GAM frequency, $\omega_{BAE}$ (Eq.~\ref{eq_omBAE}), 
using the (ECE) experimental measure  of  $T_e$ and the assumption that 
$T_i=0.8T_e$. 
The latter assumption, verified for shot $\#42039$  (a reference shot for 
our subsequent analysis) using the CRONOS transport code \cite{Basiuk_03} 
(which returned $T_i=0.82T_e$ for this shot), 
is reasonable in  Tore-Supra for the densities covered in 
these experiments (ie, {\it large} densities, which allow sufficient
collisions between electrons and ions to allow $T_{i}$ to be no too
far from $T_{e}$). 
What this shows, is that {\bf ions are not cold in our
experiments}.
More precisely, an average over 10ms windows of the spectrum and values of 
$\omega_{BAE}$ has been used to avoid the details of the sawtooth dynamics 
(which could modify $\delta \hat W_\text{MHD}$  in the BAE dispersion 
relation), and simply portray the main tendencies of the modes frequency.

Fig.~\ref{fig_FcySpectrum} shows a good agreement between the experimental data 
and $\omega_{BAE}$, with an error which does not exceed 
$10 \%$  in most cases and may be attributed to a 30\% incertainty on the 
knowledge of $T_e/T_i$. In particular, the increase of the mode frequency with
temperature is well reproduced. 
{\bf This validates the assumption that the observed modes belong the BAE/GAM 
family}.

Note however that the fact that most experimental points are below $\omega_{BAE}$ 
(as was also found in \cite{Buratti_05} and \cite{Annibaldi_07}) may suggest the 
necessity to add some corrections to the BAE/GAM formula~\ref{eq_omBAE}: 
Alfv\'enic \cite{Heidbrink_93}, diamagnetic or finite Larmor radius corrections 
\cite{Annibaldi_07}, or the effect of energetic particles \cite{Fu_08} 
(entering the BAE dispersion relation in $\delta \hat W_\text{MHD}$ for example). 
To investigate the role of such effects, the distance of the experimental points
to $\omega_{BAE}$  is plotted in Fig.~\ref{AnalyzedFcySpectrum_simple}, as a
function of the Alfv\'en velocity $v_A$,
the normalized energetic ion pressure $\beta_h=2\mu_0n_hT_h/B_0^2$ 
(The estimate of $\beta_h$ used for this experimental work will be described 
more clearly in the next chapter.).
No obvious contribution of $v_A$ or $\beta_h$ is unveiled by this analysis. 
In particular, it was shown in \cite{Fu_08} that a new branch of the GAM family, 
called {\bf Energetic Geodesic Acoustic Mode}, and characterized by a much lower 
frequency (of about half or a third of $\omega_{BAE}$) was reached for 
sufficiently high values of $\beta_h$. No such behaviour in discriminated in 
Fig.~\ref{AnalyzedFcySpectrum_simple},  
and we can consequently exclude the possibility that the  modes
studied in here are purely energetic modes, similar to EGAMs.\\

\subsubsection{GAMs or BAEs?}
There remains to distinguish between BAEs and GAMs. 
As explained earlier, the toroidal mode number of acoustic modes cannot be 
identified in Tore-Supra. However, following section 
\ref{ssection_DistinguishingBAEGAM}, we can find some clues indicating that
the observed are BAEs rather than GAMs. 

First, Tore-Supra reflectometry measurements are performed in the tokamak
equatorial plane, where GAM fluctuations are expected to be negligible.

Secondly, in all the conducted shots, modes were found to be localized 
with a  (possibly off-axis) maximum inside the $q=1$ surface. This suggests
a role of the rational $q$ surfaces (possibly resonant surface).
Moreover, we know that GAM localization depends on the existence of an extremum
of $\omega_{BAE}(r)$. Accordingly, in experiments showing GAMs with a global 
structure or simulations of GAMs , the existence 
of GAMs was related to either a non-monotic $q$ profile \cite{Boswell_06}, 
or non-monotonic temperature profile \cite{Huysmans_95}.
Such non-standard conditions are not expected in our experiments, which 
correspond in most cases to discharges with on axis ICRH heating 
(and without any off-axis current drive).

Finally, as will be clearer in the next chapter, the excitation of an
$\sf n=0, m=0)$ mode by energetic particles 
is a difficult process, and most experimental observations of GAMs 
with a global structure performed until now, required specific heating conditions 
(counter injection of Neutral Beam Ions \cite{Nazikian_08}
or high field ICRH heating \cite{Boswell_06}).
In our experiments, the mode excitation did not require such
constraining  heating characteristics. 

As a conclusion, {\bf Tore-Supra acoustic modes are identified as BAEs.}

\section{Summary}

{
\it 
In this chapter, we derived the dispersion relation and structure of 
Beta Alfv\'en Eigenmodes, in the absence of energetic particles, using 
a kinetic model for the plasma, which retained high order Finite Larmor
Radius effects and low order transit effects, corresponding to 
the approximation $\omega \gg v_{ti}/qR_0$, but no diamagnetic effects.

\begin{itemize}
\item Our derivation of the {\bf BAE dispersion relation} 
  carried out using a decomposition of the mode structure
  onto its poloidal Fourier components returned the same dispersion relation
  as the one found in Ref.~\cite{Zonca_99} using the ballooning formalism,
  if the latter is taken in the high q-limit, without diamagnetic effects.

\item The calculation of the {\bf mode structure} was carried out explicitely, 
  displaying the mode two-scale radial structure, where one may distinguish
  an {\bf inertial region} dominated by strong radial gradients and a 
  {\bf smoother region corresponding to MHD scales}.
  In particular, it allowed to determine the typical size of the inertial 
  region (correctly assessed with the kinetic formalism)
  as well as the geometric constants of relevance, 
  which enter the dispersion relation, both for modes 
  with an even shape and modes with an odd shape around the mode inertial 
  layer.
  
\item Next, we compared Beta Alfv\'en Eigenmodes with Geodesic Acoustic Modes.
  We showed that the {\bf eigenmode equation of both modes is degenerate}, 
  because of
  a formal similarity between the electrostatic limit and the approximate
  cancellation the parallel electric field for BAEs, the so-called MHD
  property.

\item As a consequence of the BAE/GAM degeneracy, the two modes are not easy
  to distinguish in Tore-Supra.
  Thus, we studied the features that can be analyzed in experiment for that 
  purpose and concluded that {\bf the modes found in Tore-Supra are BAEs}.
\end{itemize}

Let us now determine the effects of a fast particle population on
the modes stability.
}


\begin{savequote}[20pc]
\sffamily
Les th\'eories ont caus\'e plus d'exp\'eriences que les
exp\'eriences n'ont caus\'e de th\'eories.
\qauthor{Joseph Joubert (1754-1824)}
\end{savequote}




\chapter{Linear stability of Beta Alfv\'en Eigenmodes in the presence 
of energetic particles}
\label{chapter_LinearStability}

Let us finally approach the problem of the role of energetic particles in the 
modification of the stability properties of a plasma.
We already mentioned the possibility of introducing a contribution of 
the energetic particles in the mode dispersion relation to build up the 
{\bf fishbone-like dispersion relation} in Eq.~\ref{eq_NNFishboneDispRel}, 
where we essentially replaced the traditional $\delta W_\text{MHD}$ of the MHD 
energy principle, by $\delta W_f + \delta W_k$, without much justification.
It is tempting to make the same manipulation in the BAE dispersion relation 
Eq.~\ref{eq_BAEDispersionRelation} which obviously couples an inertial part
and energy potential energy $\delta \hat W$. 
We already anticipated some modifications of the mode behavior involved by such
an additional contribution in the previous chapters, 
such as the destabilization of MHD modes by the energetic particles, as well as 
the emergence of new modes (such as the EGAMs mentioned in subsection 
\ref{ssection_BAEIdentificationToreSupra}) when the energetic particle 
contribution becomes important.
The aim of this chapter is to study the {\bf linear stability of BAEs in the
presence of energetic particles}.\\

We start this chapter with an analytical computation of the BAE fishbone-like
dispersion relation in a relevant way for the study of Tore-Supra modes, using 
a {\bf perturbative approach to add damping mechanisms neglected so far, 
  and the effect of energetic particles}. 
For acoustic modes $\sim v_{ti}/R_0$, resonances with passing
thermal ions ${\bf n*}\cdot\m{\Omega_i}\sim v_{ti}/qR_0 $, 
the so-called {\bf ion Landau damping}, are expected to be the main
damping process, whereas the energetic particle drive can provide the
driving source.
The perturbative treatment of damping follows the small order expansion
of the transit effects performed in the previous
chapter, and it makes some sense if  
$ \omega^2_{BAE} = (T_i/m_i)(7/2 \tau_e +1)\gg (T_i/m_i)(1/q^2)$.
For stability studies (that is to say, when studying the closeness to
{\bf marginal stability}, where damping and drive compensate), it is
consistant to make use of a perturbative approach for the energetic
population as well.
Nevertheless, the form of the energetic particle contribution provided
below can be extended to the general case.
From the competition of driving and damping mechanisms, it is possible
to deduce the {\bf BAE stability threshold}.\\

Next, we compare the calculated threshold with experiments. In this work, 
{\bf experiments were conducted in order to allow for a systematic 
and statistical analysis of the experimental conditions and tunable
parameters which favor the BAE destabilization, in the presence of a 
population of ICRH heated ions}. 
Earlier stability analysis were conducted for Toroidal Alfv\'en Eigenmodes 
(TAEs) \cite{Betti_92,Kerner_94} or fishbones \cite{Nabais_05,Zonca_07}. 
For BAEs, the major point is the role of Landau damping, which can be
important and get close to the validity limit of the perturbative
approach, for the type of experiments performed.
Landau damping was even thought to prevent the apparition of modes in the 
acoustic frequency range. And indeed, in the tokamak
JET, the observation of global GAMs was reported to be possible only when 
thermal ions were ``cold'', hence in a regime of low Landau damping 
\cite{Boswell_06}. 
A legitimate question is whether the same is required for the excitation
of BAEs.
A second interesting point for BAEs is the {\bf localization of
  damping}, which depends on the mode structure and can differ significantly
from TAEs.

In the experiments presented here, various global and local parameters have 
been varied to determine the conditions for BAE excitation and the precise role 
of these parameters in the excitation. 
In a first part, the role of {\bf global macroscopic parameters} is studied, 
based on a statistical analysis of various shots designed to keep homothetic 
plasma profiles. Next, attention is given to the {\bf effects of a $q$ profile
and fast ion profile modification, in order to estimate in particular
the role of the fast population anisotropy}.

\section{Computation of the fishbone-like dispersion relation relevant to 
Tore-Supra Beta Alfv\'en Eigenmodes}
\label{section_BAEFishboneLikeDispersionRelation}
\subsection{The fishbone-like dispersion relation applied to Beta Alfv\'en 
Eigenmodes}

In the absence of energetic particles and neglecting damping mechanisms, 
the BAE dispersion relation has been found in 
Eq.~\ref{eq_BAEDispersionRelation}, to be divided into an inertia related
part, and an incompressible MHD-like region. This separation mimics the 
traditional MHD energy relation Eq.~\ref{eq_dWMHD}, 
but contains an additional spatial information, according to which
inertial terms are concentrated where high radial gradient are 
present, that is, inside a layer of typical size (see Eq.~\ref{eq_InertiaLength})
\begin{equation}
L_I = \sqrt{2Q}/|k_\theta s| \sim
\left(\frac{\rho_i}{k_{\theta} s}
\frac{\omega_{BAE}}{\omega_A}\right)^{1/2}.
\label{eq_InertiaLengthbis}
\end{equation}

The additive form of the Lagrangian Eq.~\ref{eq_OmElectromagLagrangian}, which 
is simply related to the traditional MHD energy relation
(by Eq.~\ref{eq_LagrangianAnddWMHD}, 
$-1/2\mathcal{L}_\omega  = -(\omega^2/2)\int d{\bf x}^3nm_i|\xi|^2 
+ \delta W_\text{MHD}\ $)
tells us that if we now want to add energetic particles related effects, it 
makes sense to simply add the energetic particle Lagrangian
$-1/2\mathcal{L}_{h\omega}$ to the MHD dispersion relation.

In principle of course, the addition of an extra population of particles could 
modify both the inertia and MHD structure, as well as the various orderings we 
made use of for the calculations of the previous chapter. However, an estimate 
of $L_I$ given in Eq.~\ref{eq_InertiaLengthbis} returns values of the order of 
a few mm in Tore-Supra, whereas the gyroradius of suprathermal ions (typically 
with energies between 100 keV and 1MeV) is of the order of $\rho_h\sim$10mm.
For this reason, energetic ions are not expected to interact with the mode
inertial structure, as was also argued in Ref.~\cite{Tsai_93}, and their 
contribution can simply be added to the larger scale structure.

Traditionally, the non-resonant contribution of energetic particles is summed up 
to the thermal plasma MHD part, to form a real energy potential 
$\delta W_f = \delta W_ \text{MHD} - (1/2)\mathcal{L}_{s\omega}^{non-res}$
which contains all the {\it fluid}-like contributions,
whereas the resonant ({\it kinetic}) contributions from the energetic particles
is computed separately,  $\delta W_{k}= -(1/2)\mathcal{L}_{h\omega}^{res}$.
Following Eq.~\ref{eq_KineticdW}, an expression for the resonant Lagrangian 
of a species $s$ can be taken to be
\footnote{Note that the separation between between a  non-resonant 
contribution and a resonant contribution is not univocal, and that a different
cut-off could have been made.}
\begin{equation}
\mathcal{L}^{res}_{s\omega} 
=  \sum_{\mathbf{n}=(0,{\sf n}_2,{\sf n}_3)}
\int d^3{\bf x}d^3{\bf p}
\ e_s^2 \frac{\omega\partial_{E|\mu} 
+ {\sf n}_3 \Omega_{\ast s}/T_s }
{\omega-\mathbf{n} \cdot \mathbf{\Omega}_s} F_{s}
\left|\left(\frac{\mathbf{v}_{gs}\cdot\nabla\psi_{\omega}}{-i\omega}
+\mathcal{E}_{\omega}\right)_{\mathbf{n}} \right|^2
\label{eq_ResonantLagrangian}
\end{equation}
and now have the three announced contributions of the fishbone-like dispersion 
relation Eq.~\ref{eq_NNFishboneDispRel}, 
an inertia ( / or generalized inertia) term which we can note $\delta I$, 
a fluid-like contribution $\delta W_f$
and a kinetic energetic particle term, $\delta W_k$.

For consistency with the previous notation and with the literature 
\cite{Zonca_07},
we use a normalized form of the generalized MHD relation, consistent with the
normalizing constants of Eq.~\ref{eq_BallooningNormalization} and 
Eq.~\ref{eq_AmplitudeNormalization}, that is, we multiply the MHD relation by 
\begin{equation}
\hat{C}=\frac{\mu_0}{B^2_0}\frac{R|k_\theta|}{2\pi r_s}q^2
\left(\frac{\omega^2B_0^2s^2}{k_\theta^2\Psi_0^2}\right),
\label{eq_FinalNormalization}
\end{equation}
to obtain the normalized fishbone-like dispersion relation
\begin{equation}
\delta {\hat W} 
\equiv -\delta\hat{I}+\delta\hat{W}_{f}+\delta \hat{W}_k=0.
\label{eq_NormalizedFishboneDispRel}
\end{equation}
where in particular, 
$\delta\hat{W}_f=\delta \hat{W}_\text{MHD}
-(\hat{C}/2)\mathcal{L}_{h\omega}^{non-res}$, and $\delta \hat W_\text{MHD}$
corresponds to the quantity used in Eq.~\ref{eq_BAEDispersionRelation}.
Again, the quantities of Eq.~\ref{eq_FinalNormalization}, 
$k_\theta={\sf m}/r, s, q$,
need to be assessed  close to the main resonant surface of the BAE, $r=r_s$.
$\Psi_0/s$ is a characteristic value of the mode amplitude,  whose relation to 
the mode MHD structure has been calculated in  Eq.~\ref{eq_OddIdeal} 
and~\ref{eq_EvenIdeal} for modes with a given parity. 
In particular, for the tearing parity, the last parenthesis of  
Eq.~\ref{eq_FinalNormalization} can be rewritten $1/\delta \xi_0$  where 
$\delta \xi_0$ represents the jump of the radial MHD displacement close to the 
mode inertial region.\\

Let us calculate now determine the different terms involved in 
Eq.~\ref{eq_NormalizedFishboneDispRel} more precisely.

\subsection{The energetic particle term}
\label{ssection_EnergeticParticleTerm}
In this section, we compute the resonant Lagrangian 
\ref{eq_ResonantLagrangian} for a population of energetic particles ($s=h$), 
relevant to a Tore-Supra ICRH heated plasma.

\subsubsection{Energetic particle equilibrium distribution functions}

When ICRH heating is used in the hydrogen minority scheme, hydrogen ions are 
accelerated in the  {\it perpendicular} direction, where the resonance condition 
$\omega_{ICRH} = \Omega_{c,h} = e B_\eq/m_h$ or 
$ R = eB_0R_0/(m_h\omega_{ICRH})$ is met, that is on a {\it vertical surface of 
constant $R$}. 
To the lower approximation (and in particular in the absence of MHD
activity), these hot ions are thought to release their energy on
electrons, via collisions, such that an equilibrium is reached when
the local ICRH input power, $P_{ICRH}(r)$ is balanced by the energy
released
on electrons $\propto \nu_{he}n_hT_h$, with  $\nu_{he}$ the collision frequency
between the heated particles and the electrons. 
Our formalism assumes the existence of an {\it equilibrium}
distribution function for the hot particles, which makes sense as long
as the growth rate of the considered MHD modes $\gamma$ 
is smaller than $\nu_{he}$ and than the typical particle
characteristic frequency. In our range of parameters (typically close
to threhold, for stability analysis), it is a reasonable assumption,
$\gamma \ll \omega_r<\sim \Omega_{h,1...3}$
 (for $\omega=\omega_r+i\gamma$).

Following the heating, a tail of hot hydrogen ions is expected, and it has
been found to be well approximated by a Maxwellian of temperature 
$T_h>T_i\sim T_e$ \cite{White_89}. 
Moreover, at the location where resonance occurs, 
the energy of the resulting population of hot particles is dominated
by its perpendicular energy, such that the equilibrium invariant 
$\mu/E \sim 1/B_\eq$ is almost
independent on the hot particle.

Consistently, a traditional analytic fit for a population of ICRH heated ions is 
the {\bf anisotropic Maxwellian},
\begin{equation}
F_h(r, \lambda, {\sf E}) = \tilde{n}(r) f_\lambda(\lambda-\lambda_0) 
e^{-{\sf E}/T_h (r)}
\label{eq_AnisotropicMaxwellian}
\end{equation}
where $f_\lambda$ is a highly picked function in 0 often taken to be a 
$\delta$-function \cite{White_89}, and $\lambda_0$ is a constant defining the
resonant surface $R/R_0 =\lambda_0$.
As required, $F_h$ has been expressed as a function of {\bf three equilibrium 
invariants}, here chosen to be the radial coordinate $r$ (invariant with the 
assumption of small orbit width), the energy {\sf E}, and 
$\lambda = \mu B_0/{\sf E}$.
The word {\bf anisotropic} refers to the dependence in $\lambda$ (or $\mu$)
of the distribution function. 

Taking $f_\lambda$ such that $\int d\lambda f_\lambda = 1$, it is
possible to rewrite the equilibrium distribution function 
\ref{eq_AnisotropicMaxwellian}
as a function of the more traditional characteristic quantities of
a distribution function. 
From Appendix \ref{appssection_NormAnisotropicMaxwellian}, it comes
\begin{eqnarray}
\tilde{n}(r) = 
\frac{n_{h}(r)}{(2 \pi m_h T_h)^{3/2}}
\frac{2}{< \bar{\Omega}_b(\lambda,r)>_\lambda}
\quad \text{ with }
<...>_\lambda 
= \int_0^{1+r/R}d\lambda f_\lambda.
\end{eqnarray}

\subsubsection{Derivation of the energetic particle term}
When a mode oscillates with a real frequency $\omega_r$ 
($\omega = \omega_r + i\gamma $), different populations of particles can 
resonate with it. As a consequence, $\delta \hat{W}_k$ is 
the sum of many resonances,
which are seen from Eq.~\ref{eq_ResonantLagrangian} to be of the form 
$\omega - {\bf n^*}\cdot{\bf \Omega}_s=0$ with 
${\bf n^*}=(0,{\sf n}_2,{\sf n}_3)$.
The latter terms come from the the projection of the fields structure, typically   
expressed in the $(r, \theta, \varphi)$ coordinate system, onto the particle  
relevant coordinates, the action-angle variables, and they can be an infinite
number. However, {\bf only a few of them are at the same time relevant to
the population of particles in place, and with a non-negligible field component}.

\begin{figure}[ht!]
\begin{center}
\begin{minipage}{\linewidth}
\begin{center}
\includegraphics[width=0.5\linewidth]{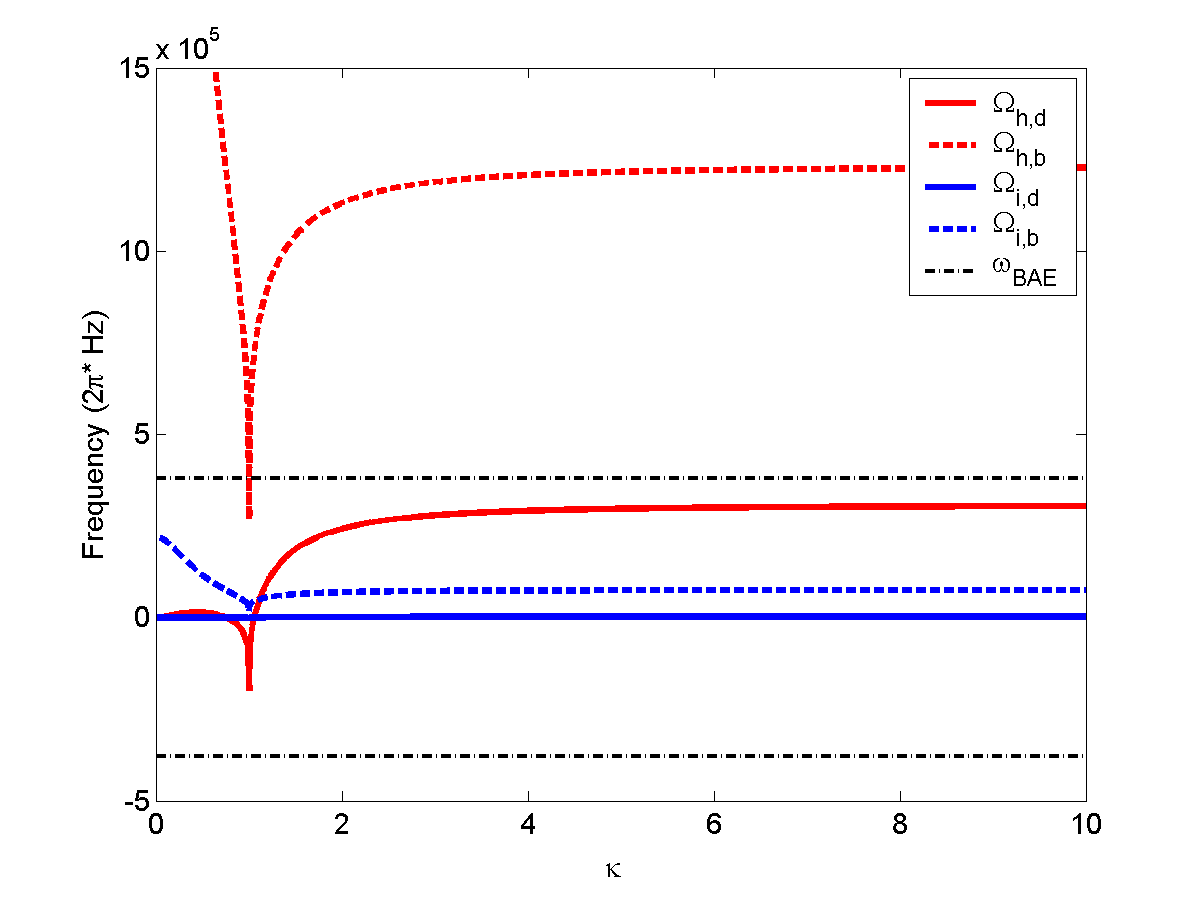}
\caption
[\footnotesize Typical characteristic equilibrium frequencies 
of hot ions relevant to Tore-Supra parameters.]
{\label{fig_EquilibriumFcies}
\footnotesize
Equilibrium characteristic frequencies, calculated assuming
$R_0=2.45$m, $q=1$, $s=0.1$, $r=0.14$m,
$400$keV hot hydrogen ions, and 3keV thermal ions.
Those quantities are given as function of the pitch angle
$\kappa^2 = 2(r/R)\lambda/\left(1-(1-r/R)\lambda\right)$,
such that $\kappa>1$ corresponds to trapped particles, and
$\kappa<1$ to passing  particles.}
\end{center}
\end{minipage}
\end{center}
\end{figure}

First, considering a BAE with a single finite {\sf n} number, it can be 
shown that the only field components which do not cancel verify 
${\sf n}_3\neq{\sf n}$ (more details are given in Appendix 
\ref{appssection_ActionAngleProjection}).
To determine the relevant values of ${\sf n}_2$, the BAE frequency
$\omega_{BAE}$, estimated to be an approximation for $\omega_r$,  is plotted 
in Fig.~\ref{fig_EquilibriumFcies}, 
along with the typical numerical values of the equilibrium characteristic 
frequencies $(\Omega_{s,b},\Omega_{s,d})$ relevant to ICRH heated hydrogen ions
(with typical energies between 100keV and a few hundreds keV) 
and thermal ions, 
using the expressions of the {\it normalized} drift and bounce frequencies
(Eq.~\ref{eq_NormBounceDrift}) given in Appendix \ref{appsection_KbKd}.
It comes from Fig.~\ref{fig_EquilibriumFcies} that the most effective resonance
with hot ions are of the form $\omega - {\sf n}\Omega_{h,d} = 0$, because 
$\Omega_{h,b}\gg\omega_{BAE}$. 
For $ \omega_r {\sf n}>0$, trapped ions are the only candidates for 
this excitation. Instead, for $\omega_r {\sf n}<0$, resonance may be achieved by 
all types of particles presenting a drift reversal. For low shear, it means that 
passing ions (taking ${\sf n}_2=-{\sf n}$, $q\approx 1$) and marginally trapped 
ions can resonate with the mode. 
In the following, only the case with $ \omega_r {\sf n}>0$ is considered. 
Indeed, it can be argued that for standard profiles with a negative radial gradient
(as expected in our experimental conditions), a mode with $\omega_r {\sf n}<0$ 
is damped and not excited  by the energetic population. This argument will be 
made clearer in the following paragraphs.

Following these remarks, it appears meaningful to simply keep resonances 
of the form  $({\sf n}_1={\sf n}_2=0$, ${\sf n}_3={\sf n})$ and 
{\bf fast trapped ions}. As an addition, {\bf the MHD-constraint}, 
$\mathcal{E}_\omega = 0$, which is verified out of the inertial region
in the absence of the fast ions, 
is assumed to be valid when fast ions are added as well.
In the approximation of {\bf thin banana} ($k_r \delta_b \ll 1$ with 
$\delta_b$ the particle banana widths) for the energetic ions,
which is partly justified in the MHD region, and after some
calculations given in Appendix~\ref{appsection_dWk}, the 
energetic contribution reduces to
\begin{eqnarray}
\delta\hat{W}_k &=& -2^{3/2}\pi^2{\sf m}q{(r_s)}^2m_h^{3/2}
\frac{\mu_0}{B_0^2}\times\\
&& \nonumber
\int^a_0 \frac{1}{r_s^2}r dr|\bar{\xi}|^2
\int^{1+r/R}_{1-r/R} d\lambda \bar{\Omega}_d^2{\bar\Omega_b}
\int^{+\infty}_0 \frac{d{\sf E}}{{\sf E}-{\sf E}_{res}}{\sf E}^{5/2}
\left({\sf E}_{res}
\partial_{\sf E|\mu}-\frac{1}{\bar{\Omega}_d}R_0\partial_r\right)F_{h}
\label{eq_HalfExpanded_dWk}
\end{eqnarray}
where $\bar{\xi} = \xi/ (k_\theta\Psi_0/ B_0 \omega s)$ with
$\xi=k_{\theta}\psi_\omega^{\sf m}/B_0\omega$,
is the normalized MHD displacement, associated with the field main poloidal 
component.
${\sf E}_{res}$ is the necessary energy to meet the resonant condition
$\omega - {\sf n}\Omega_{h,d}=0$,
\begin{equation}
{\sf E}_{res} = \frac{r}{{\sf n}q(r)}\frac{e_hB_0R_0}{\bar{\Omega}_d} \omega.
\end{equation}
Eq.~\ref{eq_HalfExpanded_dWk} is in agreement with Eq.~8 of Ref.~\cite{Chen_84}.\\

It is already possible to make a few physical comments based on the form of 
Eq.~\ref{eq_HalfExpanded_dWk}, to get a first feeling of the role of an 
energetic population in the drive of a mode.
Assuming a mode of frequency $\omega = \omega_r +i\gamma$ to be purely 
oscillating ($\gamma=0$),  resonance occurs for ${\sf E} = {\sf E}_{res}(\omega_r)$,
which is at the origin of a discontinuity in Eq.~\ref{eq_HalfExpanded_dWk}.
Solving this discontinuity is traditional for the treatment of wave-particle
interaction.

First, we note that the discontinuous function $1/({\sf E}-{\sf E}_{res})$ can 
be replaced by the well-defined {\it principal value} distribution 
$\mathcal{P}\cdot\int_0^{+\infty} d{\sf E}/({\sf E}-{\sf E}_{res})$ to give
a meaning to Eq.~\ref{eq_HalfExpanded_dWk}.
Nevertheless, 
in order for $\delta \hat{W}_k (\omega)$ to be analytic in the complex 
plane ($\omega \in \mathds{C}$), the proper substitution is  
$1/({\sf E}-{\sf E}_{res}) \longrightarrow 
\mathcal{P}\cdot\int_0^{+\infty} d{\sf E}/({\sf E}-{\sf E}_{res})
\pm i\delta$, where the sign preceeding $\delta$ needs to be defined.
The traditional way to determine this sign is to make use of an argument of 
{\bf causality} \cite{NicholsonBook}. 
Looking at a fluctuation field $\mathcal{F}_\omega = \mathcal{F}(\omega)$, 
as the Laplace transform of a time dependent field 
$\mathcal{F}(t) = 
\int_{\omega = \omega_r + i\gamma} d\omega_r \mathcal{F}(\omega)\exp(-i\omega t)$, 
cancellation of the signal at $t=-\infty$ (the origin of time) requires $\gamma>0$.
This means that the function $\mathcal{F}_\omega$ needs to have a physical 
meaning in the upper half plane only. Hence, if one wants to define an analytic
function in the whole complex plane, one first needs to define such dispersion
relation in the upper half plane and continue it in the lower half plane, 
analytically.
In our case, the integration over {\sf E} is well defined in the upper half
plane for $\omega = \omega_r + i\gamma$, $\gamma>0$. When now $\gamma$ is sent to 
0, it comes
\begin{equation}
\frac{1}{{\sf E}-{\sf E}_{res}(\omega_r + i\gamma )}\\
\begin{array}{c}
\gamma\ll\omega_r\\
\longrightarrow
\end{array}\\
\frac{1}{{\sf E}-{\sf E}_{res}(\omega_r) }  
+\frac{i{\sf E}_{res}(\gamma)}{({\sf E}-{\sf E}_{res}(\omega_r))^2} \\ 
\end{equation}
The imaginary part is positive, and so should be the sign in front of the
delta function.

As expected, this resonant behavior is at the origin of a {\bf wave-particle 
energy transfer}. 
Indeed, we know from Eq.~\ref{eq_JE},
that for $\omega \approx \omega_r$,
$\omega_r\im(\delta\hat{W}_k(\omega_r)$ is positively proportional to the 
energy transfer from  the hot particles to the wave 
($-2\omega_r\im{\mathcal{L}_{s\omega}}$), and it is the resonance which allows
this imaginary part to be finite.

More precisely, we observe 
that the amplitude and sign of the energy transfer depend on the existence of 
{\bf gradients of the equilibrium distribution function}, $\partial_r F_h$ and 
$ \partial_{\sf E}F_h $
\footnote{Note that the lack of gradient along the third invariant
$\partial_\mu $ simply comes from the fact that our definitions are 
gyroaveraged.}.
For ${\sf E}_{res} \sim T_h$
(that is for the type of hot population expected to resonate with the mode), 
the {\bf contribution of the radial gradient is dominant}. Typically, 
the ratio of the first to the second term in the second parenthesis of 
Eq.~\ref{eq_HalfExpanded_dWk}, is of the order of 
$\omega/\Omega_{*h} \sim ({\sf E}_{res}/T_h)(R_0/L_{ph}) \sim 1/\epsilon$
(neglecting anisotropy as a first approximation), 
where $L_{ph}$ is a typical radial gradient of the hot particle distribution 
function. For this reason, the localization extent of a heating mechanism
is of importance in the drive of a mode.

In particular, keeping the hot ion radial gradient induced drive only, 
it follows that the {\bf the sign of the energy transfer from the
hot particles to the wave is given by}
$\m{ -\sgn({\sf m}\omega_r\partial_r F_h) = \sgn ({\sf n}\omega_r\partial_rF_h)
}$. We recover the announced idea that only waves with 
$\omega_r {\sf n}>0$ can be excited.

A second result announced in the previous chapter is the difficulty to excite
GAMs. For $\sf n=0$ modes, radial gradient induced excitation 
as well as resonance with the particle drift motion are not possible.
However, because of anisotropy, it was shown in Ref.~\cite{Boswell_06}, that
the energy derivative ${\sf E}_{res}\partial_{\sf E|\mu}$ 
could be larger than $O(1)$ (For clarity, the contribution of anisotropy to the 
energy derivative is the second term of the brackets in 
Eq.~\ref{eq_fullydevelopeddWk}), and such excitation is not inconsistent with 
a resonance with passing hot ions.
It remains that excitation of GAMs could only be found in very special 
conditions.
\\

To make comparisons with Tore-Supra experiments, it is useful to apply 
further simplification.
First, we assume the hot ion distribution to be an {\bf anisotropic 
Maxwellian}, that is, of the form defined in Eq.~\ref{eq_AnisotropicMaxwellian}.

Secondly, in the absence of the observed modes toroidal and poloidal mode 
numbers, we assume the mode MHD structure to be a {\bf kink with 
$({\sf m, \sf n})=(-1,1)$ poloidal component}.

As explained earlier, a kink mode is an incompressible MHD instability 
driven by the equilibrium current.
In a tokamak and for discharges with a standard monotonic $q$-profile, 
$({\sf m, \sf n})=(-1,1)$ kink modes are common modes, which are in 
particular at the origin of sawtooth crashes.
Such modes are usually recognized to have an odd step-like structure 
\cite{Bussac_75}
with a discontinuity localized at the mode resonant surface, 
ie $\m{\xi} = \m{\xi}_0$ for radii $q(r)<1$ 
and $\m{\xi} = \m{\xi}_0$ for $q(r)>0$.

We know from the previous chapter that the BAE inertial region can connect
with both an even or an odd step-like structure, and we showed that even modes 
with a frequency which is close to the BAE accumulation point tend to be 
very localized. Because our experiments 
(Fig.~\ref{fig_SpectrumAndRadialStruc41925}) show rather extended structures,
in agreement with a localization inside the $q=1$ surface,
it is reasonable to expect the BAE to have an odd structure and an inertial
layer localized on the $q=1$. 
A second argument to justify this choice is the remark that sawtooth 
oscillations are found in our experiments 
(periodically excited before the sawtooth crashes). 
It means that in the types of equilibria set up in our experiments, the 
kink typical  structure is to be associated with a small
{\it incompressible} $\delta W_{MHD}$, which again is in agreement with the 
fact that BAE oscillations are found close to the BAE accumulation point.
Finally, observations of ${\sf n}=1 $ modes oscillating with an acoustic 
frequency have been performed in the JET tokamak, where the toroidal mode 
number could be measured \cite{Zonca_07}. It is reasonable to postulate 
a similar physics here.

With the above assumptions, the energetic particle term takes the form
\begin{eqnarray}
\delta\hat{W}_k &=& 
-\pi q{(r_s)}^2{\sf m}
\int_0^a \frac{1}{r^2_s}r dr|\bar{\xi}|^2\beta_h
\int_{1-r/R}^{1+r/R} d\lambda\frac{\bar{\Omega}_b\bar{\Omega}_d^2}
{<\bar{\Omega}_b>_\lambda} \times \\
\nonumber&& \quad\quad\quad
\left\{
-\frac{{\sf E}_{res}}{T_h}Z_5f_{\lambda}
-\frac{{\sf E}_{res}}{T_h}Z_3\lambda f'_{\lambda}\right.\\
&&\nonumber\quad\quad\quad
\quad\quad\quad\quad
\left. -\frac{1}{\bar{\Omega}_d} 
\left[\frac{R\partial_r n_h}{n_h} Z_5 
 -\frac{R_0\partial_r <{\bar \Omega}_b>}{<{\bar\Omega}_b>} Z_5
 +\frac{R_0\partial_r T_h}{T_h}\left(Z_7-\frac{3}{2}Z_5\right)
 \right]f_{\lambda} \right\}
\label{eq_fullydevelopeddWk}
\end{eqnarray}
with $\beta_h = 2\mu_0n_hT_h/B_0^2$, 
and
$Z_p(y_0)=(1/\sqrt{\pi})\int^{+\infty}_{-\infty}dyy^{p+1}\exp(-y^2)/(y^2-y_0^2)$,
for $y^2_0={\sf E}_{res}/T_h$. 
In particular, the function $Z$, such that $Z(y_0)=y_0Z_{-1}(y_0)$, is the 
{\it Fried et Conte function}, and 
\begin{equation}
\begin{array}{rclrcl}
Z_3(y_0)&=&\frac{1}{2}+y_0^2+y^3_0Z(y_0),\\
Z_5(y_0)&=&\frac{3}{4}+\frac{y_0^2}{2}+y_0^4+y_0^5Z(y_0),&
Z_7(y_0)&=&\frac{15}{8}+\frac{3}{4}y_0^2+\frac{y_0^4}{2}+y_0^6+y_0^7Z(y_0).
\end{array}
\label{eq_FriedEtConte}
\end{equation}

\subsection{Inertia and Landau damping}
A partial computation of inertia, $\delta\hat I$, was done in the 
previous chapter where resonance related energy transfers were absent.
In this approximation, $\delta\hat{I}$ was found to be purely real
and of the form
\begin{eqnarray}
\label{RealInertia}
\nonumber
\delta \hat{I}
&=& \frac{\hat{C}}{2}\int_{inertia}d^3{\bf x} 
\frac{ne^2}{T_{i}}\frac{\omega_A^2}{\omega^2}
\psi^{\sf m\ast}_\omega
\left[(qR_0)^2\rho_i^2\nabla_\|\partial_r^2\nabla_\|
+ \Lambda^2\rho_i^2\partial_r^2
+ \left(\frac{Q}{k_\theta\rho_is}\right)^2\rho_i^4\partial_r^4 \right]
\psi_\omega^{\sf m} 
\label{eq_InertiaEquationForm}
\\
\label{eq_RealInertiaLimits}
&=&\left\{ 
\begin{array}{ll}
i\Lambda|s(r_s)| & \mbox{for large } \Lambda^2 \\
-2\sqrt{Q}\
\frac{\Gamma\left(3/4-\Lambda^2/4Q \right)}
     {\Gamma\left(1/4-\Lambda^2/4Q\right)}
|s(r_s)|
& \mbox{for }  \Lambda^2\longrightarrow 0
\end{array}\right.
\end{eqnarray}
where $\Lambda^2$ and $Q^2$ two real functions of $\omega$,
representing generalized inertia and high order finite Larmor radius effects, 
given by Eqs.~\ref{eq_Coeff4thOrderLambda}, \ref{eq_Coeff4thOrderSigma} and
\ref{eq_QDefinition}.
When they are purely real, $\Lambda^2$ and $Q^2$ can be also
interpreted as a  measure of the {\bf inertia length}.
Indeed, when $\Lambda^2$ is large such that the term in $Q^2$ is
negligible in Eq.~\ref{eq_InertiaEquationForm}, the balance of field
line tension and generalized inertia returns a typical inertia length,
$L_{I}\sim|\Lambda|/|k_{\theta}s(r_s)|$.
When $\Lambda^2 \rightarrow 0$, the case considered in chapter
\ref{chapter_BAEdescription}, high order FLR come into play and are
able to set up
a new relevant inertia length, given in Eq.~\ref{eq_InertiaLengthbis},
$L_I \sim \sqrt{Q}/|k_\theta s(r_s)|$.
\\

However, as mentioned in the chapter introduction, resonance (and
resonance induced energy transfers) with thermal ions can be expected
in the acoustic frequency range, as also suggested by 
Fig.~\ref{fig_EquilibriumFcies},
for very passing particles ($\kappa^2 \ll 1$  in this figure).
As will be shown, resonance with passing thermal ions, can lead to
a damping characterized by a resonance of the form 
$\omega - {\bf n}^*\cdot\m{\Omega}_{i} = \omega-k_\|v_\|$, which has
been given the name of {\bf Landau damping}.
Landau damping explains the existence of a threshold for BAE
excitation, even for modes which are in the compressibility
induced BAE gap (represented in Fig.~\ref{fig_CS42039_wCompressibility} for a 
Tore-Supra discharge) and escape continuum damping, 
or for modes which are close to the BAE
gap edge (ie, close to $\omega_{BAE}$) and likely to exhibit an undamped
kinetic structure.

A precise calculation of $ \delta {\hat I}$ including a non-perturbative 
treatment of Landau damping can be found in Refs.~\cite{Zonca_99,Annibaldi_07}.
In this analysis, a much simpler perturbative model is used, valid for
large $\omega qR/v_{ti}$, and more tractable for statistical
comparisons with experiments.
With this approximation, the {\it real} dispersion relation given in 
Eq.~\ref{eq_InertiaEquationForm}
is lower order, and Landau damping can be introduced at the same order
as the previous transit effects, corresponding to the real part of the
resonant term.\\

For passing ions, $\Omega_{i,2} \approx v_\|/qR_0$, $\Omega_{i,3} 
\approx v_\|/R_0$.
Relevant resonances are with the sidebands, 
${\sf n_2}\Omega_{i,2} + {\sf n}_3 \Omega_{i,3} 
= ({\sf  m}/q+{\sf n})(v_\|/R_0)
= k_{\|}^{\sf m\pm1}v_\|
\approx \pm v_{\|}/qR_0$, close to the main mode resonant surface.
The computation of Landau damping which results from the latter
resonances follows the same logics as described for the resonances
with energetic ions. It is developed in Appendix \ref{appsection_Inertia},
starting from the resonant lagrangian \ref{eq_ResonantLagrangian}.
It returns
\begin{equation}
\label{eq_LandauDamping}
\begin{array}{rcl}
\frac{\hat{C}}{2}{\rm Im}(\mathcal{L}^{res}_{i\omega}) (\omega) 
&=& \frac{\hat{C}}{2}\int d^3{\bf x}
\frac{ne^2}{T_{i}}(\rho_i\partial_r\psi_\omega^{\sf m})^2 \delta_I 
\\
\mbox{with } \delta_I 
&=&\frac{\sqrt{\pi}}{2\sqrt{2}} \left(q\frac{v_{ti}}{R\omega}\right)
e^{-\frac{1}{2}\left(\frac{q R \omega}{ v_{ti}}\right)^2}
\left[ 2 + 2\left(\left(\frac{qR\omega}{v_{ti}}\right)^2 
+ 2\tau_e\right) 
+\left(\left(\frac{qR\omega}{v_{ti}}\right)^2
+2\tau_e\right)^2 \right].
\end{array}
\end{equation}

Let us analyze the expression \ref{eq_LandauDamping}.
It is already clear that the calculated resonant behavior implies an
energy tranfer from the wave to the particles because
$\sgn(\omega_r \im(\mathcal{L}^{res}_{i\omega}))>0$. A wave with a positive
energy is consequently {\bf damped}.

For $\omega\approx \omega_r$ in the acoustic frequency range, $\delta_I$ is of 
the order of 1 (at the limit of the perturbative treatment) and should not vary
significantly for high density ICRH experiments, because the BAE frequency is 
proportional to $v_{ti}/R_0$. 
A more important role may be expected from the mode radial gradient included in 
the term $ (\rho_i \partial_r \psi_\omega^{\sf m})^2$.
The presence of this radial gradient gives a major importance to the inertial 
region characterized by high radial gradients, and explains why {\bf Landau 
damping may be considered to be mainly an inertial phenomenon} and integration 
reduced to the inertia region $\int \longrightarrow \int_{inertia}$.
Note however that this reduction is consistent with the ordering of 
$\delta {\hat W}_k$ (calculated in the MHD region), only in
the perturbative frame, that is for $\delta_I < 1$. Beyond this limit, an argument
of localization of the Landau resonance condition should be invoked instead, to 
restrict Landau damping to the inertial region. 
It results that Landau damping can be put in the form
\begin{equation}
\frac{\hat C}{2}{\rm Im}(\mathcal{L}^{res}_{i\omega})
\approx\frac{\omega^2}{\omega_A^2}\delta_I(r_s)
\frac{2\pi}{k_{\theta}}\int_{inertia}dr(\partial_r{\bar\xi})^2
\approx \frac{\omega^2}{\omega_A^2}\delta_I(r_s)\frac{c_0}{k_{\theta}L_{I}}
\end{equation} 
where $c_0$ is a constant of order 1 depending on the details of
the inertia structure, and the integration is approximated by $c_0/2\pi L_I$.
In this form, the role of the mode inertia length $L_I$ is clearly put forward.
In particular, the precise computation of high order finite Larmor
radius effects, 
which is necessary for a correct estimate of $L_I$ when $\Lambda^2\rightarrow0$ 
finds a justification here.

We may even be more precise, 
noticing that the addition of the resonant inertia imaginary part
Eq.~\ref{eq_LandauDamping} to the real inertia Eq.~\ref{RealInertia}
allows the direct inclusion of Landau damping  in the ``generalized inertia'':
\begin{equation}
\re(\Lambda^2)\longrightarrow \Lambda^2=\re(\Lambda^2)
+i(\omega^2/\omega_A^2)\delta_I(r_s).
\label{eq_ComplexifiedLambda}
\end{equation}
This way, a direct estimate of
the inertial energy $d{\hat I}$ is given by formulas 
\ref{eq_RealInertiaLimits}, where now $\Lambda$ is to be understood 
as the square root of a complex function of $\omega$. 
In particular, a direct assessement of the $c_0$ constant above follows,
\begin{eqnarray}
 {\rm Im}(\delta \hat{I})
&\approx& \left\{ 
\begin{array}{lll}
 \frac{1}{2}\frac{\omega^2}{\omega_A^2}
\delta_I(r_s)\frac{1}{k_{\theta}L_{I}},
&k_{\theta}s(r_s)L_{I}=|\re(\Lambda)|,
&{\rm for\,large\,\Lambda^2\, and \,} \omega<\omega_{BAE}  \\
 \frac{\sqrt{\pi}}{2}
\frac{\omega^2}{\omega^2_A}\delta_I(r_s)\frac{1}{k_{\theta}L_I},
&k_\theta s(r_s)L_I=|\sqrt Q|,
&{\rm\,for\, \Lambda^2\longrightarrow 0^+}
\end{array}\right..
\label{eq_InertiaWLengths}
\end{eqnarray}

\begin{figure}[ht!]
\begin{center}
\begin{minipage}{\linewidth}
\begin{center}
\includegraphics[width=0.5\linewidth]{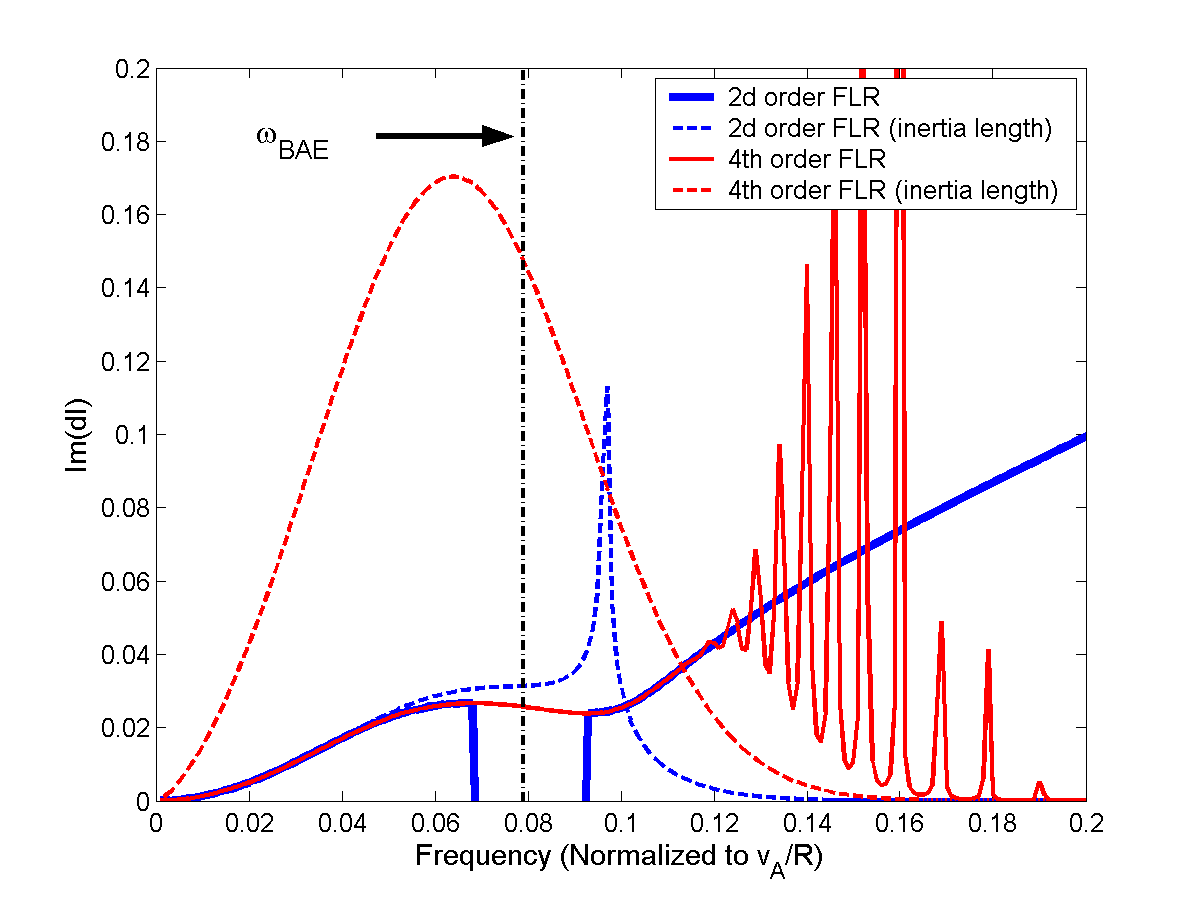}
\caption
[\footnotesize
Landau damping of BAEs.]
{\label{fig_Inertia}
\footnotesize 
Inertia localized damping, $\im(\delta\hat I)$, as a
function of the real frequency $\omega_r$ for the different models 
given by  formulas \ref{eq_RealInertiaLimits} 
and \ref{eq_InertiaWLengths}.
Below $\re(\Lambda^2)=0$ represented by the divergence of the blue
dotted curve, damping is to be attributed to Landau damping, whereas 
continuum damping dominates for $\re(\Lambda^2)>0$.
}
\end{center}
\end{minipage}
\end{center}
\end{figure}

The estimate of Landau damping for the various models, that is 
formulas  \ref{eq_RealInertiaLimits} with the {\it complexified inertia} 
and formulas \ref{eq_InertiaWLengths}, is plotted in Fig.~\ref{fig_Inertia}.
The expressions of Landau damping with the inertia length are observed
to be good approximations in their 
validity areas. The one including 2D order FLR effects  diverges
where $\Lambda^2=0$ (slighty shifted from $\omega_{BAE}$ due to the
high-q correction).
More interestingly, 
the role of high order finite Larmor radius effects clearly appears. 
Close to $\Lambda^2=0$, the  inclusion of 4th order finite Larmor
radius effects is shown to be necessary to compute Landau
 damping appropriately.
Because it is the frequency region where experimental points have
been found, the {\bf 4th order FLR model is  used in our subsequent 
analysis}, that is, 
the second equation of Eq.~\ref{eq_RealInertiaLimits}, with the complexified 
inertia Eq.~\ref{eq_ComplexifiedLambda}.

\subsection{The fluid energy}
The incompressible MHD energy corresponding to a kink structure has
in toroidal geometry has been calculated in Ref.~\cite{Bussac_75} for
discharges with a monotonic q-profiles. It reads
\begin{equation}
\delta\hat W_f = 3\pi(1-q(0))
\left(\frac{13}{144}-\beta^2_p(r_s)\right)\frac{r_s^2}{R_0^2}.
\label{eq_Bussac}
\end{equation}

\section[Threshold simplification for experimental comparison]
{Simplication of the threshold for comparison with experiments}
\label{section_ThresholdSimplification}
The BAE excitation {\bf threshold} can now be deduced from
\ref{eq_NormalizedFishboneDispRel}. 
Close to threshold ($ \omega=\omega_r+i\gamma$ with 
$|\gamma| \ll|\omega_r|$), the dispersion relation can be expanded
to the first order: 
$
\delta {\hat W}(\omega_r+i\gamma) = 0 
=i\gamma (\partial/\partial{\omega_r}){\delta\hat {W}}(\omega_r)
+\delta{\hat W}(\omega_r)
$
to the first order.
It follows, 
\begin{equation}
\gamma\frac{\partial}{\partial \omega_r}\re(\delta\hat{W}) 
=-\im(\delta{\hat W})
\end{equation}
where we know that the right
hand side is related to the wave-particle energy transfer 
($P_{wave\rightarrow s}=2\omega_r {\rm
  Im}(\mathcal{L}^{res}_{s\omega})$ ), and hence not surprisingly
determines if the wave is damped  ($\gamma < 0 $) or excited  
($\gamma > 0$). 
For a wave with {\it positive energy} (see section 
\ref{ssection_PositiveEnergyWaveDensity}),  
$ \omega_r\partial {\rm Re} (\delta {\hat W}) /\partial\omega_r <0$ 
and in agreement with the causality constraint, 
the excitation condition simply becomes 
$\omega_r {\rm Im} (\delta{\hat W}(\omega_r))>0$. 
Since $\sgn({\rm Im}(\delta \hat{W})) = -\sgn({\rm Im}(\mathcal{L}_\omega))$, 
this simply means that instability occurs for a positive energy
transfer from the particles to the wave.

We now assume ${\rm Re}(\delta {\hat W}) \gg {\rm Im} (\delta
{\hat W})$, consistently with the above {\it perturbative} calculation of
Landau damping, such that the lower order solution of 
$\delta{\hat W} (\omega_r)=0$
is well approximated by $\omega_{BAE}$. Hence, excitation occurs for
$ {\rm Im}(\delta {\hat W}_k)(\omega_{BAE})>{\rm Im}(\delta{\hat I})
(\omega_{BAE})$.
Under the simplifying assumption that the mode is more localized than 
the fast ion distribution, such that global characteristic values of
the fast ion distribution ($T_h, n_h$...) may be considered instead of 
precise radial distributions,
and considering the effect of the radial pressure gradient  to be
dominant in Eq.~\ref{eq_fullydevelopeddWk}, 
this condition reduces to 
\begin{equation}
\frac{\beta_h}{\beta_i}\left(\frac{R_0}{L_{ph}}\right)
\left(\frac{L_I}{r_s}\right)>
\frac{c_0(7/2+2\tau_e)\delta_I}
{\pi^{3/2}{\Omega}_{h,d}(\lambda_0)\delta_h (T_h/{\sf E}_{res})}.
\label{eq_Threshold}
\end{equation}
Here $L_{ph}$ is again a typical radial gradient length of the hot ion 
distribution.
$\delta_h$ is a function of $(T_h/{\sf E}_{res})$ which takes the form
$\delta_h(T_h/{\sf E}_{res})=
\mathcal{P}(\sqrt{{\sf E}_{res}/T_h})\exp(-{\sf E}_{res}/T_h)$, 
with $\mathcal{P}$ a polynomial derived from the imaginary part of
the last bracket of \ref{eq_fullydevelopeddWk}.
When only the contributions of the radial density and temperature gradients
are kept, $\mathcal{P}$ is equal to 
\begin{equation}
{\mathcal P}(\sqrt{\frac{{\sf E}_{res}}{T_h}}) =
\left(\frac{{\sf E}_{res}}{T_h}\right)^{5/2} \left\{
L_{ph}\frac{\partial_r n_h}{n_h}|_{r_s}
+ L_{ph}\frac{\partial_r T_h}{T_h}|_{r_s}
\left(\frac{{\sf E}_{res}}{T_h}-\frac{3}{2}\right)
\right\}.
\label{eq_ThresholdPolynom}
\end{equation}
$\delta_h$ makes the excitation easier  when  $T_h<\sim{\sf E}_{res}$ 
(The optimal values of $T_h/{\sf E}_{res}$ is 0.24 
 for the case of a dominating temperature gradient, as also illustrated in 
Fig.~\ref{fig_ModeExistence}) 
and it is typically of order $1$ when the optimal value of $E_{res}/T_h$ is 
reached.

The form of Eq.~\ref{eq_Threshold} is similar to the TAE excitation
threshold calculated in \cite{Betti_92} but it includes the additional role of 
the inertia length $L_I$, which is to be related to the fact that the 
frequencies of interest $\omega \sim \omega_{BAE}$ are small compared to
$\omega_A$. Interestingly, it appears that for $ T_h/{\sf E}_{res}$ slightly
below $1$ (achievable in Tore-Supra) and for $\omega\sim\omega_{BAE}$
(leading to $\delta_I \sim 1$), the right hand side of
\ref{eq_Threshold} is of order $1$. Hence, it seems possible to access the 
 {\bf threshold in Tore-Supra experimental conditions} described
subsection \ref{ssection_ToreSupraAcousticModes}.\\

To allow for an easy comparison of the computed threshold with
experimental observations, we now want to {\bf link the parameters of 
Eq.~\ref{eq_Threshold} with measurable quantities}. 
{\it When} it is possible to assume the plasma shape, the shape of the fast ion
distribution ($R_0/L_{ph}$ in particular), the  $q$ profile ($r_s$ in
particular) and $T_{e}/T_{i}$ to be constants, scalings can be used:
\begin{equation}
\begin{array}{lll}
L_I\propto\sqrt{Q}, &\omega_{BAE}\propto\sqrt{T_e}, \\ 
n_hT_h \propto P_{ICRH}/\nu_{he} &\text{with }
\nu_{he}\propto n_0/T_e^{3/2}
\end{array}
\end{equation}
where the second line follows our previous description of the ICRH 
induced equilibrium distribution function given in 
section~\ref{ssection_EnergeticParticleTerm}  
and $\nu_{he}$  is again the ion-electron collision frequency.
Taking the hot particle distribution to be mainly made out of hydrogen
ions, it also comes that $n_h\propto f_{min}n_0$, and  
Eq.~\ref{eq_Threshold} becomes
\begin{equation}
c_1 n_0^{1/4}\frac{P_{ICRH}}{n_0^{2}}\frac{T_{e}}{B_0}
>c_3/
\delta_h\left(c_2 \frac{1}{f_{min}}\frac{P_{ICRH}}{n_0^{2}}\frac{T_e}{B_0}\right)
\label{eq_ModifiedThreshold}
\end{equation}
where we conserved the right and left hand sides of
\ref{eq_Threshold}, such that $c_1, c_2, c_3$ are unambiguously
defined as the proportionality coefficients between the physical
quantities of Eq.~\ref{eq_Threshold} and the measurable quantities: 
$T_e$ (taken at the center), $n_0$ (the central density) 
and $f_{min}$, respectively  measured by ECE, reflectometry
and neutrals detection diagnostics. 

In this formula, all tunable
parameters are found to have an ambiguous role in the excitation,
depending on the distance to the resonance condition $T_h/E_{res}\sim 1$.
In particular, the {\bf B-field, known to favor the excitation of
energetic particle modes such as precessional fishbones (characterized
by an excitation condition of the form $\beta_h>\beta_{h0}$
\cite{White_89}) has a reduced role on the BAE excitation threshold}.
More precisely, if we make use of the rough scaling law $ n_0T_e\sim B_0
f(n_0, P_{ICRH})$ with $f$ a given function,
$B_0$ disappears in the threshold
for fixed values of $n_0$ and $P_{ICRH}$. Note that 
the same is not true for TAEs, for which a similar derivation shows a 
non-negligible but ambiguous impact of the $B$-field. 
In particular, it is easy to see that taking 
$\omega\propto B_0/\sqrt{n_0}$ (for TAEs)
in the left hand side of \ref{eq_Threshold}
makes the resonance condition $T_h/{\sf E}_{res}\sim 1$ harder to reach for
higher $B_0$.
One should notice however that the role of $B_0$ in the left
hand side of Eq.~\ref{eq_ModifiedThreshold}  {\it depends on the model}
chosen for the inertia. Again, a correct description of
the inertia length appears important to understand the mode excitation.
This idea will be further discussed.

\section{A solver for the dispersion relation}
\label{section_DispersionRelationSolver}
In order to evaluate and solve the fishbone-like dispersion relation, we 
wrote a {\sf c++} program using the {\sf Matpack} library 
\cite{Matpack}
for the computation of integrals and special functions.
The program takes as input equilibrium parameters calculated from the CRONOS
and PION codes, to calculate the various terms of the fishbone like
dispersion relation.

The program can assess the dispersion relation for a chosen value
of the frequency $\omega_0$ and deduce the mode stability, 
or find the roots of the dispersion relation in a circle of
center $\omega_0\in\mathds{C}$ and radius $|\delta\omega|$ given 
by the user, using Davies method \cite{Davies_86}.
The input and output parameters of the solver are given in 
Tab.~\ref{tab_FPDispRel}
. 
\begin{table}[h!]
\begin{center}
\begin{tabular}{lll}
\hline\hline
\bf Input  & $\bullet$ Equilibrium data reconstructed by 
& -$B_0$, $R_0$, $q(r), \tau_e(r_s), n(r_s)$ \\
&CRONOS or experimental diagnostics& \\
&$\bullet$ Hydrogen distribution function & -$n_h, T_h, \lambda_0$\\
&computed by PION and projected & -$\Delta\lambda$ (the typical width of 
$f_\lambda$)\\
&onto an anisotropic Maxwellian&is set up by the user\\
& $\bullet$ User defined mode characteristics & 
-$(\sf m, n), \omega_0 + \delta \omega$\\
&&-Radial Structure: Kink/Numerical \\
& $\bullet$ Model & -Inertia: w/wo high order FLR\\
&& -$\delta\hat{W}_k$: w/wo anistropy drive $f_\lambda'$\\
&& -Value of $\delta \hat{W}_\text{MHD}$\\
\hline
\bf Output & $\bullet$ Simple estimate of stability & 
$\delta \hat{W}_k(\omega_0)$, $\delta \hat{I}(\omega_0)$,\\
&&$\re(\omega_0)\im(\delta\hat {W}(\re{\omega_0}))>0 $ ? \\
&$\bullet$ Search of  dispersion relation roots & 
Roots\\
&in circle of center $\omega_0$ and radius $|\delta
\omega|$& \\
\hline\hline
\end{tabular}
\end{center}
\caption{\label{tab_FPDispRel}
\footnotesize
Inputs and outputs of the dispersion relation root solver.
}
\end{table}

\section[Comparison with Tore-Supra experiments]
{Comparison of the theoretical threshold with experiments carried out
  in the Tore-Supra tokamak}
\label{section_ComparisonWithExperiments}

\subsection{Objectives of the experimental campaign }
To analyze the BAE stability and in particular estimate the
relevance of the thresholds described by Eqs.~\ref{eq_Threshold} and 
\ref{eq_ModifiedThreshold},
experiments were performed to determine the experimental excitation thresholds 
of BAEs in Tore-Supra, for {\bf various equilibrium parameters}. 
Two aims were given to this study:
\begin{itemize}
\item Identify the role of {\it global} parameters, 
  $B_0, n_0, f_{min}$, when profiles are kept homothetic.
\item Analyze some effects related to a profile modification, for
 example a modification of the $q$-profile.
\end{itemize}

\subsection{Experimental settings}
\subsubsection{Scenari and parameters}
\begin{figure}[ht!]
\begin{center}
\begin{minipage}{\linewidth}
\begin{center}
\includegraphics[width=0.6\linewidth]{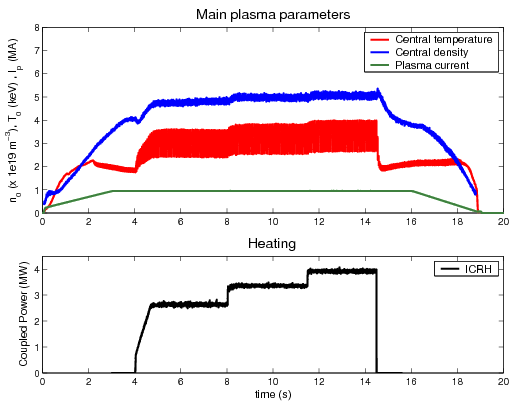}
\caption{\label{fig_Param41925}
\footnotesize Time evolution of the main plasma parameters in shot \#41925.}
\end{center}
\end{minipage}
\end{center}
\end{figure}

Experimental conditions were varied around the {\bf reference parameters}: 
$B_0=3.8$T, $I_p=1$MA, $n_0=5e19$m$^{-3}$, $P_{ICRH}=2.5$MW,
$f_{min}=4\%$, for which earlier observations of BAEs could be performed.
To determine the thresholds for BAE excitation, 
ICRH power was increased step by step with an attempt to keep
the other parameters fixed (In reality, these parameters cannot be
kept fixed as explained in subsection \ref{ssection_ExpDesign}, 
which limits the range of parameters which can be explored), using
long enough steps to allow for equilibrium parameters to reach steady-state
$\sim 3.5 s$.
Fig.~\ref{fig_Param41925} displays a typical scenario used during our
set of experiments.

First, equilibrium conditions were modified, with an attempt to keep
profiles homothetic. For this, pure ICRH discharges were used, keeping
the ratio $I_p/B_0$ constant to limit modifications of the $q$-profile. 
Three values were considered for the field, $B_0=2.8/ 3.2/ 3.8$T,
three (target) values for the central density $n_0 =4.5 / 5/ 5.5 e^{19}$m$^{-3}$. 
The minority fraction,  $f_{min}$ was varied from 2.5\% to 10\%, 
whereas $P_{ICRH}$ was increased from $0$ to $5.5MW$ with 0.5MW steps.
Next, additional experiments were conducted to determine a possible
role of the shape of the $q$ profile, controlled by some input of LH
power (already indicated to make possible the generation
of a localized current), $P_{LH} =0.0/1.2/2.2$MW.

\subsubsection{Diagnostic set up}
Various fluctuation diagnostics have been set up for the experiments:
reflectometry, ECE fluctuation diagnostics, interferometry, soft X
rays, designed to measure fluctuations in the last half second of each
ICRH step.
Even if some observations could be performed on the electron
temperature (after a heavy post-treatment though, see Fig.~\ref{fig_Psd42039}), 
the best observations were achieved with reflectometry which proved
high sensitivity. 
For this reason, most of the following analysis is based on reflectometry.
The two modes of the reflectometry diagnostics (described in 
subsection \ref{ssection_Diagnostics})
were used 
to obtain the long time evolution of the mode during a whole sawtooth period,
as well as its radial structure and to allow for the mode detection over a 
broad radial range.

\subsection{General qualitative features of the mode stability}
\label{ssection_GeneralExperimentDescription}
Before entering quantitave considerations, let us describe
some features of the occurence of BAEs in our discharges, qualitatively.
\subsubsection{BAE stability and sawtooth dynamics}
\begin{figure}[ht!]
{\footnotesize Shot \#42039, $t \sim9.5s$: 
$B_0=3.80T$, $n_0=4.45e19m^{-3}$, $I_p=0.95MA$, 
$P_{LH}=0.0MW$, $P_{ICRH}=3.1MW$, $f_{min}=8\%$}
\begin{center}
\begin{minipage}{0.49\linewidth}
\begin{center}
\includegraphics[width=\linewidth]{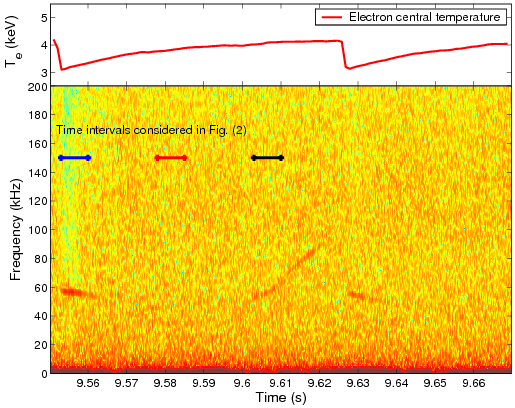}
\caption
[\footnotesize 
Reflectometry spectrogram at the plasma center for shot $\#$42039 
(t $\sim 9.5s$)]
{\label{fig_Specgram42039}
\footnotesize 
Reflectometry spectrogram at the plasma center for shot 
$\#$42039. In this picture, the time interval [9.56 9.62] corresponds to 
a typical sawtooth period.}
\end{center}
\end{minipage}\hfill
\begin{minipage}{0.49\linewidth}
\begin{center}
\vspace*{1.cm}
\includegraphics[width=\linewidth]{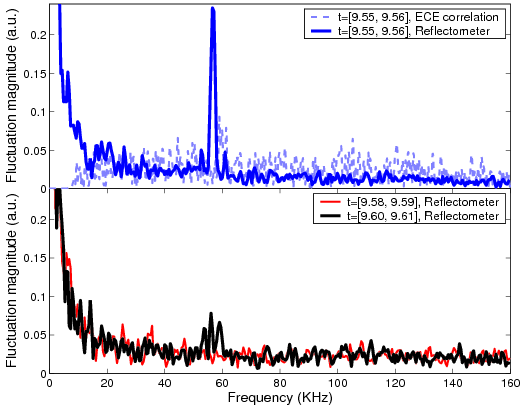}
\caption
[\footnotesize
Detailed fluctuation spectra during a sawtooth period.]
{
\label{fig_Psd42039}
\footnotesize
Fluctuation spectra corresponding to the time intervals shown in 
Fig.~\ref{fig_Specgram42039}, taken at the plasma center.
The plotted ECE correlation spectrum results from the correlation of
the 10 first ms of each sawtooth period between $t=9s$ and $10s$ 
(necessary to discrimate the mode) and the filtering of the 
perturbations below 10 kHz.
}
\end{center}
\end{minipage}
\end{center}
\end{figure}

We already mentioned the existence of sawtooth oscillations in all of
our discharges, that is, oscillations of the central temperature
characterized by a sudden crash and a slower restoration of about 20s
(Fig.~\ref{fig_Specgram42039} and \ref{fig_SpectrumAndRadialStruc41925}).
The reason for these crashes is that our equilibria are
inside the limit where pure MHD incompressible kink instabilities are unstable 
(that is $\delta \hat W_f <0$ in Eq.~\ref{eq_Bussac}). Kink
instabilities (which can be identified easily on fluctuation
diagnostics, with a real frequency of around $3$kHz) grow such that
central confinement gets lost: it is
the sawtooth crash. Following this drop of the central
temperature, the plasma gets MHD stable ($\delta \hat W_f >0$) and the
central temperature can be restored until MHD marginal stability
$\delta \hat{W}_f=0$ is reached once again.

As a consequence, despite our attempts to reach steady-state conditions for each
ICRH step, sawtooth oscillations does not allow to rigorously talk
about a constant central temperature, or a constant density. Moreover,
the presence of BAEs was found to be influenced by the sawtooth oscillations.
This influence is well illustrated in Fig.~\ref{fig_Specgram42039} and 
\ref{fig_Psd42039}, which are typical observations we made. 
As observed in these figures, the mode is more intense at the
beginning and at the end of the sawtooth period, where the mode
frequency also appears to grow (sometimes faster than the temperature
relaxation time).

For the statistical analysis of the conditions for BAE
destatbilization, we will not be considering the details of the sawtooth 
dynamics. In particular, the so-called central temperature will be a time
average of the central temperature, whereas the presence of BAEs will
have to be understood as the presence of the mode at the end of each
sawtooth period. Because our statistical analysis makes use of ``scalings'', 
this choice seems reasonable.

For more detailed analysis (the detailed analysis carried out for 
shot \#42039 in the following), 
the temperature and $q$ profiles computed by CRONOS will be used, 
which corresponds to the equilibria prior to the sawtooth crash.

\subsubsection{BAE stability and LH Heating}
\begin{figure}[ht!]
{\footnotesize 
Shot \#42806: $B_0=3.86T$, $n_0=4.65e19m^{-3}$, $I_p=0.95MA$, 
$P_{LH}=0.0MW$, $P_{ICRH}=1.7MW$, $f_{min}=5.5\%$\\
Shot \#42815: $B_0=3.86T$, $n_0=4.75e19m^{-3}$, $I_p=0.95MA$, 
$P_{LH}=2.2MW$, $P_{ICRH}=1.7MW$, $f_{min}=4.7\%$}
\begin{center}
\begin{minipage}{\linewidth}
\begin{center}
\includegraphics[width=0.5\linewidth]{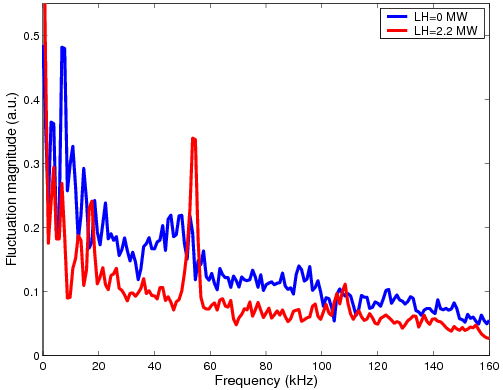}
\caption{\label{fig_PsdWLH}
\footnotesize Modification of the BAE density fluctuation 
spectrum with the introduction of LH power. }
\end{center}
\end{minipage}
\end{center}
\end{figure}

A second striking feature which appeared during our experiments is the
role of LH power. As shown in Fig.~\ref{fig_PsdWLH}, 
the mode amplitude is signicantly enhanced in the
presence of LH heating ($P_{LH} = 2.1MW$). This phenomenon was found to be
particularly obvious close to marginal stability, suggesting a reduction of
the mode excitation threshold with the introduction of LH heating. 
In some shots, the modes also appeared to be radially extended in the
presence of LH power.

Even if LH power implies some additional heating of the ion population, it is 
not expected to produce ions with sufficient energies to excite acoustic modes.
Hence, these observations suggest that the addition of LH power modifies 
{\it equilibrium} properties in a way which favors the mode excitation.\\

\subsection{Single case analysis}
\label{eq_SingleCaseAnalysis}
We first start the comparison with a particular case analysis to verify 
the above claim that the threshold condition Eq.~\ref{eq_Threshold}
is accessible for relatively standard parameters, using the numerical tool
described in section \ref{section_DispersionRelationSolver}.

\begin{figure}[ht!]
{\footnotesize Shot \#42039, $t \sim 7s$: 
$B_0=3.80T$, $n_0=4.37e19m^{-3}$, $I_p=0.95MA$, 
$P_{LH}=0,0MW$, $P_{ICRH}=2.3MW$, $f_{min}=8\%$ \\
Shot \#42039, $t \sim 14s$: 
$B_0=3.80T$, $n_0=4.54e19m^{-3}$, $I_p=0.95MA$, 
$P_{LH}=0.0MW$, $P_{ICRH}=4.6MW$, $f_{min}=8\%$
}
\begin{center}
\begin{minipage}[t]{0.465\linewidth}
\begin{center}
\includegraphics[width=\linewidth]{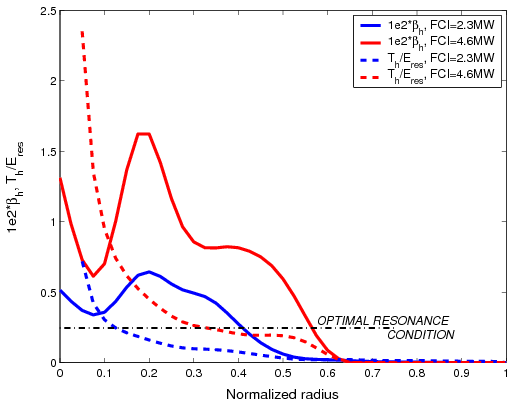}
\caption
{\label{fig_ThEres42039}
\footnotesize Radial profiles of the fast ion normalized pressure
and resonance parameter $T_h/E_{res}$, for different power input.}
\end{center}
\end{minipage}\hfill
\begin{minipage}[t]{0.524\linewidth}
\begin{center}
\includegraphics[width=\linewidth]{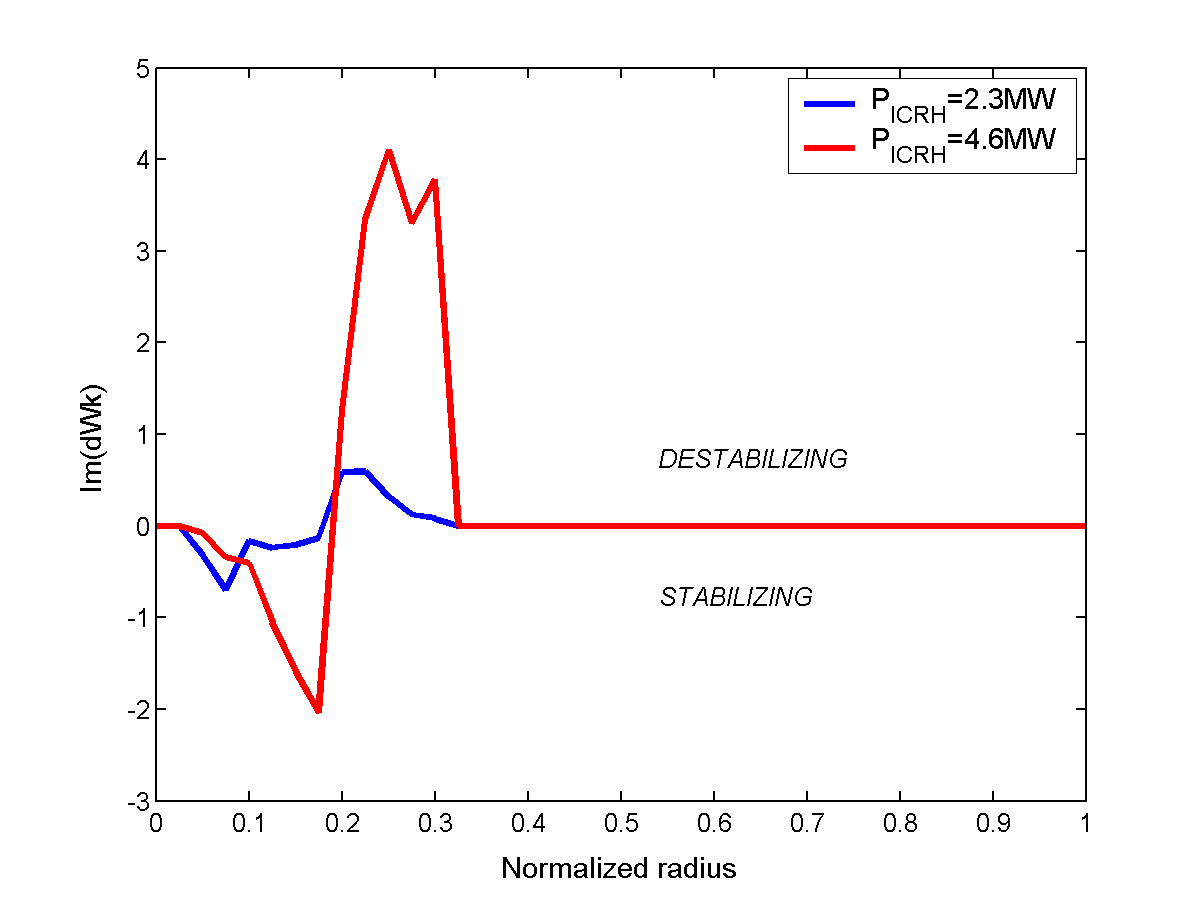}
\caption
[\footnotesize
Radial profile of the fast ion energy drive for shot \#42039 
($t\sim 7s$).]
{\label{fig_idWk42039}
\footnotesize Radial profile of the fast ion induced
wave-particle energy transfer, ${\rm Im}(d{\hat W}_k({\bar r}))$, for 
different power input. 
${\rm Im}(d{\hat W}_k({\bar r}))$ verifies 
$\int_{\bar r}d{\bar r}{\rm Im}(d{\hat W}_k({\bar r})) 
= {\rm Im}(d{\hat W}_{k})$ with
${\bar r}$ the normalized radius.}
\end{center}
\end{minipage}
\end{center}
\end{figure}

The studied case is shot \#42039, analyzed at two different times, 
$t\sim 7s$ and $ t\sim 14 s$, corresponding to two different levels of
ICRH power, but similar equilibrium parameters.
The main experimental parameters of the shot are given in 
Fig.~\ref{fig_idWk42039}. The q-profile was found from a CRONOS simulation
to be almost parabolic, with a $q=1$ surface localized at a normalized 
radius of 0.35, and a central value consistent with q=0.7
(within the code errorbars).
From an experimental point of view, modes were observed with reflectometry
in both cases, and the first set of experimental conditions ($t\sim 7s$)
was found to be close to marginal stability 
(when increasing the power lever step by step).

The features of the fast ion distribution calculated by PION are given
in Fig.~\ref{fig_ThEres42039}, where it appears that relatively large 
values of the fast ion pressure can be reached (for comparison, 
$\beta_i\sim 0.2-0.3\%$), and that the resonant condition
$ T_h/E_{res}\sim 1$ (more precisely optimal for $T_h/E_{res}\sim 0.25$ for a 
dominant temperature gradient as illustrated in Fig.~\ref{fig_ModeExistence})
is accessible, which makes the resonant drive reasonable.
As also appear in this figure, the localization of fast particles is
slightly off-sxis.
Now, from the comparison of the two experimental conditions, it appears that 
the introduction of ICRH power can have an effect on the two free parameters 
involved in the threshold Eq.~\ref{eq_Threshold}: the fast ion pressure $\beta_h$
and the resonance condition which is better met a higher power
($t\sim 14s$)
between the normalized radii 0.2 and 0.35, where the fast ion pressure profile 
presents a negative slope.

The corresponding profiles for the drive are given in 
Fig.~\ref{fig_idWk42039}. The drive appears to be very sensitive 
to the gradients of the fast particle radial pressure, as expected.
Only regions characterized by a negative radial gradient of the fast ion 
population 
(see the form of $\beta_h$ in Fig.~\ref{fig_ThEres42039}) lead to a 
drive, whereas positive gradients lead to damping.
Consequently, a stronger off-axis heating could be expected to be stabilizing
for the modes described here, even for large values of $P_{ICRH}$.
Finally, in agreement with the shapes of the fast particle distribution 
functions, the drive at
higher power is three times as strong as in the lower power case.
\\

Let us now compare drive and damping, calculated with the previously defined
expressions of $\delta\hat{I}$ (with 4th order FLR), and of $\delta\hat{W}_k$.
(without anisotropy
\footnote{Anisotropy is usually taken to be a limited effect and to be relevant
  only with very particular heating conditions \cite{Boswell_06, Nazikian_08}
  With the $\delta$-function used as a model for the fast ion distribution, 
  we do not want to introduce an articially strong anisotropy induced 
  excitation.}
). 
For the marginal conditions ($t\sim7s$) and the experimental value of the mode 
oscillation frequency $\omega_r^\text{exp}/2\pi=55kHz$, it comes
\begin{equation}
\im\delta {\hat I}(\omega_r^\text{exp})=0.024 \\
\im\delta {\hat W}_k(\omega^\text{exp}_r)=0.026
\label{eq_dIdWk42039t7}
\end{equation}
which seems to suggest a good agreement between theory and experiment.
Indeed, from this calculation, the mode appears to be rightfully excited, and
close to  marginal stability.

Note that the use of the  dispersion relation solver of section of 
\ref{section_DispersionRelationSolver} as a root solver is unfortunately
much less conclusive at the moment.
For a choice of $\delta\hat{W}_\text{MHD}=0$ (consistent with the
assumption of an $(-1,1)$ kink structure for the BAE, a structure which is 
obviously close to marginal MHD stability as proved by the presence of 
sawtooth oscillations), the calculated solution of the dispersion relation 
$(\omega_r^\text{sol}, \gamma^\text{sol})$ is
\begin{equation}
\omega_r^\text{sol}/2\pi = 66 $kHz$ \\
\gamma^{sol}/\omega_r^\text{sol}=-1.6\times10^{-3},
\end{equation}
which implies that the mode is damped, and that some corrections to the 
dispersion relation needs to be provided to avoid the difference between the
computed and the experimental values of the oscillation frequency
(In particular, the perturbative treatment of Landau damping is questionable).
Nevertheless, the root solver, being currently still under a testing phase 
(for dispersion relations with multiple roots), caution is necessary at the 
moment.

In any case, the comparison between theory and experiments confirms the 
closeness to marginal stability for the shot \#42039 at $t\sim 7s$.
This case will be considered as a reference in the following.


\subsection{Statistical analysis of the role of global parameters}

\subsubsection{Procedure}
In this part, we compare the observation of BAEs with the threshold
calculated in Eq.~\ref{eq_ModifiedThreshold}, taking the marginal
experimental conditions described above as a reference to define
the constants $c_1$, $c_2$ and $c_3$, defined there.
Such analysis can be seen as an experimental proof-check of the 
perturbative treatment of the excitation and damping of BAEs.

In the following, $c_1$ and $c_2$ are adjusted to match the experimental 
data of shot $\#42039$ at time $t\sim7s$ with the computed value of the
fast ion distribution calculated by the code PION at the normalized radius
of maximum excitation, 0.25. 
$c_3$ is made consistent with the ratio of $\delta\hat{I}$ and 
$\delta\hat{W}_k$ computed for shot $\#42039$, that is
\begin{equation}
c_3 = c_1n_0^{1/4}\frac{P_{ICRH}}{n_0^2}\frac{T_e}{B_0}\delta_h
\times \frac{\delta\hat{I}}{\delta \hat{W}_k}
\label{eq_c_3}
\end{equation}
where $\delta\hat{I}$ and $\delta\hat{W}_k$ are the values given 
in Eq.~\ref{eq_dIdWk42039t7}.\\

The theoretical threshold is compared with the experimental mode onset, 
defined as the existence of a time
coherent signal/noise value exceeding  2, on the reflectometry diagnostic.

\subsubsection{Results}
\begin{figure}[ht!]
  \centering 
  \includegraphics[width=0.8\linewidth]{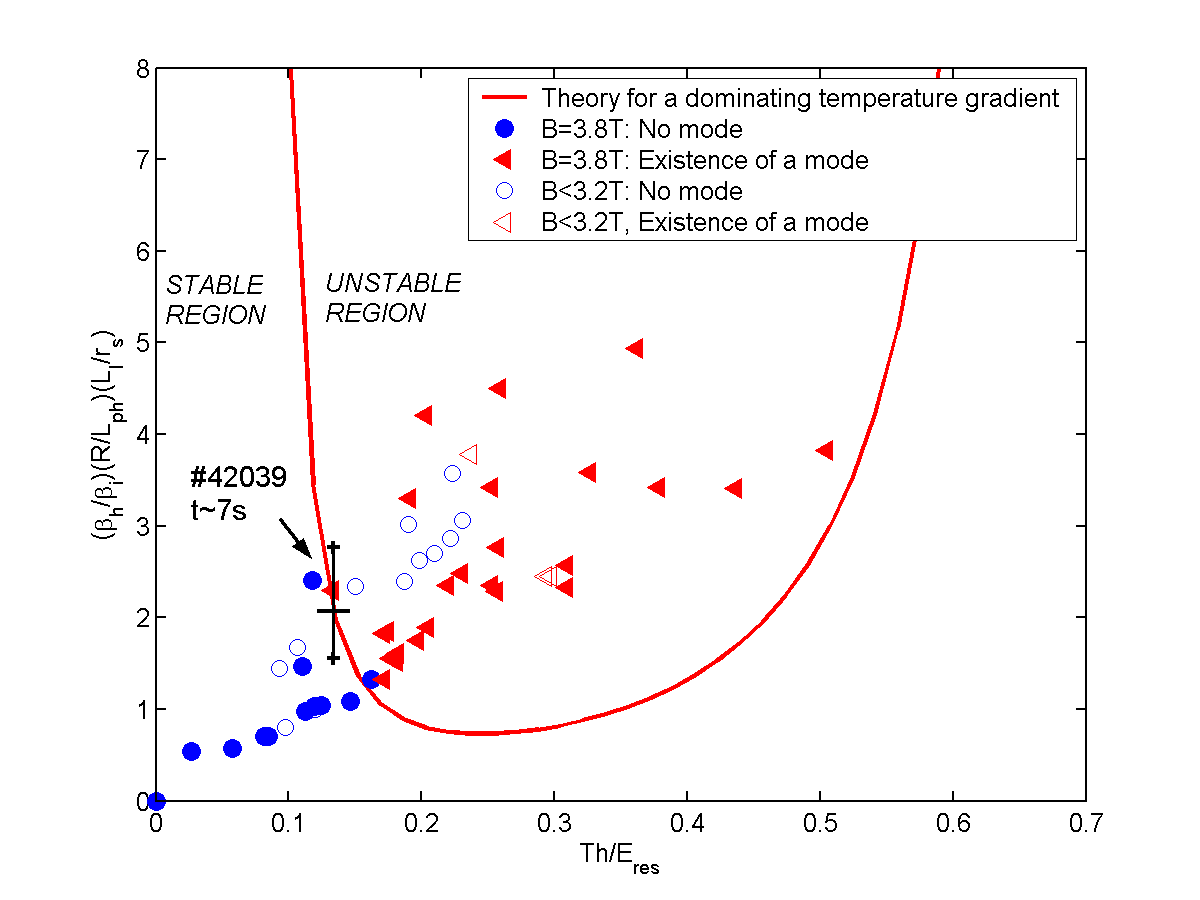}
  \caption
  [\footnotesize
  BAE apparitions in Tore-Supra discharges, compared to the theoretical
    prediction.
    ]
    {\footnotesize \label{fig_ModeExistence}
    BAE apparitions in Tore-Supra discharges, compared to the theoretical
    prediction.
    A ``mode observation''corresponds to a coherent signal/noise value 
    exceeding  2.
    The pointed dot corresponds to the  case $\#42039, t\sim 7s$.}
  
\end{figure}

The results of the mode investigation are given in 
Fig.~\ref{fig_ModeExistence}.
On this plot, the onset of modes  has been reported as a function of the 
two degrees of freedom involved in \ref{eq_ModifiedThreshold}, based on the 
scalings used to establish this formula and on the assumption that 
$c_1$, $c_2$ and $c_3$ do not significantly vary during the experiments.
The horizontal error-bar stands for the relative mismatch between the observed 
localization of the threshold (close to the dot representing shot 
$\#42039$ at time $t\sim 7s$) and the numerical computation of it
derive from Eq.~\ref{eq_dIdWk42039t7}.
Though quite small, it is observed to induce a possibly large error in the 
vertical direction due to the sensitivity of $\delta_h (T_h/E_{res})$ in 
Eq.~\ref{eq_c_3}.
The theoretical threshold has been plotted assuming a dominating 
radial temperature  gradient $L_{ph}=L_{Th}$. 
Keeping only the radial temperature gradient in formula
\ref{eq_ThresholdPolynom}, the threshold curve reads
\begin{equation}
\mathcal {Y}=c_3/\delta_h(\mathcal{X} = T_h/E_{res})
= c_3/(E_{res}/T_h)^{5/2}(E_{res}/T_h-1.5)e^{-E_{res}/T_h}
\end{equation}
where $(\mathcal{X},\mathcal{Y})$ stand for the abcissa and ordinate
of the graph.
From the full PION computation of the fast ion distribution carried out
for shot  \#42039, the consideration of a dominating temperature
gradient is reasonable. Interestingly, {\bf the vertical shape of the plotted 
theoretical curves close to $T_h/E_{res}\sim 0.1$ implies that a sufficiently 
high hot ion temperature is the major necessary condition for the mode onset} 
whatever the hot ion pressure.\\

As can be seen in this figure, a good agreement appears between
theory and experiment for the fixed large  magnetic field $B_0=3.8T$, 
which gives some confidence in the computed orders of magnitude of Landau 
damping and excitation. 
The agreement is particularly striking considering the rather large extent 
of the investigated values of the minority fraction. 
The role of the latter parameter is indeed well illustrated  by the reference 
shot $\#42039$ at time $ t\sim7s$,
experimentally observed to be marginally unstable in spite of a 
relatively large power input $ P_{ICRH}=2.3MW$ but a high minority 
fraction  $f_{min}=8\%$.
For similar equilibrium parameters and for more standard minority
fractions $f_{min}\sim4\% $, modes could be observed with a power input 
$P_{ICRH}=1.5MW$.
Hence, the {\bf perturbative framework appears to have some relevance, at least
to estimate the qualitative relative effect of  global macroscopic 
parameters}.

For lower field, excitation is seen in Fig.~\ref{fig_ModeExistence} to be 
harder than expected by theory. As explained 
in section \ref{section_ThresholdSimplification}, the model used in this 
threshold assessment  does not imply a clear role of the magnetic
field, in contradiction with the experimental observation.
This may suggest the relevance of using a different model for
the inertia length. In particular, because the latter (very small) length may 
evolve significantly in the nonlinear regime, an improvement of the
stability analysis may be considered towards a nonlinear description of the
BAE structure. However, considering the important shift of the threshold for
lower fields, found experimentally in Fig.~\ref{fig_ModeExistence}, 
this model modification is likely to be insufficient 
and other possibilities such as a harder mode detection for lower
magnetic field may provide simpler explanations.

\subsection{Detailed additional analysis}
Though less systematically studied, some interesting phenomena 
were put forward in our experiments, enhancing the role of 
additional parameters neglected so far. In the following, attention
is given to the role of the shear and  fast ion distribution
anisotropy.\\

\begin{figure}[ht!]
\begin{center}
\begin{minipage}{0.5\linewidth}
\begin{center}
\includegraphics[width=1.1\linewidth]{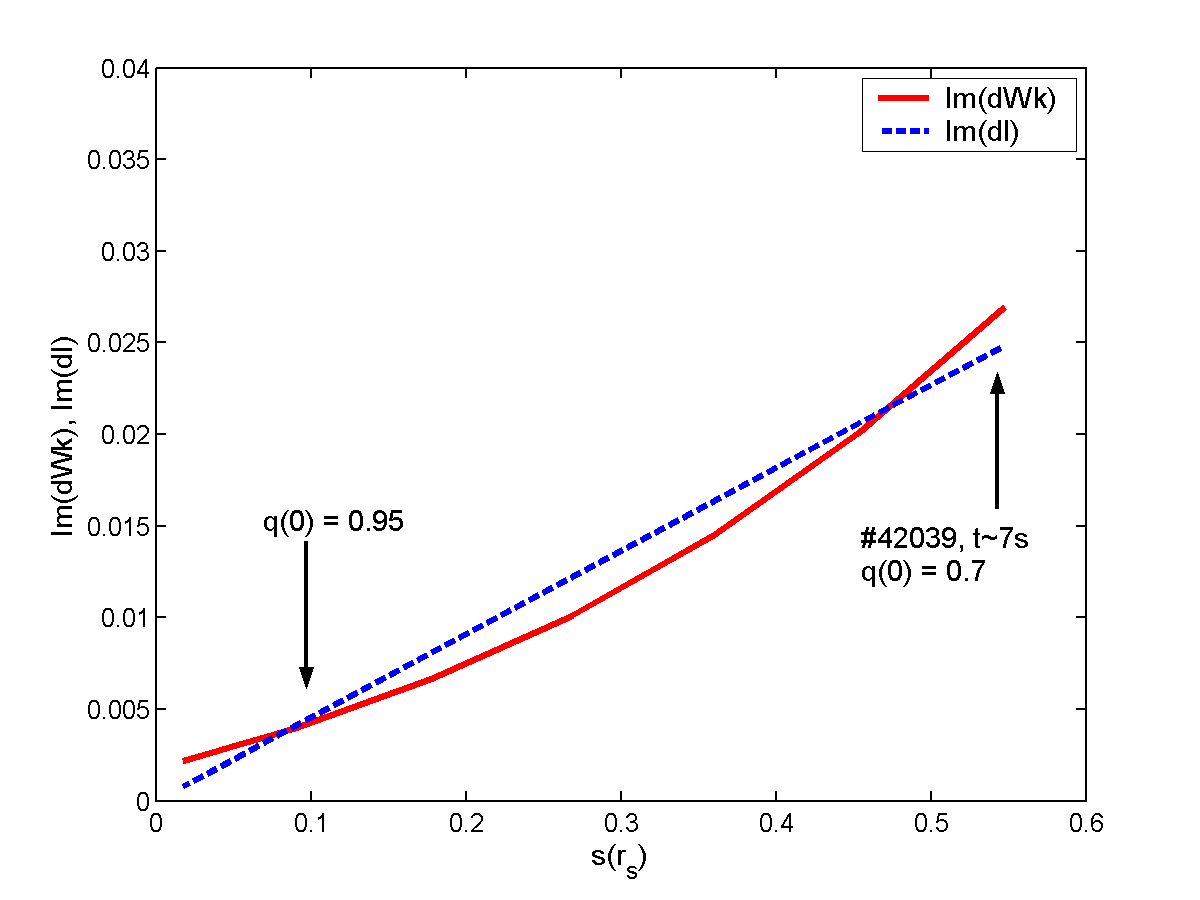}
\caption
[\footnotesize 
Effect of modification of the shear on the drive and damping.]
{\label{fig_Shear_Obs}
\footnotesize 
Effect of modification of the shear (taken at the resonance layer)
on the drive and damping.
In the main picture, fixed parameters correspond to shot\#42039, with 
$P_{ICRH}=2.3MW$. The second picture is a speculative schematic of a 
possible configuration, commented in the core of the paper.}
\end{center}
\end{minipage}\hfill
\begin{minipage}{0.49\linewidth}
\vspace*{-0.2cm}
\begin{center}
\includegraphics[width=\linewidth]{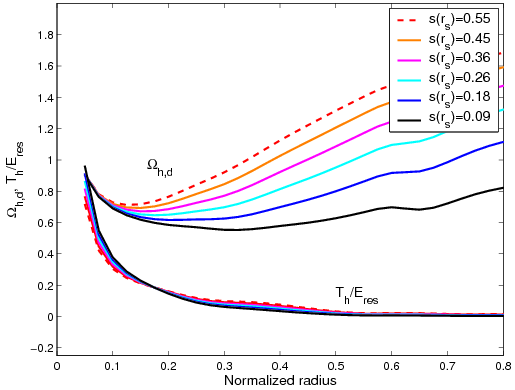}
\caption
[\footnotesize
Effect of a modification of the shear on the drift frequency $\Omega_{h,d}$
and resonance parameter $T_h/E_{res}$ radial profiles.]
{\label{fig_Shear_Explanations}
\footnotesize 
Effect of a modification of the shear on the drift frequency $\Omega_{h,d}$
and resonance parameter $T_h/E_{res}$ radial profiles.
}
\end{center} 
\end{minipage}
\end{center}
\end{figure}

The modification of the shear $s$ is the best candidate to 
explain the drop of the excitation threshold in the presence of LH
heating, described in  subsection~\ref{ssection_GeneralExperimentDescription}. 
Indeed, the level
of LH heating is not expected to heat the fast ions to
relevant energies for resonance with an acoustic mode. Hence, the noted 
threshold reduction is more likely to be linked to a modification of  
equilibrium parameters. Besides, this interpretation is in agreement 
with the discontinuous presence of the mode  in the course of the
sawtooth period shown in \ref{fig_Specgram42039}.
To better understand this role, we computed numerically the values of 
the excitation and damping for different artificial $q$ profiles.
Starting from the parameters of shot  $\#42039, t\sim 7s $, 
we tested various parabolic $q$ profiles with a fixed $q=1$ surface, 
and different central values and computed the resulting values of the
drive and damping. The results of this analysis are given in 
\ref{fig_Shear_Obs} and \ref{fig_Shear_Explanations}.
As can be seen in Fig.~\ref{fig_Shear_Obs}, a flatter $q$ profile tends to 
decrease both the excitation $d{\hat W}_k$ and damping $d{\hat I}$. 
This behaviour 
of the damping is easy to understand noticing that $L_I \propto
\sqrt{s(r_s)}^{ \ -1}$ and hence $\delta I\propto \sqrt{s(r_s)}$. 
To understand the origin of the drop of $d{\hat W}_k$,
we plotted in Fig.~\ref{fig_Shear_Explanations}, the parameters of 
Eq.~\ref{eq_Threshold} which are sensitive to a $q$ profile 
modification.
The major parameter modified here is obviously the bounce frequency, 
which is an affine function of $s$.
The simultaneous drop of the drive $\im (d{\hat W}_k)$ and of the damping 
$\im (d{\hat I})$ with
a flattened $q$ profile does not allow to explain the decrease of
the BAE excitation threshold non ambiguously. 
However, as can be seen in Fig.~\ref{fig_Shear_Obs}, the graphs representing
the drive and the damping are characterized by different curvatures.
$\delta {\hat I}$ appears to be concave and $\delta {\hat W}_k$ to be
convex, which is consistent with the interpretation of these behaviours. 
This situation enables the two curves to cross twice, meaning that an 
increase of the  shear may induce either an easier or a harder mode
excitation, depending on the starting point.
Such a situation can be postulated to explain the mode onset at
the beginning ($s\sim 0$ at the center) and at the end of a sawtooth  
period.\\

The fast ion anisotropy may also play a role in the stability
analysis. First, as can be seen in (\ref{fullydevelopeddWk}), 
anisotropy may induce an energy gradient and hence a mode excitation. 
With  the parameters of shot  \#42039 at time $t\sim 7s$ and
taking a $\delta$-function to model the fast ion pitch angle
$f_\lambda=\delta (\lambda-\lambda_0)$, 
anisotropy is seen to imply a factor of 2 on the mode drive. Although the 
$\delta$-function may seem to be a poor model, this important factor may
reveal a certain  role of this excitation mechanism.

\begin{figure}[ht!]
\begin{center}
\begin{minipage}[b]{0.5\linewidth}
\begin{center}
\includegraphics[width=\linewidth]{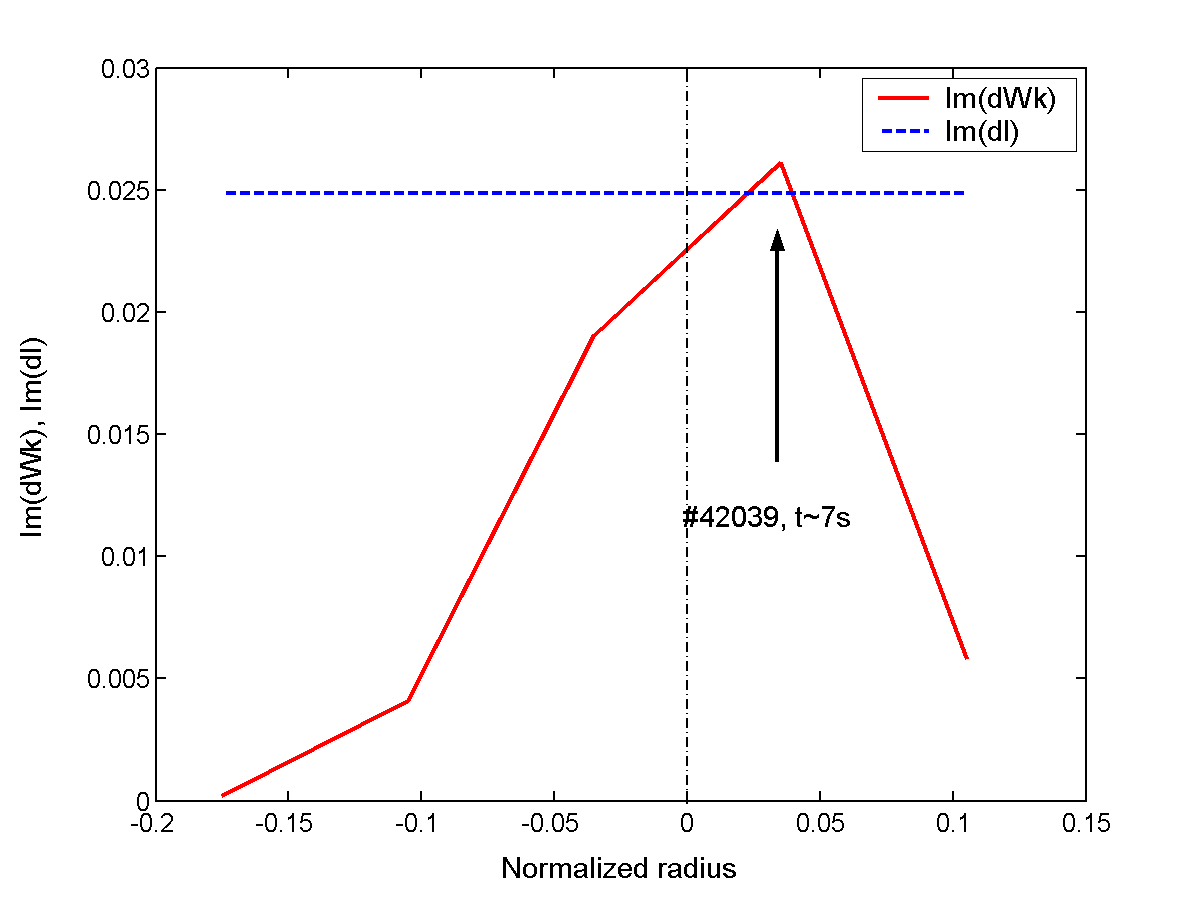}
\caption
[\footnotesize
Effect of modification of the anisotropy parameter $\lambda_0$ on the 
drive and damping.]
{\label{fig_ICRHLocalization_Obs}
\footnotesize
Effect of modification of the anisotropy parameter $\lambda_0$ on the 
drive and damping.
$\lambda_0$ gives some hint into the localization of the ICRH heating.}
\end{center}
\end{minipage}\hfill
\begin{minipage}[b]{0.48\linewidth}
\begin{center}
\includegraphics[width=\linewidth]{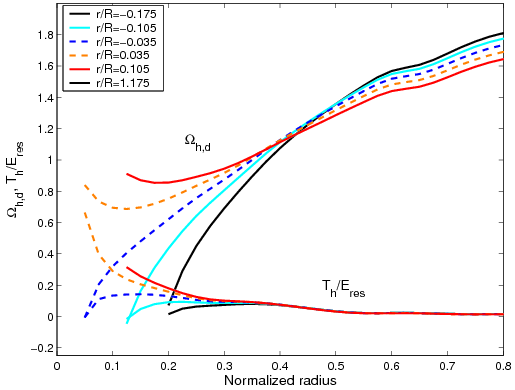}
\caption{\label{fig_ICRHLocalization_Explanations}
\footnotesize 
Effect of modification of the anisotropy parameter $\lambda_0$ on the 
drift frequency $\Omega_{h,d}$ and resonance parameter $T_h/E_{res}$ 
radial profiles.} 
\end{center}
\end{minipage}
\end{center}
\end{figure}

The fast ion anisotropy also gives some importance to the heating
localization. As an addition to the modification of the characteristic
radial gradients of the fast ion population noted above, a
modification of the heating localization  changes the particle 
pitch-angle $\lambda_0$. To assess this phenomenon, we plotted in
\ref{fig_ICRHLocalization_Obs} the values of $\im(\delta \hat{W}_k)$ 
for different values of $\lambda_0$ (and  artificially  fixed radial 
gradients) starting from the parameters of shot \#42039 at marginal 
stability.
As can be seen in Fig.~\ref{fig_ICRHLocalization_Obs}, a lower drive may
be expected for a high field side heating localization. The
explanation of this behaviour is given in
Fig.~\ref{fig_ICRHLocalization_Explanations}, where it appears that the
resonance condition is better met when fast particles are
more strongly trapped and $\bar{\Omega}_{h,d}$ subsequently enhanced.
Though not fully realistic, 
Figs.~\ref{fig_ICRHLocalization_Obs} and \ref{fig_ICRHLocalization_Explanations}
show a non-negligible role of the particles pitch-angle and of the radial
variation of the precession frequency $\Omega_{h,d}$, often taken to be
constant in the theoretical assessment of the energetic particle drive.
This idea of a role the  heating localization was suggested to us
by experimental observations, but a cleaner experimental investigation
of this phenomenon still needs to be carried out.

\section{Summary}
{
\it
In the present chapter, 
the BAE excitation threshold was calculated using analytic expressions for 
the energetic particle drive and for ion Landau damping, and it was compared
with experiments conducted on the Tore-Supra tokamak both qualitatively
and quantitatively.

\begin{itemize}
\item In the perturbative framework, the calculation of the threshold
was found to involve different parameters relative to the energetic particle
distribution as well as to equilibrium parameters. These parameters 
could be reduced to {\bf two degrees of freedom} which depend on experimental
macroscopic tunable parameters (the density, the magnetic field, the minority 
fraction and level of ICRH heating), under the assumption that radial profiles 
are homothetic, from one experiment to the other.

\item
This made possible a statistical comparison of the calculated BAE threshold with 
experiments. This comparison returned a fair agreement between theory and 
experiment, which confirms the {\bf orders of magnitude of the computed damping 
and excitation mechanisms} involved in the BAE stability, and the rough
{\bf global role of the above macroscopic tunable parameters}.
The {\bf necessity to meet the  resonant condition appeared to be the most 
determining factor for the mode destabilization}, in agreement with both theory 
and experiment.
\item
More surprisingly, the conducted analysis  highlighted an ambiguous {\bf role 
of the shear} in the BAE excitation, which may give some explanation to the
discontinous onset of the mode in the course of a sawtooth period.

\item
Nevertheless, the threshold analysis as well as the theory/experiment comparison 
also pointed out some missing physics of the threshold calculated here.

First, the strong role of the inertial layer (narrow) size in the assessment of 
Landau damping suggests that the mode stability can be modified when moving to 
the {\bf nonlinear framework}, because strong  gradients are usually not 
expected to survive nonlinearly.

Next, our model cannot  explain the higher threshold for BAE destabilization
at lower $B_0$-field. Neither can it explain the increase of the BAE frequency 
at the end of the sawtooth periods. The {\bf perturbative treatment of Landau
damping}, used here, should definitely be seen as weak point, considering the
orders of magnitude involved. Moreover, {\bf diamagnetic effects}, neglected in
this analysis, have recently been postulated for the increase of the BAE 
frequency at the end of the sawtooth period, 
observed in the Asdex Upgrade tokamak  as well \cite{Lauber_09}.
Although the latter effects are a priori negligible in our experimental
conditions, it may be interesting to assess such a claim.

Thus, both a non-perturbative treatment of Landau damping
and diamagnetic effects are currently being implemented in the BAE 
dispersion relation, following the derivations given in Ref.~\cite{Zonca_99}.
Further theory-experiment comparisons are also planned in the next experimental
campaigns.
\end{itemize}

One interesting point suggested by the {\bf analytic} calculation of the BAE 
threshold is the possibility of a {\bf nonlinear} modification of stability.
This is the motivation for the preliminary  nonlinear developments
conducted in next chapter, where indeed one process for the apparition of 
subcritical activity will be proposed.

However, one should note that the analysis to come is not directly suggested 
by the {\bf theory-experiment comparison} given here, which does not not 
advocate for the existence of a subcritical activity.
}

\begin{savequote}[20pc]
\sffamily
On doit exiger de moi que je cherche la v\'erit\'e, mais 
non que je la trouve.
\qauthor{Denis Diderot (1713-1784)
}
\end{savequote}




\chapter{Towards a nonlinear description}

In this chapter, we address the problem of the {\it nonlinear}
description of Beta Alfv\'en Eigenmodes, in order to get some ideas
about their {\bf saturation mechanisms} and the related 
{\bf transport of energetic particles}.
The goal of the following presentation is not to provide a full picture 
of the nonlinear evolution of BAEs. It is an attempt to enlight some 
difficulties related to the descripton of nonlinear {\it kinetic} behaviors, 
analyze the applicability of the most classical interpretation of fast 
particle saturation, that is the {\it nonlinear trapping} briefly outlined 
in subsection \ref{ssection_NonlinearPresentation}, 
and finally investigate some intrinsic difficulties of BAEs.
More specifically, two ideas are focused on.
First, we consider the possible existence of {\bf metastable saturated modes}, 
related to the nonlinear trapping of resonant particles. 
Next, some thoughts are given to the effects of a nonperturbative treatment
of damping and excitation.

In the first section, 
we explain how it is possible to extend the nonlinear energy principle 
displayed in the previous chapters. 
Next, we come back to the theory of {\it nonlinear trapping} in
order to determine its meaning and validity for BAEs. 
Finally, the analysis of metastable modes is carried out in section
\ref{section_SelfConsistentDamping}, 
whereas some observations and ideas related to the consideration
of non-perturbative damping and excitation are developed in section
\ref{section_NonPerturbativeCase}.

\section{Formulation of a nonlinear energy principle}
\label{section_NLEnergyPpe}

In linear analysis, perturbed solutions are {\it eigenfunctions} of 
eigenvalue $\omega$,
proportional $\propto \exp(i\omega t)$.
For such types of perturbations, we displayed in subsection 
\ref{ssection_EnergylikeRelation} of chapter \ref{chapter_GyrokineticToMHD},
an energylike dispersion relation, derived from the linear expansion of the 
low-frequency Maxwell equation, 
\begin{eqnarray}
  \nonumber
  0 = \mathcal{L_{\omega}}(\phi_\omega,{\bf A}_\omega, 
  \phi_\omega^\dagger, {\bf A}_\omega^\dagger)
  &=& -\int d^{3}\textbf{x}
  \frac{\mathbf{B}_{\omega}^{\dagger} \mathbf{B}_{\omega} }{\mu_0}
  + \sum_{s} \int d^{3}\textbf{x}
  \left({\bf j}_{s\omega}\cdot\mathbf{A}_{\omega}^{\dagger}-\rho_{s\omega}
    \phi_{\omega}^{\dagger} \right)\\
  &=& -\int d^{3}\textbf{x}
  \frac{\mathbf{B}_{\omega}^{\dagger} \mathbf{B}_{\omega} }{\mu_0}
  - \sum_s \int d^{3}\textbf{x}f_{s\omega} h_{s\omega}^{\dagger}.
  \label{eq_OmElectromagLagrangian_bis}
\end{eqnarray}
In linear studies, such a {\it dispersion relation} is sufficient to 
determine stability, accessed when solving for the roots 
$(\omega_r, \gamma)$ of the energylike relation.\\

When moving to the nonlinear frame, some additional difficulties  arise.
\begin{itemize}
\item
{\bf Several eigenmodes can couple together.}
Nonlinearities can couple multiple eigenfrequencies, and consequently,
the complete electromagnetic Lagrangian should be written as a sum of 
several modes of various eigenfrequencies.

This difficulty is real, but it does not prevent the use of energylike
relations of the form $\mathcal{L}_\omega=0$. Indeed, it is easy to see
that the linearity of Maxwell equations and of the Fourier transform,
still gives a sense to the energylike relations
\ref{eq_OmElectromagLagrangian_bis}, where
the fields $\mathcal{F}_\omega$ are taken to be  Fourier-like transforms 
of the nonlinear solution, $\mathcal{F}_\omega = 
(T/2\pi)\int_{0}^{2\pi/T} dt  \mathcal{F}(t)\exp (i\omega t)$
with $T$ larger than any time period of interest.
Making use of an energylike relation of the form 
\ref{eq_OmElectromagLagrangian_bis},
\begin{equation}
\mathcal{L}_\omega \equiv \mathcal{L}(\omega) = 0
\label{eq_SimpleEnergyRelation}
\end{equation}
is simply equivalent to the choice of one single oscillation frequency 
$\omega$ and it is valid nonlinearly, 
though now the nonlinear particle nonlinear responses need to 
be incorporated.

\item {\bf The energylike relation is in general a time dependent equation.}
The nonlinear behavior of a "mode", cannot be reduced to a single 
frequency $\omega$.
Starting for example from a linear eigenfrequency 
$\omega_0= \omega_{r0}+\gamma_0$ ($\gamma_0>0$), a mode can reach saturation
$\gamma = 0$, or its oscillation frequency can change. The shape of the 
eigenfunctions can also evolve.
In other words, the energy relation \ref{eq_SimpleEnergyRelation} is in 
general of the form
\begin{equation}
\mathcal{L}(\omega_{r0}+\delta\omega_r(t) + i\gamma(t), t) = 0
\label{eq_SimpleTimeDependentEnergyRelation}
\end{equation}
For example, if the mode structure is conserved, 
and there is a smallness parameter $\bar{\epsilon}$ such that 
$\delta\omega_r = O(\bar{\epsilon}\omega_{r0})$, 
$\mathcal{L}=\mathcal{L}_\text{(0)} + \mathcal{L}_\text{(1)}$ with
$\mathcal{L}_\text{(1)}=O(\bar{\epsilon}\mathcal{L}_0)$,
Eq.~\ref{eq_SimpleTimeDependentEnergyRelation} can be expanded in the form 
\begin{equation}
(\delta \omega_r +i\gamma)\partial_\omega\mathcal{L}_\text{(0)}(\omega_{r0}) 
= -\mathcal{L}_\text{(1)}(\omega_{r0})
\label{eq_SimpleEnergyEquationExpanded}
\end{equation}
which can be seen as a time dependent energy equation on the mode
real amplitude $\mathcal{A}$ if $\delta\omega_r=0$,
\begin{equation}
\omega_{r0}\partial_\omega\mathcal{L}_\text{(0)}(\omega_{r0})
\frac{d}{dt}\mathcal{A}
= -\omega_{r0}\mathcal{L}_\text{(1)}(\omega_{r0},\mathcal{A}).
\label{eq_SimpleEnergyEquation}
\end{equation}
Moreover, such an equation may need to be coupled to a time dependent
equation for the nonlinear particle responses.
\item {\bf The system may not remain  close to threshold, or close to an 
attractor.}
In order to display the energy equation \ref{eq_SimpleEnergyEquation},
we needed to assume that the system remained close to a given frequency 
point $\omega_{r0}$ (the threshold frequency for example, but it should also 
be possible to consider a nonlinear attractor as well). 
In particular, when $\omega_{r0}\partial_\omega\mathcal{L}$ is almost real,
such an assumption is equivalent to say that the mode energy density is 
assumed constant for the studied process. However, such an expansion makes
sense only if $|\gamma\partial_\omega\mathcal{L}|$ remains small, which may 
not be the case.

Some numerical situations have been found where a restoring force constrained
a system to remain close to the linear threshold, even for a strongly driven
mode \cite{Vann_07}.
However, in some experiments, strong frequency sweeping was found, with 
$\delta\omega_{r}\sim{\omega}_{r0}$ (see Fig.~\ref{fig_FcySweeping}), 
\begin{figure}[ht!]
\begin{center}
\begin{minipage}{\linewidth}
\begin{center}
\includegraphics[width=0.4\linewidth]{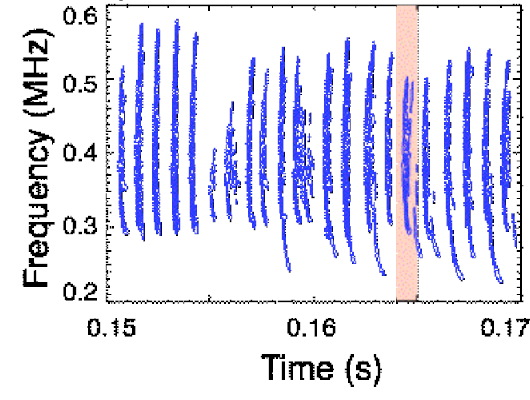}
\caption
[\footnotesize Frequency spectrum of sweeping modes measured in NSTX]
{\label{fig_FcySweeping}
\footnotesize
Frequency spectrum of sweeping modes measured in NSTX 
\cite{Fredrickson_06}.}
\end{center}
\end{minipage}
\end{center}
\end{figure}
which does not seem to fit the framework.
In particular, we may expect $\partial_\omega\mathcal{L}$ to be huge when 
{\it resonances} are included in the derivative.\\
\end{itemize}

A particular framework which avoids most of these difficulties is the
{\bf kinetic perturbative framework}, 
which is traditionally used for the study of the kinetic nonlinear 
saturation of gap modes (see section \ref{ssection_NonlinearPresentation} 
and Refs.~\cite{BerkBreizman_99}).

In this framework and in analogy with our previous computation of the linear
stability analysis, a well defined linear eigenmode with a real
eigenfunction  and a real frequency $\omega_{r0}$ exists, which is 
{\it perturbatively} damped and driven by
wave-particle resonances. Nonlinearly, it is assumed that this linear
structure is not modified, and in particular that the system remains 
close to the threshold linear eigenfrequency, 
$\delta\omega_r + i\gamma \ll \omega_{r0}$. In this case, an expansion
of the form \ref{eq_SimpleEnergyEquationExpanded} is possible, where 
$\mathcal{L}_\text{(0)}$ contains the {\it linear} thermal response assumed 
unchanged, and $\mathcal{L}_\text{(1)}$ contains the damping and driving 
{\it resonant} processes, treated nonlinearly. 

Moreover, when time decoupling is possible between the particle response
and the wave response, as it is the case in the most basic 
(but most studied) version of {\it nonlinear trapping}, 
Eq.~\ref{eq_SimpleEnergyEquationExpanded} does not even need to be expanded 
in the form of a time-dependent equation, or to be coupled with additional
equations \cite{BerkBreizman_99}.

Such a framework is useful and can find some applicability, 
but it cannot catch MHD nonlinearities (see section 
\ref{ssection_NonlinearPresentation}) corresponding to the nonlinear 
modification of $\mathcal{L}_\text{(0)}$, strongly driven modes 
(EPMs) or experimental situations characterized by a strong frequency 
sweeping. 

It is worth noting that some other approaches exist and/or could be 
developed: the use of a set of initial value time dependent equations 
\cite{Breizman_09, Zonca_07b}, the use of {\it conjugate problems}
where resonances are internalized in order to avoid strong derivatives
in Eq.~\ref{eq_SimpleEnergyEquationExpanded} and make possible the use 
of expansions close to attractors \cite{Brizard_94b}, the use of alternate 
conservation laws, such as the conservation of potential vorticity
\cite{Diamond_09} as an addition to energy conservation laws.
It can indeed be expected that a single equation cannot cover all the degrees
of freedom of the time evolution, in particular if an expansion of the form
Eq.~\ref{eq_SimpleEnergyEquationExpanded} is not possible.
\\

In the following, and for this preliminary analysis of nonlinear theories,
most ideas will refer to this limited kinetic perturbative framework.
More precisely,
our focus will be on the study of a scenario, where an energylike relation 
of the form 
\begin{equation}
(\delta \omega_r +i\gamma)\ 
\omega_{r0}\partial_\omega\mathcal{L}_\text{(0)}(\omega_{r0}) 
= -\omega_{r0}\mathcal{L}_{i\text{(1)}} 
-\omega_{r0}\mathcal{L}_{h\text{(1)}}
\label{eq_PerturbativeNLFramework}
\end{equation}
is valid, where 
$\omega_{r0}\partial_\omega\mathcal{L}_\text{(0)} 
= \omega_{r0}\partial_{\omega_r}\re\mathcal{L}_\text{(0)}$ 
is  {\it real} and similar to the energy density defined in section 
\ref{ssection_PositiveEnergyWaveDensity}
of a well defined linear eigenmode (MHD nonlinearities are neglected),
$ -\omega_{r0}\mathcal{L}_{i\text{(1)}}(\omega_{r0}) $ and
$-\omega_{r0}\mathcal{L}_{h\text{(1)}}(\omega_{r0})$ respectively refer to the
perturbative contribution of the resonant thermal ions and hot ions.

This picture is the same as the one we developed for the linear analysis,
where we simply calculated the imaginary part of the resonant contribution
in section \ref{section_ThresholdSimplification} 
(with the normalization \ref{eq_FinalNormalization}) 
implied by the linear response of the particles.
The aim is now to see how the situation is changed in the presence of a
nonlinear response of the particles.

For simplicity, the  results of the previous linear analysis will be referred 
to using simple notations
\begin{equation}
\begin{array}{lcl}
\gamma_l &=& -\im  \left( \mathcal{L}_{i(1)}(\omega_{r0})\right)
/\partial_{\omega_r}\mathcal{\re L}_{\text{(0)}}(\omega_{r0}) > 0\\
\gamma_d &=&  +\im \left( \mathcal{L}_{h(1)}(\omega_{r0})\right)
/\partial_{\omega_r}\re\mathcal{L}_{\text{(0)}}(\omega_{r0})> 0
\end{array}
\end{equation}
with the linear particle response included.
Until now, our linear analysis simply consisted in determining the sign
of $\gamma = \gamma_l-\gamma_d$.



\section{Application of the nonlinear trapping theory to Beta Alfv\'en
  Eigenmodes and limits}
\label{section_NLTrappingToBAE}
Let us now focus on the nonlinear response of the resonating populations, 
in the light of the {\bf nonlinear trapping theory}, mainly developed in 
Refs.~\cite{BerkBreizman_90}, \cite{BerkBreizman_98} and \cite{BerkBreizman_99},
which is currently the most used theory for the study of gap modes. 

We start with a short review of it (in its most basic formulation), 
designed to enhance its validy limits and to set up some notation for 
the subsequent analysis.

\subsection{The nonlinear trapping model}
\label{ssection_NLTrappingModel}
Let us consider a population of particles of distribution function $F$, and
a perturbation which can be reduced to {\bf one single resonance} in the 
particle action-angle space, that is to say to a  hamiltonian 
of the form
\begin{equation}
H = H_\eq({\bf J}) 
+ h({\bf J},t)\cos({\bf n }\cdot\m{\alpha}-\omega_{r0}t 
-\int dt \delta  \omega_r (t) \ )
\label{eq_PerturbedHamiltonianSingleResonance}
\end{equation}
where a nonlinear time evolution of the oscillation frequency 
$\delta\omega_r(t)$ is allowed. 

Linearly, resonant particles belong to a  surface defined by 
$\omega_{r0}={\bf n}\cdot\m{\Omega}_\eq({\bf J}) 
\equiv{\bf n}\cdot\partial_{{\bf J}}H_\eq$.
If, close to a point of the resonant curve ${\bf J}_R$, the Hamiltonian
can be approximated by
\begin{equation}
H = H_\eq({\bf J}_R) + \partial_{\bf J}H_\eq\cdot\delta {\bf J}
+ h({\bf J}_R,t)\cos({\bf n }\cdot\m{\alpha}-\omega_{r0}t 
-\int dt \delta  \omega_r),
\text{with } {\bf J} = {\bf J}_0 + \delta {\bf J}
\label{eq_HamiltonianExpansionCloseResonance}
\end{equation}
(the ${\bf J}$-dependence of $h$ needs to be low), major geometric 
difficulties can be locally overlooked. The dynamics is reduced to 
a 2D problem characterized by the variables
\begin{equation}
q = {\bf n }\cdot\m{\alpha}-\omega_{r0}t-\int dt \ \delta  \omega_r ,\\
p = {\bf n}\cdot\partial_{\bf J} H -\omega_{r0} -\delta \omega_r
\approx {\bf n}\cdot\partial_{\bf J}H_\eq\cdot\delta {\bf J}-\delta\omega_r
\label{eq_pqIsland}
\end{equation}
which are easily seen to be conjugate variables for the Hamiltonian
\begin{equation}
\mathcal{H} = \frac{1}{2}p^2 -{\sf C}h(t)\cos q +\dot{\delta \omega}_r\ q, 
\label{eq_IslandHamiltonian}
\end{equation}
with ${\sf C}={\bf n}\cdot\partial_{\bf J}{\bf \Omega}_0({\bf J}_R)\cdot{\bf n}$ 
the Hamiltonian curvature at resonance, in the $\bf n$-direction.

Physically, $q$ is obviously similar to an angle variable. Noticing that 
the variations of the action variables $\dot{\delta{\bf J}} = -h\sin(q){\bf n}$
simply occur in one direction defined by ${\bf n}$ (in the action three 
dimensional phase-space) and could simply be described by a scalar 
$\delta J$ such that $\delta {\bf J} = \delta J{\bf n}$, 
the variable $p$ can be simply related to $\delta J$ from its definition
in Eq.~\ref{eq_pqIsland}, $p = {\sf C}\delta J-\delta\omega_r$. 
Hence, $p$ can be seen as a measure along {\bf n} in the action space.\\

\begin{figure}[ht!]
\begin{center}
\begin{minipage}{\linewidth}
\begin{center}
\includegraphics[width=0.4\linewidth]{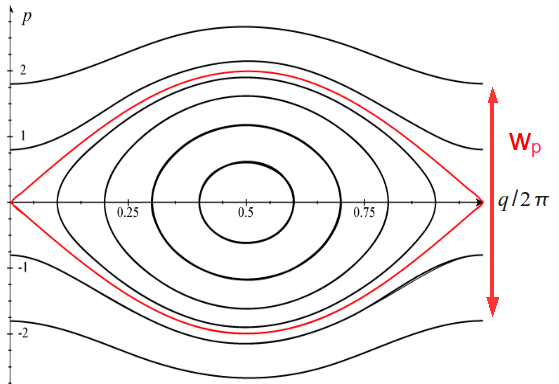}
\caption
[\footnotesize
Equipotentials of the characteristic Hamiltonian of a pendulum]
{\label{fig_PhaseSpaceIsland}
\footnotesize
Equipotentials of the characteristic Hamiltonian of a pendulum
(Eq.~\ref{eq_IslandHamiltonian} without explicit time dependence)}
\end{center}
\end{minipage}
\end{center}
\end{figure}

In the absence of frequency sweeping $\dot{\delta\omega}_r=0$ and for a slow
time dependence of the perturbation amplitude ($h$ almost constant),
Eq.~\ref{eq_IslandHamiltonian} is the traditional Hamiltonian describing 
the motion of a pendulum and leading to the creation of a {\bf phase-space island}. 
Indeed, it is easy to see that under these conditions, particles are 
constrained to follow the energy equipotentials (the constant-$\mathcal{H}$ 
lines), represented in Fig.~\ref{fig_PhaseSpaceIsland} for the Hamiltonian 
\ref{eq_IslandHamiltonian}. 
This picture shows the creation of a $(q,p)$ phase-space island where some 
particles are trapped into the potential well created by the wave. Typical 
quantities describing this island can be derived from 
Eq.~\ref{eq_IslandHamiltonian}
\begin{equation}
\begin{array}{llcl}
\text{- the island width in the $p$ direction:}         & 
w_p &=& 2\sqrt{Ch}\\
\text{- the bounce frequency of very trapped particles in the well:}&
\omega_{B} &=& \sqrt{{\sf C} h}
\end{array}
\end{equation}

More generally, such a {\bf nonlinear trapping} of the resonant particles 
inside the wave potential well makes sense if the typical bounce frequency of 
a particle inside such a well is large compared to the other typical frequencies 
of the problem: $\nu_\text{eff}$, the frequency of {\bf diffusive processes} 
(collisions, turbulent diffusion...) which can decorrelate the particles from 
the wave dynamics, or the characteristic frequencies of the mode evolution.  
The latter processes include in particular the {\bf evolution of the mode 
amplitude} related to $\dot{\omega_{B}}$ or the {\bf rate of frequency sweeping},
related to $\dot{\delta\omega_r}$.

It is possible to be somehow more explicit, noticing that if the bounce 
motion is well decoupled from the other time scales, the quantity 
$\mathcal{J}(\mathcal{H},t) = (1/2\pi)\oint dq \ p(q, \mathcal{H},t)$,
where the integral is taken along a bounce motion, is an 
{\it adiabatic invariant}, according to hamiltonian theory.
This invariance of $\mathcal{J}$ can be seen as the condition to be in the 
nonlinear trapping regime.
Calculating the time dependence of $\mathcal{J}$, it comes
$\dot{\mathcal{J}}=\tau(\dot{\mathcal{H}}-\langle\dot{\mathcal{H}}\rangle)$,
where
$\tau = (1/2\pi)\oint dq/p$
and
$\dot{\mathcal{H}} = q \ddot{\delta\omega} -\dot{ (\omega^2_b)} \cos q$.
Hence, sufficient conditions to be in the trapping regime appear to be
\begin{equation}
\ddot{\delta \omega}_r\ll \omega_B^3\ ,\\ 
\dot{\omega_b}\ll\omega_B^2\\ 
\text{and}\\
\nu_\text{eff}\ll\omega_B,
\label{eq_NonlinearTrappingValidityLimits}
\end{equation}\\

As a result of this trapping, the distribution function gets flat in the 
$p$ direction, which simply stands for the creation of a 
{\bf phase-space coherent structure} (the island), as already mentionned 
and illustrated in Fig.~\ref{fig_BOTSaturation}. As such, resonant 
particles can no longer drive or damp the wave resonantly, unless some 
collisions destroy the coherent structure 
(the upper figure in Fig.~\ref{fig_BOTSaturation}), or the 
structure moves in phase-space, leading to the frequency sweeping 
(the lower figure in Fig.~\ref{fig_BOTSaturation}). 

The nonlinear response of particles trapped into a potential well can be 
calculated from the full Boltzmann equation
\begin{equation}
\frac{dF}{dt} = \mathcal{Q} + \mathcal{C}\cdot F
\label{eq_Boltzmann_bis}
\end{equation}
where $\mathcal{Q}$ is again a source, and we choose to include 
$\mathcal{C}$ any background dissipative process which can break the 
island structure (not only collisions). Note that contrary to linear 
theories, where these dissipative effects are usually neglected 
($\nu_\text{eff}\ll\gamma_L,\gamma_d \sim  \omega_B$), 
nonlinear theories require their 
consideration, because nonlinear trapping can reduce driving and damping 
rates to the same order as dissipative processes.\\

Such a calculation has been done in Ref.~\cite{BerkBreizman_99} 
{\bf for a driving resonant species, 
and introduced in a perturbative energylike relation of the form 
\ref{eq_PerturbativeNLFramework}, \bf where a fixed damping rate 
$\gamma_d$ was considered}, with the additional assumption that 
dissipative effects are negligible in the computation of the 
linear drive which is correct for
\begin{equation}
(\partial_pF_\eq / F_\eq)^{-1} \gg \nu_\text{eff}, 
\label{eq_NoCollisionsInLinearlyTheory}
\end{equation}
following Ref.~\cite{BerkBreizman_90}.
This calculation shows that the drive is modified nonlinearly and depends
on both  $\nu_\text{eff}$ and $\delta\omega_r$.
{\bf Different saturation regimes have been identified:}
\begin{itemize}
\item {\bf A simple saturation at the given frequency} occurs when 
$\nu_\text{eff}$ is "large enough". Such saturation mechanisms are well 
known in plasma physics and result from a nonlinear reduction of the 
linear drive $\gamma_l$.

The corresponding reduction factors have been calculated explicitely for 
different dissipative processes \cite{BerkBreizman_90, Garbet_09}:
\begin{equation*}
\begin{array}{rll}
\rightarrow \text{if } \mathcal{Q} + \mathcal{C}\cdot F &= 
-\nu(F-F_\eq)
\text{ is a Krook operator}, &\nu_\text{eff} = \nu \\ 
\text{then}&\gamma_l \text{ is nonlinearly replaced by }& 
2.0\ \nu^*\gamma_l\\
\rightarrow \text{if } \mathcal{Q} + \mathcal{C}\cdot F &
= -D_p\partial_{pp}(F-F_\eq)
\text{ is a diffusive operator}, &\nu_\text{eff} = 2D_p/w_p^2 \\ 
\text{then}&\gamma_l \text{ is nonlinearly replaced by }& 
3.5\nu^*\gamma_l\\
\end{array}
\end{equation*}
where $\nu$ in the Krook operator is a collision frequency, 
whereas $D_p$ stands for a diffusion coefficient along the $p$ coordinate, and
$\nu^* = \nu_\text{eff}/\omega_B \ll 1$ for nonlinear trapping theories to 
apply.

Note that {\bf the calculation of the limit $\m{\nu^*\rightarrow 1}$ is also 
tractable}, and it is a classical results of  {\it neoclassical theory}.
When $\nu^*$ becomes large, it can be shown that $\gamma_l$ is nonlinearly 
conserved. In other words, {\bf the "reduction factor" is equal to 1.} 
\cite{Garbet_09}. In the following, we will refer to this limit as the 
{\bf quasi-linear regime}, but this terminology should not be mixed with the
quasilinear approximation used in turbulence studies, where the quasilinear regime
refers a to regime where several islands overlap such that a stochastization 
of the particle trajectories is reached. 
Nevertheless, in analogy to the latter regime, "our" quasi-linear approximation
assumes the existence of a {\bf decorrelation process} which is faster
than the trapping frequency
($\nu$ in this case, turbulent diffusion in the case of the 
the other regime).

\item {\bf A sweeping of coherent structure associated to a frequency 
sweeping}
occurs at low $\nu_\text{eff}$ \cite{BerkBreizman_99}, with a characteristic
time evolution $\propto \sqrt{t}$.

\item {\bf Intermediate oscillatory, chaotic, or excitation-relaxation 
regimes}
have also been prediced and identified numerically
\cite{BerkBreizman_92, BerkBreizman_98, Vann_02}.
\end{itemize}

Finally, the nonlinear trapping model displays several interesting 
behaviors, which are thought to reproduce several experimental observation 
of gap modes \cite{Heidbrink_08}. 
Nevertheless, it is useful to keep in mind some of its validity limits:
the consideration of {\bf one single resonance}, 
and the necessity for the {\bf perturbation to be smoothly dependent on 
the equilibrium invariants} in 
Eq.~\ref{eq_HamiltonianExpansionCloseResonance},
and finally the necessity of a {\bf time 
scale separation between the kinetic nonlinear response of resonant 
particles and the wave time evolution}
(Eqs.~\ref{eq_NonlinearTrappingValidityLimits}).

\subsection{Application to Beta Alfv\'en Eigenmodes}
\label{ssection_NLToBAE_OrdersMagnitude}
Let us determine the relevant resonances and time scales involved in the 
dynamics of Beta Alfv\'en Eigenmodes, to determine the validity of the 
nonlinear trapping model.

In the following, we look at the case of energetic and thermal ions 
separately, 
neglecting in a first approximation their non-resonant response.

\subsubsection{Resonance with energetic ions}
{\it Linearly}, we know from the previous chapters that the resonant 
behavior of particles can be described by the resonant Lagrangian 
Eq.~\ref{eq_ResonantLagrangian}.
For energetic ions,  we already explained in section 
\ref{ssection_EnergeticParticleTerm} that it makes sense to focus on one 
single resonance given by $({\sf n}_1=0, {\sf n}_2=0, {\sf n}_3={\sf n})$.
Keeping this single term only and recalling that $\mathcal{E}_\omega=0$
in the MHD region where resonance with energetic ions is effective,
the particle resonant response is seen to be the same as the one derived 
from the equivalent perturbed Hamiltonian
\begin{eqnarray}
\tilde{h}_h = 
-\int\frac{d\alpha_1}{2\pi}\int \frac{d\alpha_2}{2\pi}
\int \frac{d\alpha_3}{2\pi}
\left(e_h\frac{{\bf v}_{gh}\cdot\nabla\psi_\omega}{-i\omega}\right)
e^{-i{\sf n}_3\alpha_3}.
\end{eqnarray}
With this single resonance considered, the problem is reduced to the form 
of Eq.~\ref{eq_PerturbedHamiltonianSingleResonance}.

{\it Nonlinearly}, we assume that 
Eq.~\ref{eq_PerturbedHamiltonianSingleResonance} is still valid, such that 
the previous analysis can be applied.
This implies in particular that we consider the linear structure of the 
wave to be conserved nonlinearly, or in other words that we neglect MHD 
nonlinearities.
With this assumption and following the nonlinear trapping model described
above, an island is formed in the 2D plane 
$(q={\sf n}\alpha_3,p= {\sf n}\Omega_{h,d}-\omega_{r0} 
\approx {\sf C}_he_h\delta\Psi)$, that is, approximately in the {\bf 
radial direction}.
Hence, the relevant frequencies associated with this behavior and a 
typical value of the resonance invariants ${\bf J}_R$, 
\begin{eqnarray}
\nonumber
\omega^2_{Bh} 
&=& {\sf C}_h\tilde{h}_h({\bf J}_R)\\
&&\nonumber
\text{with the curvature }
{\sf C}_h = {\sf n}^2\partial_{J_3}\Omega_{h,d}({\bf J}_R) 
\sim \frac{{\sf n}^2q}{e_h r B_0}\frac{1}{L_{Th}}\Omega_{h,d}({\bf J}_R),\\
&&\nonumber
\text{and the perturbed Hamiltonian calculated in Appendix 
\ref{eq_EquivalentHotHamiltonian}} \\
&& \tilde{h}_h({\bf J}_R) = e_h\psi_\omega^{\sf m}({\bf J}_R).\\
\nonumber 
\nu^\text{eff}_h
&=& \frac{1}{2}\frac{D_{ph}}{(w_{ph}/2)^2}
= \frac{1}{2}\frac{D_{ph}}{\omega^2_{Bh}}
\sim \frac{1}{2}\frac{D_0}{\omega^2_{Bh}}
\frac{({\sf n}\Omega_{h,d})^2}{L^2_{Th}} 
\end{eqnarray}
with $D_0$ the {\bf spatial  diffusion} coefficient ($m^2s^{-1}$), 
$L_{Th}$ a typical gradient of the equilibrium fast ion population.

\subsubsection{Resonance with thermal ions}
We can make a similar assessment for the thermal ions.
In this case, several resonances can be of importance, and we kept the two
resonances of the form $\omega = \pm v_\|/qR_0$ in the linear computation 
of Landau damping.
The previous framework assumes these two resonances to be well separated 
in phase space, such that they can be considered one by one.
Taking for example the resonance of the form $\omega=v_{\|}/qR_0$, or 
$({\sf n}_1=0, {\sf n}_2= {\sf m}+1, {\sf n}_3 = {\sf n})$, the equivalent 
Lagrangian can be derived from Eq.~\ref{eq_EquivalentThermalHamiltonian} 
of the appendices.
If we choose to expand the perturbed hamiltonian around the point 
representing purely passing particles ($\mu=0$) and at the location 
where Landau damping is stronger, that is at the mode resonant surface, 
it comes
\begin{equation}
\tilde{h}_i({\bf J}_R) 
= e\left(\frac{1}{2}\left(\frac{\omega q R_0}{v_{ti}}\right)^2+\tau_e\right)
\left(\frac{v_{ti}}{R_0\omega}\right)
\rho_i\partial_r\psi_\omega^{\sf m} \ \ .
\label{eq_ThermalIonsEquivalentHamiltonian}
\end{equation}\\

Let us calculate the Hamiltonian curvature ${\sf C}_i$  at this point.
For passing particles and in the large aspect ratio limit, the equilibrium
motion invariants \ref{eq_GeneralInvariants} can be approximated by
\begin{equation}
\begin{array}{lcl}
&\Omega_2 \approx \frac{v_\|}{qR_0},& J_2 \approx e\Phi({\Psi}),\\
&\Omega_3 \approx \frac{v_\|}{R_0}\, ,&
J_3 = e\Psi + m_iR_0v_\|=e{\Psi}(J_2)+m_iR_0v_\| \ .
\end{array}
\label{eq_EquilibriumFciesPassing}
\end{equation}
The Hamiltonian curvature at this point follows 
\begin{eqnarray}
{\sf C}_i = {\bf n}\cdot\partial_{\bf J} (k^{\sf m+1}_\|v_\|) 
= \frac{(k_\|^{\sf m+1})^2}{m_i}
- \frac{\sf (m+1)}{m_iq^2R_0^2}\ \frac{\sf m+1}{r}sqR_0
\frac{\rho_i}{r}\frac{\omega qR_0}{v_{ti}}
\approx \frac{1}{m_iq^2R_0^2} \ , 
\end{eqnarray}
where the different quantities need to be assessed at the resonant surface
$r_s$, corresponding to the direction of the new canonical system of 
coordinates $(q, p={\sf C}_i\delta J)$.
More precisely, coming back to the {\it natural} space-velocity
space, and making an 
expansion around the resonant point ${\bf J}_R$ using 
Eqs.~\ref{eq_EquilibriumFciesPassing},
${\bf J} = {\bf J}_R + \delta J {\bf n}$, we can relate the $\delta J$
variations to the variations of the more natural variables $\delta r$ and
$\delta v_\|$
\begin{eqnarray}
\frac{\delta r}{a} 
&=& -\frac{q}{r^2eB_0}\delta (e\Psi)
= -\frac{q}{r^2eB_0}\left(-\frac{\sf n_2}{q(J_2)}\delta J\right)
= \frac{\sf m+1}{r^2eB_0}\ \delta J\\
\frac{\delta v_\|}{v_{ti}} 
&=& 
\frac{\delta(m_i R_0 v_\|)}{m_iR_0v_{ti}} 
= \frac{1}{m_iR_0v_{ti}}
\left( {\sf n}_3 + \frac{{\sf n}_2}{q(J_2)} \right)\delta J 
= \frac{1}{m_iR_0v_{ti}}\frac{1}{q}\delta J
\end{eqnarray}
It follows that the direction described by $p=C\delta J$ has components 
both in the {\it radial} and {\it velocity} directions. Nevertheless, at 
the typical macroscales involved, the  {\bf alignement along the velocity 
direction dominates}.\\

We can now display the characteristic frequencies involved. 
In velocity space, {\bf collisions} are the relevant dissipative process.
It comes 
\begin{eqnarray}
\nonumber
\omega^2_{Bi} 
&=& \left(\frac{v_{ti}}{qR_0}\right)^2 \frac{h_i({\bf J}_R)}{T_i}\\
&&\nonumber\\
\nonumber 
\nu^\text{eff}_i
&=& \frac{1}{2}\frac{D_{pi}}{(w_{pi}/2)^2}
= \frac{1}{2}\frac{D_{pi}}{\omega^2_{Bi}}
\sim
\frac{1}{2}\nu_{ii}
\frac{v_{ti}^2}{q^2R_0^2}\ 
\frac{1}{\omega^2_{Bi}}
\end{eqnarray}
with $\nu_{ii}$ is the ion-ion collision frequency.

\subsubsection{Numerical application}
Let us assess these various frequencies for Tore-Supra relevant 
parameters for the type of shots described in chapter 
\ref{chapter_LinearStability}:
$R_0=2.45$m, $R_0/L_{ph} = 30$,
and the mode structure is assumed to verify 
$\xi \sim 1$mm (the MHD displacement), 
$\rho_i k_r\sim(\rho_ik_\theta)^{1/4}$ in the inertial region, 
$\omega_{r0}/2\pi\sim (v_{ti}/R_0)/2\pi \sim  (v_{ti}/qR_0)/2\pi \sim 5.10^4$ Hz.

For the dissipative processes, we take a background spatial diffusion
of $D_0 = 0.1$m$^2$s$^{-1}$ and a collision frequency $\nu_{ei}=10 Hz$. 
It comes
\begin{eqnarray}
\text{For energetic ions: }\frac{2\pi}{\omega_{Bh}} \sim 2.10^{-4} s, &
\nu^*_h = \frac{\nu^\text{eff}_h}{\omega_{Bh}}\sim 0.02\\
\text{For thermal ions: \:\:\:}\frac{2\pi}{\omega_{Bi}} \sim  1.10^{-4} s, &
\nu^*_i = \frac{\nu^\text{eff}_i}{\omega_{Bi}}\sim 0.3 &
\text{ if } \rho_ik_\theta\sim 0.01 \\
\frac{2\pi}{\omega_{Bi}} \sim  6.10^{-5} s, &
\nu^*_i = \frac{\nu^\text{eff}_i}{\omega_{Bi}}\sim 0.02 &
\text{ if } \rho_ik_\theta\sim 0.001
\end{eqnarray}

\subsection{Validity analysis}
From the previous paragraph, the consideration of 
{\bf one single resonance} fully makes sense for the energetic ions, and 
it seems reasonable to separate the two resonances defined by
$v_\| = \omega_{r0}qR_0$ and $v_\| = -\omega_{r0}qR_0$. 
Indeed, a breaking of the latter condition would mean
that the thermal ion population is strongly modified by Landau damping,
and we could expect the nonlinear eigenmode to have much stronger 
nonlinear features (not the long-lived stable frequency oscillations 
observed in Tore-Supra).

Moreover, the condition of {\bf nonlinear trapping} seems valid for both 
the resonant energetic ions and the resonant thermal ions. 
First, $\nu_h^*, \nu_i^* \ll 1$. Secondly, the nonlinear time scale 
evolution which could possibly compete with the trapping frequencies 
$\omega_{Bi}$ and $\omega_{Bh}$ needs to be faster than $10^{-4}s$, and
we did not observe such very fast time scale evolutions in our 
experiments.\\

These remarks seem to advocate for the use of the nonlinear trapping model, 
and it may be tempting to apply the results briefly outlined above.
However some caution is necessary for the study of BAEs:

\begin{itemize}
\item The local expansion around ${\bf J}_R$ in 
  Eq.~\ref{eq_HamiltonianExpansionCloseResonance} may be questionned, because 
  it implies the disappearing of some geometry, and in particular the complete 
  access to the mode radial structure.

  For energetic ions which resonate with the mode in the BAE MHD region 
  (characterized by smooth radial gradients), this may not be an issue. 
  For resonant thermal ions, which 
  mainly resonate in the BAE inertial layer, the equivalent Hamiltonian is 
  proportional to $\rho_i\partial_r\psi_\omega^{\sf m}$ 
  (see Eq.~\ref{eq_ThermalIonsEquivalentHamiltonian}) 
  and strongly varies in the inertial region, to match the smoother MHD region.
  In particular, if a radial component of the perturbed hamiltonian needs
  to be kept, a coupling of several dimensions occurs: the problem can not be
  reduced to 2  dimensions only.
  
  Note that a response to this argument could be that a smoother structure can 
  be expected for the mode in the nonlinear regime 
  (either due to MHD nonlinearities or resonant flattening...)
  than the one described in chapter \ref{chapter_BAEdescription}. Such argument
  may be relevant, but it does not allow a  direct application of the above 
  theory.

\item We claimed the validity of nonlinear trapping based on the remark that 
  Tore-Supra experiments did not display fast dynamics, such as fast sweeping.
  
  However, some other regimes could be possible where the nonlinear distortion  
  of the energetic particle drive is so strong that sweeping occurs on time 
  scales of the order of the bounce frequency, as claimed in 
  Ref.~\cite{Zonca_07b}.

\item So far, the kinetic nonlinear saturation of Alfv\'en gap modes has been
  mainly studied using a fixed damping $\gamma_d$, such that saturation occurs
  due to the reduction of the kinetic resonant drive only. 
  For BAEs, which are both {\it damped} and {\it driven} via resonant processes, 
  it may be interesting to determine the effects of a simultaneous nonlinear 
  evolution of drive and damping.

\item Finally, it has to be noted that most of the current results concerning the
  nonlinear saturation of modes driven by fast particles, have been derived in 
  the perturbative framework, described in \ref{section_NLEnergyPpe}. It is in 
  particular true for the different saturation regimes outlined at the end  of 
  subsection \ref{ssection_NLTrappingModel}, though some efforts are currently 
  done to avoid this approximation \cite{Breizman_09}.

  This perturbative framework gives sense to the artificial separation between
  the main bulk of the thermal ions made in Eq.~\ref{eq_PerturbativeNLFramework}, 
  and the assumed small group of resonant
  damping thermal ions (thought to be negligible for the mode linear structure).
  However, should the validity of this framework break, new regimes could be found
  and a strong impact on the BAE structure may be expected because of the 
  possibility of resonance with thermal ions.
\end{itemize}

In the next two sections, we attempt to provide some response to the last 
two points.

\section{Effects of a self-consistent nonlinear damping}
\label{section_SelfConsistentDamping}

\subsection{Motivation: the possibility of subcritical behaviors}
\label{ssection_SubcriticalMotivation}
The most simple resonant saturation mechanism described by the nonlinear 
trapping theory with a fixed damping $\gamma_d$, relies on the idea that the 
linear drive is {\it nonlinearly} subject to a {\bf reduction}: 
\begin{equation}
\gamma_l \rightarrow c_0\nu_l^*\gamma_l\ 
\\ \text{ with } c_0\text{  a multiplying factor of order 1,}
\end{equation}
where $\nu_l^*<1$ is a decreasing function of the mode amplitude.
We indicated in subsection \ref{ssection_NLTrappingModel} that $\nu_l^*$ 
could be calculated as a function of the driving species bounce frequency 
$\omega_{Bl}$ ($l$ is used here to refer to the driving species)
which can be seen as a measure of the mode amplitude,  
$\omega_{Bl} = \sqrt{C_lh_l}$.

Hence, above the linear threshold $\gamma_l>\gamma_d$, saturation can occur
for the amplitude $\omega_{Bl}$ verifying 
$c_0\nu_l^*(\omega_{Bl})\gamma_l=\gamma_d$.\\

If damping is no longer taken fixed, but assumed to result from a resonant 
process as well, we may wonder if a similar reduction factor $\nu^*_d$ also 
applies for the damping, and if, in this case, saturation is  still possible.

The existence of these nonlinear reduction factors  rises in particular the
question of the possibility of {\bf subcritical modes}.
Indeed, if $\nu^*_l> \nu^*_d$,
one may question the existence of linearly stable modes $\gamma_l<\gamma_d$ 
which become unstable nonlinearly, ie: $\nu^*_l\gamma_l>\nu^*_d\gamma_d$,
because of the nonlinear reduction of the damping rate.

If such subcritical modes exist and correspond to the explanation given above,
they should be considered an issue for the stability of BAEs. Indeed,
our rough computation of the reduction factors corresponding to ion Landau 
damping and hot ion drive (in subsection \ref{ssection_NLToBAE_OrdersMagnitude})
returned
\begin{eqnarray}
 \nu^*_d = \nu^*_i  \sim  \nu^*_h = \nu^*_l
\end{eqnarray}
which suggests that subcritical behaviors could be relevant to experimental
conditions for some particular range of parameters.\\

In the following, we verify the possibility of a subcritical activity using a 
simple electrostatic 2D numerical model, and compare the behavior of these modes 
with the idea of distinct reduction factors  developed above.
Our focus will be on the behavior of subcritical modes with a well defined 
frequency, characterized by a "simple saturation". 
Such modes will be referred to as {\bf metastable modes}, and may be relevant to 
Tore-Supra experiments.

As will be obvious from the simulations to be presented, subcritical regimes
characterized by a sweeping/chirping oscillation frequency also exist. 
They will be studied in a different PhD work.

\subsection{Model}

\subsubsection{The Bump On Tail (BOT) problem and its analogy to the BAE problem}
In order to get some insight into the simultaneous nonlinear evolution of two 
competing resonant phenomena, we make use of a simple variation of the 
traditional {\bf electrostatic 2D  Bump-on-Tail (BOT) problem} \cite{ONeil_65}, 
{\bf to which we add a self-consistent resonant damping}. 

The 2D electrostatic BOT problem is a model for an electroneutral plasma, where
an electron population characterized by an equilibrium small bump in velocity 
space (see the red curve in  Fig.~\ref{fig_F1F2}) 
is computed using the 2D Boltzmann 
equation, whereas the ion population is considered fixed. 
In such plasmas, waves oscillating at the {\bf electron  plasma frequency}
\begin{equation}
\omega_{pe} = \sqrt{n_ee^2/(m_e\epsilon_0)},
\end{equation}
the so-called Langmuir waves can develop, 
and the velocity bump can provide a linear resonant drive for them if resonance
occurs in the positive slope of the bump.
In order to model the competition of two resonant mechanisms, we take advantage of 
a similar model with a population presenting  a bump on tail ($F_1$), 
and add a second species ($F_2$)  to this picture, 
intended to present a negative slope at the localization of resonance in order to 
provide damping (the blue curve in Fig.~\ref{fig_F1F2}).
The idea is to substitute the BAE problem by a simpler 2D problem, with the correspondence
\begin{equation}
\begin{array}{lcl}
\text{BAE}  & \rightarrow &\text{Langmuir wave} \\
\text{Energetic particles}     & \rightarrow & \text{Bump of the first population}  \\
\text{Damping thermal ions} & \rightarrow & \text{Second population}
\text{Bump of first population}.
\end{array}
\end{equation}

More precisely, we make use of two populations of particles of distribution functions
$F_1$ and $F_2$ modelled using 2D Boltzmann equations with simple Krook collision 
operator, and related using Poisson equation for the computation of the 
electrostatic field:
\begin{equation}
\begin{array}{rcl}
\partial_tF_1  + v\partial_xF_1 + (e_1E/m_1)\ \partial_vF_1 
= -\nu_1 (F_1 - F_{1\eq}),\\
\partial_t F_2 + v\partial_xF_2 + (e_2E/m_2)\ \partial_vF_2
= -\nu_2(F_2 - F_{2\eq}),\\
-\partial_x E  
= (1/\epsilon_0) \int dv \ 
[e_1(F_1 - F_{1\eq}) + e_2(F_2-F_{2\eq})],
\end{array}
\label{eq_ExpandedBOTModel}
\end{equation}
and we assume the first population to present a bump $F_\text{1\eq}$.\\

Why is this reduced 2D problem relevant to the BAE problem?
Under the approximation of the nonlinear trapping theory developed above, 
we know that the BAE model can be reduced to a 2D Boltzmann equation
for each of the two resonant species involved. Next, their contribution to the
wave can be taken into account using an energylike equation, which is nothing 
but an expression of Maxwell equation. Hence, the problem presented above 
makes sense for the BAEs as well (even if the two resonant processes at stake
do not take place in the same phase-space direction), 
as long as no interaction between the two 
populations involved is needed outside of Maxwell equations (for example, 
collisions between the two populations).
The Hamiltonian curvature, ${\sf C}_s =k^2/m_s $ in the electrostatic problem
(for a wave number $k$ and a species $s$), 
simply needs to be replaced by the BAE relevant curvatures, and the Krook 
collision frequency by the relevant dissipative processes $\nu_\text{eff}$ 
given earlier in subsection \ref{ssection_NLToBAE_OrdersMagnitude}.

A weaker feature of this model is that we chose to fully separate the damping 
mechanism from the bulk (in order to recover the traditional features of the 
fixed $\gamma_d$ model), whereas the  thermal ions are the damping species
in the case of BAEs. In a perturbative approach (where the resonant thermal ions
are not assumed to contribute significantly to the wave structure itself) 
however, such a separation makes some sense.

\subsubsection{Numerical implementation}
The model \ref{eq_ExpandedBOTModel} was implemented using the 
{\sf {\bf C}ALV{\small i}} plateform developed by the INRIA-Calvi Team, which 
gathers several pre-compiled {\sf Fortran} routines behind a {\sf Python} 
interface, designed to offer multiple schemes to solve the 2D Vlasov equation 
\cite{Calvi_09}.

Simarly as in Ref.~\cite{Vann_02}, we can notice that taking 
some reference quantities $n_0, T_0, e_0, m_0$  
($T_0$ a temperature of the order of the bulk plasma,
$n_0$ a density  of the order of the bulk density,
$m_0=m_1, e_0 = e_1$)
and using the normalized variables
\begin{eqnarray}
&&\begin{array}{lclclclclclclcl}
  x   &\rightarrow& (x/\lambda_{D0})   &,&
  t   &\rightarrow& \omega_{p0}t       &,&
  v   &\rightarrow& v/V_0 \\
  F_i &\rightarrow& (V_0/N_0)F_i &,&
  n_s &\rightarrow& n_s/N_0 &,&
  \omega_{ps} &\rightarrow& \omega_{ps}/\omega_{p0}
  \\
  E   &\rightarrow& \sqrt{\epsilon_0/(N_0T_0)}\ E &,&
\nu_1 &\rightarrow& \nu_1 / \omega_{p0}
\end{array}\\
&&\text{ with }
\lambda_{D0} = \sqrt{\epsilon_0T_0/(n_0e_0^2)},
V_0 = \sqrt{T_0/m_0},
\omega_{p0} = \sqrt{n_0e_0^2/(m_0\epsilon_0)}\\
&&\nonumber
\text{(some reference Debye length, thermal velocity and plasma frequency), }
\end{eqnarray}
the BOT problem can be rewritten with typical quantities of the order 
of 1. Such normalizations were used in the code and in the results displayed 
below.\\

Finally, the implemented model is the following
\begin{eqnarray}
\partial_tF_1 + v\partial_xF_1 + E\partial_vF_1
&=& -\nu_1 (F_1 - F_{1\eq})
\label{eq_Vlasov1}\\
\partial_tF_2 + v\partial_xF_2 
+ \left[\frac{e_2}{e_1}\right]\left[\frac{m_1}{m_2}\right]E\partial_vF_2
&=& -\nu_2 (F_2 - F_{2\eq})
\label{eq_Vlasov2}\\
-\partial_x E &=& \int dv(F_1-F_{1\eq}) + \frac{e_2}{e_1}\int dv(F_2-F_{2\eq})
\label{eq_Poisson2species}
\end{eqnarray}
with equilibrium distribution functions of the form
\begin{eqnarray}
\text{with \,\,}
F_{1\eq}(v) &=& \frac{n_\text{b}}{\sqrt{2\pi}v_{tb}}
\exp\left[-\frac{1}{2}\left(\frac{v}{v_{tb}}\right)^2\right]
+\frac{n_\text{l}}{\sqrt{2\pi}v_{tl}}
\exp\left[-\frac{1}{2}
\left(\frac{v-v_\text{0}}{v_{tl}}\right)^2\right]
\\
F_{2\eq}(v) &=& \frac{n_{d}}{\sqrt{2\pi}v_{td}}
\exp\left[-\frac{1}{2}\left(\frac{v}{v_{td}}\right)^2\right]
\end{eqnarray}
and the possibility to enforce an initial perturbation of the first species,
of amplitude $\alpha$ and wave number $k$
\begin{equation}
F_1(x,v, t=0) = F_{1\eq}(v)(1+\alpha\cos(kx)).
\end{equation}

Noticing that the problem could in fact be devided into three species, 
the bulk  $b$, the driving species $l$ and the damping $d$ species, the 
alternate notation 
$\nu_b = \nu_l = \nu_1$, $\nu_d=\nu_2$ 
$e_d/e_l=e_2/e_1$, 
$m_d/m_l = m_2 / m_1$,
is used from now on.

\subsubsection{The adiabatic/perturbative approximation}
\begin{figure}[ht!]
\begin{center}
\begin{minipage}{\linewidth}
\begin{center}
\includegraphics[width=0.65\linewidth]{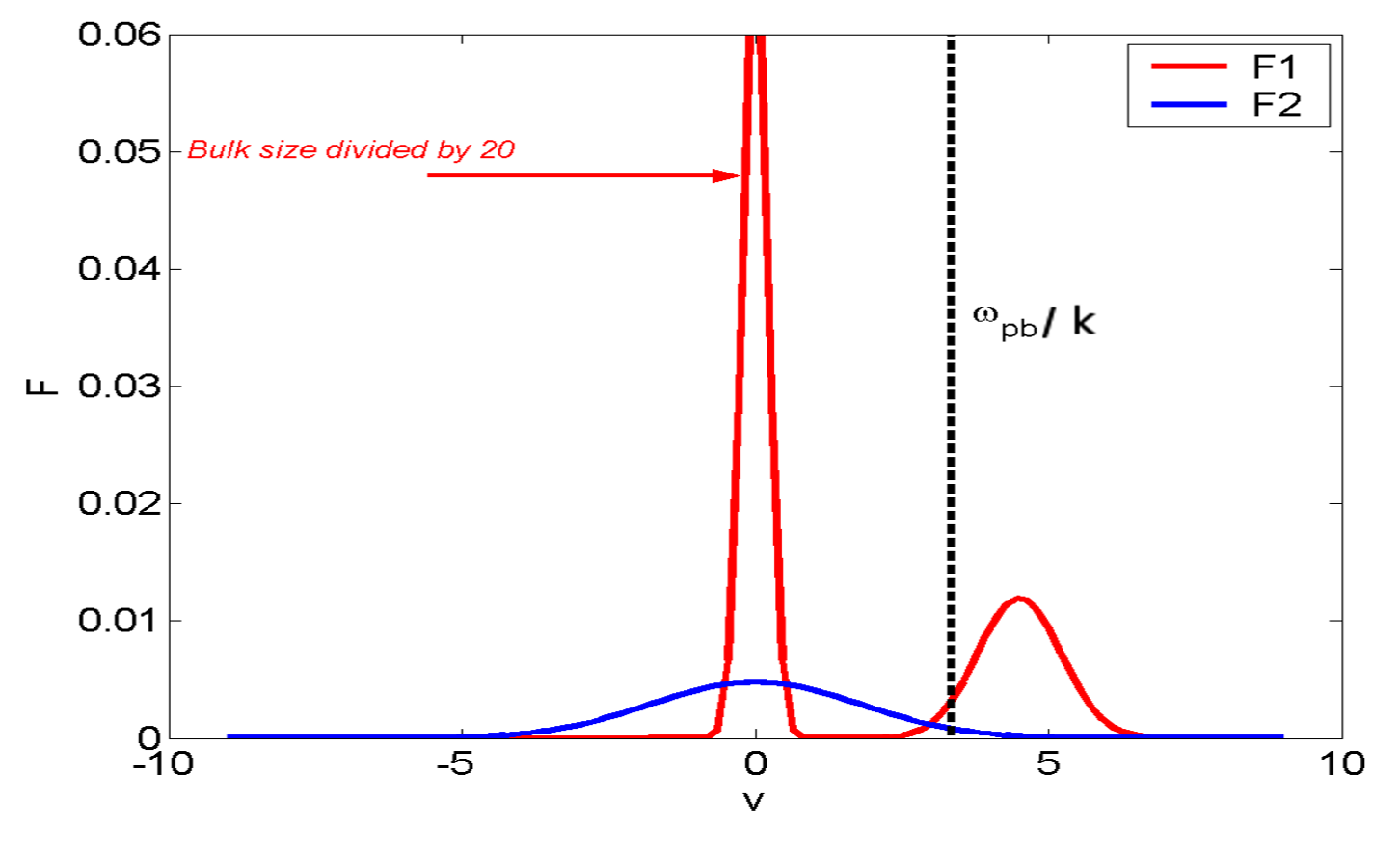}
\caption
[\footnotesize
Shape of the equilibrium distribution functions, for typical parameters
of the simulations.]
{\label{fig_F1F2}
\footnotesize
Shape of the equilibrium distribution functions, for typical parameters
of the simulations.
More precisely, the fixed parameters of q.~\ref{eq_FixedParametersInBOT}
are used with $n_d=0.03$.}
\end{center}
\end{minipage}
\end{center}
\end{figure}

In order to allow for some analytic comparisons of the simulated results with
the ideas developed in subsection \ref{ssection_SubcriticalMotivation}, a 
simple {\bf perturbative approach} is used  in the following,  
and the bulk plasma is taken to be fully {\bf adiabatic} in order to avoid 
any ambiguity on the species responsible for the mode damping.
More precisely, we take $n_{b}\gg n_{l}\sim n_{d}$,
and $kv_{tb} \ll\omega_{pb}$, where $\omega_{pb}$ 
is the normalized plasma frequency of the bulk species.
Such approximations are illustrated in Fig.~\ref{fig_F1F2}
and allow for a  non-ambiguous definition of the linear drive $\gamma_L$ 
and damping $\gamma_d$. 

As an addition, we consider limited dissipations
$\nu_b = \nu_l \ll \omega_{bp}, (k\partial_vF_{l\eq}/F_{l\eq})^{-1}$
$\nu_d \ll (k\partial_vF_{d\eq}/F_{d\eq})^{-1}$. 
\\

The calculation of the particle linear responses from the linear
expansion of Boltzmann equations \ref{eq_Vlasov1}, \ref{eq_Vlasov2}
for perturbation of frequency $\omega$ and wave number $k$, 
and their introduction  in  Poisson equation \ref{eq_Poisson2species}
is a tractable problem which returns the general dispersion relation
\begin{equation}
-k^2 = \sum_{s = b,l,d} \frac{\omega^2_{ps}}{v_{ts}^2}
\left(1+\frac{\omega+i\nu_s-\delta_{s=l}\ kv_0}{\sqrt{2}\ kv_{ts}}\right)
Z \left(\frac{\omega+i\nu_s-\delta_{s=l}kv_0}{\sqrt{2}\ kv_{ts}}\right).
\label{eq_BOTLinearDispRel}
\end{equation}
where $Z$ is the plasma dispersion function, also called the Fried Et Conte 
function and already defined before Eq.~\ref{eq_FriedEtConte}.

In general however, it is not possible to define independently the linear
drive $\gamma_l$ and the linear damping $\gamma_d$, because the addition/removal 
of a non-perturbative species can significantly modify the oscillation 
frequency, and hence of the localization and rate of resonant energy 
transfers.
In the perturbative adiabatic framework, this is possible.
When the bulk density dominates, the dispersion relation is almost real.
At the lower order, the adiabatic approximation $\omega \gg kv_{tb}$
associated to the limited dissipation returns:
\begin{equation}
\omega^2 = \omega^2_{pb}.
\end{equation}
To the next order, the bump is clearly found to induce a drive, and the 
second species a damping. Assessing the first order modification of $\gamma$
implied by the introduction of these two species, it comes
\begin{eqnarray}
\gamma_l &=& 
\frac{-1}{2}\frac{\omega_{pb}}{k^2}\left(\frac{\omega^2_{pl}}{v^2_{tl}}\right)
\left(1 + \frac{\omega_{pb}-kv_0}{\sqrt{2}kv_{tl}}\right)
\im Z\left(\frac{\omega_{pb}-kv_0}{\sqrt{2}kv_{tl}}\right)
\\
\gamma_d &=& 
\frac{-1}{2}\frac{\omega_{pb}}{k^2}\left(\frac{\omega^2_{pd}}{v^2_{td}}\right)
\left(1 + \frac{\omega_{pb}}{\sqrt{2}kv_{td}}\right)
\im Z\left(\frac{\omega_{pb}}{\sqrt{2}kv_{td}}\right).
\end{eqnarray}

\subsubsection{Fixed and free parameters}
The aim of the subsequent analysis  is to study the stability of the BOT
problem. For that purpose, several stability diagrams are possible.
In order to recover the features of the BOT problem with a fixed damping, 
we choose to vary parameters which are already present there
and are known to be crucial for stability as well as for the determination 
of the relevant saturation regime \cite{BerkBreizman_99}: 
{\bf the collisionality of the first species} $\nu_1$
and {\bf the damping rate} $\gamma_d\propto\omega^2_{pd}\propto n_d$.
An advantage of such stability diagrams is that they were already produced 
for the BOT with a fixed damping in Refs.~\cite{Vann_02,Lesur_09}.

Consequently, in all the following simulations, we keep fixed:
\begin{equation}
\begin{array}{ll} 
\text{ -the wave number, } &k=0.3\\ 
\text{ -the equilibrium distribution function } F_{1\eq}, 
&n_b=1.0, v_{tb}=0.3, n_l=0.03,\\
& v_{tl}=1.0, v_0=4.5\\ 
\text{ -and some parameters of the damping species, }
&v_{td}=2.5, e_d/e_l=1.0.
\end{array}
\label{eq_FixedParametersInBOT}
\end{equation}

Stability diagrams are ploted varying $\nu_b=\nu_l$ and $n_d\propto\gamma_d$,
whereas the mass ratio $m_d/m_l$ and the collisionality of the damping species
$\nu_d$, which are both involved in the nonlinear dynamics (as will be 
clear in Eq.~\ref{eq_reductionFactor2Species}), 
are changed from one stability diagram to another. 
Finally, to validate the metastable character of the found mode, 
the amplitude of the initial perturbation is also varied, and expressed in 
terms of $\omega_{Bb}$ in the whole description.
\\

This independent variation of the collisionality, density, and mass may 
thought to be somehow artificial (in particular because these quantities are 
physically related). But it is justified recalling that the aim of this analysis
is not to provide a better understanding of Langmuir waves, but to catch a piece
of physics which may take place in different phase-space directions. 
In particular, $\nu$ is not simply intended to represent collisions, but any
dissipative process.
\subsection{Recovering the known saturation regimes of the perturbative 
nonlinear trapping model with a fixed damping}
\begin{figure}[ht!]
  \begin{center}
    \begin{minipage}{0.52\linewidth}
      \begin{center}
	\vspace{-1.3cm}
        \includegraphics[width=\linewidth]{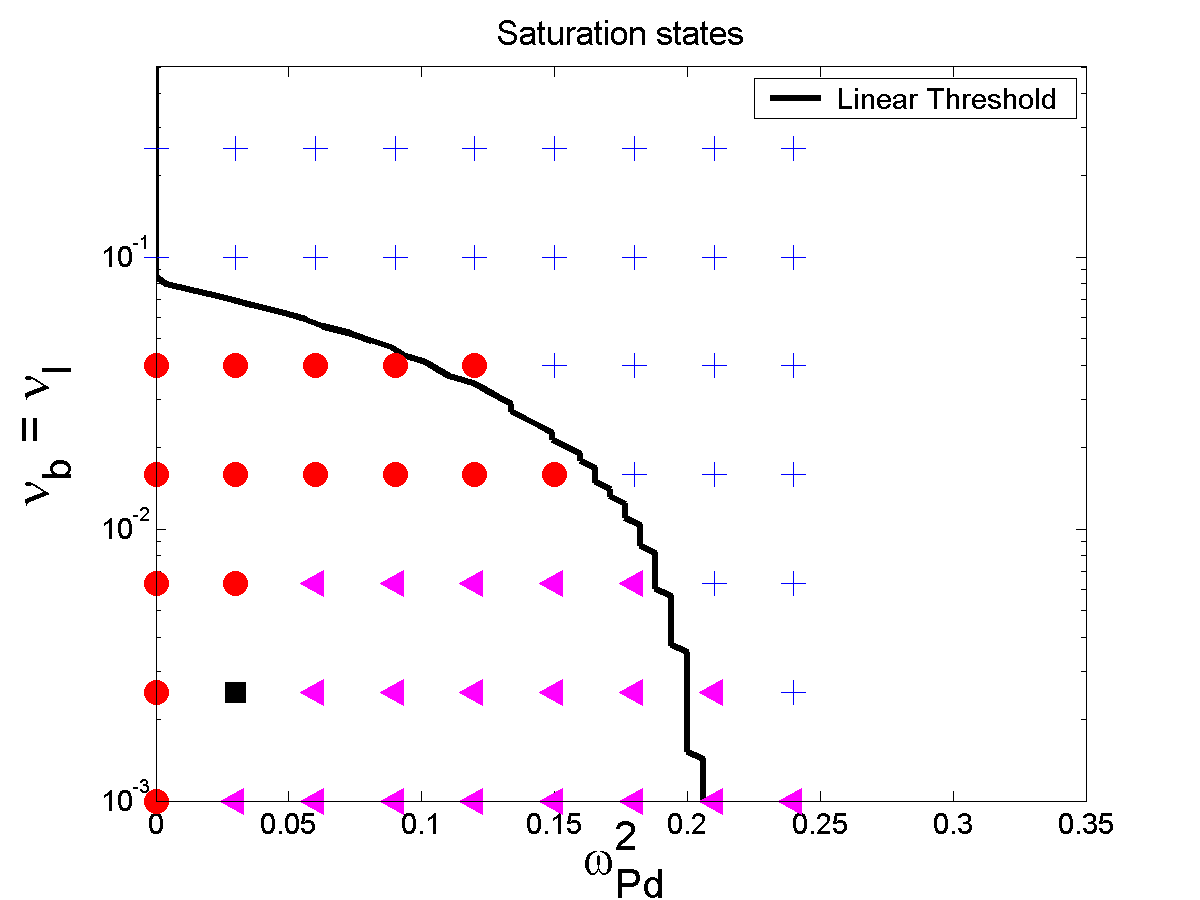}
	\caption[\footnotesize Nonlinear saturation diagram without
        metastability.]
        {\label{fig_states_runs2}
          \footnotesize
          Stability diagram without metastability, obtained for the parameters
          $m_d/m_l=2.0$, $\nu_d=0.01$, $\omega_{Bb}(t=0)=0.1$.
          Four behaviors are distinguished: 
          simple saturations (red dots), 
          chaotic regimes    (pink triangles), 
          damped modes       (blue crosses), 
          and intermediate oscillatory behaviors (black squares).
        }
      \end{center}
    \end{minipage}
    \begin{minipage}{0.47\linewidth}
      \begin{center}
        \includegraphics[width=\linewidth]{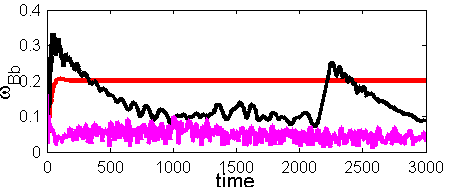}
        \includegraphics[width=0.91\linewidth]{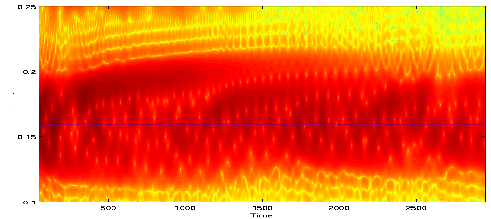}	
        \caption[\footnotesize 
        Some examples of nonlinear saturation regimes.]
        {\label{fig_Examples_runs2}
	  \footnotesize 
          Some examples of nonlinear saturation regimes, extracted from the
          stability diagram Fig.~\ref{fig_states_runs2}.
          First, the field amplitude (expressed in terms of $\omega_{Bb}$) 
          has been represented, corresponding to three different
          types of saturation regimes: 
          simple saturation, 
          chaotic behavior, or 
          relaxation-excitation dynamics.
          Next, a spectrogram of the field signal is represented showing the 
          existence of frequency sweeping at low collisionality.
        }
      \end{center}
    \end{minipage}
  \end{center}
\end{figure}

A first example of such type of diagram is given in Fig.~\ref{fig_states_runs2}
corresponding to $m_d/m_l=2.0$, $\nu_d=0.01$, $\omega_{Bb}(t=0)=0.1$.
It has been represented along with the linear stability threshold, numerically 
derived from the numerical resolution of Eq.~\ref{eq_BOTLinearDispRel} using
the root solver described in section \ref{section_DispersionRelationSolver}.\\

In this picture, different regimes have been identified, also illustrated 
in Fig.~\ref{fig_Examples_runs2}.
\begin{itemize}
\item {\bf Simple saturation regimes} correspond to field perturbations
  characterized by a well defined oscillation frequency, 
  whose amplitude converges to a fixed value (see Fig.~\ref{fig_Examples_runs2})
\item {\bf Damped modes} correspond to a decrease of the field amplitude, much
  below 10\% of the initial amplitude.
\item {\bf Chaotic regimes} refer to behaviors characterized by some 
  indetermination of the oscillation frequency and associated to bounces of the 
  perturbed field amplitude (see Fig.~\ref{fig_Examples_runs2}).
  It includes in particular the existence frequency shifts, associated to the
  creation of phase-space structures (as in the spectrogram of 
  Fig.~\ref{fig_Examples_runs2}) \cite{BerkBreizman_99}.
\item {\bf Intermediate regimes} which do not clearly fall into the previous 
  categories have been distinguished. Most of them present an oscillatory 
  behavior of the field amplitude, or an excitation-relaxation dynamics. 
  Similar oscillations and excitation-relaxation phenomena have  identified and 
  explained in Refs.~\cite{BerkBreizman_92, BerkBreizman_98}.
\end{itemize}

Finally, the model presented here is found to reproduce the same regimes as 
those predicted in standard theories of nonlinear trapping with a fixed damping.
The stability diagram presented in Fig.~\ref{fig_states_runs2} reproduces the
same qualitative features of the diagrams of Refs.~\cite{Vann_02, Lesur_09}, 
where similar parameters have been used for species 1
(though not a fully adiabatic model in Ref.~\cite{Vann_02}, 
and a collisionality which only applies to the bump in Ref.~\cite{Lesur_09}).

From a stability point of view, modes (/pseudo modes) are mainly found inside 
the linear stability region, which seems to advocate for the validity of the 
linear analysis.


\subsection{Metastability}
The validity of the linear stability analysis breaks down when the 
characteristic parameters of the damping species are varied. 

\subsubsection{Existence of metastable modes}
When the collisionality $\nu_d$ or the mass of the damping species 
$m_d/m_l$ are reduced, some modes appear outside of the linear stability
regions, which can only live if the initial perturbation is large enough.

\begin{figure}[ht!]
\begin{center}
\begin{minipage}{\linewidth}
\begin{center}
\includegraphics[width=0.5\linewidth]{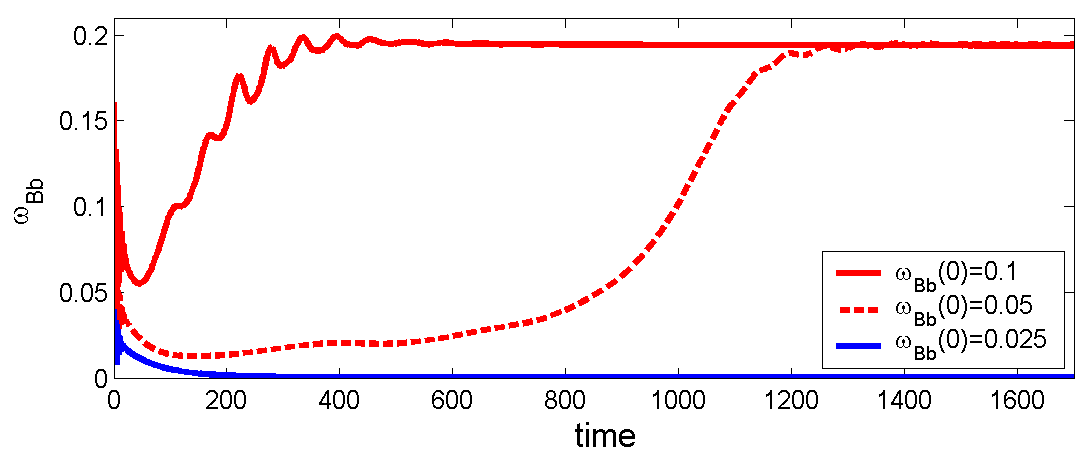}
\caption{\label{fig_OneMetastableModeWReducedAmplitudes}
  \footnotesize 
  Time evolution of the amplitude of a metastable mode, for various
  initial conditions.
}
\end{center}
\end{minipage}
\end{center}
\end{figure}

\begin{figure}[ht!]
\begin{center}
\begin{minipage}{\linewidth}
\begin{center}
\includegraphics[width=\linewidth]{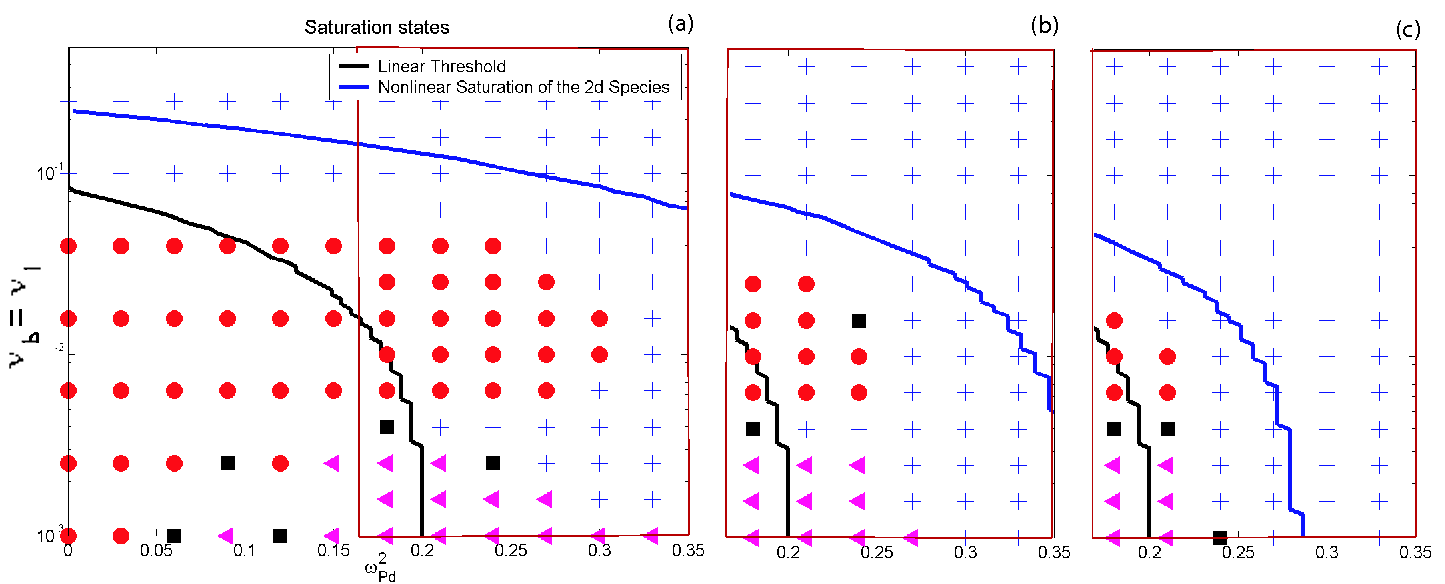}
\caption[\footnotesize
Stability diagrams corresponding to various amplitudes of the initial 
perturbation]
{\label{fig_MetastabilityWReducedAmplitudes}
\footnotesize 
Stability diagrams corresponding to $m_d/m_l=0.5$, $\nu_d=0.005$, for various
magnitudes of the initial perturbation: $\omega_{Bb}(t=0)=0.1$ in Fig.~(a), 
$\omega_{Bb}(t=0)=0.05$ in Fig.~(b), $\omega_{Bb}(t=0)=0.025$ in Fig.~(c).}
\end{center}
\end{minipage}
\end{center}
\end{figure}

An example of such mode found for
$m_d/m_l=0.5$, $\nu_d=0.005$, $\omega^2_{pd}=0.21$, $\nu_b=\nu_l=0.016$, is
given in Fig.~\ref{fig_OneMetastableModeWReducedAmplitudes}, where we represented 
the time evolution of its amplitude.
The localization of this mode in a stability diagram is given in 
Fig.~\ref{fig_MetastabilityWReducedAmplitudes}, and clearly shows that it is 
localized outside of the stability window.
This idea is confirmed by Fig.~\ref{fig_OneMetastableModeWReducedAmplitudes}. 
In this figure, the existence of the mode is seen to depend on the amplitude of 
the initial perturbation.
In any cases, the amplitude presents an initial decreasing phase,
which can be interpreted as a linear (stable) phase, and possibly grow in a later
phase to reach a finite saturation level if the initial perturbation is large
enough. This behavior indicates that the simulated mode is {\bf metastable}.\\

The stability diagram corresponding to $m_d/m_l=0.5$, 
$\nu_d=0.005$, $\omega_{Bb}(t=0)=0.1$, can be simulated and it is represented in 
Fig.~\ref{fig_MetastabilityWReducedAmplitudes}. 
It shows the existence of several modes outside of the linear stability region.
Again, when the initial perturbation amplitude is decreased, these subcritical
modes disappear. 

More precisely, we plotted in 
Fig.~\ref{fig_MetastabilityWReducedAmplitudes}, the curve 
$\omega_{Bd}(t)=-\gamma$ ($\gamma$ being the linear growth rate).
If the simulated modes are metastable, a condition for their existence is that
they reach the nonlinear regime. In particular, if our intuition of metastability
developed in subsection \ref{ssection_SubcriticalMotivation} is correct, the
damping species needs to reach the nonlinear regime to survive. 
A measure of the necessary
time to reach the nonlinear regime is the bounce frequency $\omega_{Bd}$
(its inverse gives the necessary time for a particle trajectory to be
significanlty modified by the wave).
Hence, in the  subcritical region  (where $\gamma<0$), the curve 
$\omega_{Bd}(t)=-\gamma$, gives some indication of the possibility for a mode
to reach the nonlinear regime before being significanlty damped.

This curve somehow follows the boundary of the metastable region, and confirms
the idea that the observed subcritical activity depends on the possibilty for the 
damping species to reach the nonlinear regime.\\

Note that the runs carried out in this analysis also clearly show the existence
of subcritical chaotic behaviors. However, we will not be considering them in the
following.

\subsubsection{Interpretation of metastability}
Let us now determine if the observation of metastable modes is in agreement 
with the interpretation given in subsection 
\ref{ssection_SubcriticalMotivation}, 
that is to say with the idea that it results from 
{\bf different nonlinear reduction factors}.
In the following, the consequences of such an interpretation are drawn and 
compared with the results of 6 stability diagrams, obtained using an initial 
perturbation $\omega_{Bb} = 0.1$, and the 6 possible combinations of the damping
species parameters: $\nu_d = 0.001, 0.005, 0.01$ and $m_d/m_l = 0.5, 2.0$.\\

The reduction factors $\nu^*_l$ and $\nu^*_d$ 
(or more precisely $2.0\nu^*_l$ and $2.0\nu^*_d$ )
can be easily calculated for the 
BOT problem, for which the curvature of the Hamiltonian is equal to 
${\sf C}_s = k^2/m_s$ for a species $s$. Using $\omega_{Bb}$ as a reference 
amplitude, it comes
\begin{equation}
\nu^*_l = \frac{\nu_l}{\omega_{Bb}},\\ 
\nu^*_d = \frac{\nu_d}{\omega_{Bb}}\sqrt{\frac{m_d}{m_l}}
       =  \frac{\nu_l}{\omega_{Bb}}
       \left(\frac{\nu_d}{\nu_l}\sqrt{\frac{m_d}{m_l}}\right).
\label{eq_reductionFactor2Species}
\end{equation}
In the framework of the nonlinear trapping theory, the role of the damping
species collisionality (compared to $\nu_l$) and of the mass ratio $m_d/m_l$ 
which were both decreased when moving from Fig.~\ref{fig_states_runs2}
to Fig.~\ref{fig_MetastabilityWReducedAmplitudes} is clear: their drop 
leads to a smaller nonlinear reduction factor and agrees with the idea that 
metastability results from the nonlinear reduction of damping.

Using the perturbative framework summarized by 
Eq.~\ref{eq_PerturbativeNLFramework} and the nonlinear trapping reduction,
one could expect
\begin{equation}
\text{ metastability for }   \nu^*_l\gamma_l > \nu^*_d\gamma_d \\
\text{ and saturation for } \nu^*_l\gamma_l = \nu^*_d\gamma_d.
\label{eq_NLTrappingEquation}
\end{equation}
Unfortunately, because both sides of the latter equality are proportional to 
$1/\omega_{Bb}$, such a model does not lead to saturation. 
This is in contradiction with our simulations 
which clearly displayed well defined saturation levels, which could be 
reproduced under variations of the initial conditions (see for example the 
saturation levels of the two metastable modes in 
Fig.~\ref{fig_OneMetastableModeWReducedAmplitudes}). This suggests that some
corrections are needed.\\

The assessment of the reduction factors based on the simulated saturation 
levels for our 6 sets of parameters 
(and for both linearly unstable modes and metastable modes) 
returns $\nu^*_l \in [0, 0.25]$ and $\nu^*_d\in[0, 0.20]$.
Two remarks follow.

\begin{itemize}
\item The computed values of $\nu^*_l, \nu^*_d$ show that the use of the 
  nonlinear trapping model is  legimate, but that some large $\nu^*$ effects
  play a role.
  In subsection \ref{section_NLTrappingToBAE}, we indicated  that for large 
  $\nu^*$ 
  (in particular for a negligible mode amplitude $\omega_{Bb}\rightarrow 0$), 
  the resonant response almost 
  remained linear (it is the {\it quasilinear response}, of "reduction factor" 
  1.0). 
  Hence the idea to fit the two regimes to take into account some large-$\nu^*$ 
  effects. 
  This can be done making an inverse
  average of the multiplying factors involved in the nonlinear trapping and 
  quasilinear regimes. In other words, a more reasonable reduction factor may be
  \begin{equation}
    \frac{2.0\nu^*}{1.0+2.0\nu*} \\
    \begin{array}{ll}
      \rightarrow 2.0\nu^* &\text{ for small }\nu^*\\
      \rightarrow 1.0 &\text{ for a null amplitude of the mode}
    \end{array}
  \end{equation}
\item In the energy balance  Eq.~\ref{eq_PerturbativeNLFramework}, the 
  contribution of the background dissipation was not considered, which
  is valid if the values of the driving and damping species exceed any lower 
  background dissipation.
  Linearly, such an assumption can be reasonable ($\gamma_l = 0.12$ with our 
  parameters), but this fully breaks down when moving to the nonlinear regime,
  where both the drive and damping are multiplied by their reduction factor
  $\nu^* \ll 1$.

  The effect of the bulk collisions on the energy
  balance Eq.~\ref{eq_PerturbativeNLFramework} can be calculated in a simple way,
  noticing that they simply imply a frequency shit $\omega\rightarrow\omega+i\nu_b$ 
  is the BOT dispersion relation, compared to the non-collisional case. 
  In other words, writing $\mathcal{L}_b$ the contribution of the bulk, it comes
  \begin{equation}
    \mathcal{L}_b(\omega) = \mathcal{L}_{\eq}(\omega-i\nu_b)
    = \partial_\omega\mathcal{L}_{\eq}(\omega_{r0})(\omega - i\nu_b-\omega_{r0})
    = i\partial_\omega\mathcal{L}_{\eq}(\omega_{r0})(\gamma -\nu_{b})
  \end{equation}
  The background damping is simply  given by the dissipation frequency
  $\nu_b$.
\end{itemize}
Finally the two previous remarks suggest a saturation for
\begin{equation}
\frac{2.0\nu_l}{\omega_{Bl}+2.0 \nu_l}\gamma_l = 
\frac{2.0\nu_d}{\omega_{Bd}+2.0 \nu_d}\gamma_d + \nu_b\\
\text{ or again,}
\end{equation}
\begin{equation}
0.25*\nu_b\omega^2_{Bb} 
-0.5
\left(\nu_l(-\nu_b+\gamma_l) 
+ \nu_d\sqrt{m_d/m_l}(-\nu_b-\gamma_d)) \right) 
\omega_{Bb}
- \nu_l\nu_d(-\nu_b+\gamma_l-\gamma_d)=0
\label{eq_SecondOrderEquation}
\end{equation}

If this model is valid, the existence of saturated modes depends on the existence
of positive roots for the mode amplitude $\omega_{Bb}$.
The present equation is a second order equation,
and we can notice that the product of its two roots is directly related to the 
condition for linear stability, whereas the sum of them is the same as the criterion
given in Eq.~\ref{eq_NLTrappingEquation}
(modulo the corrections resulting from the introduction of bulk dissipation).

\begin{figure}[ht!]
\begin{center}
\begin{minipage}{0.49\linewidth}
\begin{center}
\includegraphics[width=1.1\linewidth]{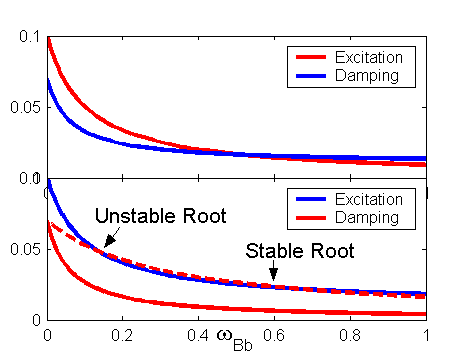}
\caption{\label{fig_SolutionExistence}
\footnotesize Illustration of the different situations offered by 
the proposed nonlinear model, which may lead to a saturation.}
\end{center}
\end{minipage}
\begin{minipage}{0.49\linewidth}
\begin{center}
\includegraphics[width=1.1\linewidth]{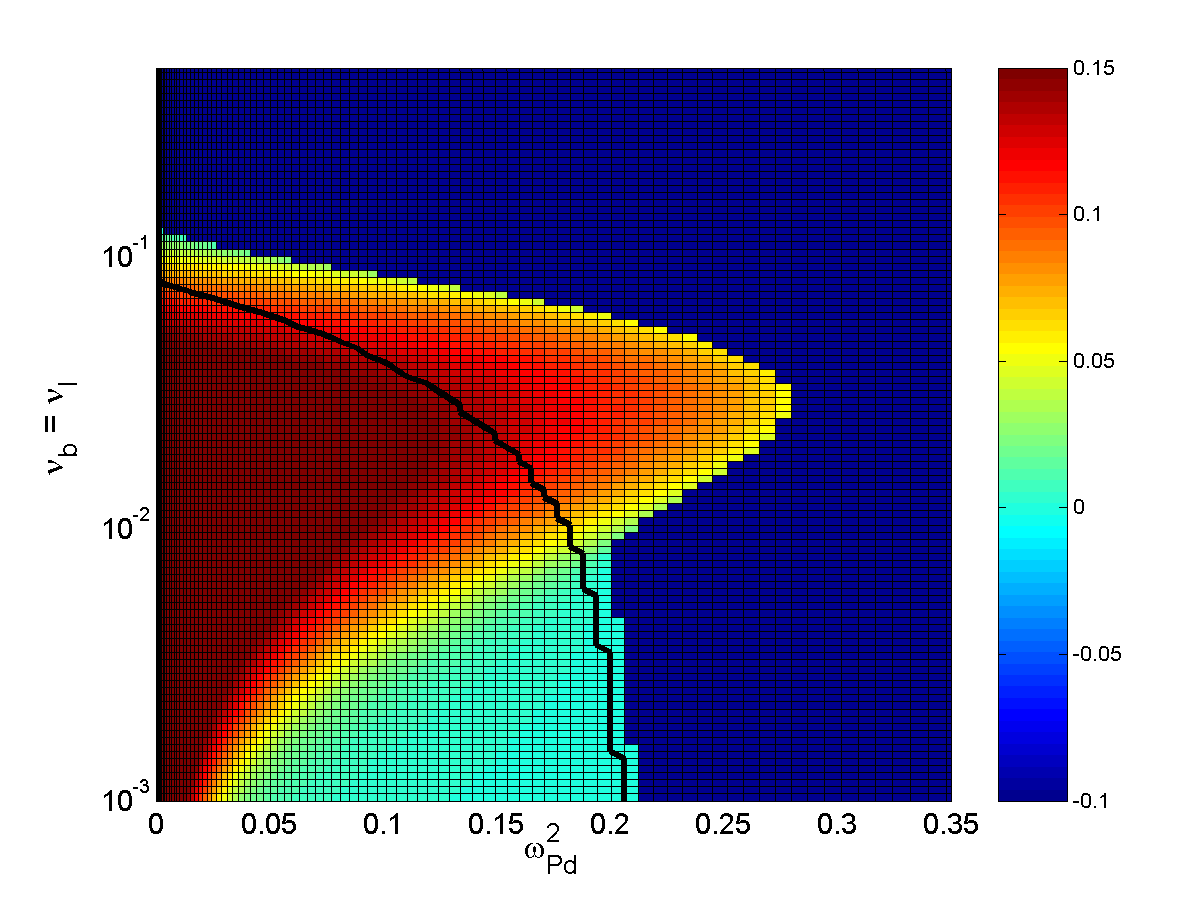}
\caption{\label{fig_HigherRootNLAmplitude}
  \footnotesize
  Stable saturation level predicted by the proposed nonlinear model, 
  for the stability diagram of parameters   $m_d/m_l=0.5$, $\nu_d=0.01$.
}
\end{center}
\end{minipage}
\end{center}
\end{figure}

It follows that {\bf linearly unstable modes can allways reach saturation}.
The situation is different in the subcritical region of course.
But {\bf metastability is possible}. When solutions exists, {\bf they are
(both) positive if the generalized version of Eq.~\ref{eq_NLTrappingEquation} is 
verified}
\begin{equation}
\nu_l(-\nu_b+\gamma_l) > \nu_d\sqrt{m_d/m_l}(-\nu_b-\gamma_d)) 
\end{equation} 
In this case however, only  one of the two roots is stable, as illustrated
in Fig.~\ref{fig_SolutionExistence}, where we represented the nonlinear 
driving rate and the dissipation rate as a function of the mode amplitude.
The latter stable roots are represented in Fig.~\ref{fig_HigherRootNLAmplitude}
(when they exist).

\subsubsection{Model verification}
\begin{figure}[h!]
\begin{center}
\begin{minipage}{0.49\linewidth}
\begin{center}
\includegraphics[width=1.1\linewidth]{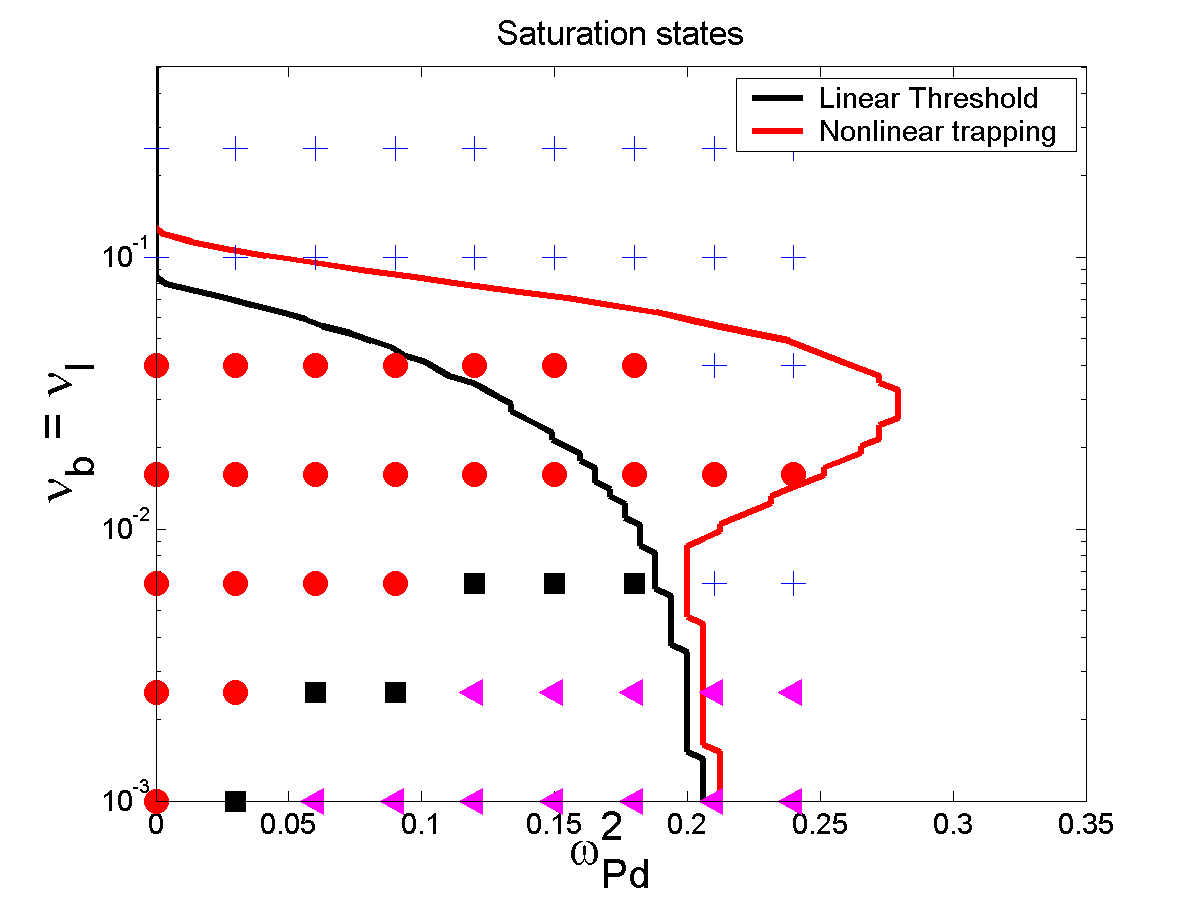}
\caption
{\label{fig_states_runs3_halfnude}
\footnotesize 
  Stability diagram without metastability, obtained for the parameters
  $m_d/m_l=0.5$, $\nu_d=0.01$, $\omega_{Bb}(t=0)=0.1$.  
}
\end{center}
\end{minipage}
\begin{minipage}{0.49\linewidth}
\begin{center}
\includegraphics[width=1.1\linewidth]{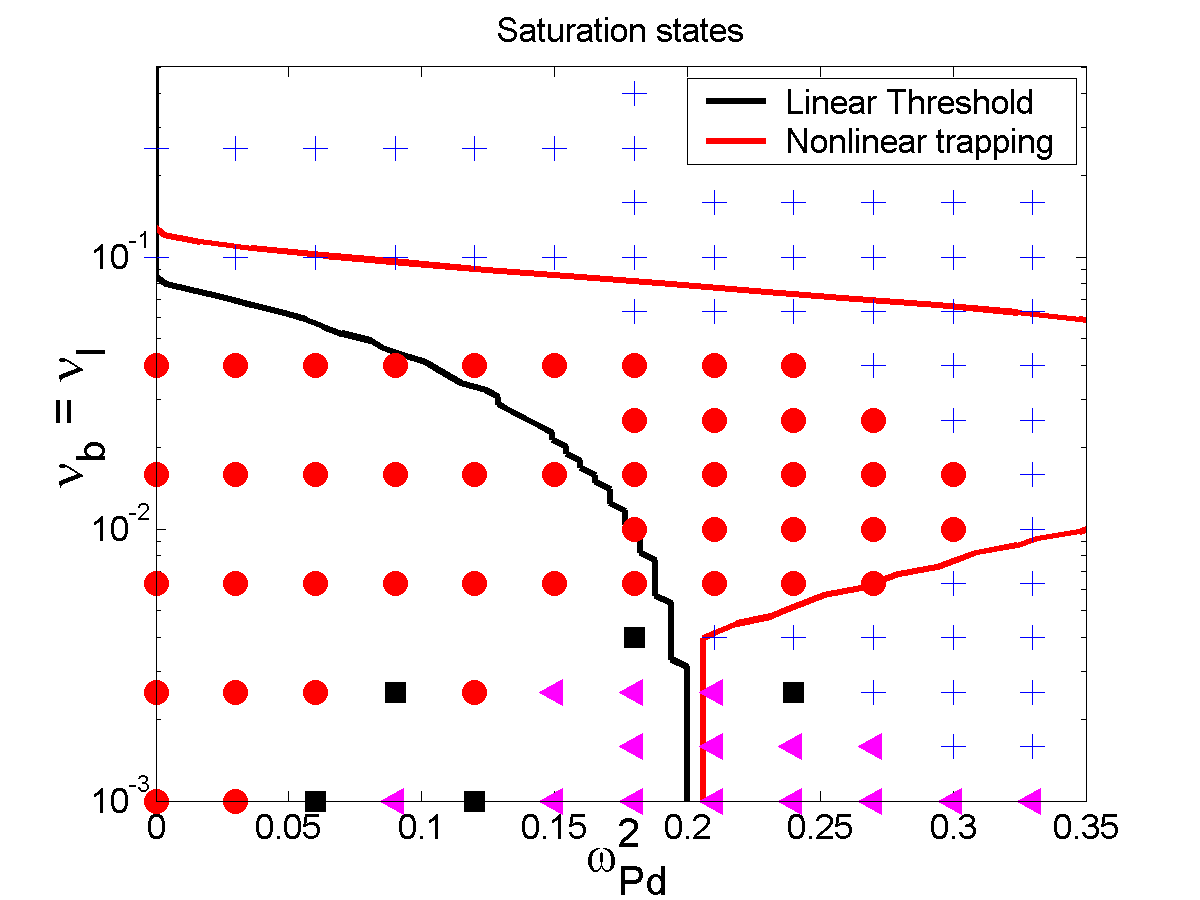}
\caption{\label{fig_states_runs4_halfnude}
\footnotesize 
  Stability diagram without metastability, obtained for the parameters
  $m_d/m_l=0.5$, $\nu_d=0.005$, $\omega_{Bb}(t=0)=0.1$.  
}
\end{center}
\end{minipage}
\end{center}
\end{figure}

\begin{figure}[h!]
\begin{center}
\begin{minipage}{\linewidth}
\begin{center}
\includegraphics[width=0.7\linewidth]{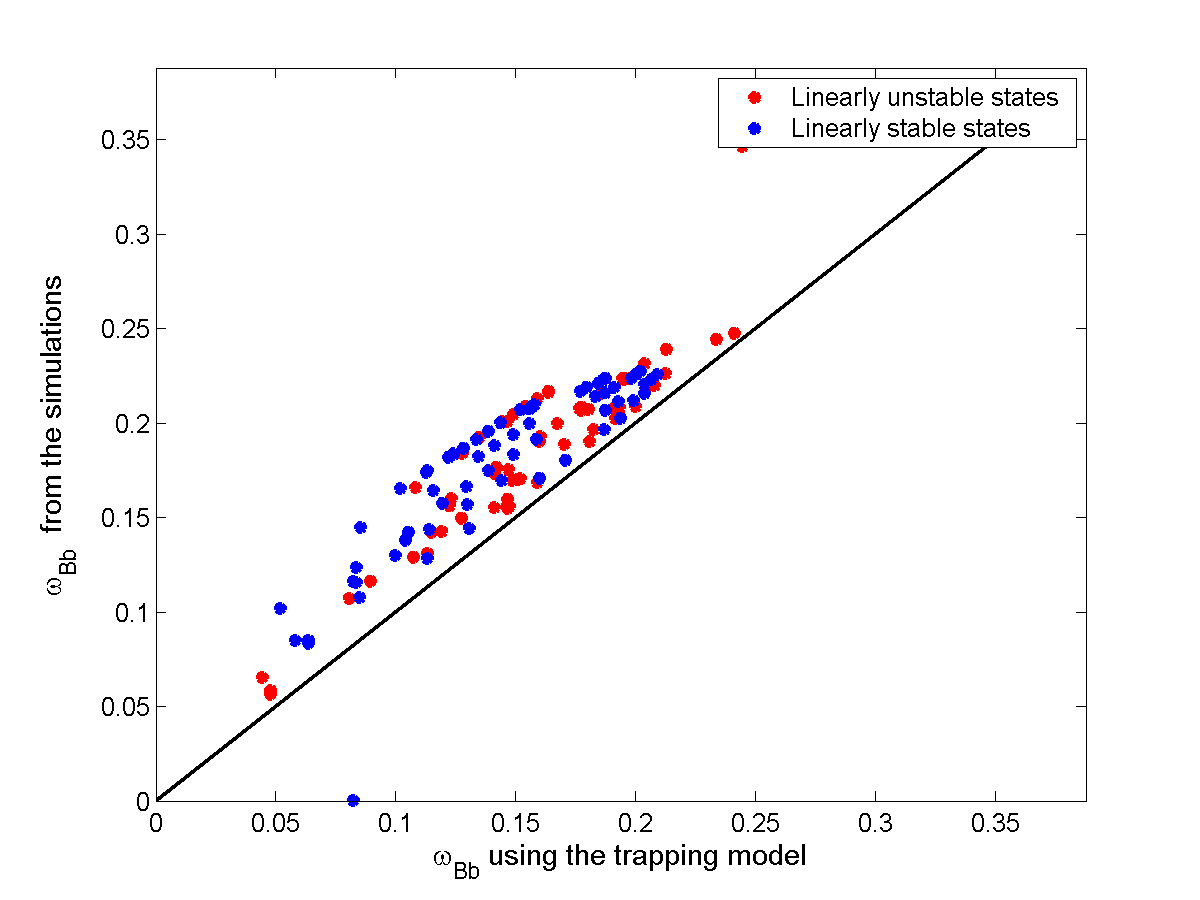}
\caption{\label{fig_AllSaturationLevels}
\footnotesize 
Simulated nonlinear saturation levels compared to prediction, for 
linearly unstable and metastable modes.}
\end{center}
\end{minipage}
\end{center}
\end{figure}

In the next three plots, the existence criteria  
(Figs.~\ref{fig_states_runs3_halfnude}) and saturation levels
(\ref{fig_AllSaturationLevels} where our 6 sets of parameters are 
represented)  derived from 
Eq.~\ref{eq_SecondOrderEquation} are compared with the simulations.

A clear agreement appears, which confirms the possibility and relevance of 
metastability, due to  kinetic nonlinear saturation.
In particular, this study confirms that dissipative effects become 
particularly important in nonlinear regimes, even when they are negligible
in a linear analysis.

Because of the order of magnitudes computed in subsection
\ref{ssection_NLToBAE_OrdersMagnitude}, metastability via kinetic resonant
saturation should be seen as a 
serious candidate for the existence of subcritical activity.


\section{Breaking of the condition perturbative-adiabatic approach...
towards a BAE relevant description}
\label{section_NonPerturbativeCase}

In this section, we simply want to make a few concluding comments on the
perturbative-adiabatic approach, a lot used in past theories
(which we also used in this thesis...), starting from a BAE relevant 
non-adiabatic run, with:
\begin{equation}
\begin{array}{lll}
\text{A non-perturbative bump, with} & n_l = 0.15, v_{tl}={1.0} \\
\text{A non adiabatic species - damping species, with } &
m_d / m_l = 0.5, n_d = 1.0, \\
&v_{td} = 1.2 = \omega_{pd}/k,  \nu_d = 0.01.\\
\end{array}
\end{equation}
and the initial condition $\omega_{Bd}(t=0) = 0.14$.  In particular, the 
bulk is not adiabatic $\omega_{pd}/kv_{td} = 2.7  (\sim \sqrt{7/2 + 2})$.

In this figure, various cases corresponding to different values of the 
bump collisionality are represented. The time evolution of the mode amplitude
is shown as well as the frequency spectrum for some of these cases, where the 
blue lines represent the frequency $\omega_{pb}=\omega_{pd}$.
\begin{figure}[ht!]
\begin{center}
\begin{minipage}[t]{0.49\linewidth}
\begin{center}
\includegraphics[width=\linewidth]{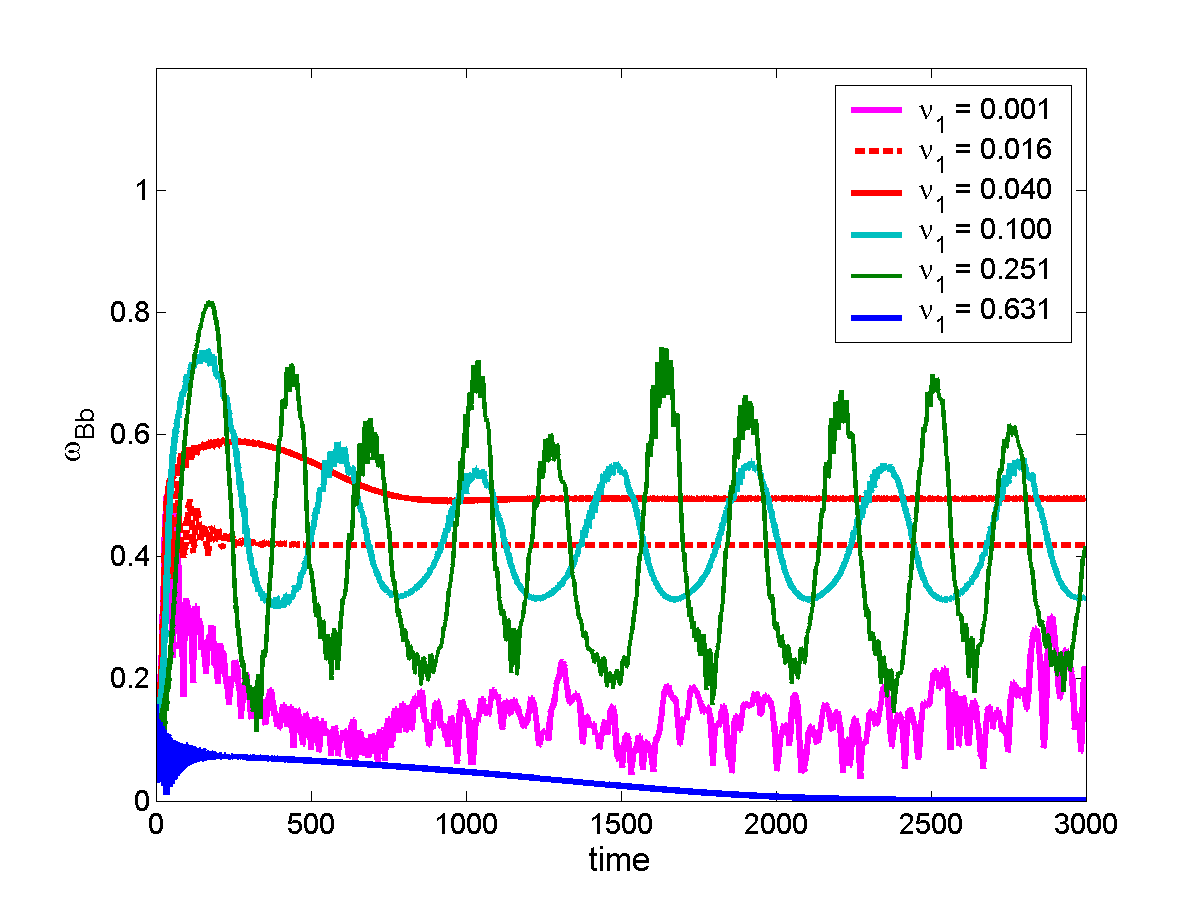}
\end{center}
\end{minipage}
\begin{minipage}[]{0.49\linewidth}
\begin{center}
\includegraphics[width=\linewidth]{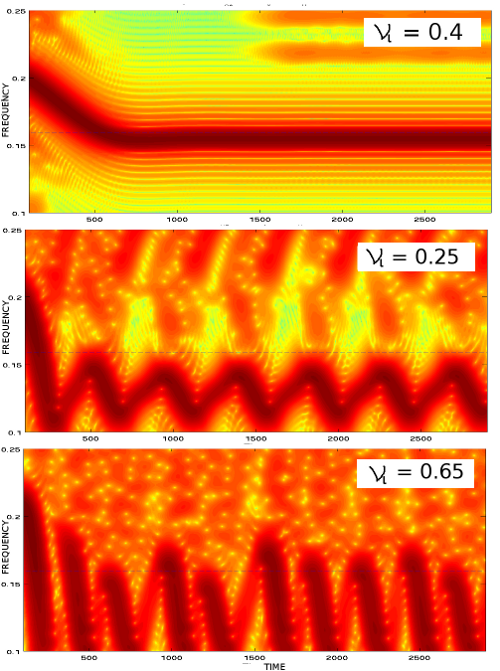}
\end{center}
\end{minipage}
\end{center}
\caption[
\footnotesize
Some nonlinear states obtained from a non-adiabatic model for the bulk
plasma.]
{\label{fig_NonAdiabatic_states}
\footnotesize
Some nonlinear states obtained from a non-adiabatic model for the bulk
plasma, and for different values of the driving species collisionality.
}
\end{figure}

What this set of figures shows is that moving to the non 
perturbative/non-adiabatic case may lead to very different saturation regimes,
as the one described above, and traditionally considered in theory/experiment
comparisons. 
The oscillatory regimes shown in these figures are different and topologically 
separated from the chirping or chaotic structures mentionned earlier.
They seem to correspond to a cycle between two stable solutions.

A second point is that nonlinear saturation may not be such a smooth, 
stabilizing process, as suggested by the current the nonlinear
trapping theory presented here. In this case, sweeping is clearly 
associated to a {\bf growth}.\\

A lot remains to be understood by the author of the present manuscript. 
But our aim is definitely to go further in this direction.
This was simply to finish with nice pictures...

\section{Summary}
{\it
In this chapter, we analyzed some difficulties related to the nonlinear 
description of energetic particles with the attempt to catch their relevance
for Beta Alfv\'en Eigenmodes, 
and we analyzed one particular problem in details: 
the possible existence of metastable modes due to resonant saturation.

\begin{itemize}
\item 
  We found that the time scales involved in the modes observed in Tore-Supra
  were consistent with the approximations of the nonlinear trapping theory.
\item
  We raised the question of a possible {\bf nonlinear modification of 
    stability}, 
  and suggested one process leading to metastable modes:  
  the possibility for the driving and damping mechanisms at stake, to
  saturate at different levels, because kinetic resonant nonlinearities.
  
\item 
  We developed a 2-D model (based on pre-compiled routines), 
  catching the kinetic features of the wave-particle interaction, to 
  check the proposed idea of metastability. {\bf Simulations confirmed the 
  proposed mechanism, both qualitatively and quantitatively}.

\item Again, the analysis puts forward the necessity to avoid the 
perturbative treatment of energetic particles.
\end{itemize}
}

\begin{savequote}[10pc]
\sffamily
\qauthor{}
\end{savequote}




\chapter{Conclusion}
One of the major scientific goals for ITER is to reach and explore the 
burning plasma regime, in which significant amounts of energy are 
generated by the D­T fusion reactions. Such a regime is characterized 
by a large population of fusion-produced alpha particles, born with a
largely suprathermal energy, 3.5 MeV. 
As a consequence of their interaction with the plasma, such energetic
particles may not only be redistributed and lost, they can also endanger
the whole plasma stability and impact on turbulence and plasma confinement
properties.\\

In this thesis, we studied one type of interaction which can take place between
energetic particles and the main plasma: {\bf the destabilization of a meso-scale
acoustic mode by energetic particles, the Beta Alfv\'en Eigenmode (BAE)}.
For this, we made use of a framework which applies more generally to  
a large class of energetic particle phenomena, with an effort to keep the
presentation general whenever the particular physics of BAEs was not involved.

The first part of the work was devoted to the building of the latter 
framework, a variational gyrokinetic framework, based on previous formulations 
\cite{Edery_92}. 
The effort done here was to define the physical meaning of the instabilities
such dispersion relation can display, offer a clean and compact formulation
of the gyrokinetic equation relevant for both geometric and particle variables
and appropriate for the modelling of resonant behaviors, and finally draw the 
link between well-established formulations used in this field: 
a traditional set of gyrokinetic equations used for the study of shear Alfv\'en
waves, and the Magneto-HydroDynamic formalism. Finally, the applicability 
of such a variational framework to nonlinear kinetic regimes was assessed.

The second part was devoted to the characterization of BAEs, with two goals.
A first goal was to determine the eigenmode properties which can be of 
interest in the destabilization of the mode. For this, we proposed a 
derivation of the BAE dispersion relation using a Fourier decomposition, and 
limited to a perturbative computation of transit effects, which could recover 
previous derivations, conducted in the so-called ballooning representation. 
We also calculated the BAE structure explicitely.
A second goal was to distinguish BAEs from other modes belonging to the 
acoustic frequency range, and we showed in particular what were the 
similarities and differences between  BAEs and the so-called Geodesic Acoustic
Modes, in the light of our derivation. This analysis allowed us to identify
Tore-Supra modes as BAEs.

Finally, we entered the core of the analysis in a third part, that is the 
problem of the destabilization of BAEs by a population of energetic particles.
Such a problem can be seen as a competition between the mode thermal damping, 
called Landau damping, and the drive by energetic particles making necessary
the determination of a threshold for BAE destabilization.
We first started such a determination with a perturbative linear analysis
designed to offer an intuitive understanding of the experimental tunable 
parameters at stake in the threshold, 
and we compared this analysis with experiments carried out in the Tore-Supra
tokamak.
The latter comparison displayed qualitative and quantitative agreement, 
allowing to build some confidence in the mechanisms
at stake in the BAE destabilization as well as the on the rough role of 
global macroscopic parameters involved. We could also provide some 
explanation for more detailed behaviors such as an obvious role of
the shear in the destabilization. However, some caution is needed for the
latter interpretations because of two aspects: the use of a perturbative
approach and the apparition of a set of phenomena which cannot fit our 
framework, such as the frequency upshift of the BAE frequency during the 
course of a sawtooth period or the harder observations of BAEs at low magnetic
field. 

Among other interpretations, the mismatch between linear theory and 
experiments could be associated with a nonlinear modification of stability. 
In the thesis, we proposed one scenario, appropriate for BAEs, which makes 
possible the existence of metastable modes and we verified it numerically.\\

Though it can provide some intuitive, tractable understanding of the physics
at stake, the most questionable feature of the thesis is the use of the 
perturbative approach.

In the near future, we would like to improve the linear model used in the
thesis to include diamagnetic effects \cite{Lauber_09} 
as well as a non-perturbative treatment of Landau damping based on 
Ref.~\cite{Zonca_98}, to make a test of the relative role of each component
for Tore-Supra parameters, and provide some explanations for the shift
between the BAE basic formula and the experiments.

Another point which needs to be addressed, is the reason for the stable
frequency behavior of BAEs in Tore-Supra, compared to other tokamak devices
characterized by frequency sweeping \cite{Zonca_09}, which is not fully
clarified at the moment. Following the
nonlinear sweeping model predicted in Ref.~\cite{BerkBreizman_99}, 
phase-space structures associated frequency chirping occurs for negligible 
dissipative phenomena, and the Tore-Supra shots we conducted are indeed highly
collisional. However, there may exist other situations leading to
frequency sweeping. Some more analysis needs to be made for this issue.

Finally, the investigation of a nonlinear modification of stability displayed
interesting and conclusive results, which could be relevant in tokamak
conditions. Again, a non-pertubative treatment may be more realistic and could
lead not only to a nonlinear modification of the amplitude of the driving and 
damping mechanisms at stake, but also to a strong modification 
of the mode structure itself.
Also, such nonlinear effects need an experimental investigation.\\

On a broader scale, we can note that recent research in the field of energetic 
particles has definitely fallen into nonlinear studies, which are
necessary for the computation of energetic particle losses.
It may be important to fit in a consistant picture the different 
regimes considered for the nonlinear evolution: the so-called MHD
nonlinearities and kinetic resonant nonlinearities, low scale and large
scale  chirping regimes possibly associated to a wave growth, and to add
some effects of geometry, which are often overlooked.

Ideally, such studies should go in the direction of a control of the unwanted
phenomena related to fast particles.


\publications
\subsubsection{Publications}
$\bullet$ 
C.~{Nguyen}, X.~{Garbet} and A.-I.~Smolyakov.
\newblock {Variational derivation of the dispersion relation of kinetic coherent  modes in the acoustic frequency range in tokamaks},
\newblock {\em Physics of Plasmas, 15, 1}, 2008.\\
$\bullet$ 
C.~{Nguyen}, X.~{Garbet}, R.~{Sabot}, L.-G. {Eriksson}, M.~{Goniche}, P.~{Maget}, V.~{Basiuk}, J.~{Decker}, D.~{Elb\`eze}, G.~T.~A. {Huysmans}, A.~Macor, J.-L. S\'egui and M.~Schneider.
\newblock {Excitation of Beta Alfv\'en Eigenmodes in Tore-Supra},
\newblock {\em Plasma Physics and Controlled Fusion, 51, 095002}, 2009.\\
$\bullet$ 
C.~{Nguyen}, X.~{Garbet}, V.~Grandgirard and M.~Lesur.
\newblock{On the possibility of a nonlinear modification of the stability of fast particle driven modes},
\newblock{\em To be submitted}, 2009.\\ 
$\bullet$ 
A.-I.~Smolyakov, C.~Nguyen and X.~Garbet.
\newblock{Kinetic theory of electromagnetic geodesic acoustic modes}, 
\newblock{\em Plasma Physics Controlled Fusion, 50, 115008}, 
2008.\\
$\bullet$
A.-I.~Smolyakov, C.~Nguyen and X.~Garbet.
\newblock{Electromagnetic effects on Geodesic Acoustic Modes And Beta Alfv\'en Eigenmodes},
\newblock{\em Submitted to Nuclear Fusion}, 2009.\\
$\bullet$
A.~Macor, M.~Goniche, J.-F.~Artaud, J.~Decker, D.~Elb\`eze, X.~Garbet, G.~Giruzzi, G. T.~Hoang, P.~Maget, D.~Mazon, D.~Molina, C.~Nguyen, Y.~Peysson, R.~Sabot and J.-L. S\'egui.
\newblock{Redistribution of Suprathermal Electrons due to Fishbone Frequency Jumps},
\newblock{\em Physical Review Letters, 102, 155005}, 2009.\\
$\bullet$ 
R.~Sabot, A.~Macor, C.~Nguyen, J.~Decker, D.~Elbeze, L.-G.~Eriksson, X.~Garbet, M.~Goniche, G. T. A.~Huysmans, Y.~Ladroit, P.~Maget and J.-L.~S\'egui.
\newblock{Observation of acoustic and sub-acoustic fast particles driven modes in Tore-Supra},
\newblock{\em Nuclear Fusion, 49, 085033}, 2009.\\
$\bullet$
X.~Garbet, G.~Dif-Pradalier, C.~Nguyen, Y.~Sarazin, V.~Grandgirard and P.~Ghendrih.
\newblock{Neoclassical equilibrium in gyrokinetic simulations},
\newblock{\em Physics of Plasmas, 16, 062503}, 2009.\\ 

\subsubsection{Orals}         
$\bullet$
American Transport Task Force workshop (Denver, 2008). \\
$\bullet$
IAEA Technical Meeting on Energetic Particles (Kiev, 2009).\\
$\bullet$
Invited to European Fusion Theory Conference (Riga, 2009).\\
$\bullet$
Invited to European Physical Society (Dublin, 2010).

\bibliographystyle{unsrt}
\bibliography{Thesis.bib}

\begin{thebibliography}{100}

\bibitem{Rider_95}
T.~H. Rider.
\newblock {\em Fundamental Limitations on Plasma Fusion Systems Not in
  Thermodynamical Equilibrium}.
\newblock PhD thesis, MIT, 1995.

\bibitem{Jarboe_94}
T.~R. {Jarboe}.
\newblock {Review of spheromak research}.
\newblock {\em Plasma Physics and Controlled Fusion}, 36:945--990, June 1994.

\bibitem{Lyon_90}
J.~F. Lyon, G.~Grieger, F.~Rau, A.~Iiyoshi, A.~P. Navarro, L.~M. Kovrizhnykh,
  O.~S. Pavlichenko, and S.~M. Hamberger.
\newblock {Stellerators}.
\newblock {\em Nuclear Fusion}, 30:1695--1715, 1990.

\bibitem{Duong_93}
H.~H. {Duong}, W.~W. {Heidbrink}, E.~J. {Strait}, T.~W. {Petrie}, R.~{Lee},
  R.~A. {Moyer}, and J.~G. {Watkins}.
\newblock {Loss of energetic beam ions during TAE instabilities}.
\newblock {\em Nuclear Fusion}, 33:749--765, May 1993.

\bibitem{White_95}
R.~B. {White}, E.~{Fredrickson}, D.~{Darrow}, M.~{Zarnstorff}, R.~{Wilson},
  S.~{Zweben}, K.~{Hill}, Y.~{Chen}, and G.~{Fu}.
\newblock {Toroidal Alfv{\'e}n eigenmode-induced ripple trapping}.
\newblock {\em Physics of Plasmas}, 2:2871--2873, August 1995.

\bibitem{Angioni_08}
C.~Angioni, A.~G. Peeters, and G.~V. Pereverzev.
\newblock Gyrokinetic simulation of alpha particle transport and consequences
  on iter transport modelling.
\newblock 13th EU-US TTF Workshop and 1st EFDA Transport Topical Group meeting,
  2008.

\bibitem{NicholsonBook}
D.~R. Nicholson.
\newblock {\em Introduction to Plasma Theory}.
\newblock John Wiley and Sons, 1983.

\bibitem{HazeltineBook}
R.~D. Hazeltine and J.~D. Meiss.
\newblock {\em Plasma Confinement}.
\newblock Addison-Wesley Publishing Company, Redwood City, 2007.

\bibitem{BiskampBook}
D.~Biskamp.
\newblock {\em Nonlinear Magnetohydrodynamics}.
\newblock Cambridge University Press, 1997.

\bibitem{Brizard_07}
A.~J. {Brizard} and T.~S. {Hahm}.
\newblock {Foundations of nonlinear gyrokinetic theory}.
\newblock {\em Review of Modern Physics}, 79:421--468, April 2007.

\bibitem{GoldsteinBook}
H.~Goldstein, C.~P. Poole, and J.~L. Safko.
\newblock {\em Classical Mechanics}.
\newblock Addison Wesley, 3rd edition, 2001.

\bibitem{Littlejohn_83}
R.~G. Littlejohn.
\newblock Variational principles of guiding center motion.
\newblock {\em Journal of Plasma Physics}, 29:111--125, 1983.

\bibitem{WhiteBook}
R.~B. White.
\newblock {\em The theory of Toroidally Confined Plasmas}.
\newblock World Scientific Publishing Company; 2 Revised edition, 2006.

\bibitem{Kaufman_72}
A.~N. {Kaufman}.
\newblock {Quasilinear Diffusion of an Axisymmetric Toroidal Plasma}.
\newblock {\em Physics of Fluids}, 15:1063--1069, June 1972.

\bibitem{GarbetHDR}
X.~Garbet.
\newblock Instabilit\'es, turbulence et transport dans un plasma magn\'etis\'e.

\bibitem{Zonca_07}
F.~{Zonca}, P.~{Buratti}, A.~{Cardinali}, L.~{Chen}, J.-Q. {Dong}, Y.-X.
  {Long}, A.~V. {Milovanov}, F.~{Romanelli}, P.~{Smeulders}, L.~{Wang}, Z.-T.
  {Wang}, C.~{Castaldo}, R.~{Cesario}, E.~{Giovannozzi}, M.~{Marinucci}, and
  V.~{Pericoli Ridolfini}.
\newblock {Electron fishbones: theory and experimental evidence}.
\newblock {\em Nuclear Fusion}, 47:1588--1597, November 2007.

\bibitem{Hasegawa_74}
A.~{Hasegawa} and L.~{Chen}.
\newblock {Plasma Heating by Alfv{\'e}n-Wave Phase Mixing}.
\newblock {\em Physical Review Letters}, 32:454--456, March 1974.

\bibitem{Vlad_99}
G.~{Vlad}, F.~{Zonca}, and S.~{Briguglio}.
\newblock {Dynamics of Alfv{\'e}n waves in tokamaks}.
\newblock {\em Nuovo Cimento Rivista Serie}, 22:1--97, July 1999.

\bibitem{Hasegawa_76}
A.~{Hasegawa} and L.~{Chen}.
\newblock {Kinetic processes in plasma heating by resonant mode conversion of
  Alfven wave}.
\newblock {\em Physics of Fluids}, 19:1924--1934, 1976.

\bibitem{Vaclavik_91}
J.~{Vaclavik} and K.~{Appert}.
\newblock {Theory of plasma heating by low frequency waves: magnetic pumping
  and Alfv\'en resonance heating}.
\newblock {\em Nuclear Fusion}, 31:1945--1997, 1991.

\bibitem{Grad_69}
H.~{Grad}.
\newblock {Plasmas}.
\newblock {\em Physics Today}, 22(12):34--36, December 1969.

\bibitem{Heidbrink_08}
W.~W. {Heidbrink}.
\newblock {Basic physics of Alfv{\'e}n instabilities driven by energetic
  particles in toroidally confined plasmas}.
\newblock {\em Physics of Plasmas}, 15(5):055501--+, May 2008.

\bibitem{Appert_82}
K.~{Appert}, R.~{Gruber}, F.~{Troyon}, and J.~{Vaclavik}.
\newblock {Excitation of global eigenmodes of the Alfven wave in Tokamaks}.
\newblock {\em Plasma Physics}, 24:1147--1159, September 1982.

\bibitem{Appert_84}
K.~{Appert}, J.~{Vaclavik}, and L.~{Villard}.
\newblock {Spectrum of low-frequency, nonaxisymmetric oscillations in a cold,
  current-carrying plasma column}.
\newblock {\em Physics of Fluids}, 27:432--437, February 1984.

\bibitem{Appert_85}
K.~{Appert}, G.~A. {Collins}, F.~{Hofmann}, R.~{Keller}, A.~{Lietti}, J.~B.
  {Lister}, A.~{Pochelon}, and L.~{Villard}.
\newblock {Observations of toroidal coupling for low-n Alfv{\'e}n modes in the
  TCA tokamak}.
\newblock {\em Physical Review Letters}, 54:1671--1674, April 1985.

\bibitem{Berk_01}
H.~L. {Berk}, D.~N. {Borba}, B.~N. {Breizman}, S.~D. {Pinches}, and S.~E.
  {Sharapov}.
\newblock {Theoretical Interpretation of Alfv{\'e}n Cascades in Tokamaks with
  Nonmonotonic q Profiles}.
\newblock {\em Physical Review Letters}, 87(18):185002--+, October 2001.

\bibitem{Breizman_03}
B.~N. {Breizman}, H.~L. {Berk}, M.~S. {Pekker}, S.~D. {Pinches}, and S.~E.
  {Sharapov}.
\newblock {Theory of Alfv{\'e}n eigenmodes in shear reversed plasmas}.
\newblock {\em Physics of Plasmas}, 10:3649--3660, September 2003.

\bibitem{Kieras_82}
C.~E. {Kieras} and J.~A. {Tataronis}.
\newblock {The shear Alfv{\'e}n continuous spectrum of axisymmetric toroidal
  equilibria in the large aspect ratio limit}.
\newblock {\em Journal of Plasma Physics}, 28:395--+, 1982.

\bibitem{Cheng_86}
C.~Z. Cheng and M.~S. Chance.
\newblock Low-n shear alfv\'en spectra in axisymmetric toroidal plasmas.
\newblock {\em Physics of Fluids}, 29(11):3695--3701, 1986.

\bibitem{Berk_91}
H.~L. {Berk}, J.~W. {Vandam}, Z.~{Guo}, and D.~M. {Lindberg}.
\newblock {\em {Continuum damping of low-n toroidicity-induced shear Alfven
  eigenmodes}}.
\newblock September 1991.

\bibitem{Chen_07}
L.~{Chen} and F.~{Zonca}.
\newblock {Theory of Alfv{\'e}n waves and energetic particle physics in burning
  plasmas}.
\newblock {\em Nuclear Fusion}, 47:727--+, October 2007.

\bibitem{Goniche_08}
M~Goniche, G.~T.~A. Huysmans, F.~Turco, P.~Maget, J.~L. Segui, J.~F. Artaud,
  G.~Giruzzi, F.~Imbeaux, P.~Lotte, D.~Mazon, and D.~Molina.
\newblock {Identification of fast particle triggered modes by means of
  correlation electron cyclotron emission on Tore-Supra.}
\newblock {\em Fusion Science and Technology}, 53(1):88--96, 2008.

\bibitem{Manhajan_94}
S.~M. {Mahajan}.
\newblock {Spectrum of Alfv{\'e}n waves, a brief review}.
\newblock {\em Physica Scripta Volume T}, 55:160--+, 1994.

\bibitem{Cheng_85}
C.~Z. {Cheng}, L.~{Chen}, and M.~S. {Chance}.
\newblock {High-n ideal and resistive shear Alfv{\'e}n waves in tokamaks}.
\newblock {\em Annals of Physics}, 161:21--47, April 1985.

\bibitem{Chen_94}
L.~{Chen}.
\newblock {Theory of magnetohydrodynamic instabilities excited by energetic
  particles in tokamaks}.
\newblock {\em Physics of Plasmas}, 1:1519--1522, 1994.

\bibitem{Zonca_92}
F.~{Zonca} and L.~{Chen}.
\newblock {Resonant damping of toroidicity-induced shear-Alfv{\'e}n eigenmodes
  in tokamaks}.
\newblock {\em Physical Review Letters}, 68:592--595, February 1992.

\bibitem{Fu_08}
G.-Y. {Fu}.
\newblock {Energetic-Particle-Induced Geodesic Acoustic Mode}.
\newblock {\em Physical Review Letters}, 101(18):185002--+, October 2008.

\bibitem{Zonca_96}
F.~{Zonca}, L.~{Chen}, and R.~A. {Santoro}.
\newblock {Kinetic theory of low-frequency Alfv{\'e}n modes in tokamaks}.
\newblock {\em Plasma Physics and Controlled Fusion}, 38:2011--2028, 1996.

\bibitem{Betti_92}
R.~{Betti} and J.~P. {Freidberg}.
\newblock {Stability of Alfv{\'e}n gap modes in burning plasmas}.
\newblock {\em Physics of Fluids B}, 4:1465--1474, June 1992.

\bibitem{Lauber_03}
P.~Lauber.
\newblock {\em Linear Gyrokinetic Description of Fast Particle Effects on the
  MHD Stability in Tokamaks}.
\newblock PhD thesis, Technische Universit\"at M\"unchen, 2003.

\bibitem{Zonca_98}
F.~{Zonca}, L.~{Chen}, R.~A. {Santoro}, and J.~Q. {Dong}.
\newblock {LETTER TO THE EDITOR: Existence of discrete modes in an unstable
  shear Alfv{\'e}n continuous spectrum }.
\newblock {\em Plasma Physics and Controlled Fusion}, 40:2009--2021, 1998.

\bibitem{Chavdarovski_09}
I.~{Chavdarovski} and F.~{Zonca}.
\newblock {Effects of trapped particle dynamics on the structures of
  low-frequency shear Alfv\'{e}n continuous spectrum}.
\newblock {\em to be published in Plasma Physics and Controlled Fusion}, 2009.

\bibitem{Rosenbluth_83}
M.~N. {Rosenbluth}, S.~T. {Tsai}, J.~W. {van Dam}, and M.~G. {Engquist}.
\newblock {Energetic Particle Stabilization of Ballooning Modes in Tokamaks}.
\newblock {\em Physical Review Letters}, 51:1967--1970, November 1983.

\bibitem{Porcelli_91}
F.~{Porcelli}.
\newblock {Fast particle stabilisation}.
\newblock {\em Plasma Physics and Controlled Fusion}, 33:1601--1620, November
  1991.

\bibitem{White_89}
R.~B. White, M.~N. Bussac, and F.~Romanelli.
\newblock High-$\beta{}$, sawtooth-free tokamak operation using energetic
  trapped particles.
\newblock {\em Physical Review Letters}, 62(5):539--542, Jan 1989.

\bibitem{Chen_84}
L.~{Chen}, R.~B. {White}, and M.~N. {Rosenbluth}.
\newblock {Excitation of Internal Kink Modes by Trapped Energetic Beam Ions}.
\newblock {\em Physical Review Letters}, 52:1122--1125, 1984.

\bibitem{Biglari_91}
H.~{Biglari} and L.~{Chen}.
\newblock {Unified theory of resonant excitation of kinetic ballooning modes by
  energetic ions and alpha particles in tokamaks}.
\newblock {\em Physical Review Letters}, 67:3681--3684, December 1991.

\bibitem{Zonca_EPS07}
F.~{Zonca} and L.~{Chen}.
\newblock {The general fishbone-like dispersion relation: a unified description
  for shear Alfv\'en Mode excitations}.
\newblock {\em Proceedings of the 34th EPS Conference on Plasma Physics,
  Warsaw, Poland}, 2007.

\bibitem{Berk_05}
H.~L. {Berk}, C.~J. {Boswell}, D.~{Borba}, A.~C.~A. {Figueiredo}, T.~{Johnson},
  M.~F.~F. {Nave}, S.~D. {Pinches}, S.~E. {Sharapov}, and J.~{EFDA
  contributors}.
\newblock {Explanation of the JET n = 0 chirping mode}.
\newblock {\em Nuclear Fusion}, 46:888--+, October 2006.

\bibitem{ONeil_65}
T.~{O'Neil}.
\newblock {Collisionless Damping of Nonlinear Plasma Oscillations}.
\newblock {\em Physics of Fluids}, 8:2255--2262, December 1965.

\bibitem{Fu_92}
G.~Y. {Fu} and C.~Z. {Cheng}.
\newblock {Excitation of high-n toroidicity-induced shear Alfv{\'e}n eigenmodes
  by energetic particles and fusion alpha particles in tokamaks}.
\newblock {\em Physics of Fluids B}, 4:3722--3734, November 1992.

\bibitem{Zonca_95}
F.~{Zonca}, F.~{Romanelli}, G.~{Vlad}, and C.~{Kar}.
\newblock {Nonlinear Saturation of Toroidal Alfv{\'e}n Eigenmodes}.
\newblock {\em Physical Review Letters}, 74:698--701, January 1995.

\bibitem{Odblom_02}
A.~{{\"O}dblom}, B.~N. {Breizman}, S.~E. {Sharapov}, T.~C. {Hender}, and V.~P.
  {Pastukhov}.
\newblock {Nonlinear magnetohydrodynamical effects in precessional fishbone
  oscillations}.
\newblock {\em Physics of Plasmas}, 9:155--166, January 2002.

\bibitem{BerkBreizman_90}
H.~L. {Berk} and B.~N. {Breizman}.
\newblock {Saturation of a single mode driven by an energetic injected beam.
  III. Alfv{\'e}n wave problem}.
\newblock {\em Physics of Fluids B}, 2:2246--2252, September 1990.

\bibitem{BerkBreizman_99}
H.~L. {Berk}, B.~N. {Breizman}, J.~{Candy}, M.~{Pekker}, and N.~V.
  {Petviashvili}.
\newblock {Spontaneous hole-clump pair creation}.
\newblock {\em Physics of Plasmas}, 6:3102--3113, August 1999.

\bibitem{Pinches_06}
S.D. Pinches, V.G. Kiptily, S.E. Sharapov, D.S. Darrow, L.-G. Eriksson, H.-U.
  Fahrbach, M.~Garcia-Munoz, M.~Reich, E.~Strumberger, A.~Werner, the ASDEX
  Upgrade~Team, and JET-EFDA Contributors.
\newblock Observation and modelling of fast ion loss in jet and asdex upgrade.
\newblock {\em Nuclear Fusion}, 46(10):S904--S910, 2006.

\bibitem{Hsu_92}
C.~T. {Hsu} and D.~J. {Sigmar}.
\newblock {Alpha-particle losses from toroidicity-induced Alfv{\'e}n
  eigenmodes. Part I: Phase-space topology of energetic particle orbits in
  tokamak plasma}.
\newblock {\em Physics of Fluids B}, 4:1492--1505, June 1992.

\bibitem{Sharapov_00}
S.~E. {Sharapov}, B.~{Alper}, D.~{Borba}, L.-G. {Eriksson}, A.~{Fasoli}, R.~D.
  {Gill}, A.~{Gondhalekar}, C.~{Gormezano}, R.~F. {Heeter}, G.~T.~A.
  {Huysmans}, J.~{Jacquinot}, A.~A. {Korotkov}, P.~{Lamalle}, M.~J.
  {Mantsinen}, D.~C. {McDonald}, F.~G. {Rimini}, D.~F.~H. {Start}, D.~{Testa},
  P.~R. {Thomas}, and {JET Team}.
\newblock {Energetic particle physics in JET}.
\newblock {\em Nuclear Fusion}, 40:1363--1381, July 2000.

\bibitem{Sigmar_92}
D.~J. {Sigmar}, C.~T. {Hsu}, R.~{White}, and C.~Z. {Cheng}.
\newblock {Alpha-particle losses from toroidicity-induced Alfv{\'e}n
  eigenmodes. Part II: Monte Carlo simulations and anomalous alpha-loss
  processes}.
\newblock {\em Physics of Fluids B}, 4:1506--1516, June 1992.

\bibitem{Candy_97}
J.~{Candy}, D.~{Borba}, H.~L. {Berk}, G.~T.~A. {Huysmans}, and W.~{Kerner}.
\newblock {Nonlinear interaction of fast particles with Alfv{\'e}n waves in
  toroidal plasmas}.
\newblock {\em Physics of Plasmas}, 4:2597--2611, July 1997.

\bibitem{Berk_95}
H.~L. {Berk}, B.~N. {Breizman}, and M.~{Pekker}.
\newblock {Numerical simulation of bump-on-tail instability with source and
  sink}.
\newblock {\em Physics of Plasmas}, 2:3007--3016, August 1995.

\bibitem{Zonca_05}
F.~Zonca, S.~Briguglio, L.~Chen, G.~Fogaccia, and G.~Vlad.
\newblock Transition from weak to strong energetic ion transport in burning
  plasmas.
\newblock {\em Nuclear Fusion}, 45(6):477--484, 2005.

\bibitem{Zonca_07b}
F.~Zonca, P.~Buratti, A.~Cardinali, L.~Chen, J.~Q. Dong, Y.~X. Long, A.~V.
  Milovanov, F.~Romanelli, P.~Smeulders, L.~Wang, Z.~T. Wang, C.~Castaldo,
  R.~Cesario, E.~Giovannozzi, M.~Marinucci, and V.~Pericoli Ridolfini.
\newblock Electron fishbones: Theory and experimental evidence, 2007.

\bibitem{Mazzucato_98}
E.~{Mazzucato}.
\newblock {Microwave reflectometry for magnetically confined plasmas}.
\newblock {\em Review of Scientific Instruments}, 69:2201--2217, June 1998.

\bibitem{Basiuk_03}
V.~{Basiuk}, J.~F. {Artaud}, F.~{Imbeaux}, X.~{Litaudon}, A.~{B{\'e}coulet},
  L.-G. {Eriksson}, G.~T. {Hoang}, G.~{Huysmans}, D.~{Mazon}, D.~{Moreau}, and
  Y.~{Peysson}.
\newblock {Simulations of steady-state scenarios for Tore Supra using the
  CRONOS code}.
\newblock {\em Nuclear Fusion}, 43:822--830, September 2003.

\bibitem{Eriksson_93}
L.-G. {Eriksson}, T.~{Hellsten}, and U.~{Willen}.
\newblock {Comparison of time dependent simulations with experiments in ion
  cyclotron heated plasmas}.
\newblock {\em Nuclear Fusion}, 33:1037--1048, July 1993.

\bibitem{Troyon_84}
F.~{Troyon}, R.~{Gruber}, H.~{Saurenmann}, S.~{Semenzato}, and S.~{Succi}.
\newblock {MHD-Limits to Plasma Confinement}.
\newblock {\em Plasma Physics and Controlled Fusion}, 26:209--215, January
  1984.

\bibitem{Edery_92}
D.~{Edery}, X.~{Garbet}, J.-P. {Roubin}, and A.~{Samain}.
\newblock {Variational formalism for kinetic-MHD instabilities in tokamaks }.
\newblock {\em Plasma Physics and Controlled Fusion}, 34:1089--1112, 1992.

\bibitem{Berk_88}
H.~L. {Berk} and D.~{Pfirsch}.
\newblock {Relation of wave energy and momentum with the plasma dispersion
  relation in an inhomogeneous plasma}.
\newblock {\em Physics of Fluids}, 31:1532--1543, June 1988.

\bibitem{Brizard_92}
A.~{Brizard}.
\newblock {Hermitian structure for linearized ideal MHD equations with
  equilibrium flows}.
\newblock {\em Physics Letters A}, 168:357--362, September 1992.

\bibitem{Brizard_94}
A.~{Brizard}.
\newblock {On the relation between pseudo-Hermiticity and dissipation}.
\newblock {\em Physics Letters A}, 187:382--390, May 1994.

\bibitem{LashmoreDavies_07}
C.~N. {Lashmore-Davies}.
\newblock {Two-stream instability, wave energy, and the energy principle}.
\newblock {\em Physics of Plasmas}, 14(9):092101--+, September 2007.

\bibitem{Hasegawa_68}
A.~{Hasegawa}.
\newblock {Theory of Longitudinal Plasma Instabilities}.
\newblock {\em Physical Review}, 169:204--214, May 1968.

\bibitem{LashmoreDavies_05}
C.~N. {Lashmore-Davies}.
\newblock {Negative energy waves}.
\newblock {\em Journal of Plasma Physics}, 71:101--109, April 2005.

\bibitem{Tsai_93}
S.-T. {Tsai} and L.~{Chen}.
\newblock {Theory of kinetic ballooning modes excited by energetic particles in
  tokamaks}.
\newblock {\em Physics of Fluids B}, 5:3284--3290, 1993.

\bibitem{Breizman_05}
B.~N. {Breizman}, M.~S. {Pekker}, and S.~E. {Sharapov}.
\newblock {Plasma pressure effect on Alfv{\'e}n cascade eigenmodes}.
\newblock {\em Physics of Plasmas}, 12(11):112506--+, November 2005.

\bibitem{Zonca_99}
F.~{Zonca}, L.~{Chen}, J.~Q. {Dong}, and R.~A. {Santoro}.
\newblock {Existence of ion temperature gradient driven shear Alfv{\'e}n
  instabilities in tokamaks}.
\newblock {\em Physics of Plasmas}, 6:1917--1924, 1999.

\bibitem{Catto_81}
P.~J. {Catto}, W.~M. {Tang}, and D.~E. {Baldwin}.
\newblock {Generalized gyrokinetics}.
\newblock {\em Plasma Physics}, 23:639--650, July 1981.

\bibitem{Frieman_82}
E.~A. {Frieman} and L.~{Chen}.
\newblock {Nonlinear gyrokinetic equations for low-frequency electromagnetic
  waves in general plasma equilibria}.
\newblock {\em Physics of Fluids}, 25:502--508, March 1982.

\bibitem{Antonsen_80}
T.~M. Antonsen and B.~Lane.
\newblock Kinetic equations for low frequency instabilities in inhomogeneous
  plasmas.
\newblock {\em Physics of Fluids}, 23:1205--1214, 1980.

\bibitem{Garbet_09}
X.~Garbet, G.~Dif-Pradalier, C.~Nguyen, P.~Angelino, Y.~Sarazin,
  V.~Grandgirard, P.~Ghendrih, and A.~Samain.
\newblock {A minimal collision operator for implementing neoclassical transport
  in gyrokinetic simulations}.
\newblock {\em to be published in Physics of Plasmas}, 2009.

\bibitem{Hahm_88}
T.~S. {Hahm}.
\newblock {Nonlinear gyrokinetic equations for tokamak microturbulence}.
\newblock {\em Physics of Fluids}, 31:2670--2673, September 1988.

\bibitem{Chen_91}
L.~{Chen} and A.~{Hasegawa}.
\newblock {Kinetic theory of geomagnetic pulsations. I - Internal excitations
  by energetic particles}.
\newblock {\em Journal of Geophysical Research}, 96:1503--1512, 1991.

\bibitem{Chu_92}
M.~S. {Chu}, J.~M. {Greene}, L.~L. {Lao}, A.~D. {Turnbull}, and M.~S. {Chance}.
\newblock {A numerical study of the high-n shear Alfv{\'e}n spectrum gap and
  the high-n gap mode}.
\newblock {\em Physics of Fluids B}, 4:3713--3721, 1992.

\bibitem{Gorelenkov_07}
N.~N. {Gorelenkov}, H.~L. {Berk}, E.~{Fredrickson}, S.~E. {Sharapov}, and {Jet
  Efda Contributors}.
\newblock {Predictions and observations of low-shear beta-induced shear
  Alfv{\'e}n acoustic eigenmodes in toroidal plasmas}.
\newblock {\em Physics Letters A}, 370:70--77, 2007.

\bibitem{Huysmans_95}
G.~T.~A. {Huysmans}, W.~{Kerner}, D.~{Borba}, H.~A. {Holties}, and J.~P.
  {Goedbloed}.
\newblock {Modeling the excitation of global Alfv{\'e}n modes by an external
  antenna in the Joint European Torus (JET)}.
\newblock {\em Physics of Plasmas}, 2:1605--1613, 1995.

\bibitem{Udintsev_06}
V.~S. {Udintsev}, M.~{Goniche}, G.~{Giruzzi}, G.~T.~A. {Huysmans},
  F.~{Imbeaux}, P.~{Maget}, X.~{Garbet}, R.~{Sabot}, J.~L. {S{\'e}gui},
  F.~{Turco}, T.~P. {Goodman}, D.~{Molina}, H.~{Weisen}, and {the Tore Supra
  team}.
\newblock {LETTER TO THE EDITOR: Studies of high frequency hot ion
  instabilities by means of correlation ECE on Tore Supra}.
\newblock {\em Plasma Physics and Controlled Fusion}, 48:L33--L44, 2006.

\bibitem{Sabot_06}
R.~{Sabot}, F.~{Clairet}, and {the Tore Supra Team}.
\newblock {Recent results on turbulence and MHD activity achieved by
  reflectometry}.
\newblock {\em Plasma Physics and Controlled Fusion}, 48:B421--B432, 2006.

\bibitem{Turnbull_93}
A.~D. {Turnbull}, E.~J. {Strait}, W.~W. {Heidbrink}, M.~S. {Chu}, H.~H.
  {Duong}, J.~M. {Greene}, L.~L. {Lao}, T.~S. {Taylor}, and S.~J. {Thompson}.
\newblock {Global Alfv{\'e}n modes: Theory and experiment}.
\newblock {\em Physics of Fluids B}, 5:2546--2553, 1993.

\bibitem{Heidbrink_93}
W.~W. {Heidbrink}, E.~J. {Strait}, M.~S. {Chu}, and A.~D. {Turnbull}.
\newblock {Observation of beta-induced Alfv{\'e}n eigenmodes in the DIII-D
  tokamak}.
\newblock {\em Physical Review Letters}, 71:855--858, 1993.

\bibitem{Buratti_05}
{Buratti} P.
\newblock {Observation of high-frequency waves during strong tearing mode
  activity in FTU plasmas without fast ions}.
\newblock {\em Nuclear Fusion}, 45:1446--1450, 2005.

\bibitem{Annibaldi_07}
S.~V. {Annibaldi}, F.~{Zonca}, and P.~{Buratti}.
\newblock {Excitation of beta-induced Alfv{\'e}n eigenmodes in the presence of
  a magnetic island}.
\newblock {\em Plasma Physics and Controlled Fusion}, 49:475--483, 2007.

\bibitem{Zimmerman_05}
0.~Zimmerman and {et al.}
\newblock Proc 32nd EPS Conf. on Plasma Physics and Controlled Fusion
  (Tarragona, Spain) P4-059, 2005.

\bibitem{Nabais_05}
F.~Nabais, D.~Borba, M.~Mantsinen, M.~F.~F. Nave, S.~E. Sharapov, and Joint
  European Torus-European Fusion Development Agreement (JET-EFDA) contributors.
\newblock Fishbones in joint european torus plasmas with high
  ion-cyclotron-resonance-heated fast ions energy content.
\newblock {\em Physics of Plasmas}, 12(10):102509, 2005.

\bibitem{Zonca_09}
F.~{Zonca}, L.~{Chen}, A.~{Botrugno}, P.~{Buratti}, A.~{Cardinali},
  R.~{Cesario}, V.~{Pericoli Ridolfini}, and {JET-EFDA contributors}.
\newblock {High-frequency fishbones at JET: theoretical interpretation of
  experimental observations}.
\newblock {\em Nuclear Fusion}, 49:085009, 2009.

\bibitem{Hallatschek_01}
K.~{Hallatschek} and D.~{Biskamp}.
\newblock {Transport Control by Coherent Zonal Flows in the Core/Edge
  Transitional Regime}.
\newblock {\em Physical Review Letters}, 86:1223--1226, 2001.

\bibitem{Miyato_04}
N.~{Miyato}, Y.~{Kishimoto}, and J.~{Li}.
\newblock {Global structure of zonal flow and electromagnetic ion temperature
  gradient driven turbulence in tokamak plasmas}.
\newblock {\em Physics of Plasmas}, 11:5557--5564, December 2004.

\bibitem{Scott_05}
B.~D. {Scott}.
\newblock {Energetics of the interaction between electromagnetic ExB turbulence
  and zonal flows}.
\newblock {\em New Journal of Physics}, 7:92--+, March 2005.

\bibitem{Itoh_05}
K.~{Itoh}, K.~{Hallatschek}, and S.-I. {Itoh}.
\newblock {Excitation of geodesic acoustic mode in toroidal plasmas}.
\newblock {\em Plasma Physics and Controlled Fusion}, 47:451--458, March 2005.

\bibitem{Naulin_05}
V.~{Naulin}, A.~{Kendl}, O.~E. {Garcia}, A.~H. {Nielsen}, and J.~J.
  {Rasmussen}.
\newblock {Shear flow generation and energetics in electromagnetic turbulence}.
\newblock {\em Physics of Plasmas}, 12(5):052515--+, May 2005.

\bibitem{Sugama_06}
H.~{Sugama} and T.-H. {Watanabe}.
\newblock {Collisionless damping of zonal flows in helical systems}.
\newblock {\em Physics of Plasmas}, 13(1):012501--+, January 2006.

\bibitem{Angelino_06}
P.~{Angelino}, A.~{Bottino}, R.~{Hatzky}, S.~{Jolliet}, O.~{Sauter}, T.~M.
  {Tran}, and L.~{Villard}.
\newblock {Effects of plasma current on nonlinear interactions of ITG
  turbulence, zonal flows and geodesic acoustic modes}.
\newblock {\em Plasma Physics and Controlled Fusion}, 48:557--571, May 2006.

\bibitem{McKee_03}
G.~R. {McKee}, R.~J. {Fonck}, M.~{Jakubowski}, K.~H. {Burrell},
  K.~{Hallatschek}, R.~A. {Moyer}, and D.~L. {Rudakov}.
\newblock {Experimental characterization of coherent, radially-sheared zonal
  flows in the DIII-D tokamak}.
\newblock {\em Physics of Plasmas}, 10(1712), 2003.

\bibitem{Hamada_05}
Y.~{Hamada}, A.~{Nishizawa}, T.~{Ido}, T.~{Watari}, M.~{Kojima}, Y.~{Kawasumi},
  K.~{Narihara}, K.~{Toi}, and {JIPPT-IIU Group}.
\newblock {Zonal flows in the geodesic acoustic mode frequency range in the
  JIPP T-IIU tokamak plasmas}.
\newblock {\em Nuclear Fusion}, 45:81--88, February 2005.

\bibitem{Ido_06}
T.~{Ido}, Y.~{Miura}, K.~{Kamiya}, Y.~{Hamada}, K.~{Hoshino}, A.~{Fujisawa},
  K.~{Itoh}, S.-I. {Itoh}, A.~{Nishizawa}, H.~{Ogawa}, Y.~{Kusama}, and {JFT-2M
  group}.
\newblock {Geodesic acoustic-mode in JFT-2M tokamak plasmas}.
\newblock {\em Plasma Physics and Controlled Fusion}, 48:41--+, April 2006.

\bibitem{Melnikov_06}
A.~V.~Melnikov et~al.
\newblock {\em Plasma Physics and Controlled Fusion}, 48:S87--S110, 2006.

\bibitem{Conway_05}
G.~D. {Conway}, B.~{Scott}, J.~{Schirmer}, M.~{Reich}, A.~{Kendl}, and {the
  ASDEX Upgrade Team}.
\newblock {Direct measurement of zonal flows and geodesic acoustic mode
  oscillations in ASDEX Upgrade using Doppler reflectometry}.
\newblock {\em Plasma Physics and Controlled Fusion}, 47:1165--1185, August
  2005.

\bibitem{Shats_06}
M.~G. {Shats}, H.~{Xia}, and M.~{Yokoyama}.
\newblock {Mean E {$\times$} B flows and GAM-like oscillations in the H-1
  heliac}.
\newblock {\em Plasma Physics and Controlled Fusion}, 48:17--+, April 2006.

\bibitem{Boswell_06}
C.~J. {Boswell}, H.~L. {Berk}, D.~N. {Borba}, T.~{Johnson}, S.~D. {Pinches},
  and S.~E. {Sharapov}.
\newblock {Observation and explanation of the JET n=0 chirping mode}.
\newblock {\em Physics Letters A}, 358:154--158, October 2006.

\bibitem{Gorelenkov_09}
N.~N. {Gorelenkov}, M.~A. {van Zeeland}, H.~L. {Berk}, N.~A. {Crocker},
  D.~{Darrow}, E.~{Fredrickson}, G.-Y. {Fu}, W.~W. {Heidbrink}, J.~{Menard},
  and R.~{Nazikian}.
\newblock {Beta-induced Alfv{\'e}n-acoustic eigenmodes in National Spherical
  Torus Experiment and DIII-D driven by beam ions}.
\newblock {\em Physics of Plasmas}, 16(5):056107--+, May 2009.

\bibitem{Nazikian_06}
R.~{Nazikian}, H.~L. {Berk}, R.~V. {Budny}, K.~H. {Burrell}, E.~J. {Doyle},
  R.~J. {Fonck}, N.~N. {Gorelenkov}, C.~{Holcomb}, G.~J. {Kramer}, R.~J.
  {Jayakumar}, R.~J. {La Haye}, G.~R. {McKee}, M.~A. {Makowski}, W.~A.
  {Peebles}, T.~L. {Rhodes}, W.~M. {Solomon}, E.~J. {Strait}, M.~A.
  {Vanzeeland}, and L.~{Zeng}.
\newblock {Multitude of Core-Localized Shear Alfv{\'e}n Waves in a
  High-Temperature Fusion Plasma}.
\newblock {\em Physical Review Letters}, 96(10):105006--+, March 2006.

\bibitem{Nguyen_09}
C.~{Nguyen}, X.~{Garbet}, R.~{Sabot}, L.-G. {Eriksson}, M.~{Goniche},
  P.~{Maget}, V.~{Basiuk}, J.~{Decker}, D.~{Elb\`eze}, G.~T.~A. {Huysmans},
  A.~Macor, J.-L. S\'egui, and M.~Schneider.
\newblock {Excitation of Beta Alfv\'en Eigenmodes in Tore-Supra}.
\newblock {\em Plasma Physics and Controlled Fusion}, 51:095002, 2009.

\bibitem{Zonca_08}
F.~{Zonca} and L.~{Chen}.
\newblock {Radial structures and nonlinear excitation of geodesic acoustic
  modes}.
\newblock {\em Europhysics Letters}, 83:35001--+, August 2008.

\bibitem{Smolyakov_08a}
A.~I. {Smolyakov}, C.~{Nguyen}, and X.~{Garbet}.
\newblock {Kinetic theory of electromagnetic geodesic acoustic modes}.
\newblock {\em Plasma Physics and Controlled Fusion}, 50:115008, 2008.

\bibitem{Fitzpatrick_94}
R.~{Fitzpatrick}.
\newblock {Stability of coupled tearing and twisting modes in tokamaks}.
\newblock {\em Physics Plasmas}, 1:3308--3336, October 1994.

\bibitem{Pegoraro_86}
F.~{Pegoraro} and T.~J. {Schep}.
\newblock {Theory of resistive modes in the ballooning representation}.
\newblock {\em Plasma Physics and Controlled Fusion}, 28:647--667, April 1986.

\bibitem{Smolyakov_09}
A.~I. Smolyakov, C.~Nguyen, and X.~Garbet.
\newblock {Electromagnetic effects on geodesic acoustic and beta-induced
  Alfv\'{e}n; eigenmodes}.
\newblock {\em Submitted to Nuclear Fusion}, 2009.

\bibitem{Sabot_09}
R.~{Sabot}, A.~{Macor}, C.~{Nguyen}, J.~{Decker}, D.~{Elb\`eze}, L.~G.
  Eriksson, X.~Garbet, M.~{Goniche}, G.~T.~A. {Huysmans}, P.~{Maget}, and J.~L.
  {S\'egui}.
\newblock {Observation of acoustic and subacoustic fast particles driven modes
  in Tore-Supra}.
\newblock {\em Nuclear Fusion}, 49(8):085033--+, August 2009.

\bibitem{Nazikian_08}
R.~{Nazikian}, G.~Y. {Fu}, M.~E. {Austin}, H.~L. {Berk}, R.~V. {Budny}, N.~N.
  {Gorelenkov}, W.~W. {Heidbrink}, C.~T. {Holcomb}, G.~J. {Kramer}, G.~R.
  {McKee}, M.~A. {Makowski}, W.~M. {Solomon}, M.~{Shafer}, E.~J. {Strait}, and
  M.~A.~V. {Zeeland}.
\newblock {Intense Geodesic Acousticlike Modes Driven by Suprathermal Ions in a
  Tokamak Plasma}.
\newblock {\em Physical Review Letters}, 101(18):185001--+, October 2008.

\bibitem{Kerner_94}
W.~{Kerner}, D.~{Borba}, G.~T.~A. {Huysmans}, F.~{Porcelli}, S.~{Poedts}, J.~P.
  {Goedbloed}, and R.~{Betti}.
\newblock {Stability of global Alfven waves (TAE, EAE) in JET tritium
  discharges}.
\newblock {\em Plasma Physics and Controlled Fusion}, 36:911--923, May 1994.

\bibitem{Bussac_75}
M.~N. {Bussac}, R.~{Pellat}, D.~{Edery}, and J.~L. {Soule}.
\newblock {Internal kink modes in toroidal plasmas with circular cross
  sections}.
\newblock {\em Physical Review Letters}, 35:1638--1641, December 1975.

\bibitem{Matpack}
http://www.matpack.de/.

\bibitem{Davies_86}
B.~{Davies}.
\newblock {Locating the Zeros of an Analytic Function}.
\newblock {\em Journal of Computational Physics}, 66:36--+, September 1986.

\bibitem{Lauber_09}
P.~Lauber, M.~Br\"udgam, S.~G\"unter, M.~Curran, M.~{Garc\'ia Munoz},
  V.~Igochine, M.~Maraschek, K.~Sassenberg, and {Asdex-Upgrade Team}.
\newblock Fast particle driven modes in asdex upgrade.
\newblock IAEA Technical Meeting on energetic particles, Kiev, 2009.

\bibitem{Vann_07}
R.~G.~L. {Vann}, H.~L. {Berk}, and A.~R. {Soto-Chavez}.
\newblock {Strongly Driven Frequency-Sweeping Events in Plasmas}.
\newblock {\em Physical Review Letters}, 99(2):025003--+, July 2007.

\bibitem{Fredrickson_06}
E.~Fredrickson, N.~N. Gorelenkov, and H.L. Berk.
\newblock Proceedings of the APS, 2006.

\bibitem{Breizman_09}
B.~N. Breizman.
\newblock Nonlinear travelling waves in energetic particle phase space.
\newblock IAEA Technical Meeting on energetic particles, Kiev, 2009.

\bibitem{Brizard_94b}
A.~{Brizard}.
\newblock {Eulerian action principles for linearized reduced dynamical
  equations}.
\newblock {\em Physics of Plasmas}, 1:2460--2472, August 1994.

\bibitem{Diamond_09}
P.~H. Diamond.
\newblock Conservation of potential vorticity.
\newblock Festival de Th\'eorie, Aix-en-Provence, 2009.

\bibitem{BerkBreizman_98}
H.~L. {Berk}, B.~N. {Breizman}, and N.~V. {Petviashvili}.
\newblock {Spontaneous hole-clump pair creation in weakly unstable plasmas
  [Physics Letters A 234 (1997) 213]}.
\newblock {\em Physics Letters A}, 238:408--408, February 1998.

\bibitem{BerkBreizman_92}
H.~L. {Berk}, B.~N. {Breizman}, and H.~{Ye}.
\newblock {Scenarios for the nonlinear evolution of alpha-particle-induced
  Alfv{\'e}n wave instability}.
\newblock {\em Physical Review Letters}, 68:3563--3566, June 1992.

\bibitem{Vann_02}
R.~G.~L. Vann.
\newblock {\em Characterisation of fully nonlinear Berk-Breizman
  phenomenology}.
\newblock PhD thesis, University of Warwick, 2003.

\bibitem{Calvi_09}
http://www-math.u-strasbg.fr/calvi/?lang=en.

\bibitem{Lesur_09}
M.~Lesur, Y.~Idomura, and X.~Garbet.
\newblock {Fully nonlinear features of the energetic beam-driven instability}.
\newblock {\em Physics of Plasmas}, 16:9, September 2009.

\bibitem{Connor_78}
J.~W. {Connor}, R.~J. {Hastie}, and J.~B. {Taylor}.
\newblock {Shear, periodicity, and plasma ballooning modes}.
\newblock {\em Physical Review Letters}, 40:396--399, February 1978.

\end{thebibliography}
\appendix





\chapter{Notions of Hamiltonian mechanics}
\label{appchapter_HamiltonianNotions}
We recall in this section some fundamentals of hamiltonian mechanics necessary
for our subsequent derivation of the gyrokinetic equation. Those are
largely  described in the review paper  by Brizard  et al.\cite{Brizard_07}.
\section{Hamilton's  equations in general coordinates}
The motion of a single charged particle in an electromagnetic field can be 
described using the differential one-form particle Lagrangian: 
\begin{equation}
\underline {\hat{ \Gamma}} ({\bf x}, {\bf p})  
= {\bf p}\cdot {\sf d}{\bf x} - \hat{H}{\sf d}t 
\end{equation}
where $({\bf x}, \mathbf{p})=({\bf x}, m{\bf v} +
e{\bf A})$ are the particle canonical coordinates in 6 dimensional
phase-space, $\hat{H}$ is the electromagnetic hamiltonian $\hat{H}
= |\mathbf{p}-eA|^2/2m +e\phi$, and {\sf d} denotes an exterior
derivative. (Contrary to Brizard  et al.\cite{Brizard_07}, we simply use the
usual 6 dimensional phase-space, more practical for our subsequent
derivations.) More generally, we can use an arbitrary 6 dimensional phase-space
system of indepedent coordinates ${\bf Z}=(Z^a)_{a=1...6}$ associated with a 
Lagrangian of the form 
$ \underline{ \Gamma} ({\bf Z}, t)   
= {\bf \Gamma}({\bf Z}, t)\cdot{\sf d} {\bf Z}  - H({\bf Z}, t){\sf
  d}t$ verifying  $\underline{\Gamma} ({\bf Z}, t) 
=\underline{\hat{ \Gamma}} ({\bf x}, {\bf p})+
{\sf d}S$, with S a scalar field depending on phase space
coordinates and time. 

{\bf Hamilton's principle} asserts that the dynamically allowed paths in a
time interval $[t_1, t_2]$ with given initial and final conditions 
are stationary points of the  action integral
$\mathcal{A} = \int_{t_1}^{t_2} \underline\Gamma$ for the class of phase space
paths, ie $\delta \mathcal{A} = 0$, where :
\begin{eqnarray}
\begin{split}
 \delta \mathcal{A} = \delta \int_{t_1}^{t_2} 
\left( {\bf \Gamma}\cdot \frac{d{\bf Z}}{dt}  - H \right)dt 
=\int_{t_1}^{t_2}\left( \frac{\partial\Gamma_b}{\partial Z^a}\cdot
\frac{d Z^b}{dt}\delta Z^a + \Gamma_a\frac{d \delta Z^a}{d t}  
- \frac{\partial H }{\partial Z^a}
\delta Z^a\right) dt
\end{split}
\label{eq_HamiltonDerivation1}
\end{eqnarray}  
Integrating Eq.~\ref{eq_HamiltonDerivation1} by parts and using the
invariance of the initial and final conditions 
($\delta Z^a = 0$ at the end points), 
we get:
\begin{equation}
\delta \mathcal{A}  =
\int_{t_1}^{t_2}\left(\omega_{ab}\frac{dZ^b}{dt}-\frac{\partial
  \Gamma_a}{\partial t}-\frac{\partial H}{\partial Z^a}\right)\delta Z^a
\end{equation} 
where the   $\omega_{ab}$'s are defined as:  
\begin{equation}
\omega_{ab} = \frac{\partial \Gamma_b}{\partial Z^a} 
-  \frac{\partial \Gamma_a}{\partial Z^b}\,\,, \quad a, b = 1...6
\end{equation}
and correspond to the matrix coefficients of the
two-form $ {\bf \omega } = {\sf d} \left( {\bf \Gamma}({\bf Z}, t)\cdot{\sf
  d} {\bf Z}\right) $. 
And finally, $\delta \mathcal{A} = 0$ returns \cite{Littlejohn_83}:
\begin{equation} 
\omega_{ab}\frac{dZ^b}{dt}=\frac{\partial H}{\partial Z^a}
+\frac{\partial\Gamma_a}{\partial t}
\label{eq_eulerlagrange}
\end{equation}
For regular Lagrangian systems, ${\bf \omega}$ is invertible with
inverse $J\equiv\omega^{-1}$, and equations \ref{eq_eulerlagrange}
can be rewritten in the following form, known as Hamilton's equations:
\begin{equation}
\frac{dZ^a}{dt}=J^{ab}\left(\frac{\partial H}{\partial
  Z^b}+\frac{\partial\Gamma_b}{\partial t}\right)=[Z^a, H]+[Z^a,
  Z^b]\frac{\partial \Gamma_b}{\partial t}
\end{equation}
where Poisson brackets are defined by the relations:
\begin{eqnarray}
[f, g] &=& \frac{\partial{f}}{\partial Z^a}[Z^a,
  Z^b]\frac{\partial{g}}{\partial Z^b}\\
\ [Z^a, Z^b] &=& J^{ab}({\bf Z}) 
\end{eqnarray}
\section{Hamiltonian formulation of the Vlasov equation}
Based on an averaging of the individual particles behaviors, 
the Vlasov equation suggests a description of the collective behavior
of a dense set of charged particles, which states that the full time 
derivative of the particles phase-space distribution function $F$
cancels. Its Hamiltonian formulation reads:
\begin{equation}
\frac{dF}{dt} =\frac{\partial{F}}{\partial{t}}
+\frac{\partial{F}}{\partial{Z^a}}\frac{d
  Z^a}{dt}=\frac{\partial{F}}{\partial{t}} - [H, F]- \frac{\partial
  \bf \Gamma}{\partial t}\cdot[{\bf Z},F]
\end{equation}  
\section{Coordinate Transformations}
\label{appsection_CoordinateTransformations}
Coordinate transformations are appropriate means to simplify the six 
dimensional Vlasov equation. Gyrokinetic theory in particular, 
is based on their use.

Let's consider a time-dependent coordinate transformation  ${\bf  Z} $ to 
$\bar{\bf Z} = \bar{\bf Z}({\bf Z}, t) $.  Under such a coordinate
transformation, both the expressions of the phase-space distribution
function and of the particle Lagrangian are modified. Of course, the
new distribution function $\bar{F}$ has to verify $\bar{F}(\bar{Z}, t) 
= F(Z,t)$ to describe the same population of particles. As for the particle
Lagrangian, one more degree of freedom is offered by the variational
form of Hamilton's principle, and  the physics of the charged
particle motion is simply conserved if the new
Lagrangian verifies a gauge transformation \cite{Littlejohn_83}, ie: 
$\underline{\bar{\Gamma} }(\bar{\bf Z}, t)=\underline{\Gamma} ({\bf Z}, t) +
{\sf d}S$ with S  a scalar field depending on phase space
coordinates and time. 
This latter condition relates the hamiltonians and the so-called
symplectic components of the Lagrangian ${\bf \Gamma} = (\Gamma_a)_{a=1...6}$
of the two coordinate systems, and it shows the conservation of Poisson
brackets in a coordinate transformation.
Indeed,  $   \bar{\bf \Gamma}
\cdot{\sf d} {\bf \bar Z}  - {\bar H}{\sf d}t 
= {\bf \Gamma}\cdot{\sf d} {\bf Z}  - H{\sf  d}t +
{\sf d}G$ implies: 
\begin{eqnarray}
\bar{\Gamma}_a(\bar{\bf Z}, t) &=& \Gamma_b({\bf
  Z}, t)\frac{\partial Z_b}{\partial \bar{Z}_a} 
+ \frac{\partial G}{\partial \bar{Z}_a }\\
\bar{H}(\bar{\bf Z}, t) &=& H({\bf Z}, t) 
+ \Gamma_a\frac{\partial Z_a}{\partial t}
+\frac{\partial G}{\partial t} 
\end{eqnarray}
and hence:
\begin{equation}
\bar{\omega}_{ab} = \frac{\partial \bar{\Gamma}_b}{\partial \bar{Z}^a} 
-  \frac{\partial \bar{\Gamma}_a}{\partial \bar{Z}^b}
= \frac{\partial Z_c}{\partial \bar{Z}^a}
\left(\frac{\partial \Gamma_d}{\partial Z^c} 
-  \frac{\partial \Gamma_c}{\partial Z^d}\right)
\frac{\partial Z_d}{\partial \bar{Z}^b} 
= \frac{\partial Z_c}{\partial \bar{Z}^a}\omega_{cd}
\frac{\partial Z_d}{\partial \bar{Z}^b}
\end{equation}
which can  be inverted to yield:
\begin{equation}
\bar[ f, g\bar] = \frac{\partial{f}}{\partial \bar Z^a}
\bar{J}^{ab}\frac{\partial{g}}{\partial \bar Z^b}
= \frac{\partial{f}}{\partial \bar Z^a}
\frac{\partial \bar{Z}^a}{\partial Z_c}J^{cd}
\frac{\partial \bar{Z}^b}{\partial Z_d}\frac{\partial{g}}{\partial \bar Z^b}
= [ f, g], {\,\,\,\rm for\,\, any} f\, {\rm and } \,g
\end{equation}

For perturbative treatments ot the Vlasov equation, a particular class 
of coordinate transformations is used, the class of Lie near-identity
transformations,  defined as: 
\begin{equation}
\mathcal{T}_{\delta}: {\bf Z} \longrightarrow \bar{\bf Z} 
= \exp \left( \sum_1^{\infty}  {\bf G}_n\cdot {\sf
  d}\right){\bf Z}
\end{equation}
where  the ${\bf G}_n ({\bf Z}, t)$'s, or generator vector fields  of
the transformation, verify ${\bf G}_n = O(\delta^n)$, and $\delta
<< 1$ is a dimensionless ordering parameter. Under such a transformation, 
the conservation of
the {\it values} taken by a given scalar function f, ie : $
\bar{f}(\bar{\bf Z}, t) = f({\bf Z}, t)$,
or by a given 1-form $\underline\Gamma$, ie $\underline{\bar\Gamma} 
(\bar{\bf Z}, t)= \underline\Gamma ({\bf Z}, t)$, fully defines the new functions 
$\bar{f}$ and $\bar{\underline\Gamma}$ which can be expressed using a so-called
{ \it push-forward} operator ${\sf T}^{-1}_{\delta}$ (the inverse operator
of the so-called {\it pull-back} operator ${\sf T}_{\delta}$): ${\sf
  T}^{-1}_{\delta}$: $f \longrightarrow \bar{f} $, ${\sf
  T}^{-1}_{\delta}$: $\underline\Gamma \longrightarrow
\bar{\underline\Gamma} $. 
For near-identity transformations, Lie-transform perturbation theory
shows that \cite{Brizard_07}:
\begin{equation}
{\sf T^{\pm 1}_{\delta}} = \exp\left(\pm\sum_1^\infty\mathcal{L}_n\right)
\label{eq_pushforward}
\end{equation} 
where $\mathcal{L}_n$ is a Lie derivative:
\begin{eqnarray}  \mathcal {L}_n f &\equiv& {\bf G}_n
\cdot {\sf d} f = G^a_n \frac{\partial f}{\partial Z^a} 
{\quad\rm for\,\, a\,\, scalar\,\, function\,}  f\\ 
\mathcal {L}_n \underline\Gamma &\equiv&{\bf G}_n
\cdot {\sf d} \underline\Gamma + {\sf d} ({\bf G}_n \cdot \underline \Gamma)
= \left[G^a_n(\frac{\partial \underline\Gamma_b}{\partial Z_a} -
  \frac{\partial\underline\Gamma_a}{\partial Z_b}) +
  \frac{\partial \,G_n^a\underline\Gamma_a}{\partial Z_b}\right]{\sf d} Z^b 
{\quad \rm for\,\, a\,\, {\sf 1-}form\,\,\,}  \underline\Gamma \quad
\label{eq_pushforward_1form} \end{eqnarray} 
Note that in the latter equation Eq.~\ref{eq_pushforward_1form}, we
simply wrote : $ \underline \Gamma = \underline\Gamma_a{\sf d} Z^a$
(without separating  the symplectic components ${\bf \Gamma}\cdot{\sf d}
{\bf Z}$ and the hamiltonian part $-H {\sf d} t$). Hence,
$a$ and $b$ should be taken from 1 to 7 to describe the 7-dimensional space:
$(Z_{a=1...6}, Z_7 = t)$. This equation makes sense if we use: 
$G_n^7 = 0$, $\underline \Gamma_7 = -H,
\underline\Gamma_{a=1...6} = \Gamma_{a=1...6}$. \\
Identity \ref{eq_pushforward} simply results from the development of
the expressions  $\bar{f}(\bar{\bf Z}, t) = f({\bf Z}, t)$ and 
$\underline{\bar\Gamma} (\bar{\bf Z}, t)= \underline\Gamma ({\bf Z},
t)$. To the first order in $\delta$ (of interest to us in this
paper), it is easily shown: 
\begin{eqnarray}
\bar{f}(\bar{\bf Z}, t) &=& f(\bar{\bf Z} -  {\bf
  G}_1, t) = f(\bar{\bf Z}) - {\bf G}_1 \cdot \nabla f 
= f(\bar{\bf Z}) - {\bf G}_1 \cdot {\sf d} f 
\label{eq_TransformedDistribution}\\
\underline{\bar\Gamma} (\bar{\bf Z}, t) &=& \underline\Gamma
  (\bar{\bf Z}-{\bf G}_1, t){\sf d}  (\bar{\bf Z}-{\bf
  G}_1) = \underline \Gamma (\bar{\bf Z}) 
- \left( G^a_1\frac{\partial \underline\Gamma_b}{\partial Z_a} 
+\underline\Gamma_a \frac{  \partial G_1^a}{\partial Z_b}\right){\sf d} Z^b 
\end{eqnarray}   

Let's now see the benefit of such a  transformation for perturbative
analysis. Assume  a first order perturbation of an initial state
$\underline \Gamma = \underline\Gamma_0 +
\underline\Gamma_1$ ($\underline\Gamma_1 = O(\delta)$). 
Such a perturbation modifies the charged particle
equations of motion, and may have an impact on both the Poisson
bracket structure given by $\omega =\omega_0 + \omega_1$, 
and the Hamiltonian $H = H_0 + H_1$. Hence, it may be
desirable to use a near-identity coordinate transformation to simplify 
the new equations of motion. As explained before, the transformed 
Lagrangian has to verify  
$\underline{\bar{\Gamma} }(\bar{\bf Z}, t)
=\underline{\Gamma} ({\bf Z}, t) + {\sf d}S$, 
or $\bar{\Gamma} = {\sf T}_\delta^{-1} \Gamma + {\sf d}S$. 
If we apply  identity Eq.~\ref{eq_pushforward} and separate the hamiltonian 
part of the Lagrangian (so that  {\bf Z} now lives in the 6 dimensional
phase-space), the first order order of this equality reads:
\begin{equation}
 \bar{\bf \Gamma_1}\cdot{\sf d} {\bf Z} - \bar{H}_1 {\sf d}t 
= {\bf\Gamma}_1\cdot{\sf d} {\bf Z}  - H_1 {\sf d}t  
- {\bf G}_1\cdot \omega_0 +  \left({\bf G}_1 \cdot {\sf d} H_0 + {\bf
  G}_1\cdot \frac{\partial \Gamma_0}{\partial t }\right){\sf d}t +
{\sf d} S_1 
\end{equation} where we included in $S_1$ both the first order of $S$
and the field ${\bf G}_1 \cdot \underline \Gamma$ resulting from the
application of the Lie derivative Eq.~\ref{eq_pushforward_1form}. 
The components of this equation may be separated to give:
\begin{equation}
 G^a_1 = [S_1, Z^a]_0 + (\Gamma_{1 b} - \bar{\Gamma}_{1 b}) J_0^{ba}\\
\label{eq_TransformedSymplectic}
\end{equation}
and:
\begin{eqnarray}
\nonumber\bar{H}_1 &=& H_1 -
{\bf G}_1 \cdot {\sf d} H_0 - {\bf
  G}_1\cdot \frac{\partial \Gamma_0}{\partial t }
- \frac{\partial S_1}{\partial t}\\
\nonumber &=& H_1-[S_1, H_0] - [S_1, {\bf Z}]
\cdot \frac{\partial{\bf \Gamma}}{\partial t}
- (\Gamma_1^a - \bar\Gamma_1^a)\left([Z^a, H_0]+[Z^a, {\bf Z}]
\cdot \frac{\partial{\bf\Gamma}}{\partial t}\right)
-\frac{\partial S_1}{\partial t}\\
&=&H_1-({\bf \Gamma}_1 - \bar{\bf\Gamma}_1)\cdot \frac{d_0 \bf Z}{dt}
-\frac{d_0 \,S_1}{dt}
\label{eq_TransformedHamiltonian} 
\end{eqnarray}
where $d_0/dt$ is the full time derivative along the unperturbed
particle trajectory.  Then, two degrees of freedom clearly appear in 
the transformation: the choice of ${\bf G}_1$ and the choice of $S_1$,
which allow to fix both the new Poisson bracket structure derived from
$\bar{\Gamma_1}$ using Eq.~\ref{eq_TransformedSymplectic}, and the new  
Hamiltonian $\bar{H}_1$ using Eq.~\ref{eq_TransformedHamiltonian}. 

In modern Gyrokinetic theory, near-identity transformations are used
to simplify the Poisson bracket structure and to remove the fast
gyromotion of the particle from the new Hamiltonian. Hence, a new
particle trajectory or ``gyrocenter'' trajectory, which is independent 
from the fast gyromotion,  is determined, whereas the complexity of the 
fast gyromotion gets fully included in the coordinate transformation itself.  


\chapter{Charged particle motion} 
\label{appchapter_ChargedParticleMotion}
\section{Derivation of action-angle variables}
\label{appsection_ActionAngleVariables}
We derive in the following section the system of action-angle variables used 
in the thesis, starting from the guiding-center coordinates defined
in Ref.~\cite{Littlejohn_83}.
For this, we will make a large use of the Hamiltonian formulas recalled in 
Eq.~\ref{eq_LagrangianTransformation}, and use similar procedure and arguments
as can be found in Refs.~\cite{WhiteBook, Kaufman_72}.\\

We start from the the guiding-center Lagrangian given by Littlejohn
\cite{Littlejohn_83} presented in the thesis core
\ref{eq_GuidingCenterLagrangian}, 
\begin{equation}
\underline {\Gamma}_\text{gc} ({\bf Z}) = 
(e{\bf A}_\eq ({\bf X})+mv_{\|}{\bf b}_\eq)\cdot {\sf d}{\bf X}
+ \frac{m}{e}\mu{\sf d}\gamma
- \left(\frac{1}{2}mv^2_{\|} + \mu B_\eq({\bf X}) 
+ e\phi_\eq({\bf X})\right){\sf d}t
\end{equation}
To make use of the configuration todoidal and poloidal periodicities,
it is appropriate to write the fields using their covariant 
representation. Noticing from Eq.~\ref{eq_BDefinition} that 
${\bf B}_\eq 
=\nabla\Psi\times\nabla\varphi+\nabla\Phi\times\nabla\theta$, 
it directly comes a convenient choice for the vector potential
 ${\bf A}_\eq$
\begin{equation}
{\bf A}_\eq = \Psi\nabla\varphi + \Phi\nabla\theta .
\label{eq_Aeq}
\end{equation}
Next, we use Eq.~\ref{eq_BAxisymmetricDefinition} to write the magnetic 
field covariant representation
\begin{equation}
{\bf B}_\eq= B_{\Psi}\nabla\Psi 
+B_{\theta}\nabla\theta 
+I(\Psi)\nabla\varphi.
\end{equation}
The Lagrangian becomes
\begin{equation} 
\underline {\Gamma}_\text{gc} ({\bf Z}) 
=\left(\frac{mv_\|}{B_\eq}B_{\theta}+e\Phi\right){\sf d}\theta
+\left(e\Psi + \frac{mv_{\|}}{B_\eq}I\right){\sf d}\varphi
+\frac{m}{e}\mu{\sf d}\gamma
+\frac{mv_\|}{B_\eq}B_\Psi{\sf d}\Psi-H_{\rm gc} {\sf d} t
\end{equation}

In order to display a canonical system of variables, we need to 
remove one coordinate.
Noticing that a change of coordinate ${\bf X} \rightarrow {\bf X} + {\bf X'}$
leads to additional terms in the Lagrangian
$
\underline {\Gamma} \rightarrow 
\underline {\Gamma} 
+ {\bf A}_\eq\cdot{\sf d} {\bf X'}
+ (mv_{\|}/B_\eq) {\bf B}_\eq \cdot {\sf d}{\bf X'}
$ 
the $\Psi$-dependence is found to disappear with
\begin{equation}
\begin{array}{lll}
\text{- the second order in } \rho^ *\text{ change of coordinates: } & 
{\sf d}{\bf X'}  =  -\frac{B_{\Psi} }{B^2_\eq}{\bf B}_\eq {\sf d}\Psi\\
\text{- a choice of a Gauge for }{\bf A}_\eq \text{such that: }&
{\bf A}_\eq\cdot{\bf B}_\eq = 0
\end{array}
\end{equation}
Finally the Lagrangian is expressed in a system of canonical coordinates
$(\gamma,\theta,\varphi,P_{\gamma}, P_{\theta}, P_{\varphi})$ 
\begin{equation}
\underline{\Gamma} = P_\theta {\sf d} \theta + P_\varphi{\sf d}\varphi 
+ \frac{m}{e}\mu {\sf d}\gamma- H_\eq{\sf d} t, 
\end{equation}
with
\begin{eqnarray}
P_{\gamma}  &=& \frac{m}{e}\mu\\
P_{\theta}  &=& \frac{mv_\|}{B_\eq}B_{\theta}+e\Phi \\
P_{\varphi} &=& e\Psi + \frac{mv_{\|}}{B_\eq}I(\Psi).
\end{eqnarray}\\

Here, $P_\gamma$ is directly associated to the traditional adiabatic invariant
$\mu$,
and $P_\varphi, \dot{P}_{\varphi} = \partial_{\varphi}H_\eq$ is clearly an exact
motion invariant associated with axisymmetry (at equilibrium).
A priori, $P_{\theta}$ has no reason to be an invariant and we need additional
work to display a set of action-angle variables.
Noticing that $ H_\eq(\theta, P_{\gamma}, P_{\theta}, P_{\varphi})$, where
$H_\eq$ is an invariant at equilibrium, we can rewrite 
$P_\theta = P_\theta(\theta,H_\eq,P_\gamma,P_\varphi)$, and we
can  cancel the $\theta$-dependence using a $\theta$-averaging, 
at constant invariants, ie: along a particle trajectory.
For this, we define
\begin{equation}
J_2 = \oint \frac{d\theta}{2\pi} P_\theta(\theta,H_\eq,
P_\gamma,P_\varphi) \ 
\label{eq_J2Definition}
\end{equation}
which is an invariant by definition.\\

Let us now show that $J_2$ can be used as invariant in a set of action-angle
variables. 
The clean way to proceed is to use a coordinate transformation, as described
in Eq.~\ref{eq_LagrangianTransformation}. 
We make the transformation $(\gamma,\theta,\varphi,P_\gamma,P_\theta,P_\varphi)
\rightarrow (\alpha_1,\alpha_2,\alpha_3,J_1,J_2,J_3)$, using the following
transformation generating function
\begin{equation}
G (\gamma,\theta,\varphi,J_1,J_2,J_3) = \gamma J_1 + \varphi J_3
+\int_0^\theta d\theta' P_\theta (\theta',J_1,J_2,J_3).
\end{equation}
Such generating function implies the following requirements for the new 
coordinates
\begin{eqnarray}
P_\gamma  &=& \partial_\gamma G  = J_1;\quad  
P_\theta  = \partial_\theta G  = P_\theta;\quad
P_\varphi = \partial_\varphi G = J_3\\
\m{\alpha } &=&  \partial_{\bf J} G= (\gamma, 0, \varphi) 
+ \int_0^\theta  d\theta' \ 
\partial_{\bf J}P_\theta (\theta',J_1,J_2,J_3)
\label{eq_AngleTransformation}.
\end{eqnarray}
Hence, we obtain a set of canonical variables characterized by three motion 
invariants, $J_1$, $J_2$ and $J_3$.

We can be more explicit and derive the physical meaning of the associated 
angles using Eq.~\ref{eq_AngleTransformation}.
For this, let us define the bounce frequency,
\begin{equation}
\frac{1}{\Omega_b} = \oint \frac{d\theta}{2\pi}\frac{1}{\dot{\theta}}
= \oint \frac{d\theta}{2\pi}\frac{\partial P_\theta}{\partial H_\eq}|_{J_1, J_3}.
\end{equation}
The derivation of Eq.~\ref{eq_J2Definition} according to $J_2$ with the 
definition 
$\Omega_2 = \partial_{J_2|J_1, J_3}H_\eq$ directly shows that 
$\Omega_2 = \Omega_b$, 
and it follows from \ref{eq_AngleTransformation} that 
\begin{equation}
\alpha_2  = \Omega_b\int_0^\theta \frac{d\theta'}{\dot{\theta'}}\\
\alpha _3 = \varphi + \hat{\varphi} 
\end{equation}
where $\oint \hat{\varphi} = 0$
\begin{equation}
\Omega_3 = \Omega_b\oint\frac{d\theta}{2\pi}\frac{1}{\dot{\theta}}\ \Omega_3 
 = \Omega_b\left\{\oint\frac{d\theta}{2\pi}\frac{1}{\dot{\theta}}\ 
   \dot{\varphi} 
+ \partial_{J_3}\oint \frac{d\theta}{2\pi} 
P_\theta (\theta, J_1, J_2, J_3)\right\}
 = \Omega_b\oint\frac{d\theta}{2\pi}\frac{1}{\dot{\theta}}\ \dot{\varphi} 
\end{equation}
$(\alpha_2,\Omega_2)$ and $(\alpha_3,\Omega_3)$ are clearly found to be 
associated respectively with the bounce and toroidal drift motion of particles.

\ssssection{Approximate calculation of the bounce and drift motion} 
Let us be a little bit more explicit on the bounce and drift motion involved,
using our knowledge of the equilibrium particles drifts
\footnote{Note that these drift should normally be accessed directly from
the action-angle formulation, but that such a formulation of the drifts
is much less tractable.}.
We know from subsection \ref{ssection_ParticleTrajectories}
that the motion of a charged particle in the tokamak geometry can be 
divided into a parallel and lower order drift motion
${\bf v} = v_\|{\bf b} + {\bf v}_g$, such that
\begin{eqnarray}
\dot{\Psi}    &=& {\bf v}_{g}\cdot\nabla\Psi \\
\dot{\theta}  &=& v_\| {\bf b}\cdot\nabla\theta + {\bf v}_{g}\cdot\nabla\theta\\
\dot{\varphi} &=& v_\| q {\bf b}\cdot\nabla\theta 
+ {\bf v}_{g}\cdot\nabla\varphi
\end{eqnarray}
At the lower order, $\dot{\theta}$ is dominated by the parallel velocity and 
the bounce frequency can be rewritten
\begin{equation}
\Omega^{-1}_2 = \oint\frac{d\theta}{2\pi}\frac{1}{{\bf
    b}\cdot\nabla\theta \ v_\|}.
\end{equation}
The drift motion remains relevant in the drift frequency.
Indeed, noticing 
\begin{eqnarray*}
\dot{\varphi} - q({\bar \Psi})\dot{\theta}
&=& \frac{v_\|}{B_\eq}
(q(\Psi)-q(\bar{\Psi}))
{\bf B}_\eq\cdot\nabla\theta 
 + {\bf v}_g\cdot \nabla(\varphi - q(\bar{\Psi})\theta)\\
&\approx & q' \delta \Psi\dot{\theta} +
 {\bf v}_g\cdot \nabla(\varphi - q(\bar{\Psi})\theta)
\end{eqnarray*}
it comes
\begin{equation}
\Omega_3 = \Omega_b\oint \frac{d\theta}{2\pi}\frac{1}{\dot{\theta}} \
\dot{\varphi} = 
\Omega_b\oint \frac{d\theta}{2\pi}\frac{1}{\dot{\theta}}
\left( 
- q' \dot{\Psi}\theta +
 {\bf v}_g\cdot \nabla(\varphi - q(\bar{\Psi})\theta)
\right)
+ q(\bar{\Psi})\Omega_b\oint \frac{d\theta}{2\pi}
\label{eq_Omega3Simplified}
\end{equation}
where the last term in Eq.~\ref{eq_Omega3Simplified} cancels for trapped
particles in the small radial drift approximation, such that the particle
drift is the main contribution in $\Omega_3$. 
In this study, the particle radial drift has been neglected and the last
term of Eq.~\ref{eq_Omega3Simplified} was consequenlty rewritten
$ q(\bar{\Psi})\Omega_b\delta_{passing}$ in Eq.~\ref{eq_NormalizedBounceDrift}.

\section{Expressions of the equilibrium characteristic frequencies for some
well defined geometries}
\label{appsection_KbKd}

\subsubsection{Circular geometry}
We now derive the expression of the normalized bounce and drift frequencies,
such that
\begin{equation}
\Omega_2 = \Omega_b 
= \pm\frac{1}{qR_0}\sqrt{\frac{2{\sf E}}{m}}\bar{\Omega}_b, \quad
\Omega_3 = \Omega_d + \delta_\text{passing}q(r)\Omega_b
=\frac{q(r)}{r}\frac{E}{eB_0 R_0}\bar{\Omega}_d
+\delta_\text{passing}q(r)\Omega_b
\label{eq_NormalizedBounceDrift_bis}
\end{equation}
in a simple circular equilibrium, without Grad-Shafranov shift and in the
large aspect ratio.

In such a geometry, ${\bf b}\cdot\nabla\theta\approx 1/qR$. It directly comes 
\begin{equation}
\bar{\Omega}_b = \left(\oint\frac{d\theta}{2\pi}
\frac{1}{\sqrt{1-\lambda(1+\epsilon \cos\theta)}}\right)^{-1}
\end{equation}
with $\lambda=\mu B_0/{\sf E}, \epsilon=r/R_0$.
Using $\kappa^2 = 2\epsilon \lambda/[1-(1-\epsilon)\lambda]$, and keeping only
first order effects in $\epsilon$, it follows
\begin{equation}
\bar{\Omega}_b^{-1}=\sqrt{\frac{2\epsilon+(1-\epsilon)\kappa^2}{2\epsilon}}
\oint \frac{d\theta}{2\pi}
\frac{1}{\sqrt{1-\kappa^2\sin^2(\theta/2)}}
\end{equation}
where we recall that by convention, we take for passing particles
($\kappa<1$), $\oint = \int_0^{2\pi}$, 
and for trapped particles ($\kappa>1$), $\oint= \int^{\theta_0}_{-\theta_0}$.

Using the change of variable $\sin u = \kappa\sin(\theta/2)$ for the trapped 
particles $\theta \in [-\theta_0, \theta_0]$, the normalized bounce frequency
is given explicitely by the formulas
\begin{equation}
\bar{\Omega}^{-1}_b
=\sqrt{\frac{2\epsilon+(1-\epsilon)\kappa^2}{2\epsilon}}
\frac{2}{\pi}
\left\{\begin{array}{lll} 
 \mathds{K}(\kappa) & \text{ for circulating particles}\\
(1/\kappa)  \mathds{K}(1/\kappa) & \text{ for trapped particles}
\end{array}
\right. .
\end{equation}
where  $\mathds{K}(\kappa)$ is the firs elliptic integral of the first kind,
$\mathds{K}(\kappa) 
= \int_0^{\pi/2} \sqrt{1+\kappa^2cos\theta}^{\ -1}d\theta$, for $\kappa<1$.\\

Similarly,
\begin{equation}
 \bar{\Omega}_d 
= \bar{\Omega}_b\oint \frac{d\theta}{2\pi}
\frac{(2-\lambda B)(\cos\theta + s\theta\sin\theta)}
     {\sqrt{1-\lambda(1+\epsilon \cos\theta)}} \ ,
\end{equation}
returns
\begin{eqnarray}
\bar{\Omega}_d 
&=& \bar{\Omega}_b
\frac{\kappa^2}{\sqrt{2\epsilon+(1-\epsilon)\kappa^2}}
\frac{1}{\sqrt{2\epsilon}}
\oint\frac{d\theta}{2\pi}
\frac{\cos\theta + s\theta\sin\theta}
{\sqrt{1-\kappa^2\sin^2(\theta/2)}}
\\
&=&
\frac{\kappa^2}{2\epsilon+(1-\epsilon)\kappa^2}
\left\{\begin{array}{lll} 
1 + \frac{2}{\kappa^2}(\frac{\mathds{E}}{\mathds{K}}-1)
+ \frac{4s}{\kappa^2}\frac{\mathds{E}}{\mathds{K}} 
 & \text{ for circulating particles} \\
\\
2\frac{\mathds{E}(1/\kappa)}{\mathds{K}(1/\kappa)}-1 
+ 4s\left[\frac{\mathds{E}(1/\kappa)}{\mathds{K}(1/\kappa)}
-\left(\frac{1}{\kappa^2}-1\right)\right]
& \text{ for trapped particles}
\end{array}
\right. .
\end{eqnarray}
where $\mathds{E}$ is the elliptic integral of the second kind,
$ \mathds{E} = \int_0^{\pi/2}\sqrt{1+\kappa^2cos\theta} \ d\theta$.

\subsubsection{Circular geometry with shift} 
For the analysis performed in the thesis, we made use of slightly more
general expressions of the bounce and drift frequencies, which  we found 
in Ref.~\cite{Zonca_07} and include the existence of a Grad-Shafranov shift,
characterized by $\alpha = -q^2R_0\partial_r\beta$.

\begin{eqnarray}
\nonumber
\bar{\Omega}_b^{-1}
&=& \sqrt{\frac{2\epsilon + (1-\epsilon)\kappa^2}{2\epsilon}}\frac{2}{\pi}
\left\{\begin{array}{lll} 
 \mathds{K}(\kappa) & \text{ for circulating particles}\\
(1/\kappa)  \mathds{K}(1/\kappa) & \text{ for trapped particles}
\end{array}\right.\\
\bar{\Omega}_d &=& 
 \frac{\kappa^2}{2\epsilon+(1-\epsilon^2)\kappa^2}
 \left\{\begin{array}{lll} 
 1 + \frac{2}{\kappa^2}(\frac{\mathds{E}}{\mathds{K}}-1)
 - \frac{4\alpha}{2\kappa^2}
 \left(2(1-\frac{1}{\kappa^2})
   -(1-\frac{2}{\kappa^2})\frac{\mathds{E}}{\mathds{K}}\right)
 -\frac{\alpha}{2q^2}\\
 + \frac{4s}{\kappa^2}
 \left(\frac{\mathds{E}}{\mathds{K}} -\frac{\pi}{2\mathds{K}}\sqrt{1-\kappa^2}  
   \right)
  \text{ for circulating particles}\\
 \\
 2\frac{\mathds{E}}{\mathds{K}}-1 
 + 4s\left[\frac{\mathds{E}}{\mathds{K}}
 +\left(\frac{1}{\kappa^2}-1\right)\right]
-\frac{\alpha}{2q^2}\\
-\frac{4\alpha}{3}\left[1-\frac{1}{\kappa^2}
  +(\frac{2}{\kappa^2}-1)\frac{\mathds{E}}{\mathds{K}})
\right]
 \text{ for trapped particles}
\end{array}\right.
\label{eq_NormBounceDrift}
\end{eqnarray}


\chapter{Field orderings in the BAE inertial layer}
\label{appchapter_InertialFieldOrderings}

We derive in this appendix the orderings claimed in 
Eq.~\ref{eq_SidebandsOrdering}, starting from the
expressions of electroneutrality and vorticity derived from the 
Lagrangian  in the acoustic frequency range
\begin{equation}
\begin{split}
-\left(1+\frac{1}{\tau_e}\right) \mathcal{E}_{\omega}
   + (1 - \Gamma_0) \psi_{\omega}
   =  -\left<\left(\frac{-i\omega }{-i\omega +  v_{\|}\nabla_{\|}
     + \textbf{v}_{gi} \cdot \nabla}\right)J_0^2
   \left(\frac{\textbf{v}_{gi} \cdot \nabla}{-i\omega}
   \psi_\omega+\mathcal{E}_\omega\right)\right>
\label{eq_ElectroneutralityIntermediate}
\end{split}
\end{equation}
\begin{equation}
\begin{split}
 -\frac{v_A^2}{\omega^2}\nabla_{\|}(\rho_i^2\nabla_{\bot}^2)
   & \nabla_{\|}\psi_\omega
  + (1 - \Gamma_0)(\psi_{\omega}-\mathcal{E}_\omega)
  - \left< \left( \frac{ \mathbf{v}_{gi} \cdot\nabla}{-i\omega}\right)
    (1 - J_0^2) \right>\psi_{\omega} \\
  &= + \left<\left(\frac{\mathbf{v}_{gi}\cdot \nabla}
   {-i\omega}\right)\left(\frac{-i\omega }
   {-i\omega +  v_{\|}\nabla_{\|}
     +  \textbf{v}_{gs} \cdot \nabla}\right)J_0^2
\left(\frac{\textbf{v}_{gi} \cdot \nabla}{-i\omega}
   \psi_\omega+\mathcal{E}_\omega\right) \right>
\label{eq_VorticityIntermediate}
\end{split}
\end{equation}\\

In a cylidrical equilibrium, the curvature verifies
${\bf b}_{\eq}\times\bf{\kappa}
=-\frac{1}{R_0}\left[\sin\left(\theta\right)
\mathbf{e}_{r}+\cos\left(\theta\right)\mathbf{e}_{\theta}\right]$
and implies a coupling of poloidal components  via the
${\bf v }_{gi}\cdot\nabla$ operator such that for any poloidal
number $\sf m'$,
\begin{equation}
\left({\bf v }_{gi}\cdot\nabla \psi_{\omega}\right)^{\sf m'}
= i\sum_{\epsilon = \pm  1} \omega_{gi, \epsilon}\psi^{{\sf m'}+\epsilon}_\omega
\label{eq_CurvatureCouplingApp}
\end{equation}
where $\omega_{gi, \epsilon}$ is an operator defined as
$\omega_{gi, \epsilon} = 
\frac{1}{2}v_{gi}(+i\epsilon k_r
+\frac{{\sf m}'+\epsilon}{r}) \approx i\epsilon\frac{v_{gi} k_r}{2}
\equiv + i\epsilon \omega_{di}$.
$i k_r$ stands for the $\partial_r$ operator and 
$ v_{gi} =\frac{-1}{eB_{\eq}R}\left(m_i v_{\|}^{2}+\mu_i B_{\eq}\right)$. 

We make the expansion of the resonance operator:
$ (-i\omega) (-i\omega +  v_{\|}\nabla_{\|}
 + \textbf{v}_{gi} \cdot \nabla)^{-1}
= 1 - \frac{v_{\|}\nabla_{\|}}{-i\omega}
-\frac{\textbf{v}_{gi} \cdot \nabla}{-i\omega} +O ( (k_\perp\rho_i)^2)$, 
and assume 
$\psi^{\sf m+1}=-\psi^{\sf m-1}$, 
$\psi^{\sf m+2}=\psi^{\sf m-2}$
$\mathcal{E}^{\sf m+1}=-\mathcal{E}^{\sf m-1}$, 
$\mathcal{E}^{\sf m+2}=\mathcal{E}^{\sf m-2}$.
If follows from electroneutrality, 
expanded to the leading order in the different harmonics, that
\begin{eqnarray}
\nonumber\frac{1}{\tau_e}\mathcal{E}^{\sf m}_\omega
&=& \left<\frac{-i\omega_{di}}{\omega} \right>
(\psi^{\sf m+1}_\omega-\psi^{\sf m-1}_\omega)
+ \left(1-\Gamma_0 -2\left< \left(\frac{\omega_{di}}{\omega}\right)^2
\right>\right)\psi^{\sf m}_\omega \\
\nonumber 
&&+ \left< \left(\frac{\omega_{di}}{\omega}\right)^2\right>
(\psi^{\sf m+2}_\omega+ \psi^{\sf m-2}_\omega)\\
\nonumber 
&&+O\left((\rho_ik_\perp)\psi^{\sf m+1}_\omega,
(\rho_ik_\perp)^2\psi^{\sf m+2}_\omega, (\rho_ik_\perp)^{-2}
\Lambda^2\psi_\omega^{\sf m}, 
(\rho_ik_\perp)^3\psi_\omega^{\sf m}\right)\\
\frac{1}{\tau_e}\mathcal{E}^{\sf m+1}_\omega
&=& \left<-i\frac{\omega_{di}}{\omega} \right>
(\psi^{\sf m+2}_\omega-\psi^{\sf m}_\omega)
= O\left((\rho_ik_\perp)\psi_\omega^{\sf m}\right)\\
\nonumber \frac{1}{\tau_e}\mathcal{E}^{m+2}_\omega
&=& \left<-i\frac{\omega_{di}}{\omega} \right>(-\psi^{\sf m+1}_\omega)
+ \left(1-\Gamma_0 -2\left< \left(\frac{\omega_{di}}{\omega}\right)^2
\right>\right)\psi^{\sf m+2}_\omega
- \left< \left(\frac{\omega_{di}}{\omega}\right)^2\right>
(\psi^{\sf m+2}_\omega+\psi^{\sf m}_\omega) 
\label{eq_ElectroneutralityProjected}\\ 
&=& \nonumber 
O\left((\rho_ik_\perp)\psi^{\sf m+1}_\omega,
(\rho_ik_\perp)^2 \psi_{\omega}^{\sf m}\right)
\end{eqnarray}
where 
$\Lambda^2 = (\omega^2/\omega^2_{BAE})(1 - \omega^2/\omega^2_{BAE})$, 
$\omega^2_{BAE} = (v_{ti}/R_0)^2(7/2+2\tau_e)$ 
(the last expression of Eq.~\ref{eq_SeveralOmBAE} is recovered).
Next, noticing that for the sidebands $qRk^{{\sf m}+l}_\| = 
l+ qRk^{\sf m}_\|\sim 1$, for $l=1 \text{ or } 2$, vorticity yields
\begin{equation}
\begin{array}{lcl}
\frac{(\rho_ik_\perp)^2}{\beta} \psi^{\sf m+l}_\omega\sim
(1-\Gamma_0)\mathcal{E}_{\omega}^{\sf m+l}
+ \left<\frac{i\omega_{di}}{\omega}
\right> (\mathcal{E}^{\sf m+1}_\omega- \mathcal{E}^{\sf m-1}_\omega),
{\rm for} \,\,l= 1 , 2
\end{array}
\label{eq_VorticityProjected}
\end{equation}

From Eq.~(\ref{eq_VorticityProjected}), it appears that 
$\psi^{\sf m+1 / m+2} = O ( (\rho_ik_\perp)^4\psi^{\sf m})$,
and we can combine this result with the previous equations to show:
$ \mathcal{E}^{\sf m}_\omega 
= O((\rho_ik_\perp)^{-2}\Lambda^2\psi^{\sf m}_\omega)$,
$ \mathcal{E}^{\sf m+1}_\omega = O((\rho_ik_\perp)\psi^{\sf m}_\omega)$
and $ \mathcal{E}^{\sf m+2}_\omega 
= O((\rho_ik_\perp)^2\psi^{\sf m}_\omega)$.
With the assumption of closeness to the BAE gap edge relevant to the
consideration of fourth order terms 
Eq.~\ref{eq_ClosenessToAccumulationPoint},
the main poloidal component appears to be negligible,
$\mathcal{E}^{m}_\omega = O((\rho_ik_\perp)^5\psi^{m}_\omega)$.

\chapter{Ballooning representation}
\label{appchapter_BallooningRepresentation}


Once one has chosen a toroidal field number $\sf n$, the calculation of a 
perturbation structure is reduced to a 2D problem, ie: the determination of 
the perturbation radial and poloidal dependences.
The ballooning representation \cite{Connor_78}
makes use of the anisotropy of tokamak instabilities
$k_\perp\gg k_\|$, which is especially true for high-{\sf n} modes, to reduce
the problem to one dimension only.
It is based on an expansion with $1/{\sf n}$ as a small parameter.

The starting point is a coordinate transformation from the traditional geometric
variables $(\theta, \varphi)$ to the coordinates
\begin{eqnarray}
\alpha    =  \varphi - q(r)\theta 
&&\text{ standing for a flux surface perpendicular coordinate}\\
\vartheta \in [-\infty, +\infty] 
&&\text{ a field align coordinate}.
\end{eqnarray}
where $ \vartheta = \theta$ in $[0, 2\pi]$.


In the ballooning representation, a field is represented in the form
\begin{eqnarray}
\delta\phi(r,\theta,\phi,t)
&\rightarrow&
\hat{\phi}_{\omega}(r, \vartheta)
\exp\left\{i{\sf n}\left(\varphi-q(r)(\vartheta-\vartheta_k)\right)
-i\omega t\right\}
\end{eqnarray}
where $\vartheta_k = -(i/2\pi nq') \partial_{r|\eta, \alpha}$.

Such a representation can be obtained following the procedure
\begin{eqnarray}
\phi_\omega 
&=&\sum_{\sf m} e^{i{\sf m}\theta+i{\sf n}\varphi}         \phi_{\omega}^{\sf m}(r)
 = \sum_{\sf m} e^{i{\sf n}\alpha}e^{iqR_0k_\|^{\sf m}\theta} \phi_{\omega}(r,{\sf m})\\
&=& \sum_{\sf m} e^{i{\sf n}\alpha}e^{iqR_0k_\|^{\sf m}\theta}
\int^{+\infty}_{-\infty}d\vartheta e^{-qR_0k_\|^{\sf m}\vartheta}
\hat{\phi}_{\omega}(r, \vartheta)
\label{eq_ConnorExpansion}
\end{eqnarray}
where the Fourier transform of $\phi(r,{\sf m})$ seen as a continuous function
of $\sf m$ has been taken.\\


At the lower order of the ballooning expansion, the $r$-dependence of 
$\hat{\phi}$ can be neglected, whereas $\vartheta_k$ can be seen as a pure 
parameter (It is implicit in the lower order equations).
Three consequences follow
\begin{itemize}
\item The problem is reduced to a 1D problem in $\vartheta$, for 
$\hat{\phi}({\vartheta})$.
\item From Eq.~\ref{eq_ConnorExpansion}, and using the expansion 
$qR_0k_\|^{\sf m} = k_\theta s x$,  $\hat{\phi}(\vartheta)$ is found
to the Fourier transform of the radius dependent function 
$ \phi_{\omega{\sf m}} (x/k_{\theta }s)$.
\item The transformed equations can be accessed from the mode structure 
equation in $(r,\theta)$ space, making the following substitutions: 
\begin{eqnarray}
\nabla_\|         &\rightarrow& \frac{1}{qR_0}\partial_\vartheta \\
&&\text{ou plus general: $B/2\pi J$} \\
\partial_r        &\rightarrow& -i{\sf n}q'(\vartheta-\vartheta_k)\\
\partial_{\theta} &\rightarrow& -i{\sf n}q
\end{eqnarray}
\end{itemize}

%



\chapter{Some details on the derivation of the fishbone-like dispersion 
relation}
\label{appchapter_FishboneDispRelDetailedDerivation}
In this Appendix, we give some details on the calculation of the
integrals which appear in the fihsbone-like dispersion relation.
\section{Volume elements}
To compute the integrals of the fishbone dispersion relation, it is
useful to express the traditional canonical variables $(\bf x, p)$ as a
function of the equilibrium invariants chosen for this analysis:
 $r$, $\sf E$ and $\lambda = \mu B_0/{\sf E}$. It comes
\begin{equation}
\begin{array}{lcl}
d{\bf \Gamma} = d^3{\bf x}d^3{\bf p} 
&=& ( R_0d\varphi rd\theta dr)
\left( 2\pi m^2_s  
\sqrt{\frac{2 {\sf E}}{m_s}}  d {\sf E} \ d\left(\frac{v_\|}{v}\right)\right)\\
&=&  2\sqrt{2}\ \pi \ m_s^{3/2} (R_0d\varphi rdr)
\sqrt{\sf  E}d{\sf E}
\sum_{\pm} \ \frac{b_{\eq}(r,\theta)d\lambda}{2\sqrt{1-\lambda b_\eq(r,\theta)}}.
\end{array}
\end{equation}
Noticing that volume elements are conserved when moving from one
canonical set of coordinate, 
the same result can be recovered from the action-angle variables, 
\begin{equation}
\begin{array}{lcl}
d{\bf \Gamma} = \sum_{\pm} d^3\m{\alpha}d^{3}{\bf J} 
&=& \sum_{\pm} d^3\m{\alpha} 
\left(\frac{m_s}{e_s}d\mu\right)\left(\frac{d{\sf E}}{\Omega_b}\right)(e_sd\Psi)\\
&=& \sum_{\pm} d^3\m{\alpha}\  m^{3/2}_s\frac{1}{B_0}d\lambda
(\sqrt{\frac{\sf E}{2}}d{\sf E})
\int \frac{d\theta}{2\pi}\frac{1}{{\bf b}_\eq\cdot\nabla\theta}
\frac{1}{\sqrt{1-\lambda b_\eq(r,\theta)}}
\left(\frac{B_0 }{q} rdr\right).
\end{array}
\end{equation}
Note however that the two calculations are useful in general.

\section{Energetic particle term}
\label{appsection_dWk}
\subsection{Normalization of the anisotropic Maxwellian}
\label{appssection_NormAnisotropicMaxwellian}
Assume we have a distribution function of the form,
\begin{equation}
F_h = \tilde{n}(r)f_\lambda(\lambda)e^{-{\sf E}/T_h(r)}.
\end{equation}
For consistency, $T_h$ needs to have the dimensions of an energy and can be 
understood as a temperature.
Let us link $\tilde{n}$ to the more traditional density of the hot population
$n_h$.

By definition $n = \int d^3{\bf p}F_h$, such that the average density
corresponding to the radius r, $n(r)$, is
\begin{eqnarray}
\nonumber
n(r) &=& \int \frac{d\theta}{2\pi}d^3{\bf p}\  F_h  
     = \sqrt{2}\pi m_s^{3/2}\tilde{n}(r)
\int d\lambda 
\int \frac{d\theta}{2\pi} \frac{b_\eq(r,\theta)}{\sqrt{1-\lambda
     b_\eq(r,\theta)}}\\
&=& \sqrt{2}\pi m_s^{3/2} 
\tilde{n}(r)  
\int \sqrt{\sf  E}d{\sf E}e^{-{\sf E}/T_h(r)}
\sum_{\pm} 
\int d\lambda f_\lambda(\lambda) 
\int \frac{d\theta}{2\pi}
\frac{b_{\eq}(r,\theta)d\lambda}{\sqrt{1-\lambda b_\eq(r,\theta)}} \ .
\end{eqnarray}
$n(r)$ can be interpreted as the average density in the radial shell, when
the approximation of very thin banana is made.

Finally, the correct normalization for $F_h$ is given by
\begin{equation}
\tilde{n}(r) = 
\frac{n_{h}(r)}{(2 \pi m_h T_h)^{3/2}}
\frac{2}{< \bar{\Omega}_b(\lambda,r)>_\lambda}\ .
\end{equation}

\subsection{Projection onto the action-angle basis}
\label{appssection_ActionAngleProjection}
The main difficulty for the computation of resonances is the necessity 
to make the projection of the fields, naturally expressed in the
{\it geometric variables}, onto the {\it particle coordinates},  the
action-angle variables. This projection can {\it a priori} lead to an infinite
number of components in the action-angle space.

For the resonant excitation of BAEs, it is explained in subsection
\ref{ssection_EnergeticParticleTerm} that one main component was relevant for 
resonance: the resonance with trapped particles for the triplet 
$({\sf n_1, n_2, n_3}) = ({\sf 0, 0, n})$.

So we need to to compute
\begin{eqnarray}
- \tilde{h} = e_h
\int\frac{d\alpha_1}{2\pi}\int \frac{d\alpha_2}{2\pi}\int \frac{d\alpha_3}{2\pi}
\left(\frac{{\bf v}_{gh}\cdot\nabla\psi_\omega}{-i\omega}
+\mathcal{E}_\omega\right) e^{-i{\sf n}_3\alpha_3} 
\end{eqnarray}
for the fast trapped particles, with the additional assumption that the hot
particles simply interact with the mode in the MHD-like region where 
$\mathcal{E}_\omega=0$
(This is argued in subsection ~\ref{ssection_EnergeticParticleTerm}).

In the following, we simply use the main poloidal component
$\psi_\omega = \psi_\omega^{\sf m} \exp(i{\sf m}\theta + i{\sf n}\varphi)$, 
which is dominant for our derivation of the BAE structure.
For strong perpendicular gradients compared to parallel gradients, the results of 
the ballooning representation can be used, and return an integrand of the form
\begin{eqnarray}
{\bf v}_{gh}\cdot\nabla\psi_\omega
&=& -v_{gh}
\left(\sin\theta \partial_r + \frac{\cos\theta}{r}\partial_\theta\right)
\psi_\omega\\
&=&-ik_\theta v_{gh}
\left(s(\theta-\theta_d)\sin\theta + \frac{\cos\theta}{r}\right)
\psi_\omega^{\sf m}(r) e^{i{\sf n}(\varphi-q\theta-\alpha_3)}
e^{iqR_0k_\|^{\sf m}\theta}.
\end{eqnarray}

We may now make the projection onto the action-angle variables. For this, we
need to be able to express the angle variables as a function of the geometric 
variables $(\theta,\varphi)$. It is possible to choose a description
such that $\alpha_2$ simply depends on $\theta$, and the $\varphi$-dependence 
is fully included in $\alpha_3$. More precisely, we can write
\begin{eqnarray}
\varphi  &=& \alpha_3 + \hat\alpha + q(\bar{\Psi})\hat\theta\\
\theta   &=& \delta_\text{passing}\alpha_2 + \hat{\theta}
\end{eqnarray}
It follows that $i{\sf n}(\varphi-q\theta-\alpha_3) = i{\sf n}\hat{\alpha}$.

To compute the integrals, we now make a few assumptions.
\begin{itemize}
\item For the gyroaverage, we simply choose $J_0(k_\perp\rho_\perp) = 1$ in the 
  MHD region, and $J_0(k_\perp\rho_\perp) = 0$ in the inertia region.
\item For the $\alpha_2$ integrand, we neglect the $\theta$-dependence of
  $\hat{\alpha}$ and $\hat{\theta}$, which makes some sense for sufficiently
  trapped particles.
  
  Moreover, we use closeness to the resonant (in agreement with the ballooning
  representation), such that $qR_0k_\| = 0$.
  
  Finally, we do not take into account the banana width of the particle 
  (assumption of thin bananas), such that no $\theta$ dependence does not appear
  $\psi^{\sf m}(r)$.
  This is meaningful in the MHD region where radial gradients are smooth, such
  that the radial drift  is less important.

\item The later integration is the easiest because of the cancellation of the 
  $\varphi$ dependence in the integrand.
\end{itemize}

From the definition of the drift frequency
($\Omega_3 = \Omega_d$ for the trapped particles), where we take 
${\bf v}_{gh}\cdot\nabla\varphi = 0$ (${\bf B}_\text{\eq}$ almost in the 
$\varphi$ direction), it finally comes
\begin{equation}
- \tilde{h} = J_0(\rho_{h\perp} k_\perp)
e_h \frac{\sf m}{q}\frac{\Omega_{h,d}}{\omega}\psi_\omega^{\sf m}(r)
\end{equation}
where the $\theta_d$ dependence disappeared with parity.

$\tilde{h}$ can be understood as a modified Hamiltonian.
In particular, where resonance with the particles occurs,
\begin{equation}
{\sf m}/q   = -{\sf n}, \\ 
{\sf n}\Omega_{h3}=\omega, \\
\tilde{h} = e_h \psi_\omega^{\sf m}.
\label{eq_EquivalentHotHamiltonian}
\end{equation}

\subsection{Complete form the energetic particle term used in the thesis}
In this subsection, we simply make more explicit the derivation of 
Eq.~\ref{eq_fullydevelopeddWk} and give some details on the distribution 
function used in our analysis of the BAE linear stability.

We start from Eq.~\ref{eq_HalfExpanded_dWk} and assume the fast ion 
population to be a anisotropic Maxwellian of the form  
$F_h=\tilde{n}_h(r)f_{\lambda}(\lambda)\exp(-E/T_{h}(r)\ )$ 
where $f_\lambda$ is chosen to be a peaked function
\begin{eqnarray}
f_{\lambda}&=&\delta(\lambda-\lambda_0)  \mbox{ or }
f_\lambda=\frac{1}{\sqrt{\pi}\Delta\lambda}
\exp\left(-(\lambda-\lambda_0)^2/\Delta\lambda^2\right)\\
\tilde{n}(r) &=& 
\frac{n_{h}(r)}{(2 \pi m_h T_h)^{3/2}}
\frac{2}{< \bar{\Omega}_b(\lambda,r)>_\lambda}
\end{eqnarray}
with the notation 
$<...>_\lambda 
= \int_0^{1+r/R}d\lambda f_\lambda$. 
It directly comes
\begin{eqnarray}
\nonumber\delta\hat{W}_k=\sqrt{\pi}(q{(r_s)}^2{\sf m})
\int_0^a \frac{1}{r^2_s}r dr|\bar{\xi}|^2\beta_h
\int_{1-r/R}^{1+r/R} d\lambda\frac{\bar{\Omega}_b\bar{\Omega}_d^2}
{<\bar{\Omega}_b>_\lambda} \times\\
\int_0^{+\infty} \frac{dE}{E-E_{res}}\left(\frac{E}{T_{h}}\right)^{5/2}
e^{-E/T_h}
\left\{ -\frac{E_{res}}{T_h}f_{\lambda}
- \frac{E_{res}}{T_h}\frac{T_h}{E}\lambda f'_{\lambda}
- \frac{1}{\bar{\Omega}_d}\frac{R\partial_{r}
  F_h}{F_h}f_{\lambda} \right\}.
\end{eqnarray}
The two first terms are related to the energy gradient
(the traditional Maxwellian term and the anistropy induced energy gradient 
\cite{Boswell_06}) and the last one to the radial gradient. 
More precisely,
\begin{equation}
\frac{\partial_r F_h}{F_h}
=\frac{\partial_r n_h}{n_h}
-\frac{\partial_r <\bar{\Omega}_b>}{<\bar{\Omega}_b>}
+\left(\frac{E}{T_h}-\frac{3}{2}\right)\frac{\partial_r T_h}{T_h}.
\end{equation}
Hence, 
\begin{eqnarray}
\nonumber
\delta\hat{W}_k &=& 
\pi(q{(r_s)}^2{\sf m})
\int_0^a \frac{1}{r^2_s}r dr|\bar{\xi}|^2\beta_h
\int_{1-r/R}^{1+r/R} d\lambda\frac{\bar{\Omega}_b\bar{\Omega}_d^2}
{<\bar{\Omega}_b>_\lambda} \times \\
\nonumber&&\left\{
-\frac{E_{res}}{T_h}Z_5f_{\lambda}
-\frac{E_{res}}{T_h}Z_3\lambda f'_{\lambda}\right.\\
&&\left. 
-\frac{1}{\bar{\Omega}_d} 
\left[\frac{R\partial_r n_h}{n_h} Z_5 
 -\frac{R\partial_r <{\bar \Omega}_b>}{<{\bar\Omega}_b>} Z_5
 +\frac{R\partial_r T_h}{T_h}\left(Z_7-\frac{3}{2}Z_5\right)
 \right]f_{\lambda} \right\}
\label{fullydevelopeddWk}
\end{eqnarray}
with 
$ Z_p = Z_p (y_0)$ such that $y^2_0=E_{res}/T_h$, 
and Z, such that $Z(y_0)=y_0Z_{-1}(y_0)$, is the Fried et Conte or plasma dispersion function,
\begin{eqnarray}
\sqrt{\pi}Z_p(y_0)=\int^{+\infty}_{-\infty}dyy^{p+1}\exp(-y^2)/(y^2-y_0^2),
Z_3(y_0)=\frac{1}{2}+y_0^2+y^3_0Z(y_0),\\
Z_5(y_0)=\frac{3}{4}+\frac{y_0^2}{2}+y_0^4+y_0^5Z(y_0),
Z_7(y_0)=\frac{15}{8}+\frac{3}{4}y_0^2+\frac{y_0^4}{2}+y_0^6+y_0^7Z(y_0).
\end{eqnarray}

\section{Details for the computation of inertia}
\label{appsection_Inertia}


In this Appendix subsection, $\psi_\omega$ and $\mathcal{E}_\omega$ refer to
the full electromagnetic fields including all poloidal components,
and each poloidal component is clearly indicated by its wave numbers 
({\sf n,m'}), whereas the main mode poloidal mode number is {\sf m}.

Passing thermal ions are expected  to resonate with BAEs, which leads to
Landau damping.
To calculate the value of the energy transfer, we compute the imaginary part
of Eq.~\ref{eq_ResonantLagrangian}, considered for thermal ions ($s=i$).
For passing particles, resonances are of the form 
 ${\sf n}_2\Omega_{i,2} + {\sf n}_3\Omega_{i,3} 
\approx (v_\|/qR) ({\sf m'} + {\sf n}q) = k^{{\sf m'},{\sf n}}_{\|}v_\|$, 
for any poloidal component 
$({\sf n}_2,{\sf n}_3) = ({\sf m'},{\sf  n})$.
Hence close to the BAE inertial region, resonances are
expected to involve the mode sidebands characterized by the poloidal mode
numbers ${\sf m'}= \sf m \pm 1$ (where $\sf m$ now stands for the
main poloidal component mode number) and to be of the form
$k^{{\sf m} \pm 1, {\sf n}}_{\|} \approx \pm  1/qR$.

From Eqs.~\ref{eq_PerturbedSidebandsExpression}, we know the fields involved.
Eq.~\ref{eq_ResonantLagrangian} (or equivalent resonant Hamiltonian) reads
\begin{equation}
\left(\frac{{\bf v}_{gi}\cdot\nabla\psi_\omega}{-i\omega}
+\mathcal{E}_\omega\right)
_{(\sf m+1,n )} 
= \frac{1}{2}\frac{v_{gi}}{\omega} \partial_r\psi^{\sf m}_{\omega} 
- \frac{T_e}{T_i}\left(\frac{v_{ti}}{R\omega}\right)
  \rho_i\partial_r\psi^{\sf m}_{\omega}
\label{eq_EquivalentThermalHamiltonian}
\end{equation}
to the lower order,  with $ v_{gi} = -(m_iv_\|^2 + \mu B)/eBR$, 
$v_{ti}=\sqrt{T_i/m_i}$ the ion thermal velocity.
Hence, noticing that for thermal ions, diamagnetic effects may be neglected 
(${\sf n}\Omega_{\ast i}/\omega \ll 1$), the imaginary part of the
thermal ion resonant contribution  becomes
\begin{eqnarray}
\nonumber {\rm Im}\left(\mathcal{L}^{res}_{i\omega}\right)&=&
{\rm Im}\left(-\int d^3{\bf x} \frac{ne^2}{T_i}
\left(\rho_i\partial_r\psi^{\sf m}_\omega\right)^2
\left(\frac{v_{ti}}{R\omega}\right)^2\right.\times\\
&& \quad
\left. \sum_{\epsilon=\pm1} 
\int d^3{\bf p}\frac{F_{i}}{n} \frac{\omega}{\omega+\epsilon v_{\|}/qR} 
 \left(\frac{1}{2}\left(\frac{v_\|}{v_{ti}}\right)^2  
+ \frac{1}{4} \left(\frac{v_\perp}{v_{ti}}\right)^2 
+ \frac{T_e}{T_i}\right)^2 \right)\\
\nonumber &=&\int d^3{\bf x} \frac{ne^2}{T_i}
(\rho_i\partial_r\psi^{\sf m}_\omega)^2
\frac{\sqrt{\pi}}{2\sqrt{2}} \ q\left(\frac{v_{ti}}{R\omega}\right)
\exp\left(-\frac{1}{2}\left(\frac{q R \omega}{ v_{ti}}\right)^2\right)\times\\
&& \quad\quad\quad
\left[ 2 + 2\left(\left(\frac{qR\omega}{v_{ti}}\right)^2 
+ 2\frac{T_e}{T_i}\right)
+\left(\left(\frac{qR\omega}{v_{ti}}\right)^2+2\frac{T_e}{T_i}\right)^2
\right].
\end{eqnarray}

\end{document}